\documentclass[aps,twocolumn,superscriptaddress,showpacs,floatfix]{revtex4}
\usepackage[latin9]{inputenc}
\usepackage{amsmath}
\usepackage{amssymb}
\usepackage{graphicx}
\usepackage{esint}
\usepackage{epstopdf}
\usepackage{braket}
\usepackage[final]{feynmp}
\usepackage{hyperref}
\usepackage{currfile}

\makeatletter

\def\endfmffile{%

	\fmfcmd{\p@rcent\space the end.^^J%
			end.^^J%
			endinput;}%

	\if@fmfio

		\immediate\closeout\@outfmf

	\fi

	\ifnum\pdfshellescape=\@ne

		\immediate\write18{mpost \thefmffile}%

	\fi}

\begin{document}
\unitlength = 1mm

\title{Symmetry characterization of the collective modes of the phase diagram of the $\nu=0$ quantum Hall state in graphene: Mean-field and spontaneously broken symmetries}

\author{J. R. M. de Nova}

\affiliation{Department of Physics, Technion-Israel Institute of Technology, Technion City, Haifa 32000, Israel}
\affiliation{Departamento de Física de Materiales, Universidad Complutense de
Madrid, E-28040 Madrid, Spain}

\author{I. Zapata}

\affiliation{Departamento de Física de Materiales, Universidad Complutense de
Madrid, E-28040 Madrid, Spain}

\begin{abstract}
We devote this work to the study of the mean-field phase diagram of the $\nu=0$ quantum Hall state in bilayer graphene and the computation of the corresponding neutral collective modes, extending the results of recent works in the literature. Specifically, we provide a detailed classification of the complete orbital-valley-spin structure of the collective modes and show that phase transitions are characterized by singlet modes in orbital pseudospin, which are independent of the Coulomb strength and suffer strong many-body corrections from short-range interactions at low momentum. We describe the symmetry breaking mechanism for phase transitions in terms of the valley-spin structure of the Goldstone modes. For the remaining phase boundaries, we prove that the associated exact $SO(5)$ symmetry existing at zero Zeeman energy and interlayer voltage survives as a weaker mean-field symmetry of the Hartree-Fock equations. We extend the previous results for bilayer graphene to the monolayer scenario. Finally, we show that taking into account Landau level mixing through screening does not modify the physical picture explained above.
\end{abstract}

\pacs{11.30.Qc 71.10.-w 73.22.Pr 73.43.-f 73.43.Lp 73.43.Nq 75.30.Ds  \volumeyear{2016} \volumenumber{number}
\issuenumber{number} \eid{identifier} \startpage{1}
\endpage{}}


\date{\today}
\maketitle

\section{Introduction}\label{sec:intro}

Since the discovery of the integer \cite{Klitzing1980,Tsui1982a} and fractional quantum Hall (QH) effects \cite{Tsui1982,Stormer1983,Laughlin1983}, a large number of works have been devoted to the study of two-dimensional (2D) systems in the presence of strong perpendicular magnetic fields. Due to its chiral character, rich valley-spin structure and relatively large cyclotron frequency, graphene is a particulary interesting scenario to test QH features.

In this way, the study of QH states in graphene has been a hot topic of research in the last years \cite{McCann2006,Yang2006,Iyengar2007,Doretto2007,Bychkov2008,Jung2009,Abanin2009,Cote2010,Shizuya2010,
Gorbar2010,Gorbar2011,Gorbar2012,Kim2011,Shizuya2011,Shizuya2012,Kharitonov2012,
Kharitonov2012PRL,Kharitonov2012a,Kharitonov2012b,Lambert2013,Toke2013,Wu2014,Knothe2015,Murthy2016,Tikhonov2016,Knothe2016,Jia2017}. Among all the possible QH states, the $\nu=0$ QH state, corresponding to the charge neutrality point, has received special attention due to its intriguing strongly insulating behavior in both bilayer and monolayer graphene at zero in-plane magnetic field. In particular, the bilayer provides a more interesting scenario because of the richer structure of its zero-energy Landau level (LL) and the possibility of introducing an energy bias between the two valleys with the help of a perpendicular electric field. In fact, Ref. \cite{Kharitonov2012PRL} presented a complete characterization of the mean-field phase diagram of the $\nu=0$ QH state in bilayer graphene, taking into account the most general spin symmetric interactions (including short-range valley/sublattice asymmetric interactions) and the introduction of a voltage between the two layers. The identified phases are ferromagnetic (F), canted anti-ferromagnetic (CAF), fully layer-polarized (FLP) and partially layer-polarized (PLP) \cite{Kharitonov2012PRL}. The resulting phase diagram is analog to that of monolayer graphene \cite{Kharitonov2012} due to the identical structure of the Hamiltonian governing short-range valley/sublattice asymmetric interactions.

Complementarily, the computation of the spectrum of the collective excitations in QH integer states has been also the subject of study of an important number of works \cite{Yang2006,Iyengar2007,Doretto2007,Shizuya2010,Toke2011,Sari87,Lambert2013,Wu2014,Murthy2016}. For instance, the spectrum of inter-LL collective excitations was obtained in Refs. \cite{Iyengar2007,Shizuya2010,Sari87}. On the other hand, the spectrum of intra-LL excitations within the zero-energy Landau level was computed for the $\nu=0$ QH state \cite{Toke2011} and the rest of integer QH states \cite{Lambert2013} allowing only for long range and interlayer Coulomb interactions. In Ref. \cite{Wu2014}, the dispersion relation of the modes for the Kekul\'e distortion (KD) and CAF phases of the $\nu=0$ QH state was computed for monolayer graphene using an effective low-energy model. Also within the $\nu=0$ QH state in the monolayer scenario, the bulk and edge collective modes of the CAF and F phases were obtained in Ref. \cite{Murthy2016}.

We devote this work to the computation of the mean-field energies and the intra-LL collective modes of the whole phase diagram of the $\nu=0$ QH state of bilayer graphene, as presented in Ref. \cite{Kharitonov2012PRL}. Specifically, we compute the dispersion relation within the time-dependent Hartree-Fock approximation (TDHFA), considering the complete Hamiltonian of short-range interactions \cite{Kharitonov2012PRL} and also including explicitly the interaction of the electrons with the filled Dirac sea \cite{Shizuya2012}, devoting special attention to the characterization of the rich orbital-valley-spin structure of the modes, induced by the short-range valley-asymmetric interactions, and to the study of phase transitions in terms of these modes. In this way, the work here presented provides a natural continuation to the recent literature on the field.

In particular, we show that only singlet modes in orbital pseudospin are strongly modified by many-body effects arising from short-range interactions while the remaining modes are almost unaffected by them due to the dominant character of long range Coulomb interactions. Indeed, as these orbital-singlet modes are independent of the Coulomb interaction strength at low momentum, they also represent the lowest-energy neutral excitations of the system, playing a crucial role in understanding the stability of the different phases. Regarding the valley-spin structure of the modes, we classify them according to the conserved valley and spin quantum numbers of the total Hamiltonian. We find that, while the FLP-PLP transition is governed by a spin-singlet mode, the F-CAF transition is governed by a valley-triplet one; this contrast arises due to the different nature of the spontaneous symmetry breaking mechanism of the CAF and PLP phases. Interestingly, we prove that the remaining phase boundaries, F-FLP and CAF-PLP, present a gapless mode arising from a mean-field symmetry inherited from the full exact $SO(5)$ symmetry existing at the same boundaries for zero Zeeman energy and zero interlayer voltage \cite{Wu2014}. Moreover, we show that the CAF and PLP phases are able to present dynamical instabilities as a result of their spontaneously broken symmetries.

The strong analogy between the $\nu=0$ QH states in bilayer and monolayer graphene allows us to straightforwardly translate most of these results to the monolayer scenario, recovering essentially the same results of Ref. \cite{Murthy2016}. We also study the effects of Landau level mixing by considering the screening effect of Coulomb interaction in the large-$N$ approximation \cite{Foster2008,Basko2008,Aleiner2007,Lemonik2010,Kharitonov2012}, showing that the described orbital-valley-spin structure still holds, quantitatively (but not qualitatively) changing the dispersion relation of the modes. Finally, we relate the results presented in this work with experimental scenarios, including a computation of the mean-field transport gaps and a discussion on the collective modes detection.

We remark that the collective modes calculated in the present work are neutral and, therefore, topologically trivial \cite{Sondhi1993,Moon1995}. The charged, topologically non-trivial excitations (skyrmions) have been studied in both monolayer \cite{Yang2006} and bilayer \cite{Abanin2009} graphene, and the effect of short-range interactions \cite{Wu2014} and screening \cite{Kharitonov2012} on the non-linear sigma-model stiffness coefficients has been addressed for monolayer graphene. Here we provide the corrections due to short range interactions to the stiffness coefficients in the bilayer case as well.

The article is arranged as follows: we first introduce the effective projected Hamiltonian considered in this work in Sec. \ref{sec:basicmodel}. We re-derive in Sec. \ref{sec:HFMF} the phase diagram of Ref. \cite{Kharitonov2012PRL} using a Hartree-Fock (HF) mean-field scheme and compute the corresponding mean-field energies and transport gaps. The dispersion relation of the different collective modes, computed within the TDHFA, is shown in Sec. \ref{sec:colmod}. We translate the same calculations to monolayer graphene in Sec. \ref{sec:MLG}. Effects of LL mixing are studied in Sec. \ref{sec:renorm}. A discussion on experimental features is presented in Sec. \ref{sec:remarksexps}. Finally, the conclusions are drawn in Sec. \ref{sec:QHFMConclusions}. Technical details about the diagonalization of the HF equations and the TDHFA are given in Appendices \ref{app:magneticFF}-\ref{app:analyticalTDHFA}.

\section{Effective Hamiltonian for the $\nu=0$ quantum Hall state in bilayer graphene} \label{sec:basicmodel}

\subsection{Low-energy Hamiltonian}

In the first place, we briefly present the effective model used in this work for bilayer graphene, following Refs. \cite{McCann2006,Kharitonov2012,Kharitonov2012PRL,Kharitonov2012a,Sari87}, where the reader is referred for more details.

The effective dynamics at low energies can be described by a two-band model \cite{McCann2006}, in which the field operator for electrons has eight components and reads:
\begin{eqnarray}\label{eq:BLGfieldoperator}
\hat{\psi}(\mathbf{x})&=&\left[\begin{array}{c}
\hat{\psi}_{+}(\mathbf{x})\\
\hat{\psi}_{-}(\mathbf{x})
\end{array}\right]\\
\nonumber
\hat{\psi}_{\xi}(\mathbf{x})&=&\left[\begin{array}{c}
\hat{\psi}_{KA\xi}(\mathbf{x})\\
\hat{\psi}_{K\tilde{B}\xi}(\mathbf{x})\\ \hat{\psi}_{K'\tilde{B}\xi}(\mathbf{x})\\ \hat{\psi}_{K'A\xi}(\mathbf{x})
\end{array}\right]\equiv\left[\begin{array}{c}
\hat{\psi}_{K\bar{A}\xi}(\mathbf{x})\\
\hat{\psi}_{K\bar{B}\xi}(\mathbf{x})\\ \hat{\psi}_{K'\bar{A}\xi}(\mathbf{x})\\ \hat{\psi}_{K'\bar{B}\xi}(\mathbf{x})
\end{array}\right],~\xi=\pm
\end{eqnarray}
with $A,\tilde{B}$ the most far apart sublattices, $K,K'$ the two valleys and $\xi$ the spin polarization. We note that the two sublattices are interchanged in the $K'$ valley so, as usually done, we will refer to the corresponding subspace as $\bar{A}\bar{B}$ in order to avoid confusions. The eight components of the field operator then correspond to the total space $KK'\otimes\bar{A}\bar{B} \otimes s$, $s$ being the spin space.


The corresponding effective Hamiltonian of the system is $\hat{H}=\hat{H}_0+\hat{H}_C+\hat{H}_{sr}$. After neglecting trigonal warping effects and other small corrections \cite{McCann2006,Kharitonov2012a,Sari87,Toke2013,Cote2013}, the single-particle Hamiltonian $\hat{H}_0$ reads:
\begin{equation}\label{eq:spBHamiltonian}
\hat{H}_{0}=\int\mathrm{d}^2\mathbf{x}~\hat{\psi}^{\dagger}(\mathbf{x})\left[H_B+\epsilon_VT_{zz}-\epsilon_Z\sigma_z\right]\hat{\psi}(\mathbf{x}),
\end{equation}
where $T_{ij}=\tau_i^{KK'}\otimes\tau_j^{\bar{A}\bar{B}}\otimes\hat{\mathbf{1}}^s$ and $i,j=0,x,y,z$ with $\tau_i$, $i=x,y,z$, the usual Pauli matrices and $\tau_0=\hat{\mathbf{1}}$ while $\sigma_z$ is the corresponding Pauli matrix in the spin space. In the following, the Pauli matrices in valley or sublattice space are denoted using the letter $\tau$ and the Pauli matrices in spin space are denoted using the letter $\sigma$.

The first term between square brackets in Eq. (\ref{eq:spBHamiltonian}), $H_B$, is a $2\times 2$ matrix acting in the $\bar{A}\bar{B}$ subspace and corresponds to the kinetic energy:
\begin{eqnarray}\label{eq:BilayerdestructionHO}
H_B&=&\hbar\omega_B\left[\begin{array}{cc}
0 & a_B^2\\
(a_B^{\dagger})^2 & 0
\end{array}\right]\\
\nonumber \omega_B&=&\frac{eB_{\perp}}{m}=1.76\times 10^{11}\frac{m_e}{m}B_{\perp}[\text{T}]~\text{Hz}\\
\nonumber &=&6.28\times 10^{12} B_{\perp}[\text{T}]~\text{Hz}~,
\end{eqnarray}
$m$ being the effective mass for which we take the experimental value $m=0.028m_e$, with $m_e$ the electron mass \cite{Mayorov2011}, and $a_B$ the magnetic annihilation operator
\begin{equation}\label{eq:aniquilacion}
a_B=\frac{l_B}{\hbar}\frac{\pi_y+i\pi_x}{\sqrt{2}},~l_B=\sqrt{\frac{\hbar}{eB_{\perp}}}=\frac{25.7}{\sqrt{B_{\perp}[\text{T}]}}~\text{nm}.
\end{equation}
The operators $\pi_i=-i\hbar \partial_i+eA_i$, $i=x,y$, are the momentum components after the Peierls substitution \cite{Hofstadter1976} and the magnitude $l_B$ is referred as the magnetic length.

The second term, $\epsilon_VT_{zz}$, arises from a voltage difference between the two layers, $\epsilon_V=Ea_z/2$, with $E$ the perpendicular electric field and $a_z\approx 0.35~\text{nm}$ the separation between the layers, while the third term takes into account the Zeeman effect, $\epsilon_Z=\mu_BB, B=\sqrt{B_{\parallel}^2+B_{\perp}^2}$. Here, we assume that the total magnetic field is not necessarily along the $z$ direction, i.e., it can present a parallel component to the graphene plane, $\mathbf{B}_{\parallel}$. The polarizations $\xi=\pm$ correspond to the spin components that are antiparallel (parallel) to the total magnetic field $\mathbf{B}$, respectively.


Taking into account that only the perpendicular magnetic field affects the orbital motion and using the Landau gauge, we can write the potential vector as $\mathbf{A}(\mathbf{x})=[0,B_{\perp}x,0]$. In this particular gauge, the eigenstates of $H_B$ are characterized by the following quantum numbers: the magnetic index $n$, which is an integer number and characterizes the energy of the corresponding Landau level, $\epsilon_n$; the momentum in the $y$-direction, $k$, and the polarization in the valley-spin ($KK'\otimes s$) space, $\alpha$. Specifically, they are given by $\Psi^{0}_{n,k,\alpha}(\mathbf{x})=\Psi^{0}_{n,k}(\mathbf{x})\chi_{\alpha}$, where $\chi_{\alpha}$ is an arbitrary 4-component spinor in valley-spin space while the orbital wave function with components in the space $\bar{A}\bar{B}$ is
\begin{eqnarray}\label{eq:Landaueigenfunctions}
\nonumber \Psi^{0}_{n,k}(\mathbf{x})&=&\frac{e^{iky}}{\sqrt{L_y}}\frac{1}{\sqrt{2}}\left[\begin{array}{c}
(\textrm{sgn}\ n)\ \phi_{|n|-2}(x+kl^2_B)\\ \phi_{|n|}(x+kl^2_B)
\end{array}\right]\\
\epsilon_n&=&(\textrm{sgn}\ n)\ \sqrt{|n|(|n|-1)}\hbar\omega_B
\end{eqnarray}
for $|n|\neq 0,1$ and
\begin{equation}\label{eq:ZLL}
\Psi^{0}_{n,k}(\mathbf{x})=\frac{e^{iky}}{\sqrt{L_y}}\left[\begin{array}{c}
0\\
\phi_{|n|}(x+kl^2_B)
\end{array}\right],~\epsilon_n=0
\end{equation}
for the degenerate levels $|n|=0,1$ with zero energy (note that $n=\pm 1$ are indeed the same state). Hereafter, we refer to this manifold of states as the zero Landau level (ZLL). In the previous equations, $L_y$ is the length of the system in the $y$ direction and $\phi_n(x)$ is the usual harmonic oscillator wave function [see Eq. (\ref{eq:oscillatorwavefunctions})]. We see that the kinetic energy is degenerate in $y$-momentum and valley-spin polarization; in particular, the degeneracy in $k$ for each magnetic level $n$ is $N_B=S/2\pi l^2_B$, $S$ being the total area of the system. Interestingly, the wave functions in the ZLL only have non-vanishing components in the subspace $KK'\otimes\bar{B}\otimes s$.

Using the previous eigenfunctions, the field operator is decomposed as
\begin{equation}\label{eq:fieldoperatorLL}
\hat{\psi}(\mathbf{x})=\sideset{}{'}\sum_{n=-\infty}^{\infty}\sum_{k,\alpha}\Psi^0_{n,k,\alpha}(\mathbf{x})\hat{c}_{n,k,\alpha}
\end{equation}
where $'$ means that $n$ takes every integer value except $n=-1$.

With respect to the interacting part of the Hamiltonian, $\hat{H}_C$ represents the long range Coulomb interaction:
\begin{equation}\label{eq:coulombHamiltonian}
\hat{H}_C=\frac{1}{2}\int\mathrm{d}^2\mathbf{x}~\mathrm{d}^2\mathbf{x'}:[\hat{\psi}^{\dagger}(\mathbf{x})\hat{\psi}(\mathbf{x})]V_0(\mathbf{x}-\mathbf{x'})[\hat{\psi}^{\dagger}(\mathbf{x'})\hat{\psi}(\mathbf{x'})]:
\end{equation}
Here, $:$ denotes normal ordering of the field operators and $V_0(\mathbf{x})=e^2_c/\kappa|\mathbf{x}|$ is the Coulomb potential, with $e^2_c\equiv e^2/4\pi\epsilon_0$ and $\kappa$ the dielectric constant of the environment. Finally, for the short-range interaction Hamiltonian, $\hat{H}_{sr}$, we consider the most general expression compatible with all the symmetries of the problem \cite{Kharitonov2012}
\begin{equation}\label{eq:srHamiltonian}
\hat{H}_{sr}=\sideset{}{'}\sum_{i,j}\frac{1}{2}g_{ij}\int\mathrm{d}^2\mathbf{x}\:[\hat{\psi}^{\dagger}(\mathbf{x})T_{ij}\hat{\psi}(\mathbf{x})]^2:
\end{equation}
The $'$ in this sum denotes that we exclude the symmetric term $i=j=0$, already accounted by the long-range Coulomb interaction. These interactions are asymmetric in the valley and sublattice spaces. The origin of these short-range interactions are the Coulomb interaction between sublattice/valley spaces and the electron-phonon interactions, which we also treat as short-ranged \cite{Kharitonov2012}. For shortness, we refer in the following to the long range Coulomb interactions as simply Coulomb interactions. Except for the kinetic energy term, this Hamiltonian is formally similar to that of monolayer graphene, see Sec. \ref{sec:MLG}.

The two-band model is expected to work quite well in a wide range of magnetic fields $1~\text{T}\lesssim B_{\perp}\lesssim 30~\text{T}$, specially for the ZLL \cite{Cote2013}. For lower magnetic fields trigonal warping effects become important and for larger magnetic fields one has to use the complete four-band model, as the overlap of the wave function of the $n=1$ level with the ignored sublattices is not negligible. Other small corrections to the two-band Hamiltonian here considered break the degeneracy between the single-particle energies of the $n=0$ and $n=1$ levels \cite{Lambert2013} but they are negligible compared to the analog Lamb shift, discussed in the next section.

\subsection{Projection onto the zero Landau level}\label{subsec:projection}

We now address the study of the $\nu=0$ QH state, which corresponds to a half-filling of the ZLL and complete filling of all LLs with $n\leq -2$. For that purpose, we make an estimation of the order of magnitude of the different terms in the Hamiltonian and we compare them with the typical energy difference between LLs, $\hbar\omega_B$. For the Zeeman term, we find
\begin{equation}
\frac{\epsilon_Z}{\hbar\omega_B}=0.014\frac{B}{B_{\perp}} \ll 1
\end{equation}
for realistic values of the ratio $B/B_{\perp}$ while for the Coulomb interaction one has
\begin{equation}\label{eq:coulombfactor}
\frac{F_C}{\hbar\omega_B}=\frac{13.58}{\kappa\sqrt{B_{\perp}[\text{T}]}},~F_C\equiv \frac{e^2_c}{\kappa l_B}
\end{equation}
Usual values for the perpendicular magnetic field are $B_{\perp}\gtrsim 1~\text{T}$ and the highest available continuous magnetic field in the laboratory is $B_{\perp}\simeq 80~\text{T}$ \cite{Yang2016}, which means that the strength of the Coulomb interaction verifies $F_C\gtrsim \hbar\omega_B$ when the environment is vacuum ($\kappa=1$). The interlayer voltage can in principle vary in a wide range \cite{Lambert2013,Cote2013} but we keep its value sufficiently small here, $\epsilon_V\ll \hbar\omega_B$. Finally, a dimensional analysis of the short-range terms gives an estimation for the coupling constants $g_{ij}\sim e^2_cd/\kappa$ with $d\sim0.1~\text{nm}$ the typical length scale of the lattice. Then, the energy scale associated to the short-range interactions is $\sim\frac{e^2_c d}{\kappa l^2_B}$, which is small compared to the LL separation as
\begin{equation}\label{eq:shortrangeestimation}
\frac{e^2_c d}{\kappa l^2_B\hbar\omega_B}\sim \frac{F_C}{\hbar\omega_B} \frac{d}{l_B}\sim\frac{0.1}{\kappa} \ll 1
\end{equation}
We note that the energy associated to the short-range interactions scales linearly with the magnetic field although for low values of the magnetic field the coupling constants themselves can be renormalized \cite{Kharitonov2012,Kharitonov2012PRL}, see also Sec. \ref{sec:renorm}.

As a result of the above analysis, the only term that is not small compared to the separation between LLs is that related with Coulomb interactions. We begin by treating Coulomb interactions as weak, $F_C\ll \hbar\omega_B$; for instance, by supposing a typical value $\kappa=5$ and a magnetic field $B\sim~20~T$, $F_C/\hbar\omega_B\sim 0.5$, which can be regarded as sufficiently small to treat it perturbatively. In Sec. \ref{sec:renorm}, we address the usual situation $F_C\gtrsim \hbar\omega_B$ and explain how to deal with it.

Taking into account the previous considerations and in order to study the lowest energy excitations, we neglect LL mixing and restrict ourselves to the ZLL by projecting the full Hamiltonian into that subspace, as previously done in Refs. \cite{Toke2011,Kharitonov2012,Kharitonov2012PRL,Lambert2013}. We remark [see Eq. (\ref{eq:ZLL}) and ensuing discussion] that the states in the ZLL belong to the $KK'\otimes\bar{B}\otimes s$ space, which means that they are localized, for each valley, in one sublattice or the other and correspondingly, in one layer or the other. Thus, within the ZLL, the sublattice degree of freedom becomes equivalent to the valley degree of freedom. The resulting effective Hamiltonian for the ZLL is:
\begin{widetext}
\begin{eqnarray}\label{eq:EffectiveHamiltonian}
\nonumber\hat{H}^{(0)}&=&\int\mathrm{d}^2\mathbf{x}~\hat{\psi}^{\dagger}(\mathbf{x})\left[-\epsilon_V\tau_z-\epsilon_Z\sigma_z\right]\hat{\psi}(\mathbf{x})+\frac{1}{2}\int\mathrm{d}^2\mathbf{x}~\mathrm{d}^2\mathbf{x'}:[\hat{\psi}^{\dagger}(\mathbf{x})\hat{\psi}(\mathbf{x})]V_0(\mathbf{x}-\mathbf{x'})[\hat{\psi}^{\dagger}(\mathbf{x'})\hat{\psi}(\mathbf{x'})]:\\
&+&\sum_{i}\frac{1}{2}\int\mathrm{d}^2\mathbf{x}\ g_{i}:[\hat{\psi}^{\dagger}(\mathbf{x})\tau_{i}\hat{\psi}(\mathbf{x})]^2:+\int\mathrm{d}^2\mathbf{x}~\mathrm{d}^2\mathbf{x'}\hat{\psi}^{\dagger}(\mathbf{x})V_{DS}(\mathbf{x},\mathbf{x'})\hat{\psi}(\mathbf{x'})
\end{eqnarray}
\end{widetext}
where $\tau_{i}\equiv\tau^{KK'}_i$ and $g_{i}=g_{i0}+g_{iz}, i=1,2,3$. As we restrict to the ZLL, the kinetic energy term is suppressed. We have neglected the symmetric short-range interaction, arising from the coupling $g_{0z}$, due to its smallness compared to the symmetric Coulomb interaction \cite{Kharitonov2012}. Using symmetry considerations, it can be proven that $g_x=g_y\equiv g_{\perp}$ \cite{Aleiner2007,Kharitonov2012}, so there are only two independent coupling constants $g_{\perp},g_z$. The potential $V_{DS}(\mathbf{x},\mathbf{x'})$ represents the mean-field interaction of the ZLL with the (inert) Dirac sea compound by all the occupied states with $n\leq -2$ \cite{Shizuya2012,Shizuya2013,Toke2013}. As shown in Appendix \ref{app:SCHF}, this potential is diagonal within the ZLL; its explicit expression is given by Eq. (\ref{eq:HFprojected}). The projection onto the ZLL leads to a field operator of the form
\begin{equation}\label{eq:fieldoperatorZLL}
\hat{\psi}(\mathbf{x})=\sum_{n=0,1}\sum_{k,\alpha}\Psi^0_{n,k,\alpha}(\mathbf{x})\hat{c}_{n,k,\alpha}
\end{equation}

\section{Hartree-Fock equations and mean-field phase diagram}\label{sec:HFMF}

In order to obtain the mean-field phase diagram of the $\nu=0$ QH state at zero temperature, we use the Hartree-Fock (HF) approximation for the self-consistent single-particle wave functions, which we denote as $\Psi_{n,k,\alpha}$. The corresponding HF equations for the Hamiltonian (\ref{eq:EffectiveHamiltonian}) read (we refer the reader to Appendix \ref{app:magneticFF} for all the technical details)
\begin{widetext}
\begin{eqnarray}\label{eq:HFeqs}
\epsilon_{n,\alpha}\Psi_{n,k,\alpha}(\mathbf{x})&=&\int\mathrm{d}^2\mathbf{x'}~V_{DS}(\mathbf{x},\mathbf{x'})\Psi_{n,k,\alpha}(\mathbf{x'})-\sum_{m,p,\beta}\nu_{m,\beta}\int\mathrm{d}^2\mathbf{x'}~V_0(\mathbf{x}-\mathbf{x'})\Psi_{m,p,\beta}(\mathbf{x})\Psi_{m,p,\beta}^{\dagger}(\mathbf{x'})\Psi_{n,k,\alpha}(\mathbf{x'})\\
\nonumber&+&\sum_{i}\sum_{m,p,\beta}\nu_{m,\beta}g_{i}\left([\Psi_{m,p,\beta}^{\dagger}(\mathbf{x})\tau_{i}\Psi_{m,p,\beta}(\mathbf{x})]\tau_{i}\Psi_{n,k,\alpha}(\mathbf{x})- \tau_{i}\Psi_{m,p,\beta}(\mathbf{x})\Psi_{m,p,\beta}^{\dagger}(\mathbf{x})\tau_{i}\Psi_{n,k,\alpha}(\mathbf{x})\right)\\
\nonumber &-&\epsilon_V\tau_z\Psi_{n,k,\alpha}(\mathbf{x})-\epsilon_Z\sigma_z\Psi_{n,k,\alpha}(\mathbf{x})
\end{eqnarray}
\end{widetext}
where the indices $n,m=0,1$ label the magnetic levels, $k,p$ are the momenta in the $y$-direction and $\alpha,\beta$ represent the polarization in the valley-spin space. We have made explicit that we are dealing with an integer QH state, so every orbital $p$ is filled in the same way and then, the occupation number $\nu_{n,p,\alpha}$ of each state solely depends on $n$ and $\alpha$, $\nu_{n,p,\alpha}=\nu_{n,\alpha}$. In particular, for the $\nu=0$ QH state, only half of the ZLL is filled. Thus, for each value of the $y$-momentum $k$, only four states of the eightfold space, formed by the $0,1$ magnetic states and the valley-spin degrees of freedom, are occupied. As usual, the direct (Hartree) term for the Coulomb interaction is suppressed by the positive charge background.

An important result is that the orbital part of the self-consistent HF wave functions is equal to that of the non-interacting wave functions, given in Eq. (\ref{eq:ZLL}); see Appendix \ref{app:magneticFF} for the proof. Hence, the only remaining task is to specify the spinors $\chi_{\alpha}$. In order to minimize the dominant Coulomb interaction, the electrons occupy in the same way the valley-spin space for the two magnetic levels, i.e., $\nu_{0,a}=\nu_{0,b}=\nu_{1,a}=\nu_{1,b}=1$, with $\chi_{a,b}$ two orthogonal spinors \cite{Abanin2009,Kharitonov2012PRL} so the mean-field ground state is of the form
\begin{equation}\label{eq:2ndHFsolution}
\ket{\Psi_0}=\prod_{p}\hat{c}^{\dagger}_{0,p,a}\hat{c}^{\dagger}_{1,p,a}\hat{c}^{\dagger}_{0,p,b}\hat{c}^{\dagger}_{1,p,b}\ket{DS},
\end{equation}
with $\ket{DS}$ the Dirac sea formed by all occupied Landau levels with $n\leq-2$.

The four remaining unoccupied states of the ZLL are characterized by the spinors $\chi_{c,d}$. In this way, the occupation number only depends on the valley-spin polarization $\alpha$, $\nu_{n,\alpha}=\nu_{\alpha}$, taking the values $\alpha=a,b,c,d$ that correspond to an orthonormal basis of the valley-spin space. These spinors are computed after projecting the HF equations into the orbital part of the wave functions, obtaining closed algebraic equations in valley-spin space:
\begin{eqnarray}\label{eq:HFenergy}
\nonumber \epsilon_{n,\alpha}\chi_{\alpha}&=&\frac{F_n}{2}\chi_{\alpha}-F_nP\chi_{\alpha}-\epsilon_V\tau_z\chi_{\alpha}-\epsilon_Z\sigma_z\chi_{\alpha}\\
&+&\sum_{i}u_{i}\left([\text{tr}(P\tau_{i})]\tau_{i}-\tau_{i}P\tau_{i}\right)\chi_{\alpha}
\end{eqnarray}
where $u_{i}=g_{i}/\pi l^2_B$ and $F_n=F_{n0}+F_{n1}$, with:
\begin{eqnarray}\label{eq:Fockeigenvalues}
F_{00}=\sqrt{\frac{\pi}{2}}F_C,~F_{01}=F_{10}=\frac{1}{2}F_{00},~F_{11}=\frac{3}{4}F_{00}
\end{eqnarray}
so $F_{00}>F_{01}>F_{11}$ and then $F_0>F_1$ as
\begin{eqnarray}\label{eq:Fockenergies}
F_0=\frac{3}{2}F_{00},~F_1=\frac{5}{4}F_{00}
\end{eqnarray}
The values of the factors $F_{nm}$ are obtained by inserting the Coulomb potential in Eq. (\ref{eq:FockFouriermatrixelement}).

The term $F_{n}/2$ in Eq. (\ref{eq:HFenergy}) is the analog of the Lamb shift \cite{Shizuya2012,Shizuya2013}, arising from the interaction of the ZLL with the Dirac sea after the proper regularization of the Hamiltonian (see Appendix \ref{app:SCHF}). The term $-F_nP$ arises from the exchange Coulomb interaction, where the matrix $P$ is the projector onto the subspace formed by $\chi_{a,b}$, $P=\chi_{a}\chi^{\dagger}_{a}+\chi_{b}\chi^{\dagger}_{b}$, and hence $P\chi_{a,b}=\chi_{a,b},~P\chi_{c,d}=0$. The remaining terms are those related with the short-range interactions and the valley-spin part of the single-particle Hamiltonian. We see that the sole dependence on the magnetic level is through the Coulomb and Lamb-shift terms, while the other contributions to the energy only depend on the spinor $\chi_{\alpha}$. Note that, although the interaction with the Dirac sea favors the filling of the magnetic level $n=1$, it is still more energetically favorable to occupy in the same way the levels $n=0,1$ due to the exchange interaction. Indeed, the mean-field state (\ref{eq:2ndHFsolution}) is an {\it exact} eigenvalue of the effective Hamiltonian (\ref{eq:EffectiveHamiltonian}) when short-range interactions are neglected. As Coulomb interactions dominate over short-range interactions, this mean-field solution is expected to provide a very good approximation to the actual ground state \cite{Yang2006,Kharitonov2012}, showing the robustness of the formalism here considered.

\begin{figure*}[tb!]
\begin{tabular}{@{}cc@{}}
    \includegraphics[width=\columnwidth]{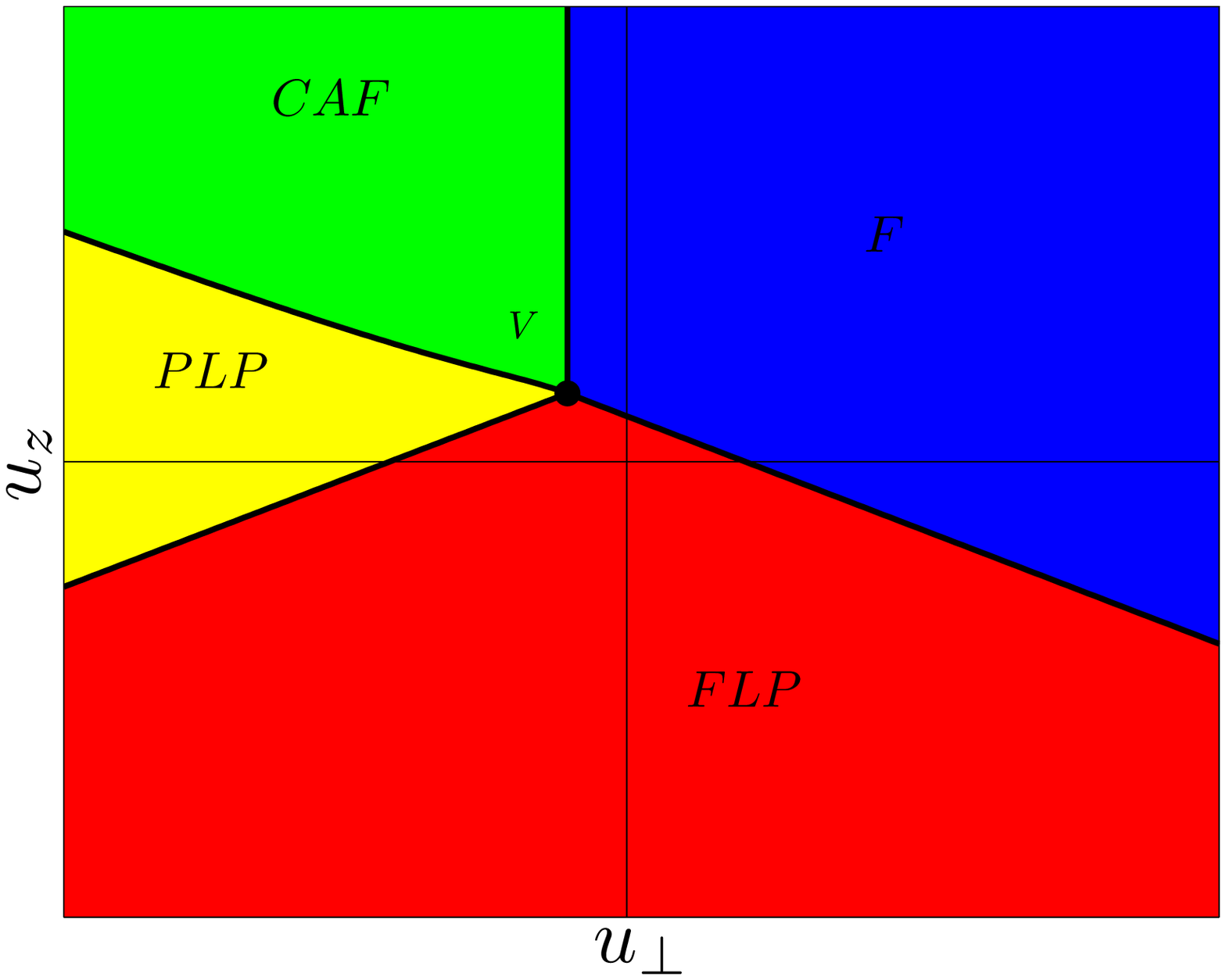} &
    \includegraphics[width=\columnwidth]{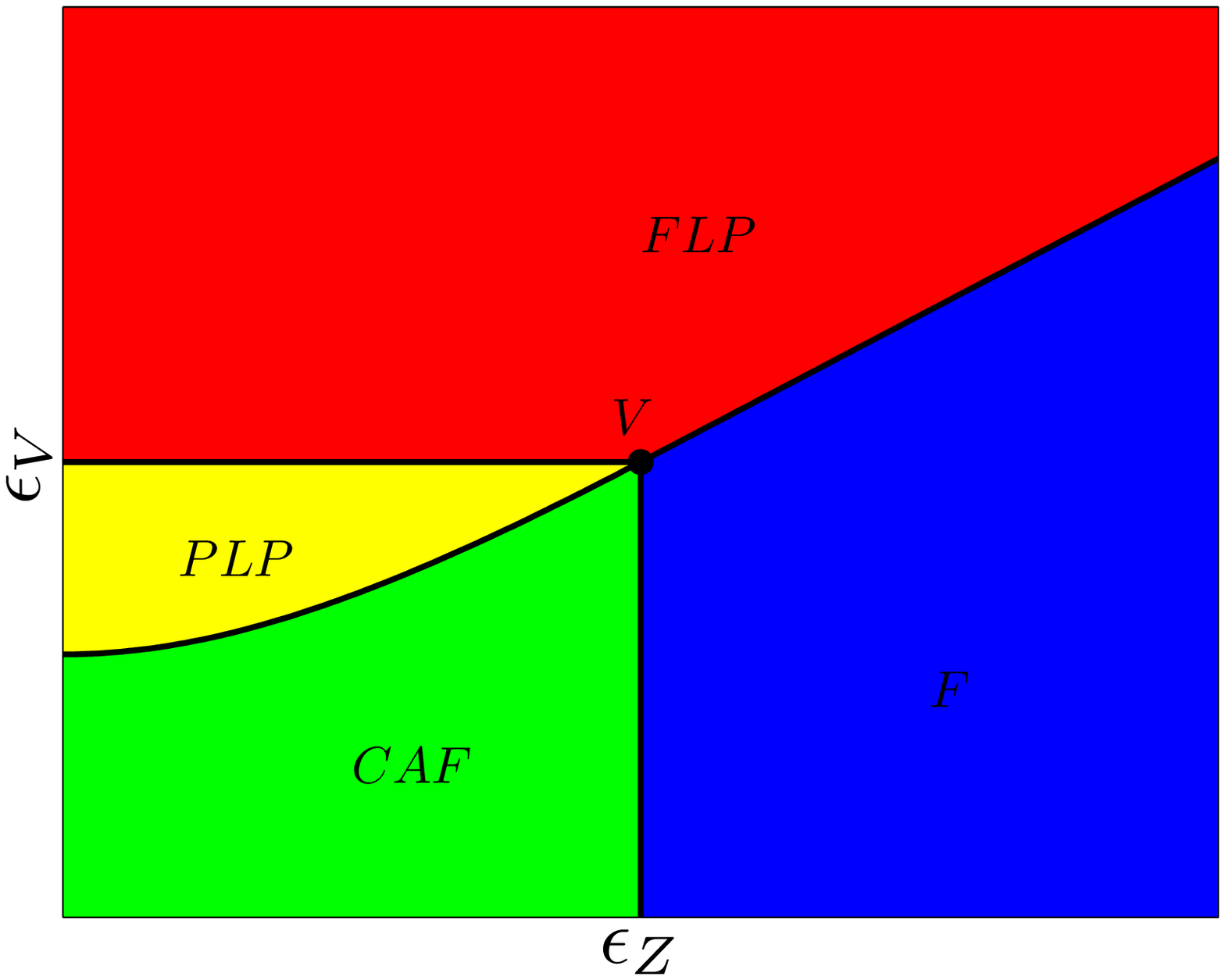} \\
\end{tabular}
\caption{Phase diagram of the $\nu=0$ QH state. Left plot: phase diagram in the parameter space $u_{\perp},u_{z}$. The point $V$ is the intersection of all phase boundaries. Right plot: expected phase diagram for fixed $u_{\perp},u_{z}$, with $u_z>-u_{\perp}>0$, as a function of the Zeeman and voltage energies, $\epsilon_Z,\epsilon_V$.}
\label{fig:PhaseDiagram}
\end{figure*}

As a result of the previous discussion, the HF energies can be written as:
\begin{equation}\label{eq:HFenergystructure}
\epsilon_{n,(a,b)}=-\frac{F_n}{2}+\epsilon_{(a,b)},~\epsilon_{n,(c,d)}=\frac{F_n}{2}+\epsilon_{(c,d)}
\end{equation}
with $\epsilon_{\alpha}$ depending only on the polarization $\alpha$. The energy of the total state (per wave vector state) is:
\begin{eqnarray}\label{eq:meanfieldenergy}
\nonumber E_{HF}&=&-\frac{(F_0+F_1)}{2}+2E(P)\\
\nonumber E(P)&=&\frac{1}{2}\sum_{i}u_{i}\left\{[\text{tr}(P\tau_{i})]^2-\text{tr}(P\tau_{i}P\tau_{i})\right\}\\
&-&\epsilon_V\text{tr}(P\tau_{z})-\epsilon_Z\text{tr}(P\sigma_z)
\end{eqnarray}
The contribution from Coulomb interaction to the total energy turns out to be degenerate and does not depend on the specific form of the occupied spinors $\chi_{a,b}$. The actual ground state of the system is determined by comparing the energies corresponding to all possible solutions to the HF equations and selecting that with lowest energy $E(P)$, in the same fashion of Refs. \cite{Kharitonov2012,Kharitonov2012PRL}. Hence, the corresponding mean-field phase diagram for the $\nu=0$ QH state is the same of those references and is represented in Fig. \ref{fig:PhaseDiagram}. The different possible phases are ferromagnetic (F), canted anti-ferromagnetic (CAF), fully layer-polarized (FLP) and partially layer-polarized (PLP). The expected phase for the $\nu=0$ QH state of bilayer graphene for $\epsilon_V=0$ and zero in-plane component of the magnetic field is the CAF phase \cite{Kharitonov2012PRL,Kharitonov2012a,Maher2013}, which implies that $u_z>-u_{\perp}>0$. The exact values of these short-range energies remain unknown but their order of magnitude is $u_z,|u_{\perp}|\sim 0.1 \hbar \omega_B$ \cite{Kharitonov2012b,Young2014}. In the following, we treat them as phenomenological inputs for the theory. All phase boundaries intersect at the critical point
\begin{equation}\label{eq:criticalpoint}
V=(\epsilon^*_{Z},\epsilon^*_V)=(-2u_{\perp},u_z-u_{\perp}).
\end{equation}
The complete phase diagram can be explored experimentally by manipulating the in-plane component of the magnetic field (which modifies the value of the total magnetic field $B$ and hence the value of $\epsilon_Z$) or the layer voltage (which modifies the value of $\epsilon_V$). This fact can be checked by looking at right Fig. \ref{fig:PhaseDiagram}, where we represent the phase diagram as a function of the Zeeman and voltage energies, $(\epsilon_Z,\epsilon_V)$, for fixed values of $u_{\perp},u_z$ such that $u_z>-u_{\perp}>0$.

The phase diagram here presented describes the $\nu=0$ QH state for $B_{\perp}\gtrsim 1$ T \cite{Kharitonov2012PRL}; the complementary phase diagram in the remaining limit of very low perpendicular magnetic fields $B_{\perp}\ll 1$ T is studied in Ref. \cite{Jia2017}.

We now briefly describe all the phases and give the expressions for the filled and empty valley-spin spinors, the matrix $P$ and the corresponding valley-spin energies $\epsilon_{\alpha}$ for each phase.

\subsection{Ferromagnetic phase}\label{subsec:Fphase}

In the F phase, all the electrons have the spin aligned with the magnetic field. The complete set of solutions involves the following 4 spinors in valley-spin space, with eigenvalues:
\begin{eqnarray}\label{eq:Fphase}
\chi_{a}&=&\ket{n_z}\otimes\ket{s_z},~\chi_{b}=\ket{-n_z}\otimes
\ket{s_z}\\
\nonumber\chi_{c}&=&\ket{n_z}\otimes\ket{-s_z},~\chi_{d}=\ket{-n_z}\otimes\ket{-s_z}\\
\nonumber\epsilon_{a,b}&=&-2u_{\perp}-u_z-\epsilon_Z\mp\epsilon_V,~\epsilon_{c,d}=\epsilon_Z\mp\epsilon_V\\
\nonumber P&=&\frac{I+\sigma_z}{2},~E(P)=-(2u_{\perp}+u_z)-2\epsilon_Z
\end{eqnarray}
Here, $\ket{\pm s_z}$ denotes the state with spin polarization $\xi=\pm$, $I$ the $4\times4$ identity matrix and $\ket{n_z}=\ket{K},\ket{-n_z}=\ket{K'}$. The F phase is always a solution of the HF equations although it does not always correspond to the actual ground-state.

\subsection{Full layer-polarized phase}\label{subsec:FLPphase}

The FLP phase is the equivalent of the F phase but in valley pseudospin, $KK'$, which means that all the electrons are concentrated on one layer (due to the equivalency of valley-sublattice-layer in the ZLL in bilayer graphene). The corresponding spinors are now:
\begin{eqnarray}\label{eq:FLPphase}
\chi_{a}&=&\ket{n_z}\otimes\ket{s_z},~\chi_{b}=\ket{n_z}\otimes\ket{-s_z}\\
\nonumber \chi_{c}&=&\ket{-n_z}\otimes\ket{s_z},~\chi_{d}=\ket{-n_z}\otimes
\ket{-s_z}\\
\nonumber\epsilon_{a,b}&=&u_z\mp\epsilon_Z-\epsilon_V,~\epsilon_{c,d}=-2u_{\perp}-2u_z\mp\epsilon_Z+\epsilon_V\\
\nonumber P&=&\frac{I+\tau_z}{2},~E(P)=u_z-2\epsilon_V
\end{eqnarray}
In analogy to the F phase, it is always a solution to the HF equations but not necessarily the ground-state. By comparing the energies of the two phases, we obtain the boundary between the F and FLP phases:

\begin{equation}\label{eq:FFLPborder}
\epsilon_V-\epsilon_Z=u_{\perp}+u_z
\end{equation}

\subsection{Canted anti-ferromagnetic phase}\label{subsec:CAFphase}

The previous phases would be the only possible phases if there were not short-range interactions, i.e., $u_{\perp}=u_z=0$. However, when taking into account these interactions, the system can exhibit canted anti-ferromagnetism or partially layer-polarization in order to minimize the interaction energy. In the CAF phase, we have that:

\begin{eqnarray}\label{eq:CAFphase}
\nonumber \chi_{a}&=&\ket{n_z}\otimes
\ket{s_a},~\chi_{b}=\ket{-n_z}\otimes\ket{s_b}\\
\nonumber \chi_{c}&=&\ket{n_z}\otimes\ket{-s_a},~\chi_{d}=\ket{-n_z}\otimes\ket{-s_b}\\
\nonumber\epsilon_{a,b}&=&-u_z-2u_{\perp}\cos^2\theta_S-\epsilon_Z\cos\theta_S \mp\epsilon_V=-u_z\mp\epsilon_V \\
\nonumber \epsilon_{c,d}&=&-2u_{\perp}\sin^2\theta_S+\epsilon_Z\cos\theta_S \mp \epsilon_V=-2u_{\perp}\mp \epsilon_V\\
\nonumber P&=&\frac{I+\cos\theta_S\sigma_z+\sin\theta_S(\mathbf{s}_{\parallel}\cdot\mathbf{\sigma})\tau_z}{2}\\
E(P)&=&-u_z-\epsilon_Z\cos\theta_S
\end{eqnarray}
where $\ket{s_{a,b}}$ are spin states with polarization given by the vectors $\mathbf{s}_{a,b}=[\pm\sin\theta_S\cos\phi_S,\pm\sin\theta_S\sin\phi_S,\cos\theta_S]$, with tilting angle $\cos\theta_S=-\epsilon_Z/2u_{\perp}$, and $\mathbf{s}_{\parallel}=[\cos\phi_S,\sin\phi_S,0]$. We choose the phases of these states in such a way that $\ket{s_b}=\sigma_z \ket{s_a}$ and $\ket{-s_b}=-\sigma_z \ket{-s_a}$ so the CAF phase continuously matches the F phase of Eq. (\ref{eq:Fphase}) for $\theta_S=0$. As the azimuth $\phi_S$ is a free parameter, the CAF phase exhibits a $U(1)$ symmetry. When this solutions exists, it always has lower energy than the regular F phase, so the condition for the presence of the CAF phase is just
\begin{equation}\label{eq:FCAFborder}
\cos\theta_S<1\Rightarrow \epsilon_Z<\epsilon_{Zc}\equiv-2u_{\perp}
\end{equation}
with $\epsilon_{Zc}$ the critical Zeeman field.

\subsection{Partially layer-polarized phase}\label{subsec:PLPphase}

In analogy with the relation between the FLP and F phases, the PLP is similar to the CAF phase but in valley space:

\begin{eqnarray}\label{eq:PLPphase}
\chi_{a}&=&\ket{\mathbf{n}}\otimes\ket{s_z},~\chi_{b}=\ket{\mathbf{n}}\otimes\ket{-s_z}\\
\nonumber \chi_{c}&=&\ket{-\mathbf{n}}\otimes\ket{s_z},~\chi_{d}=\ket{-\mathbf{n}}\otimes\ket{-s_z}\\
\nonumber\epsilon_{a,b}&=&u_{\perp}\sin^2\theta_V+u_z\cos^2\theta_V\mp\epsilon_Z-\epsilon_V\cos\theta_V=u_{\perp}\mp\epsilon_Z \\
\nonumber \epsilon_{c,d}&=&-2u_{\perp}-u_z-u_{\perp}\sin^2\theta_V-u_z\cos^2\theta_V\\
\nonumber &+&\epsilon_V\cos\theta_V\mp \epsilon_Z=-3u_{\perp}-u_z\mp \epsilon_Z\\
\nonumber P&=&\frac{I+\mathbf{n}\cdot\mathbf{\tau}}{2}\\
\nonumber E(P)&=&u_{\perp}\sin^2\theta_V+u_z\cos^2\theta_V-2\epsilon_V\cos\theta_V\\
\nonumber &=&u_{\perp}-\epsilon_V\cos\theta_V
\end{eqnarray}
with $\ket{\mathbf{n}}$ a state with valley-polarization given by $\mathbf{n}=[\sin\theta_V\cos\phi_V,\sin\theta_V\sin\phi_V,\cos\theta_V]$, $\cos\theta_V=\epsilon_{V}/(u_z-u_{\perp})$. The PLP phase also presents an $U(1)$ symmetry. In analogy with the F-CAF phase transition, whenever this solution exists, it has lower energy than the FLP phase so the existence condition for the PLP phase is
\begin{equation}\label{eq:FLPPLPborder}
\cos\theta_V<1\Rightarrow \epsilon_V<\epsilon_{Vc}\equiv u_{\perp}+u_z
\end{equation}
with $\epsilon_{Vc}$ the critical voltage energy. On the other hand, the boundary between the PLP and the CAF phases is placed at
\begin{equation}\label{eq:PLPCAFborder}
\frac{\epsilon^2_V}{u_z-u_{\perp}}+\frac{\epsilon^2_Z}{2u_{\perp}}=u_z+u_{\perp}
\end{equation}

\subsection{Transport gap}\label{subsec:Transportgap}

An experimental magnitude of interest that can be obtained within the present mean-field computation is the transport gap $\Delta_{HF}$, defined as the energy difference between the lowest energy empty state and the last filled state. From the above results, and within the convention chosen here for the occupied and empty levels, it is immediate to show that in all phases
\begin{equation}\label{eq:transportgap}
\Delta_{HF}=\epsilon_{1,c}-\epsilon_{1,b}=F_1+\Delta^{bc},~\Delta^{bc}\equiv\epsilon_{c}-\epsilon_{b},~
\end{equation}
in good agreement with Refs. \cite{Gorbar2011,Lambert2013}, where the only asymmetric interactions considered are interlayer Coulomb interactions.

\section{Collective modes}\label{sec:colmod}

\subsection{Preliminary remarks}\label{subsec:preliminar}

We proceed to compute the neutral collective modes of the previous mean-field phases using the TDHFA. The general formalism of the TDHFA and its application to integer QH states is explained thoroughly in Appendix \ref{app:TDHFA} while Appendix \ref{app:analyticalTDHFA} provides all the technical details on the results presented in this section.

Due to the particular form of the $\nu=0$ QH state, in which the unoccupied levels have different valley-spin polarization with respect to the occupied levels, the excitations correspond to valley-spin waves. Within the TDHFA, the collective modes are obtained from the eigenvalues and eigenvectors of the matrix $\tilde{X}(\mathbf{k})$, given by Eq. (\ref{eq:XYgen32}). The wave vector $\mathbf{k}$ corresponds to the momentum of the so-called magnetoexciton \cite{Kallin1984}, whose wave function is created by the action of the operator $\hat{M}^{\dagger}_{n\alpha n'\alpha'}(\mathbf{k})\equiv\hat{M}^{\dagger}_{n\lambda\xi n'\lambda'\xi'}(\mathbf{k})$ on the mean-field ground state $\ket{\Psi_0}$, with
\begin{equation}\label{eq:magnetoexcitonwavefunction}
\hat{M}^{\dagger}_{n\lambda\xi n'\lambda'\xi'}(\mathbf{k})=\sqrt{\frac{1}{N_B}}\sum_{q} e^{-iqk_xl^2_B}\hat{c}_{n,q+\frac{k_y}{2},\lambda\xi}^{\dagger}\hat{c}_{n',q-\frac{k_y}{2},\lambda'\xi'}
\end{equation}
where we have made explicit the dependence in valley $\lambda=K,K'$ and spin $\xi=\pm$ indices of the total valley-spin polarization index $\alpha$. As well known from the general theory of integer QH effect, the collective modes are expressed in terms of linear combinations of magnetoexcitons. The resulting dispersion relation is isotropic and only depends on $k=|\mathbf{k}|$, as shown in Appendix \ref{app:analyticalTDHFA}.

In order to classify these modes, we study their symmetries in the whole $8$-dimensional space $01\otimes KK'\otimes s$, formed by the tensorial product of the magnetic levels of the ZLL and the valley-spin space. For that purpose, following the notation of Ref. \cite{Lambert2013}, we define the operators:
\begin{eqnarray}
\label{eq:spin}\hat{S}_i&=&\frac{1}{2}\sum_{n,p,\lambda}\sum_{\xi,\xi'}\hat{c}^{\dagger}_{n,p,\lambda,\xi}(\sigma_i)_{\xi\xi'}\hat{c}_{n,p,\lambda,\xi'}\\
\label{eq:valleyspin}\hat{L}_i&=&\frac{1}{2}\sum_{n,p,\xi}\sum_{\lambda,\lambda'}\hat{c}^{\dagger}_{n,p,\lambda,\xi}(\tau_i)_{\lambda,\lambda'}\hat{c}_{n,p,\lambda',\xi}\\
\label{eq:orbitalspin}\hat{O}_i&=&\frac{1}{2}\sum_{p,\lambda,\xi}\sum_{n,n'}\hat{c}^{\dagger}_{n,p,\lambda,\xi}(\mu_i)_{nn'}\hat{c}_{n',p,\lambda,\xi}
\end{eqnarray}
which correspond to the components of the spin, valley and orbital pseudospin, respectively. In the above expression, $\mu_i$ are the corresponding Pauli matrices in the magnetic index, with $\mu_z=\pm 1$ for $n=1,0$.

We consider the behavior of the magnetoexcitons under transformations generated by the previous operators. For instance, for the spin operator, the commutator $[\hat{S}_i,\hat{M}^{\dagger}_{n\lambda\xi n'\lambda'\xi'}(\mathbf{k})]$ reads
\begin{eqnarray}\label{eq:magnetoexcitonSU2}
\nonumber [\hat{S}_i,\hat{M}^{\dagger}_{n\lambda\xi n'\lambda'\xi'}(\mathbf{k})]&=&\sum_{\zeta\zeta'}(G_{i})_{\zeta\zeta',\xi\xi'}\hat{M}^{\dagger}_{n\lambda\zeta n'\lambda'\zeta'}(\mathbf{k})\\
\nonumber (G_{i})_{\zeta\zeta',\xi\xi'}&=&\frac{1}{2}\left[(\sigma_i)_{\zeta\xi}\delta_{\zeta'\xi'}-\delta_{\xi\zeta}(\sigma_i)_{\xi'\zeta'}\right]~,\\
\end{eqnarray}
It is easy to prove that the matrices $G_{i}$ form a representation with spin $1/2 \otimes 1/2=1\oplus0$ of the Lie algebra of $SU(2)$. Thus, spin singlet and triplet magnetoexcitons can be constructed according to the value of the total spin $S$ and its $z$-component $S_z$, $\hat{M}^{\dagger}_{n\lambda n'\lambda',SS_z}(\mathbf{k})$. Specifically, the spin-triplet magnetoexcitons are given by
\begin{eqnarray}\label{eq:magnetoexcitontriplet}
\hat{M}^{\dagger}_{n\lambda n'\lambda'11}(\mathbf{k})&=&\hat{M}^{\dagger}_{n\lambda + n'\lambda'-}(\mathbf{k})\\
\nonumber \hat{M}^{\dagger}_{n\lambda n'\lambda'10}(\mathbf{k})&=&\frac{1}{\sqrt{2}}\left[\hat{M}^{\dagger}_{n\lambda + n'\lambda'+}(\mathbf{k})-\hat{M}^{\dagger}_{n\lambda - n'\lambda'-}(\mathbf{k})\right]\\
\nonumber \hat{M}^{\dagger}_{n\lambda n'\lambda'11}(\mathbf{k})&=&\hat{M}^{\dagger}_{n\lambda - n'\lambda'+}(\mathbf{k})
\end{eqnarray}
while the spin-singlet magnetoexciton reads
\begin{equation}\label{eq:magnetoexcitonsinglet}
\hat{M}^{\dagger}_{n\lambda n'\lambda'00}(\mathbf{k})=\frac{1}{\sqrt{2}}\left[\hat{M}^{\dagger}_{n\lambda + n'\lambda'+}(\mathbf{k})+\hat{M}^{\dagger}_{n\lambda - n'\lambda'-}(\mathbf{k})\right]
\end{equation}
Due to the formal analogy with the operators $\hat{L}_i,\hat{O}_i$, we can also build similar singlet and triplet magnetoexcitons in valley and orbital pseudospin.

The importance of the previous considerations arises from the fact that the effective Hamiltonian (\ref{eq:EffectiveHamiltonian}) commutes with the operators $\hat{\mathbf{S}}^2,\hat{\mathbf{L}}^2,\hat{S}_z,\hat{L}_z$, with
\begin{equation}
\hat{\mathbf{S}}^2=\sum_{i=x,y,z} \hat{S}_i\hat{S}_i,~\hat{\mathbf{L}}^2=\sum_{i=x,y,z} \hat{L}_i\hat{L}_i
\end{equation}
so $S,L,S_z,L_z$ are expected to be good quantum numbers that characterize the magnetoexcitons whenever $\ket{\Psi_0}$ is also an eigenstate of any of these operators. Thus, the collective modes can be labeled by a general index $\mu$, which represents a set of conserved quantum numbers in valley-spin space. Specifically, the F and FLP states have well-defined quantum numbers for $S,L,S_z,L_z$ which, as discussed in the paragraph following Eq. (\ref{eq:TDHFATontos}) and in Appendix \ref{app:analyticalTDHFA}, implies that they cannot exhibit dynamical instabilities (i.e., exploding modes with complex frequency and positive imaginary part), at least within the projected model considered in this work. On the other hand, the CAF and PLP states do not satisfy this property and, as shown in the next section, they are indeed able to display dynamical instabilities in some regions of the parameter space.

At $k=0$, apart from these valley-spin symmetries, the $z$ component of the orbital pseudospin, $O_z$, is also conserved by the Hamiltonian as it represents the angular momentum of the magnetoexciton \cite{Iyengar2007,Toke2011,Sari87}. In fact, the total orbital pseudospin $O$ is another good quantum number in this limit. As a consequence, we can provide a complete classification of the modes at $k=0$ using their total orbital-valley-spin symmetry, denoting their frequencies as $\omega^{\mu}_{OO_z}$. As shown in Appendix \ref{app:analyticalTDHFA}, the energy of the orbital-singlet modes $OO_z=00$ is completely independent of the Coulomb interaction strength. On the other hand, the orbital-triplet modes ($O=1$) depend on short-range interactions only through the valley-spin contributions of mean-field energies, but not through many-body corrections. Since Coulomb interactions are dominant, this structure implies that the triplet modes are shifted above by relatively large Coulomb energies and hence, the orbital-singlet modes are the lowest energy modes at $k=0$. In particular, the hierarchy $\omega^{\mu}_{00}<\omega^{\mu}_{11}=\omega^{\mu}_{1-1}<\omega^{\mu}_{10}$ is satisfied. It is worth noting that the triplet modes with $O_z=\pm 1$ are degenerate due to the analog Lamb shift.

For $k>0$, however, the orbital singlet and triplet magnetoexcitons are mixed and the previous classification is no longer valid. Then, in order to classify the orbital structure of the different magnetoexcitons for fixed $\mu$, we introduce the discrete index $N=0,1,2,3$ and denote the corresponding frequency as $\omega^{\mu}_N(k)$ so they respectively match the collective modes with orbital pseudospin $OO_z=00,11,1-1,10$ at $k=0$; hence, $\omega^{\mu}_{0}(k=0)=\omega^{\mu}_{00}$ and so on. The notation is chosen in such a way that, for fixed $\mu$, $\omega^{\mu}_{N}(k)< \omega^{\mu}_{N'}(k)$ for $N<N'$.

As the collective modes correspond to valley-spin waves, the behavior of the modes for low momentum is $\omega^{\mu}_{N}(k)\simeq\omega^{\mu}_{N}(0)+\rho^{\mu}_{N}k^2$, with $\rho^{\mu}_{N}$ the generalized valley-spin stiffness \cite{Kallin1984}. However, this behavior is modified as $\omega^{\mu}_0(k)\simeq v_Gk$ for the Goldstone modes of the phases with spontaneously broken symmetries.

At large momentum, $kl_B\gtrsim 1$, only Coulomb interactions are relevant since $C(\mathbf{k})$ decays as $\sim k^{-1}$ while $R(\mathbf{k})$ does it as $\sim e^{-\frac{(kl_B)^2}{2}}$, where $C(\mathbf{k})$, $R(\mathbf{k})$ are the matrices that arise due to the many-body contributions from Coulomb and short-range interactions, respectively (check Appendix \ref{app:analyticalTDHFA} for their precise definition). For $kl_B\gg 1$, the collective modes frequency approach asymptotically the mean-field particle-hole excitation energy $\hbar\omega\simeq \epsilon_{n,\alpha}-\epsilon_{n',\alpha'}$.

The above considerations imply that the modes $N=1,2,3$ are almost unaffected by short-range interactions since in both limits $kl_B\ll 1$, $kl_B\gg 1$ they solely depend on them through the valley-spin part of the mean-field contributions, which only provide a constant shift, and for $kl_B\sim 1$ Coulomb contributions are the dominant. Indeed, it is shown in Appendix \ref{app:technical} that the stiffness coefficients of the $N=1$ modes depend very weakly on $u_{\perp},u_z$ while those of the $N=3$ modes are completely independent of them. Moreover, due to the analog Lamb shift, the $N=2$ modes can be computed explicitly as they correspond to a symmetric combination of the orbital $OO_z=11,1-1$ modes as given by Eqs. (\ref{eq:N2magneto}), (\ref{eq:N2mode}), and it is seen that their dependence on $k$ only involves Coulomb interactions. As a result, we expect the dispersion relation of the $N=1,2,3$ modes to be quite independent of their valley-spin structure and then recover essentially the same results of Refs. \cite{Toke2011,Lambert2013}, where only (intra- and inter-layer) Coulomb interactions are taken into account.

The situation is quite different for the $N=0$ modes, since at $k=0$ they correspond to the orbital-singlet modes which are independent of the Coulomb strength and the only modes that experience many-body corrections coming from short-range interactions. Hence, their behavior at low momentum greatly depends on the valley-spin structure of the modes. Besides, as they present a positive stiffness (see next section), $\omega^{\mu}_{0}(k)$ increases monotonically with $k$. Hence, the orbital-singlet modes at $k=0$ are a) the lowest energy excitations of the system and b) the most sensible modes to the valley-spin structure of interactions. We therefore conclude that the orbital-singlet modes are the most natural candidates to characterize the phase transitions. We note that the orbital-singlet modes are denoted as the ``even'' modes in Ref. \cite{Toke2011}.

\subsection{Collective modes: analysis of the results}\label{subsec:numericalresults}

In this section, we compute the dispersion relation for the collective modes of the different ground states obtained in Sec. \ref{sec:HFMF}, see Eqs. (\ref{eq:Fphase}),(\ref{eq:FLPphase}),(\ref{eq:CAFphase}),(\ref{eq:PLPphase}). For each phase, we discuss the symmetries of the excitations and of the ground state, give the explicit expression for the frequencies of the orbital-singlet modes $\omega^{\mu}_{00}$ in order to study the stability of the corresponding phase and the values of the associated stiffness, $\rho^{\mu}_{0}$, and plot the dispersion relation of the different collective modes, $\omega^{\mu}_{N}(k)$. We refer the reader to Appendix \ref{app:technical} for the specific details on the calculation of the collective modes.

\begin{figure*}[tb!]
\begin{tabular}{@{}cccc@{}}
    \includegraphics[width=0.5\columnwidth]{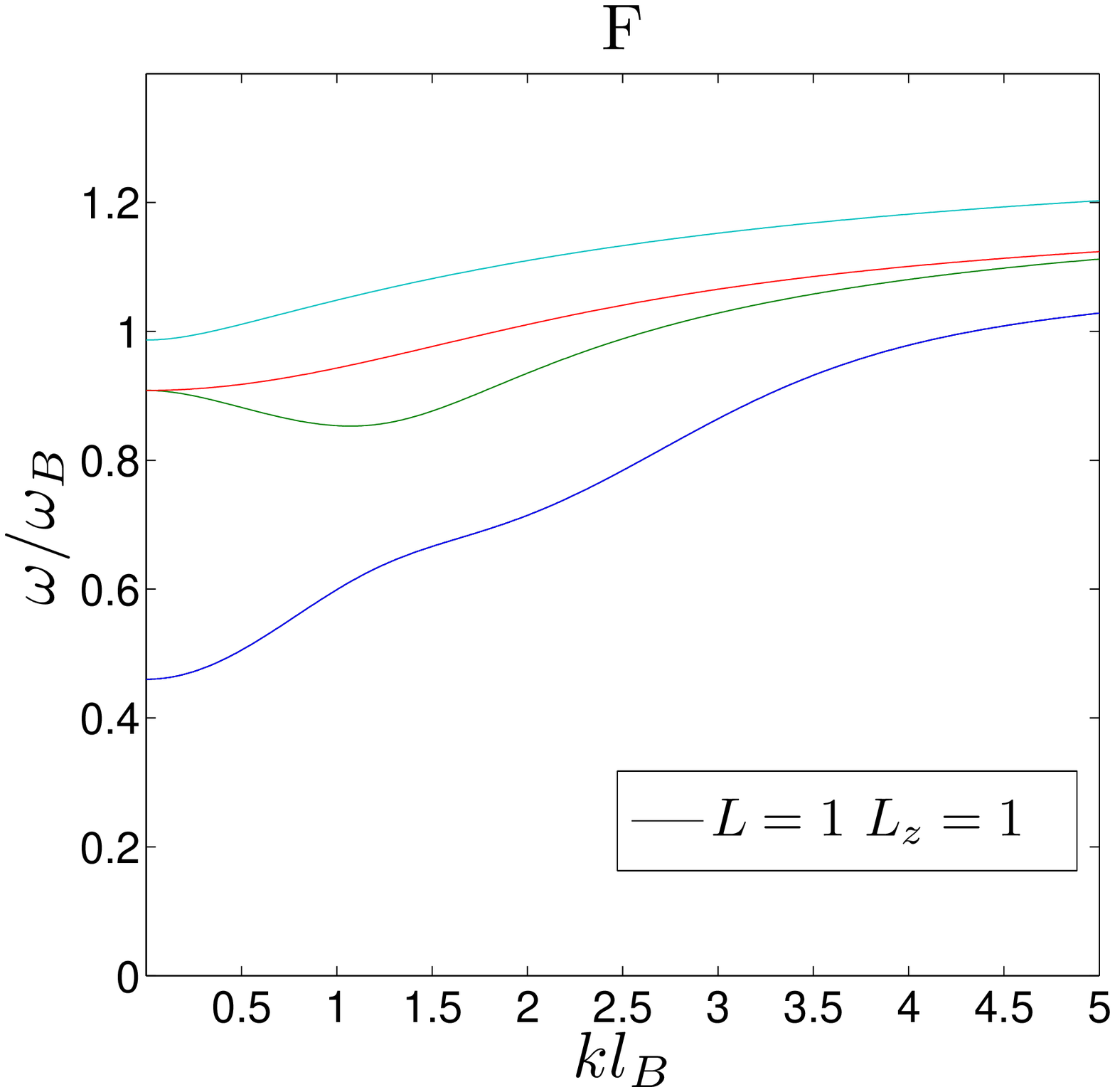} &
    \includegraphics[width=0.5\columnwidth]{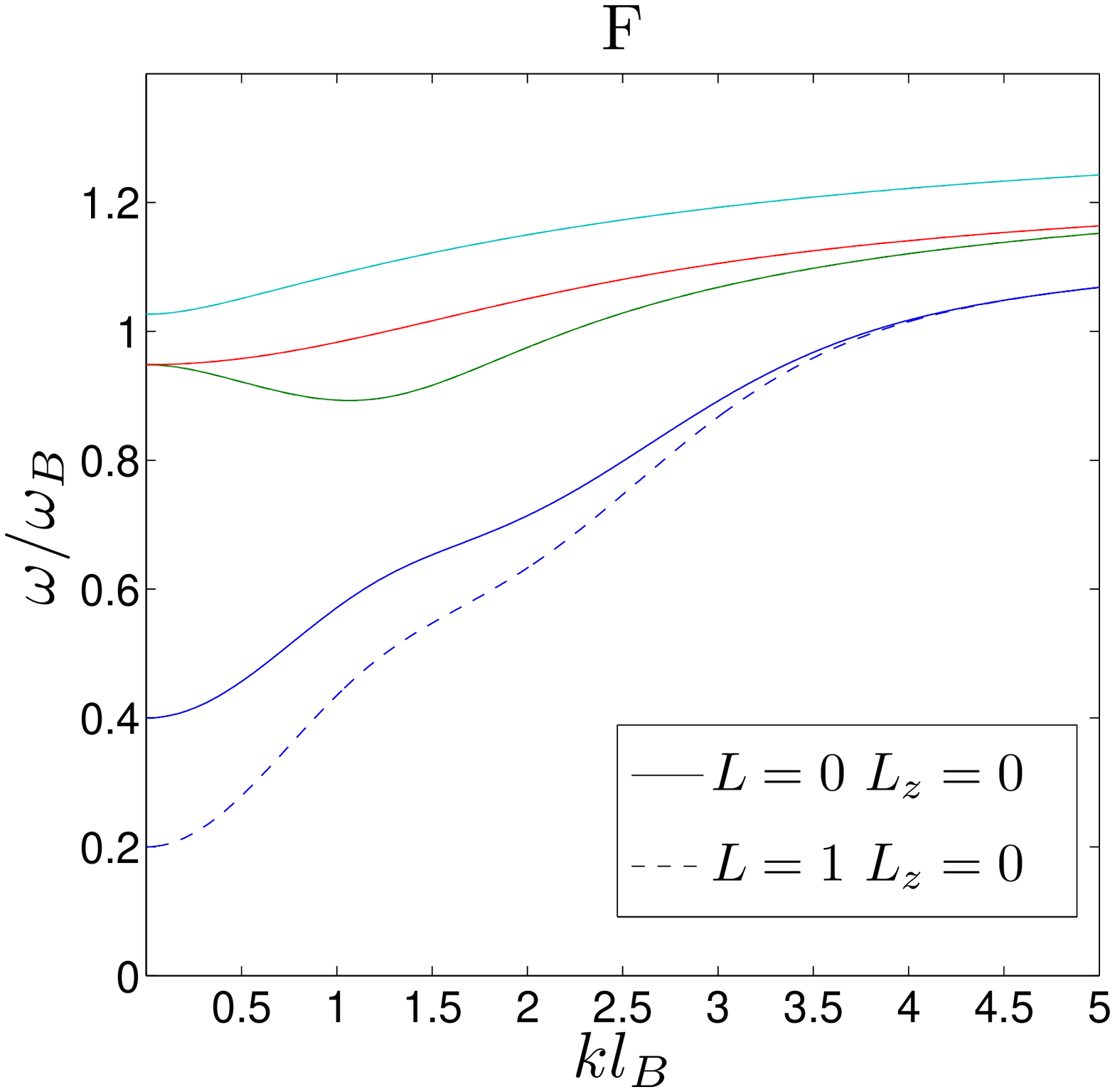} &
    \includegraphics[width=0.5\columnwidth]{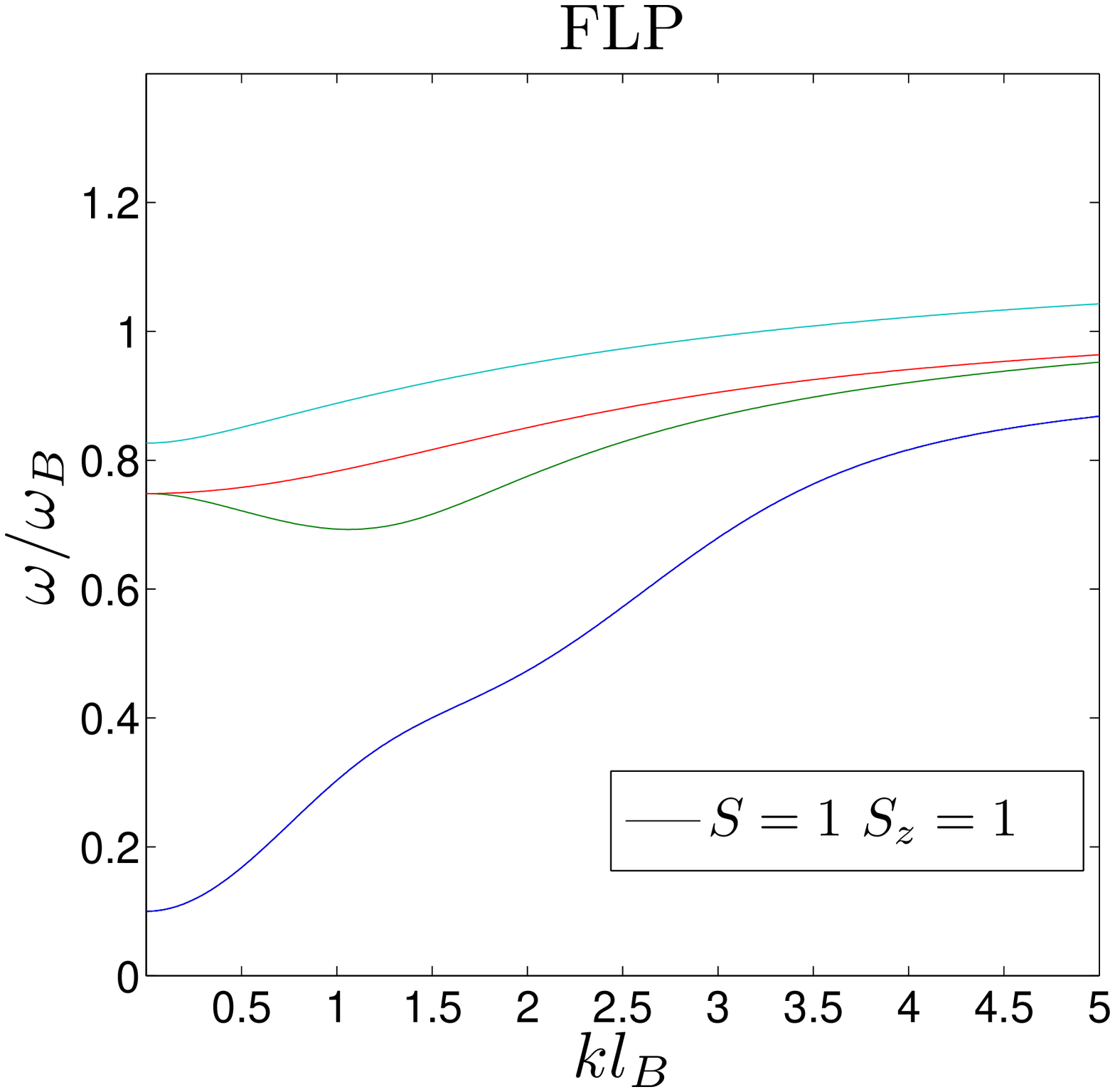} &
    \includegraphics[width=0.5\columnwidth]{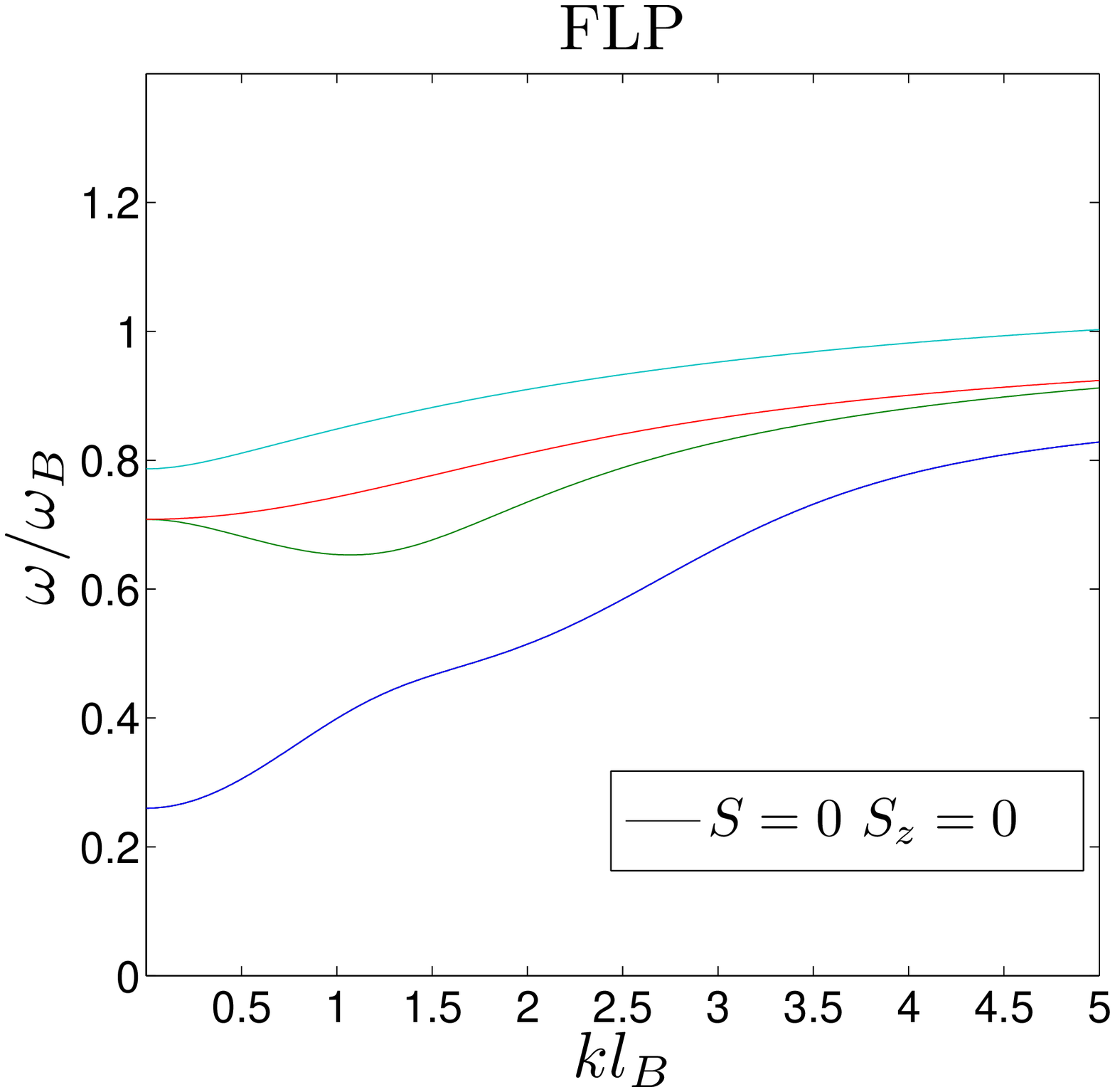} \\
    \includegraphics[width=0.5\columnwidth]{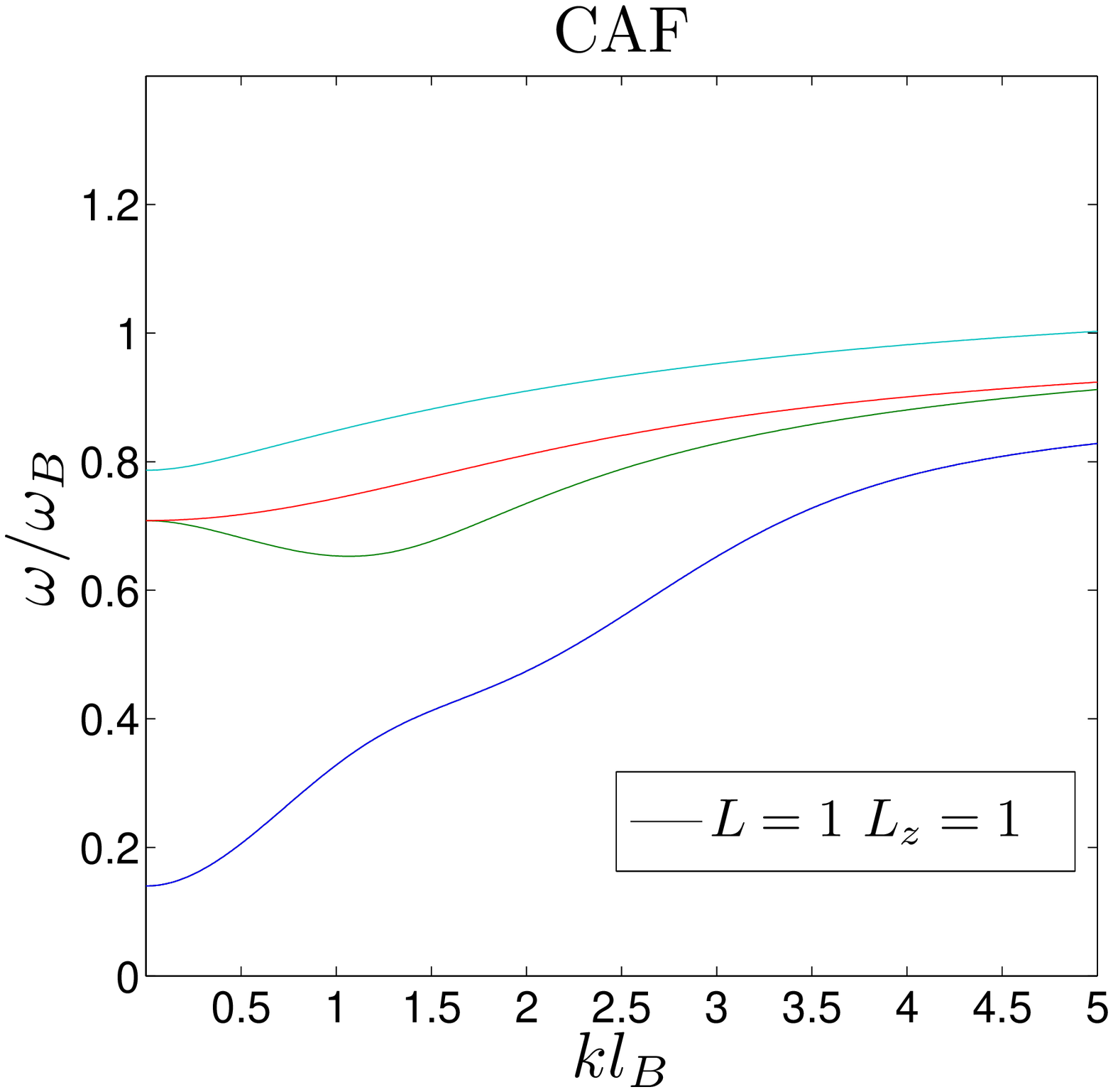} &
    \includegraphics[width=0.5\columnwidth]{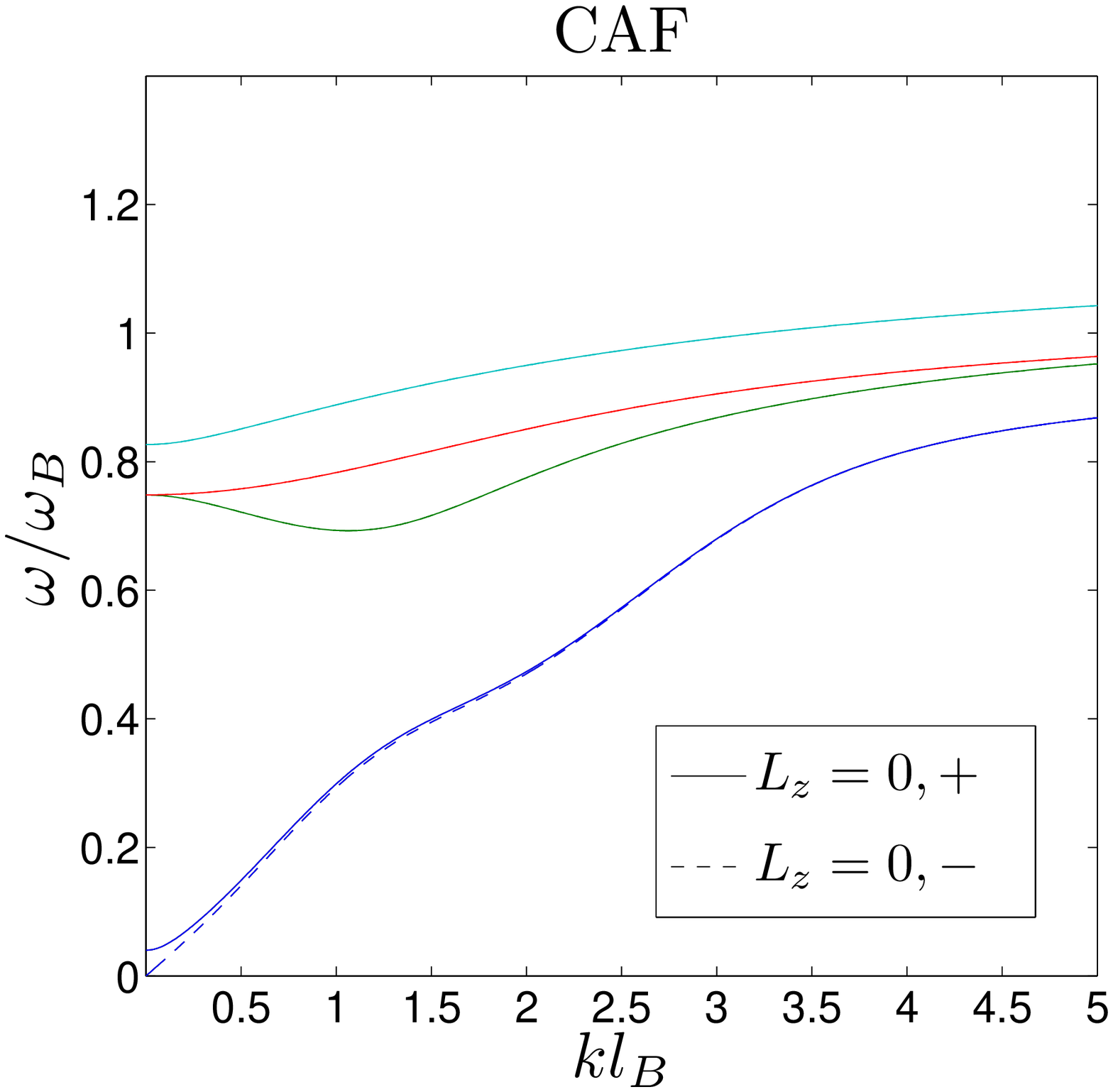} &
    \includegraphics[width=0.5\columnwidth]{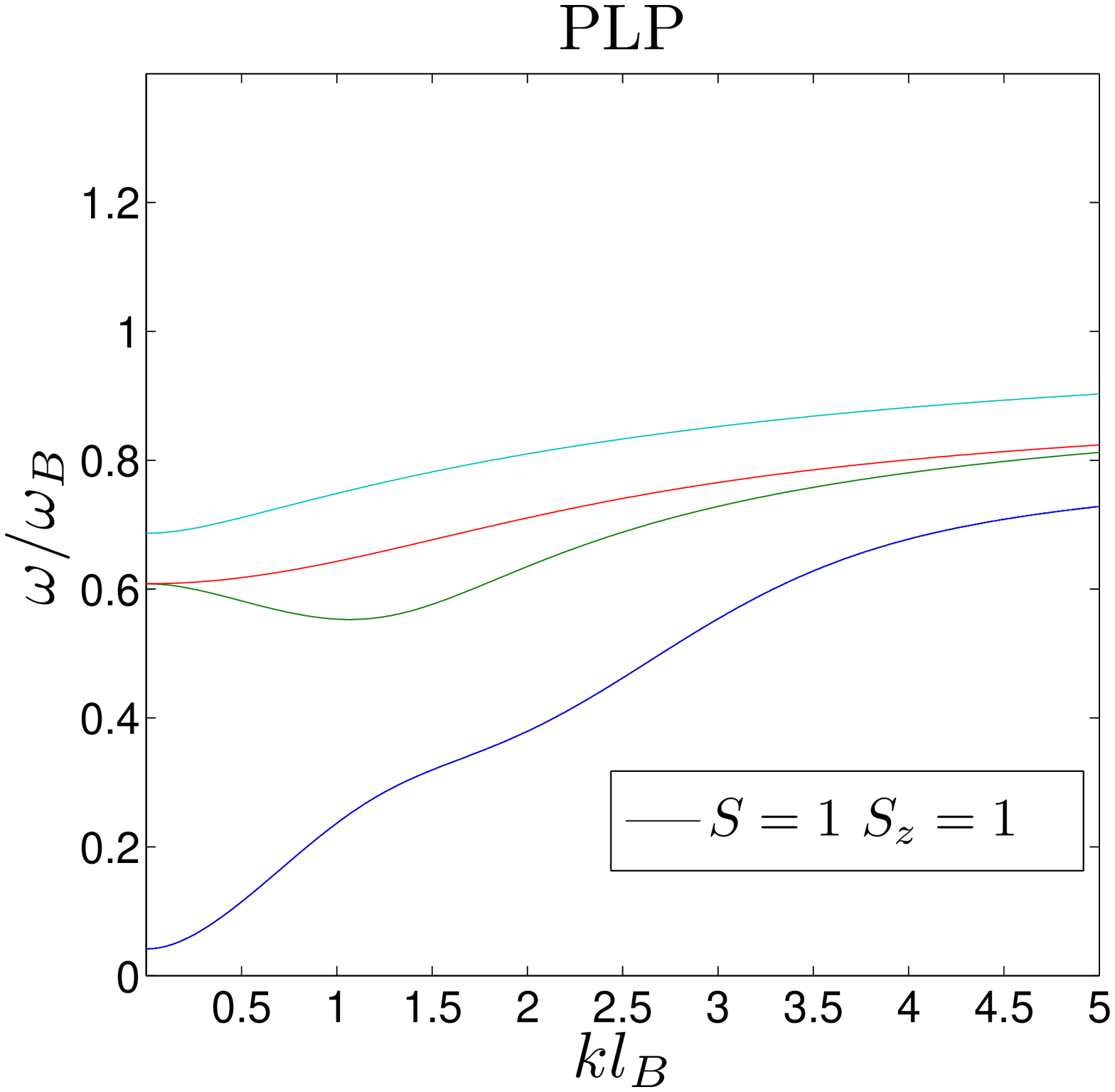} &
    \includegraphics[width=0.5\columnwidth]{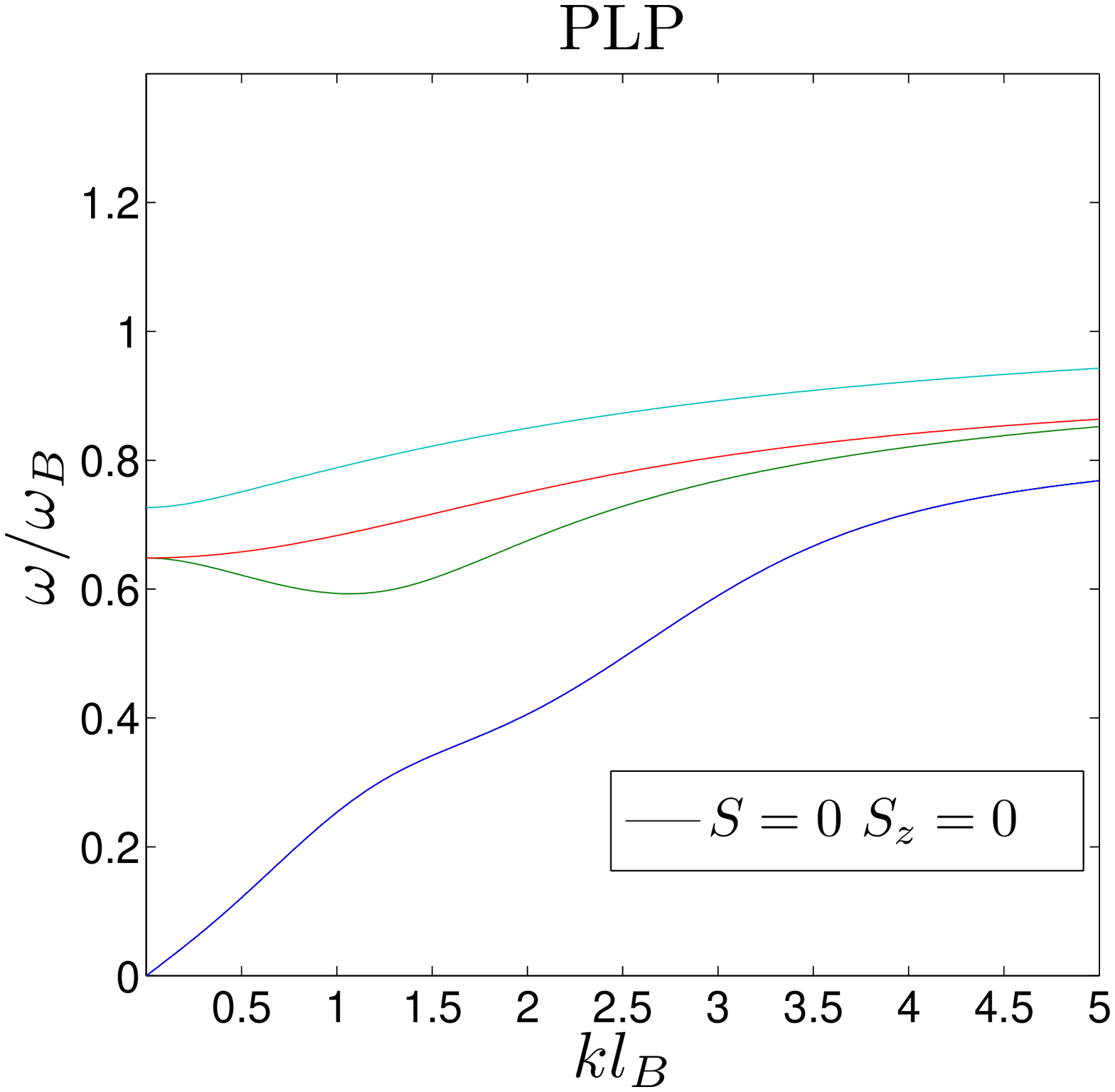} \\
\end{tabular}
\caption{Dispersion relation of the collective modes for the different phases of the $\nu=0$ QH state in bilayer graphene. We take the values $F_C=0.5\hbar\omega_B$, $u_z=0.1 \hbar \omega_B$ and $u_{\perp}=-0.05 \hbar \omega_B$ for the Coulomb and short-range energies, respectively. We remark that, when several branches are represented in the same plot, the modes with $N=1,2,3$ are so close to each other that is difficult to distinguish them. Upper left corner: dispersion relations of the  F phase with $\epsilon_V=0.02 \hbar \omega_B$ and $\epsilon_Z=0.2 \hbar \omega_B$, with the left plot devoted to the valley-triplet mode $\omega^{11}_N(k)$ and the right plot devoted to the modes with $L_z=0$, $\omega^{L0}_N(k)$, $L=0,1$ being depicted in solid (dashed) line. Upper right corner: dispersion relations of the FLP phase with $\epsilon_V=0.2 \hbar \omega_B$ and $\epsilon_Z=0.02 \hbar \omega_B$. The left plot displays the spin-triplet mode $\omega^{11}_N(k)$ and right plot the spin-singlet mode $\omega^{00}_N(k)$. Lower left corner: dispersion relations of the CAF phase with $\epsilon_V=0.02 \hbar \omega_B$ and $\epsilon_Z=0.02 \hbar \omega_B$. Left plot corresponds to the valley-triplet mode $\omega^{11}_N(k)$ and right plot to the modes with $L_z=0$, the $+(-)$ branches being depicted in solid (dashed) line. Lower right corner: dispersion relations of the PLP phase with $\epsilon_V=0.1 \hbar \omega_B$ and $\epsilon_Z=0.02 \hbar \omega_B$. Left plot corresponds to the spin-triplet mode $\omega^{11}(k)$ and right plot the spin-singlet mode $\omega^{00}(k)$. }
\label{fig:DispersionRelations}
\end{figure*}

\subsubsection{Ferromagnetic phase}\label{subsec:Fcol}

The ferromagnetic state has valley-pseudospin $L,L_z=0$ and spin $S,S_z=2N_B$, with $4N_B$ the total number of electrons in the ZLL. All magnetoexcitons carry spin $S=1, S_z=-1$ so we use the valley pseudospin in order to characterize the collective modes, $\omega^{\mu=LL_z}_N(k)$. The excitations of the ferromagnetic state consists of spin-flip excitations and full-flip excitations. The spin-flip excitations involve transitions between two different spin polarizations, keeping the same valley polarization, while in full-flip excitations both valley and spin indices are flipped at the same time. Specifically, the spin-flip excitations are characterized by the modes with $L=1,0$ and $L_z=0$. The energy of the orbital singlet associated to these modes, $\hbar\omega^{L0}_{00}$, is:
\begin{equation}\label{eq:FCAFinstability}
\hbar\omega^{L0}_{00}=2\epsilon_Z+2u_{\perp}-(-1)^L2u_{\perp}
\end{equation}
The spin-singlet mode $\hbar\omega^{00}_{00}=2\epsilon_Z>0$ is the Larmor mode \cite{Murthy2016}, while the spin-triplet mode characterizes the F-CAF phase transition, $\hbar\omega^{10}_{00}=2\epsilon_Z+4u_{\perp}=2(\epsilon_Z-\epsilon_{Zc})$; precisely, $\omega^{10}_{00}=0$ is the boundary [see Eq. (\ref{eq:FCAFborder})] between the two phases. The appearance of a such a gapless mode corresponds to the spontaneously broken $U(1)$ symmetry of the CAF phase. In the region where the CAF phase is present, the system is energetically unstable as $\omega^{10}_{00}<0$.

Regarding full-flip excitations, they are characterized by the triplet modes $L_z=\pm 1$. Indeed, their dispersion relation is the same, solely shifted by the layer voltage, $\omega^{1-1}_N(k)=\omega^{11}_N(k)+4\epsilon_V$, so we only need to compute $\omega^{11}_N(k)$ to characterize them. In particular, for the orbital-singlet mode,
\begin{equation}\label{eq:FFLPinstability}
\hbar\omega^{11}_{00}=2(u_{z}+u_{\perp}+\epsilon_Z-\epsilon_V)
\end{equation}
which is proportional to the difference between the mean-field energies of the FLP and F states. Hence, in analogy to the case of spin-flip excitations, $\omega^{11}_{00}=0$ is the boundary between the two phases.

The existence of such a gapless mode can be understood from the appearance of a mean-field symmetry right at the boundary between the FLP and the F phases. We define a mean-field symmetry as that which is not of the complete Hamiltonian but only of the mean-field HF solutions [see discussion after Eq. (\ref{eq:TDHFAGoldstone}) for more details]. Specifically, when condition (\ref{eq:FFLPborder}) is satisfied, the HF equations present a continuously degenerate mean-field ground state described by the parameters $t,\phi$ and characterized in valley-spin as
\begin{eqnarray}\label{eq:Residualphase}
\chi_{a}&=&\ket{n_z}\otimes\ket{s_z}\\
\nonumber \chi_{b}&=&\cos{t}\ket{n_z}\otimes\ket{-s_z}+e^{i\phi}\sin{t}\ket{-n_z}\otimes\ket{s_z}\\
\nonumber \chi_{c}&=&-e^{-i\phi}\sin{t}\ket{n_z}\otimes\ket{-s_z}+\cos{t}\ket{-n_z}\otimes\ket{s_z}\\
\nonumber \chi_{d}&=&\ket{-n_z}\otimes\ket{-s_z}\\
\nonumber P(t,\phi)&=&\frac{1}{2}\left(I+\sigma_z\cos^2t+\tau_z\sin^2t \right.\\
\nonumber &+& \left.\sin2t\left[\cos\phi(\Pi^x_x+\Pi^y_y)+\sin\phi(\Pi^x_y-\Pi^y_x)\right]\right)
\end{eqnarray}
with energy $E[P(t,\phi)]=E(P_{F})\cos^2t+E(P_{FLP})\sin^2t$, $P_{F,FLP}$ being the projectors of the F, FLP phases, respectively [see Eqs. (\ref{eq:meanfieldenergy}),(\ref{eq:Fphase}) and (\ref{eq:FLPphase})], and the matrices $\Pi^{i}_j$ defined as $\Pi^{i}_j\equiv \frac{1}{2}\tau_i\otimes\sigma_j$. Thus, right at the phase boundary, the total mean-field energy of the state does not depend on the parameters $t, \phi$ and a continuous mean-field symmetry arises from this fact. In particular, the parameter $t$ describes the phase transition, with $t=0$ corresponding to the F state while the FLP state is obtained for $t=\frac {\pi}{2}$.
The associated total state, denoted as $\ket{\Psi(t,\phi)}$, can be connected to the F state $\ket{F}$ by a continuous unitary transformation of the form:
\begin{eqnarray}\label{eq:SO5MF}
\nonumber \ket{\Psi(t,\phi)}&=& e^{\hat{G}(t,\phi)}\ket{F}\\
\hat{G}(t,\phi)&=& \sum_{n,p} t e^{i\phi}\hat{c}_{n,p,c}^{\dagger}\hat{c}_{n,p,b}-h.c.
\end{eqnarray}
and, in terms of the projectors $P$, $P(t,\phi)=e^{G(t,\phi)}P_{F}e^{-G(t,\phi)}$, with $G(t,\phi)$ the matrix version of the operator $\hat{G}(t,\phi)$. Note that the azimuth $\phi$ simply arises from adding a trivial rotation in valley and/or spin space to $\hat{G}(t,0)$. The generator of Eq. (\ref{eq:SO5MF}) can be further rewritten as
\begin{equation}\label{eq:SO5MFGenerator}
\hat{G}(t,\phi)=-it\left[\cos{\phi}\left(\hat{\Pi}^x_y-\hat{\Pi}^y_x\right)-\sin{\phi}\left(\hat{\Pi}^x_x+\hat{\Pi}^y_y\right)\right]
\end{equation}
The operators $\hat{\Pi}^{x}_i,\hat{\Pi}^{y}_i$, along with $\hat{S}_i,\hat{L}_z$, form a representation of $SO(5)$ that is an {\it exact} symmetry of the total Hamiltonian for $u_{\perp}+u_z=0$ and $\epsilon_Z=\epsilon_V=0$ \cite{Wu2014}. Thus, the exotic $SO(5)$ symmetry existing at the F-FLP boundary for zero Zeeman and layer voltage terms survives as a weaker mean-field version at the same phase boundary in the realistic scenario $u_z+u_{\perp}\neq0$. Remarkably, this gapless mode behaves as $\omega^{11}_0(k)\sim k^2$ for low momentum in contrast to the linear dependence found for the Goldstone modes of the phases with spontaneously broken symmetries.

The stiffness for the orbital-singlet modes is given, for dominant Coulomb interactions $F_C\gg u_{\perp},u_z$ [see Eq. (\ref{eq:stiffnessexact}) for the exact expression], by
\begin{equation}\label{eq:stiffnesspro}
\rho^{\mu}_0\simeq\left(\frac{89}{224}F_{00}-\frac{25}{49}u^{\mu}\right)l^2_B
\end{equation}
with $u^{\mu}$ a short-range coupling that depends on the valley-spin symmetry of the mode. Specifically, $u^{00}=-u_z-2u_{\perp}$, $u^{10}=-u_z+2u_{\perp}$ and $u^{11}=u^{1-1}=u_z$.

We plot in the upper left corner of Fig. \ref{fig:DispersionRelations} the dispersion relation of the full-flip (left panel) and spin-flip (right panel) excitations. In the case of spin-flip excitations, the $L=0$, $(L=1)$ branches are plotted in solid (dashed) line. As the corresponding mean-field energies are degenerate and the modes $N=1,2,3$ depend very weakly on the short-range interactions, the curves of both branches are very close to each other; indeed, this degeneracy becomes exact for the $N=2$ modes as explained in the previous section. The characteristic negative stiffness of the $N=1$ modes can also be observed in the plots [see Eq. (\ref{eq:stiffness1leading})]. Only the frequencies $\omega^{00}_{0}(k)$, $\omega^{10}_{0}(k)$ are clearly distinguished due to the many-body corrections arising from short-range interactions at low momentum.

When comparing the plots of spin-flip and full-flip excitations, we see that the curves for the $N=1,2,3$ are extremely similar. Again, this can be explained by invoking the dominant character of Coulomb interactions and the weak dependence on short-range effects. As expected, the main differences are observed once more for the lowest energy modes $N=0$. Variations of the in-plane magnetic field or the interlayer voltage only affect through a linear energy offset in $\epsilon_Z,\epsilon_V$ that depends on the valley-spin structure of each mode, as given by Eqs. (\ref{eq:FCAFinstability}), (\ref{eq:FFLPinstability}). Finally, we remark that variations of the different parameters only change quantitatively but not qualitatively the plots.

\subsubsection{Full layer-polarized phase}\label{subsec:FLPcol}

As this phase is the analog of the F phase in valley space, the results for the FLP phase are similar but interchanging $L,L_z$ and $S,S_z$. Specifically, the FLP state has spin $S,S_z=0$ and valley-pseudospin $L,L_z=2N_B$. All magnetoexcitons have valley pseudospin $L=1, L_z=-1$, so we use their spin to classify them, $\omega^{\mu=SS_z}_N(k)$. Since the interaction is spin-independent, the triplet modes present the same dispersion relation, only shifted by the Zeeman energy, $\omega^{1\pm1}_N(k)=\omega^{10}_N(k)\mp2\epsilon_Z$, hence we only need to compute $\omega^{11}_N(k)$ to characterize them.

The excitations of the FLP state correspond to valley-flip and full-flip excitations. Valley-flip excitations are characterized by the modes with $S=1,0$ and $S_z=0$. In particular, the energy of the orbital-spin-singlet mode is
\begin{equation}\label{eq:FLPPLPinstability}
\hbar\omega^{00}_{00}=2u_{\perp}-2u_{z}+2\epsilon_V=2(\epsilon_V-\epsilon_{Vc})
\end{equation}
In analogy to the case of the F-CAF transition [see Eq. (\ref{eq:FCAFinstability}) and related discussion], $\omega^{00}_{00}=0$ is the boundary between the FLP and PLP phases; the gapless character of this mode arises from the broken $U(1)$ symmetry of the PLP phase.

Regarding the F-FLP boundary, we find that it is characterized by the mode
\begin{equation}
\hbar\omega^{11}_{00}=2(\epsilon_V-\epsilon_Z-u_{\perp}-u_{z})
\end{equation}
which is just $-\hbar\omega^{S=1S_z=1}_{00}$ for the F state, see Eq. (\ref{eq:FFLPinstability}). Reasoning as before, the appearance of this gapless mode at the boundary between the two phases results from the residual mean-field symmetry shown in Eq. (\ref{eq:Residualphase}).

It is worth noting the subtle difference between the F-CAF and FLP-PLP transitions: while the former is governed by a valley {\it triplet} mode, the latter is governed by a spin {\it singlet} mode. This is due to the fact that, in the CAF phase, the occupied states $a,b$ correspond to two different spins $s_a,s_b$ for the two valley polarizations $\pm n_z$, one being the other rotated $\pi$ in the $x-y$ spin plane. In contrast, in the PLP phase, all the occupied states present the same valley polarization, specified by the vector $\mathbf{n}$, independently of their spin. This relation can be observed in the $P$ matrices: the F-CAF transition is characterized by the appearance of an operator that goes as $\sim\sigma_i\tau_z$, which has valley pseudospin $L=1,L_z=0$, while the FLP-PLP transition only varies the orientation of the projector in valley space, which has zero spin $S,S_z=0$. On the other hand, the F-FLP transition is governed by triplet modes in both spin and valley spaces. This fact can be understood from the expression of the generator $\hat{G}(t,\phi)$ of the mean-field symmetry, Eq. (\ref{eq:SO5MFGenerator}), as it has $L=1,S=1$.

The stiffness coefficients of the orbital-singlet modes are also given by Eq. (\ref{eq:stiffnesspro}) but now $u^{00}=u_z+4u_{\perp}$ and $u^{11}=u_z$, noting that $\rho^{1-1}_N=\rho^{10}_N=\rho^{11}_N$.

In the upper right corner of Fig. \ref{fig:DispersionRelations}, we represent the dispersion relation of the spin triplet (left panel) and singlet (right panel) excitations. The qualitative trends and properties of the curves are similar to those of the F state.

\subsubsection{Canted anti-ferromagnetic phase}\label{subsec:CAFcol}

\begin{figure*}[tb!]
\begin{tabular}{@{}cc@{}}
    \includegraphics[width=\columnwidth]{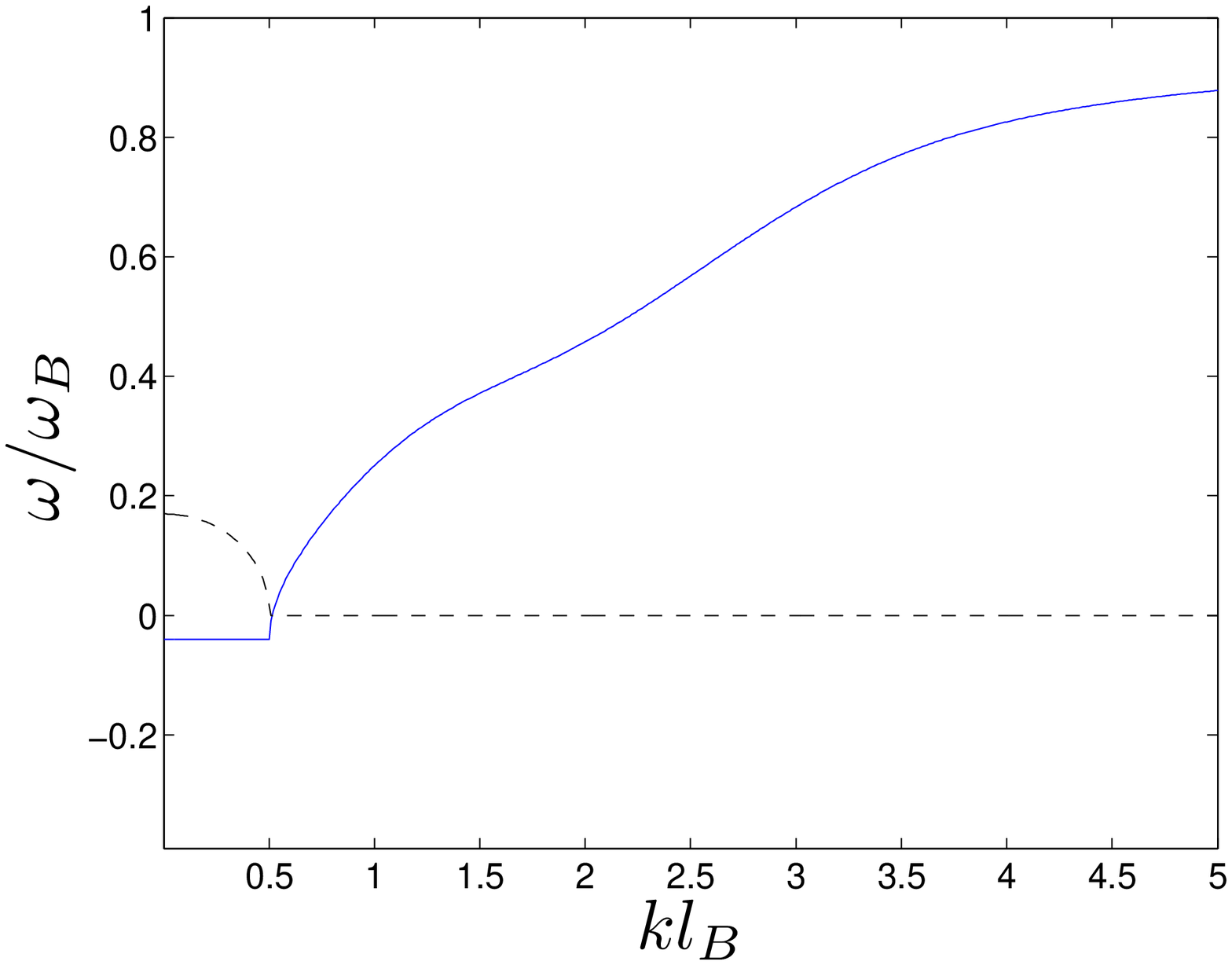} &
    \includegraphics[width=\columnwidth]{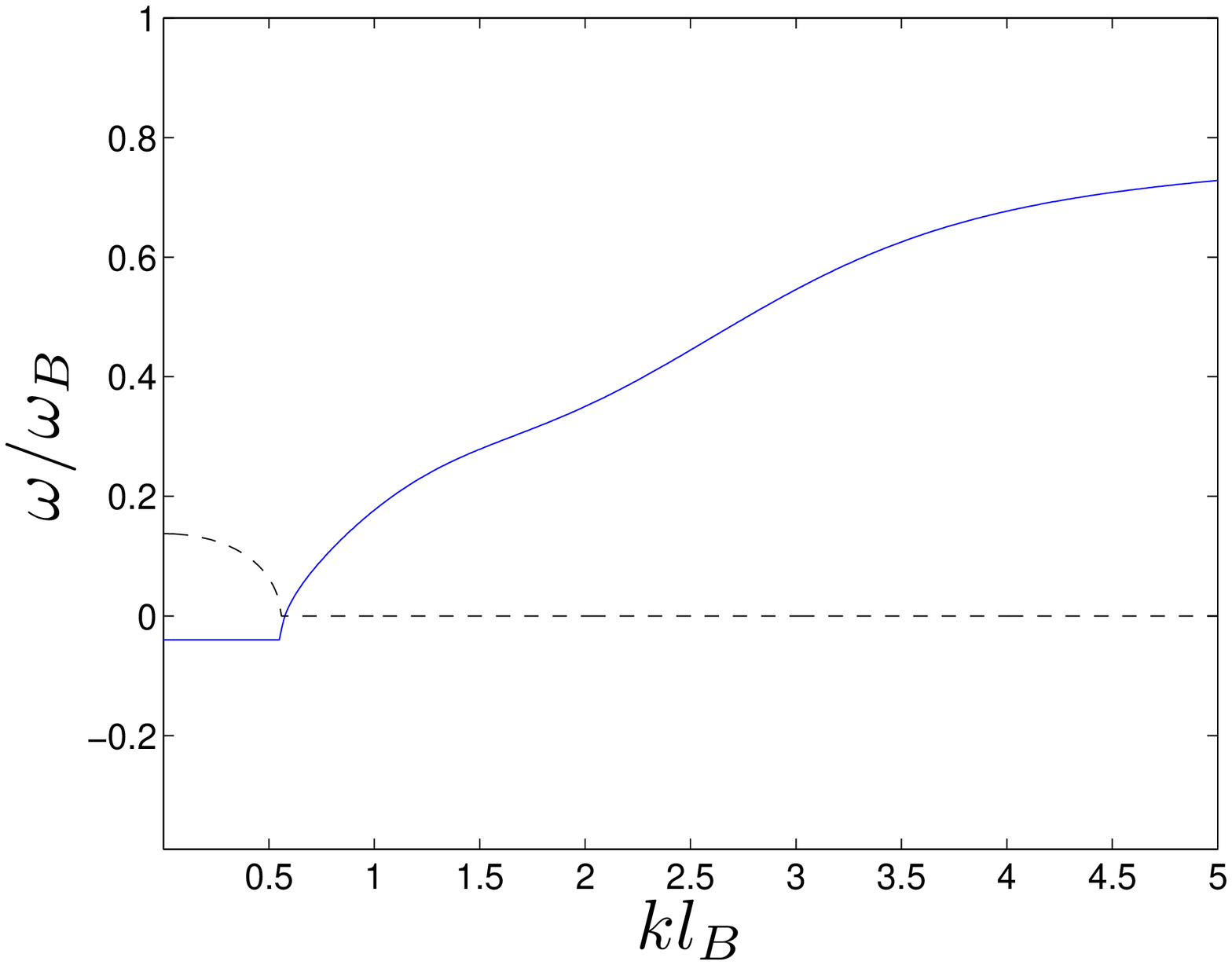} \\
\end{tabular}
\caption{Dispersion relation for complex-frequency modes, where the real (imaginary) part of the complex-frequency mode is plotted in solid (dashed) line. Left panel: plot of $\omega^{11}(k)$ for a CAF state with parameters $F_C=0.5\hbar\omega_B$, $u_z=0.05 \hbar \omega_B$, $u_{\perp}=-0.1 \hbar \omega_B$, $\epsilon_V=0.02 \hbar \omega_B$ and $\epsilon_Z=0.02 \hbar \omega_B$. Right panel: plot of $\omega^{11}(k)$ for a PLP state with parameters $F_C=0.5\hbar\omega_B$, $u_z=0.1 \hbar \omega_B$, $u_{\perp}=-0.05 \hbar \omega_B$, $\epsilon_V=0.02 \hbar \omega_B$ and $\epsilon_Z=0.02 \hbar \omega_B$.}
\label{fig:ComplexMode}
\end{figure*}

The CAF state has only well defined value of the $z$ valley pseudospin, $L_z$, so we use this quantum number to characterize the magnetoexcitons. Due to this lack of symmetry, dynamical instabilities can appear as shown in Appendix \ref{app:analyticalTDHFA}. The sector $L_z=0$ is characterized by two branches, labeled as $\omega_N^{\pm}(k)$, that correspond to generalizations of the spin-flip modes of the F phase; they match the $L=0,1$ modes for $\theta_S=0$, respectively. The energy for their orbital-singlet modes is:

\begin{equation}\label{eq:CAFspmodessimplified}
\hbar\omega^{\pm}_{00}=|2u_{\perp}|\sqrt{(1\pm\cos^2 \theta_S)^2 - \sin^4 \theta_S}
\end{equation}
In analogy to the F state, the previous equation gives the Larmor mode $\hbar\omega^{+}_{00}=|4u_{\perp}|\cos \theta_S=2\epsilon_Z>0$ and the Goldstone mode $\hbar\omega^{-}_{00}=0$. The appearance of this Goldstone mode is expected because the CAF phase spontaneously breaks the $U(1)$ symmetry of the Hamiltonian (see discussion in Appendix \ref{app:TDHFA} about Goldstone modes in the TDHFA for more details). The velocity of the Goldstone mode, given by $\omega^{-}_0(k)\simeq v_Gk$ in the limit $k\rightarrow 0$, can be computed exactly and is given by Eq. (\ref{eq:Goldstonevelocity}). In particular, in the regime $F_C\gg u_{\perp},u_z$, it is well approximated by
\begin{equation}\label{eq:Goldstonevelocitypro}
\frac{\hbar v_G}{l_B}\simeq\sqrt{2A\left[\frac{89}{224}F_{00}+\frac{25}{49}\Delta\right]},
\end{equation}
with $A=\epsilon_{Zc}\sin^2\theta_S$ and $\Delta=u_z-2u_{\perp}$. Near the phase transition, the velocity behaves as $v_G\sim\sqrt{A}\sim\sqrt{\epsilon_{Zc}-\epsilon_Z}$, in agreement with the critical behavior described in Ref. \cite{Murthy2016}. Thus, the F-CAF transition is characterized in the CAF side by a vanishing group velocity instead of a closing energy gap. There are no complex-frequency modes in the $L_z=0$ sector.

The remaining collective modes correspond to the sectors with $L_z=\pm 1$, with frequencies $\omega^{1\pm1}_N(k)$. In analogy to the F phase, they satisfy $\omega^{1-1}_N(k)=\omega^{11}_N(k)+4\epsilon_V$ so we just compute $\omega^{11}_N(k)$. We find that
\begin{equation}\label{eq:CAFpszeromode}
\hbar\omega^{11}_{00}=-2\epsilon_V + \sqrt{(2u_z-2u_{\perp}\cos^2 \theta_S)^2-4u^2_{\perp}\sin^4 \theta_S}
\end{equation}
Interestingly, there are situations in which the quantity in the square root is negative and then a dynamical instability appears. It is straightforward to show that $\omega^{11}_{00}$ is purely real whenever
\begin{equation}\label{eq:CAFdynin}
u_{\perp}+u_{z}>\frac{\epsilon^2_Z}{2u_{\perp}}
\end{equation}
is fulfilled. On the other hand, $\omega^{11}_{00}=0$ at the boundary between the PLP and CAF phases, i.e, whenever Eq. (\ref{eq:PLPCAFborder}) is satisfied. In analogy to the F-FLP transition, this gapless mode can be understood in terms of a mean-field symmetry arising right at the phase boundary as the state $\ket{\Psi(t,\phi_S,\phi_V)}$, characterized by the projector

\begin{eqnarray}\label{eq:SO5MFCAF}
P(t,\phi_S,\phi_V)&=&\frac{1}{2}\left[I
+\sigma_z s_z(t)
+\mathbf{\tau} \cdot \mathbf{n}(t) \right.
\\ \nonumber
&+& \frac{s_z(t)}{\eta(t)}
(\mathbf{s}_{\parallel}\cdot\mathbf{\sigma})
(\cos \gamma(t) \tau_z -\sin \gamma(t) \mathbf{n}_{\parallel}\cdot\mathbf{\tau})
\\ \nonumber
&-&\left.\eta(t) n(t) (\mathbf{n}_{\perp}\cdot\mathbf{\tau})(\mathbf{s}_{\perp}\cdot\mathbf{\sigma})
\right]
~,
\end{eqnarray}
with
\begin{eqnarray}
\nonumber \mathbf{n}_{\parallel}&=&[\cos\phi_V,\sin\phi_V,0],~\mathbf{s}_{\parallel}=[\cos\phi_S,\sin\phi_S,0]\\
\nonumber \mathbf{n}_{\perp}&=&\hat{\mathbf{z}} \times \mathbf{n}_{\parallel},~ \mathbf{s}_{\perp}=\hat{\mathbf{z}} \times \mathbf{s}_{\parallel}
\\ \nonumber
s_z(t)  &=& \cos^2t \cos \theta_S
\\ \nonumber
\mathbf{n}(t) &=& n(t) (\cos \gamma(t) \mathbf{n}_{\parallel} + \sin \gamma(t) \hat{\mathbf{z}})
\\
n(t) &=& \sin t \sqrt{1-\cos^2 t \cos^2 \theta_S}
\\ \nonumber
n(t) \sin \gamma(t) &=& \frac{1}{2} \sin 2t \cos \theta_V
\\ \nonumber
n(t) \eta(t) &=& \frac{1}{2} \sin 2t \cos \theta_S~,
\end{eqnarray}
has an energy $E[P(t,\phi_S,\phi_V)]=E(P_{CAF})\cos^2t+E(P_{PLP})\sin^2t$, $P_{CAF,PLP}$ being the projector of a CAF, PLP state characterized  by the angles $\theta_{S,V}$ and $\phi_{S,V}$. Therefore, as in the F-FLP case, the complete manifold of states is degenerate at the CAF-PLP boundary, giving rise to a continuous mean-field symmetry, with the parameter $t$ describing again the phase transition: the value $t=0$ yields the CAF phase while $t=\frac{\pi}{2}$ gives the corresponding PLP state at the other side of the phase transition. For simplicity, we do not give the spinors $\chi_{\alpha}(t,\phi_S,\phi_V)$ as their expression is quite cumbersome and unimportant for the present discussion. 

A summary of the stability conditions for the CAF phase is given by
\begin{widetext}
\begin{eqnarray}\label{eq:CAFinstabilitycondition}
\nonumber u_{\perp}+u_{z}&>&\frac{\epsilon^2_Z}{2u_{\perp}}+\frac{\epsilon^2_V}{u_z-u_{\perp}}>\frac{\epsilon^2_Z}{2u_{\perp}},~\omega^{11}_{00}>0\\
\frac{\epsilon^2_Z}{2u_{\perp}}+\frac{\epsilon^2_V}{u_z-u_{\perp}}&>&u_{\perp}+u_{z}>\frac{\epsilon^2_Z}{2u_{\perp}} ,~\omega^{11}_{00}<0\\
\nonumber \frac{\epsilon^2_Z}{2u_{\perp}}+\frac{\epsilon^2_V}{u_z-u_{\perp}}&>&\frac{\epsilon^2_Z}{2u_{\perp}}>u_{\perp}+u_{z} ,~\text{Im}~\omega^{11}_{00}\neq0
\end{eqnarray}
\end{widetext}
Note that $\omega^{11}_{00}>0$ as long as we stay in the region where the ground state is the CAF phase.

With respect to the stiffness coefficients, we find in the regime $F_C\gg u_{\perp},u_z$ that [see Eq. (\ref{eq:stiffnessanomalous}) for the exact expression]
\begin{equation}\label{eq:stiffnessproCAF}
\rho^{\mu}_0\simeq\frac{(u^{\mu}+\Delta)\left(\frac{89}{224}F_{00}-\frac{25}{49}u^{\mu}\right)+\frac{25}{49}A^2}{\sqrt{(u^{\mu}+\Delta)^2-A^2}}l^2_B
\end{equation}
with $u^{+}=-u_z-2u_{\perp}\cos^2 \theta_S$ and $u^{11}=u_z+2u_{\perp}\sin^2 \theta_S$. We note that the Goldstone mode has no stiffness coefficient due to its linear behavior and that $\rho^{1-1}_N=\rho^{11}_N$. Remarkably, the denominator of the above equation takes the value of the Larmor frequency for the $+$ mode and the resulting stiffness $\rho^{+}_0$ becomes relatively large due to the small value of the Zeeman energy.

In the lower left corner of Fig. \ref{fig:DispersionRelations} we plot the dispersion relation of the modes with $L_z=1$ (left panel) and $L_z=0$ (right panel). We see that a mode with zero energy that grows linearly with $k$ appears in the sector with $L_z=0$, corresponding to the Goldstone mode of the CAF state. Apart from this consideration, the qualitative form of the curves is similar to the previous cases. We note that, while a variation of the interlayer voltage only provides a trivial energy shift for the $L_z=\pm 1$ modes, the role of the in-plane magnetic field is much more critical as it determines the background mean-field state by tuning the canting angle $\theta_S$ through the Zeeman energy $\epsilon_Z$.

In left Fig. \ref{fig:ComplexMode} we consider an experimentally unrealistic case with different values of the coupling constants, $-u_{\perp}>u_z>0$, in order to study the trends of the appearance of a complex-frequency mode in the CAF phase. The plot shows the real (solid curve) and the imaginary (dashed curve) parts of the unstable mode. The decrease (and eventual vanishing) of the imaginary part of the frequency is a consequence of the exponential decay of the many-body contributions from the short-range interactions, also giving rise to the term responsible for the appearance of the instability [see Appendix \ref{app:analyticalTDHFA} for the details]. It is straightforward to show that the real part of the dynamical instability satisfies $\hbar\text{Re}[\omega^{11}_0(k)]=-2\epsilon_V$, as can be observed in the plot.

\subsubsection{Partially layer-polarized phase}\label{subsec:PLPcol}

\begin{figure}[tb!]
\includegraphics[width=\columnwidth]{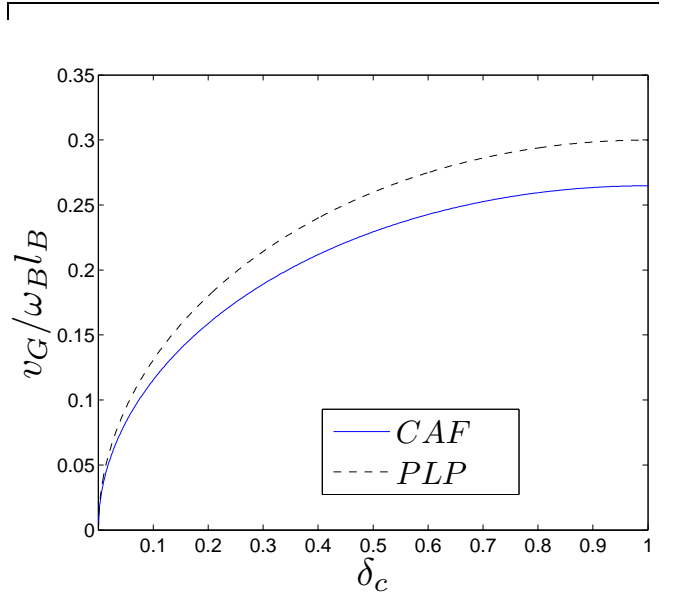}
\caption{Plot of the velocity of the Goldstone modes as a function of the parameter $\delta_c$ for the CAF phase (solid blue line) and for the PLP phase (dashed black line), with the rest of parameters keeping the same values as in Fig. \ref{fig:DispersionRelations}.}
\label{fig:Goldstone}
\end{figure}

In contrast with the CAF state, the PLP state has well defined quantum numbers in spin and valley-pseudospin. Indeed, the valley spin structure of the PLP phases is analog to that of the FLP phase but replacing $L_z$ by the component of the valley pseudospin along the direction of the vector $\mathbf{n}$, $L_n$. The excitations are also characterized by their spin, $\omega^{SS_z}_{N}(k)$, and the triplet modes satisfy $\omega^{1\pm1}_N(k)=\omega^{10}_N(k)\mp2\epsilon_Z$. However, the Hamiltonian does not commute with the operator $\hat{L}_n$ and because of this, dynamical instabilities can appear.

For the spin singlet, the orbital-singlet mode is gapless,
\begin{equation}
\omega^{00}_{00}=0~,
\end{equation}
as it is the Goldstone mode associated to the spontaneously broken $U(1)$ symmetry. Its velocity is computed in similar terms to that of the CAF phase, Eq. (\ref{eq:Goldstonevelocitypro}), but now $A=\epsilon_{Vc}\sin^2\theta_V$ and $\Delta=-u_z-4u_{\perp}$, presenting a similar critical behavior near the phase transition, $v_G\sim \sqrt{A} \sim \sqrt{\epsilon_{Vc}-\epsilon_V}$.

Considering the spin-triplet modes, the orbital-singlet frequency is
\begin{equation}\label{eq:PLPpszeromode}
\hbar\omega^{11}_{00}=-2\epsilon_Z+\sqrt{(4u_{\perp}+A)^2-A^2}
\end{equation}
It can be proven that $\omega^{11}_{00}$ is real if:
\begin{equation}\label{eq:PLPdynin}
u_{\perp}+u_{z}<\frac{\epsilon^2_V}{u_z-u_{\perp}}
\end{equation}
A zero-frequency mode $\hbar\omega^{11}_{00}=0$ appears at the boundary between the PLP and CAF phases, arising from the mean-field symmetry already discussed for the CAF phase [see Eq. (\ref{eq:SO5MFCAF}) and related]. Hence, in analogy to the F-FLP case, the CAF-PLP transition is also governed by triplet modes. Putting all together as for the CAF state:
\begin{widetext}
\begin{eqnarray}\label{eq:PLPinstabilitycondition}
\nonumber u_{\perp}+u_{z}&<&\frac{\epsilon^2_Z}{2u_{\perp}}+\frac{\epsilon^2_V}{u_z-u_{\perp}}<\frac{\epsilon^2_V}{u_z-u_{\perp}},~\omega^{11}_{00}>0\\
\frac{\epsilon^2_Z}{2u_{\perp}}+\frac{\epsilon^2_V}{u_z-u_{\perp}}&<&u_{\perp}+u_{z}<\frac{\epsilon^2_V}{u_z-u_{\perp}} ,~\omega^{11}_{00}<0\\
\nonumber \frac{\epsilon^2_Z}{2u_{\perp}}+\frac{\epsilon^2_V}{u_z-u_{\perp}}&<&\frac{\epsilon^2_V}{u_z-u_{\perp}}<u_{\perp}+u_{z} ,~\text{Im}~\omega^{11}_{00}\neq0
\end{eqnarray}
\end{widetext}
In contrast to the CAF case, for the appearance of dynamical instabilities in the PLP state it is only required a sufficiently low value of the voltage; nevertheless, this instability occurs in a region of parameter space where the CAF phase is the ground state.

For the stiffness coefficients of the spin-triplet modes, Eq. (\ref{eq:stiffnessproCAF}) is still valid for sufficiently dominant Coulomb interactions, using the value $u^{11}=u_z-A$.

In the lower right corner of Fig. \ref{fig:DispersionRelations} we plot the dispersion relation of the spin triplet (left panel) and singlet (right panel) excitations. The trends are similar to the previous cases. In particular, in analogy with the CAF case, variations in the in-plane magnetic field only vary the trivial energy shift for the $S_z=\pm 1$ modes while the influence of the interlayer voltage is much more important since it controls the value of $\theta_V$.

In right Fig. \ref{fig:ComplexMode} we study a case with sufficiently low value of the voltage so that the PLP state is dynamically unstable and hence, the triplet modes have complex frequency. The qualitative behavior resembles that of the unstable mode appearing for a CAF state, satisfying a similar relation for the real part, $\hbar\text{Re}[\omega^{11}_0(k)]=-2\epsilon_Z$.

Finally, in Fig. \ref{fig:Goldstone}, we represent the velocity of the Goldstone modes for both the CAF and PLP phases as a function of the critical parameter $\delta_c$, defined as $\delta_c\equiv(\epsilon_{Zc,Vc}-\epsilon_{Z,V})/\epsilon_{Zc,Vc}$ for the CAF and PLP phases, respectively. For low in-plane magnetic field or interlayer voltage, the velocities saturate the limit $\theta_{S,V}\rightarrow \frac{\pi}{2}$. The agreement of the approximation (\ref{eq:Goldstonevelocitypro}) for dominant Coulomb interactions with the exact result of Eq. (\ref{eq:Goldstonevelocity}) is excellent (not shown). The corresponding plot for the energy gaps of the F and FLP phases is trivial as they are proportional to $-\delta_c$, see Eqs. (\ref{eq:FCAFinstability}), (\ref{eq:FLPPLPinstability}).

\section{Monolayer graphene}\label{sec:MLG}

All of the previous work can be straightforwardly adapted to the $\nu=0$ QH state of monolayer graphene. We briefly revisit the content of Secs. \ref{sec:basicmodel}-\ref{sec:colmod}, adjusting the results to the present case.

\subsection{Effective Hamiltonian}

We start by writing the effective Hamiltonian of monolayer graphene (see Ref. \cite{Kharitonov2012} for a more complete review). The corresponding field operator is given by:
\begin{eqnarray}\label{eq:MLGfieldoperator}
\hat{\psi}(\mathbf{x})&=&\left[\begin{array}{c}
\hat{\psi}_{+}(\mathbf{x})\\
\hat{\psi}_{-}(\mathbf{x})
\end{array}\right]\\
\nonumber \hat{\psi}_{\xi}(\mathbf{x})&=&\left[\begin{array}{c}
\hat{\psi}_{KA\xi}(\mathbf{x})\\
\hat{\psi}_{KB\xi}(\mathbf{x})\\ \hat{\psi}_{K'B\xi}(\mathbf{x})\\\hat{\psi}_{K'A\xi}(\mathbf{x})
\end{array}\right]\equiv\left[\begin{array}{c}
\hat{\psi}_{K\bar{A}\xi}(\mathbf{x})\\
\hat{\psi}_{K\bar{B}\xi}(\mathbf{x})\\ \hat{\psi}_{K'\bar{A}\xi}(\mathbf{x})\\\hat{\psi}_{K'\bar{B}\xi}(\mathbf{x})
\end{array}\right],~\xi=\pm
\end{eqnarray}
where now $A,B$ are the two sublattices of the same graphene layer. Once more, the two sublattices are interchanged in the $K'$ valley so we denote the corresponding subspace as $\bar{A}\bar{B}$. The Hamiltonian is decomposed in the same fashion as in the bilayer case, $\hat{H}=\hat{H}_0+\hat{H}_C+\hat{H}_{sr}$, but the single-particle Hamiltonian $\hat{H}_0$ now reads
\begin{equation}\label{eq:MLGspBHamiltonian}
\hat{H}_{0}=\int\mathrm{d}^2\mathbf{x}~\hat{\psi}^{\dagger}(\mathbf{x})\left[H_B-\epsilon_Z\sigma_z\right]\hat{\psi}(\mathbf{x})
\end{equation}
(note that there is no layer voltage term). After an appropriated phase transformation, the matrix $H_B$ can be written as:
\begin{eqnarray}\label{eq:MLGdestructionHO}
H_B&=&\hbar\omega_B\left[\begin{array}{cc}
0 & a_B\\
a_B^{\dagger} & 0
\end{array}\right]\\
\nonumber \omega_B&=&\frac{\sqrt{2}v_F}{l_B}\simeq 5.51\times 10^{13}\sqrt{B[\text{T}]}~\text{Hz}
\end{eqnarray}
with $v_F\simeq 10^6\text{m/s}$ the Fermi velocity. The corresponding eigenfunctions and eigenvalues are similar to those of the bilayer, $\Psi^{0}_{n,k,\alpha}(\mathbf{x})=\Psi^{0}_{n,k}(\mathbf{x})\chi_{\alpha}$, with
\begin{eqnarray}\label{eq:MLGLandaueigenfunctions}
\Psi^{0}_{n,k}(\mathbf{x})&=&\frac{e^{iky}}{\sqrt{L_y}}\frac{1}{\sqrt{2}}\left[\begin{array}{c}
(\textrm{sgn}\ n)\ \phi_{|n|-1}(x+kl^2_B)\\ \phi_{|n|}(x+kl^2_B)
\end{array}\right]\\
\nonumber \epsilon_n&=&(\textrm{sgn}\ n)\ \sqrt{|n|}\hbar\omega_B
\end{eqnarray}
for $n\neq 0$ and
\begin{equation}\label{eq:MLGZLL}
\Psi^{0}_{n,k}(\mathbf{x})=\frac{e^{iky}}{\sqrt{L_y}}\left[\begin{array}{c}
0\\
\phi_{n}(x+kl^2_B)
\end{array}\right]
\end{equation}
for the ZLL, that now corresponds to just the magnetic level $n=0$.

Due to the different scaling of the cyclotron frequency with the magnetic field, the Zeeman energy now satisfies
\begin{equation}
\frac{\epsilon_Z}{\hbar\omega_B}=0.0016\frac{B[\text{T}]}{\sqrt{B_{\perp}[\text{T}]}} \ll 1
\end{equation}
while an order of magnitude analysis of the Coulomb interaction gives the dimensionless strength
\begin{equation}\label{eq:MLGcoulombfactor}
\frac{F_C}{\hbar\omega_B}=\frac{e^2_c}{\kappa l_B\hbar\omega_B}=\frac{e^2_c}{\kappa v_F}\thickapprox\frac{2.2}{\kappa}~,
\end{equation}
Note that it does not depend on the value of the magnetic field, in contrast to the case of bilayer graphene. The order of magnitude of short-range interactions is similar to the bilayer case and their typical energy also goes as $\sim F_Cd/\hbar\omega_Bl_B\ll 1$. Once more, for $\kappa=1$, Coulomb interactions are not weak. However, by taking $\kappa\sim 5$ we can get $F_C\sim 0.4\hbar\omega_B$, which can be regarded as a small value.

Under this assumption, we neglect LL mixing and project the Hamiltonian onto the ZLL. The states for the ZLL of monolayer graphene are also restricted to the $KK'\otimes\bar{B}\otimes s$ subspace, which means that they are localized, for each valley, in one sublattice or the other and hence the sublattice degree of freedom becomes equivalent to the valley degree of freedom. Reasoning in the same fashion as in the bilayer case, the resulting effective Hamiltonian is formally equal to that of Eq. (\ref{eq:EffectiveHamiltonian}) but with $\epsilon_V=0$:
\begin{widetext}
\begin{eqnarray}\label{eq:MLGEffectiveHamiltonian}
\nonumber\hat{H}^{(0)}&=&-\int\mathrm{d}^2\mathbf{x}~\epsilon_Z\hat{\psi}^{\dagger}(\mathbf{x})\sigma_z\hat{\psi}(\mathbf{x})+\frac{1}{2}\int\mathrm{d}^2\mathbf{x}~\mathrm{d}^2\mathbf{x'}:[\hat{\psi}^{\dagger}(\mathbf{x})\hat{\psi}(\mathbf{x})]V_0(\mathbf{x}-\mathbf{x'})[\hat{\psi}^{\dagger}(\mathbf{x'})\hat{\psi}(\mathbf{x'})]:\\
\nonumber &+&\sum_{i}\frac{1}{2}g_{i}\int\mathrm{d}^2\mathbf{x}\ :[\hat{\psi}^{\dagger}(\mathbf{x})\tau_{i}\hat{\psi}(\mathbf{x})]^2:+\int\mathrm{d}^2\mathbf{x}~\mathrm{d}^2\mathbf{x'}\hat{\psi}^{\dagger}(\mathbf{x})V_{DS}(\mathbf{x},\mathbf{x'})\hat{\psi}(\mathbf{x'})\\
\end{eqnarray}
\end{widetext}
The Dirac sea that creates the mean-field potential $V_{DS}(\mathbf{x},\mathbf{x'})$ is composed now by all the occupied states with $n\leq -1$.

\subsection{Hartree-Fock equations and mean-field phase diagram}

The HF equations have the same form of Eq. (\ref{eq:HFeqs}). As the $\nu=0$ QH state corresponds to half-filling of the ZLL, the electrons occupy in the same way two orthogonal spinors $\chi_{a,b}$ in valley-spin space and leave empty the remaining orthogonal spinors $\chi_{c,d}$. Then, after projecting the HF equations into the orbital part of the wave functions, we get the algebraic equation
\begin{eqnarray}\label{eq:MLGHFenergy}
\nonumber \epsilon_{0,\alpha}\chi_{\alpha}&=&\frac{F_{00}}{2}\chi_{\alpha}-F_{00}P\chi_{\alpha}\\
\nonumber &+&\sum_{i}u_{i}\left([\text{tr}(P\tau_{i})]\tau_{i}-\tau_{i}P\tau_{i}\right)\chi_{\alpha}
\\&-&\epsilon_Z\sigma_z\chi_{\alpha}
\end{eqnarray}
where now $u_{i}=g_{i}/2\pi l^2_B$. The term $F_{00}/2$ arises from the interaction with the inert Dirac sea; however, in monolayer graphene, it is just a trivial energy shift as there is only one magnetic level in the ZLL.

The previous equation presents the same valley-spin structure of Eq. (\ref{eq:HFenergy}). As a consequence, the spinorial solutions $\chi_{\alpha}$ to the HF equations are identical to those of bilayer graphene and their mean-field energies are:
\begin{equation}\label{eq:MLGHFenergystructure}
\epsilon_{0,(a,b)}=-\frac{F_{00}}{2}+\epsilon_{(a,b)},~\epsilon_{0,(c,d)}=\frac{F_{00}}{2}+\epsilon_{(c,d)}
\end{equation}
Moreover, the total energy of the ground state (per wave vector state) is:
\begin{equation}\label{eq:MLGmeanfieldenergy}
E_{HF}=-\frac{F_{00}}{2}+E(P)\\
\end{equation}
with $E(P)$ given by Eq. (\ref{eq:meanfieldenergy}). Therefore, we conclude that the mean-field phase diagram for the $\nu=0$ QH state in monolayer graphene is that of the bilayer for $\epsilon_V=0$ \cite{Kharitonov2012,Kharitonov2012PRL}. In that case, the PLP phase of the bilayer changes to a fully interlayer coherent phase (ILC) since $\theta_V=\pi/2$ and the vector $\mathbf{n}$, pointing the polarization in the valley space, is fully contained in the $x-y$ plane, $\mathbf{n}=[\cos\phi_V,\sin\phi_V,0]$. For monolayer graphene, the equivalent of the ILC phase is the Kekul\'e distortion (KD) phase: since the valley degree of freedom is equivalent to the sublattice in the zero Landau level of monolayer graphene, this state corresponds to a coherent mixture of the two sublattices. For the same reason, the FLP phase is now a charge-density wave (CDW) phase. There are experimental evidences that the phase for the $\nu=0$ QH state of monolayer graphene for zero in-plane component of the magnetic field is also the CAF state \cite{Young2014}, so $u_z>-u_{\perp}>0$ as in the bilayer case. However, as an analog of the layer voltage is lacking, in principle only the transition between CAF and F phases can be explored by changing the in-plane component of the magnetic field.

For the mean-field transport gap, we find that it satisfies a similar relation to Eq. (\ref{eq:transportgap}),
\begin{equation}\label{eq:transportgapMLG}
\Delta_{HF}=F_{00}+\Delta^{bc}
\end{equation}
\subsection{Collective modes}\label{sec:MLGcolmod}

The TDHFA developed for computing the collective modes within the ZLL of the $\nu=0$ QH state of bilayer graphene can be straightforwardly translated to the monolayer. Indeed, the valley-spin structure of the excitations is the same as in the bilayer case, while the orbital structure is trivial as it corresponds to a one-dimensional subspace spanned by the magnetic level $n=0$. Hence, excitations are characterized just by their symmetry in valley-spin space, $\omega^{\mu}(k)$, with $\mu$ labeling the same set of conserved quantum numbers described in Sec. \ref{subsec:numericalresults}. Due to the simple structure in orbital space, the collective modes for monolayer graphene can be computed analytically and their explicit expression is given at the end of Appendix \ref{app:analyticalTDHFA}. In particular, at $k=0$, $\omega^\mu(0)=\omega^{\mu}_{00}$ being $\omega^{\mu}_{00}$ the frequency of the corresponding orbital-singlet mode in the bilayer case. As a consequence, phase transitions are characterized in the same way.

\begin{figure*}[tb!]
\begin{tabular}{@{}cc@{}}
    \includegraphics[width=\columnwidth]{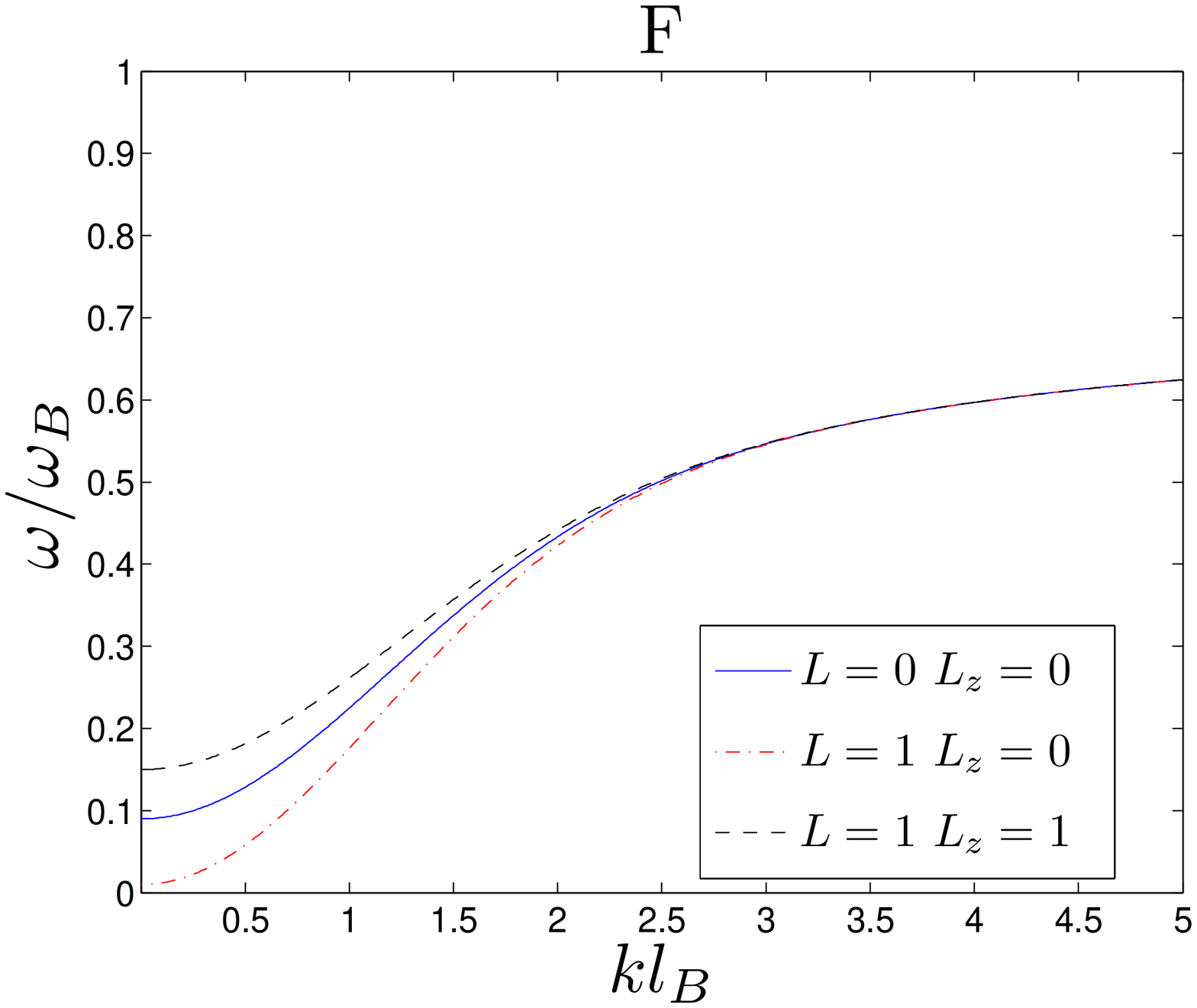} &
    \includegraphics[width=\columnwidth]{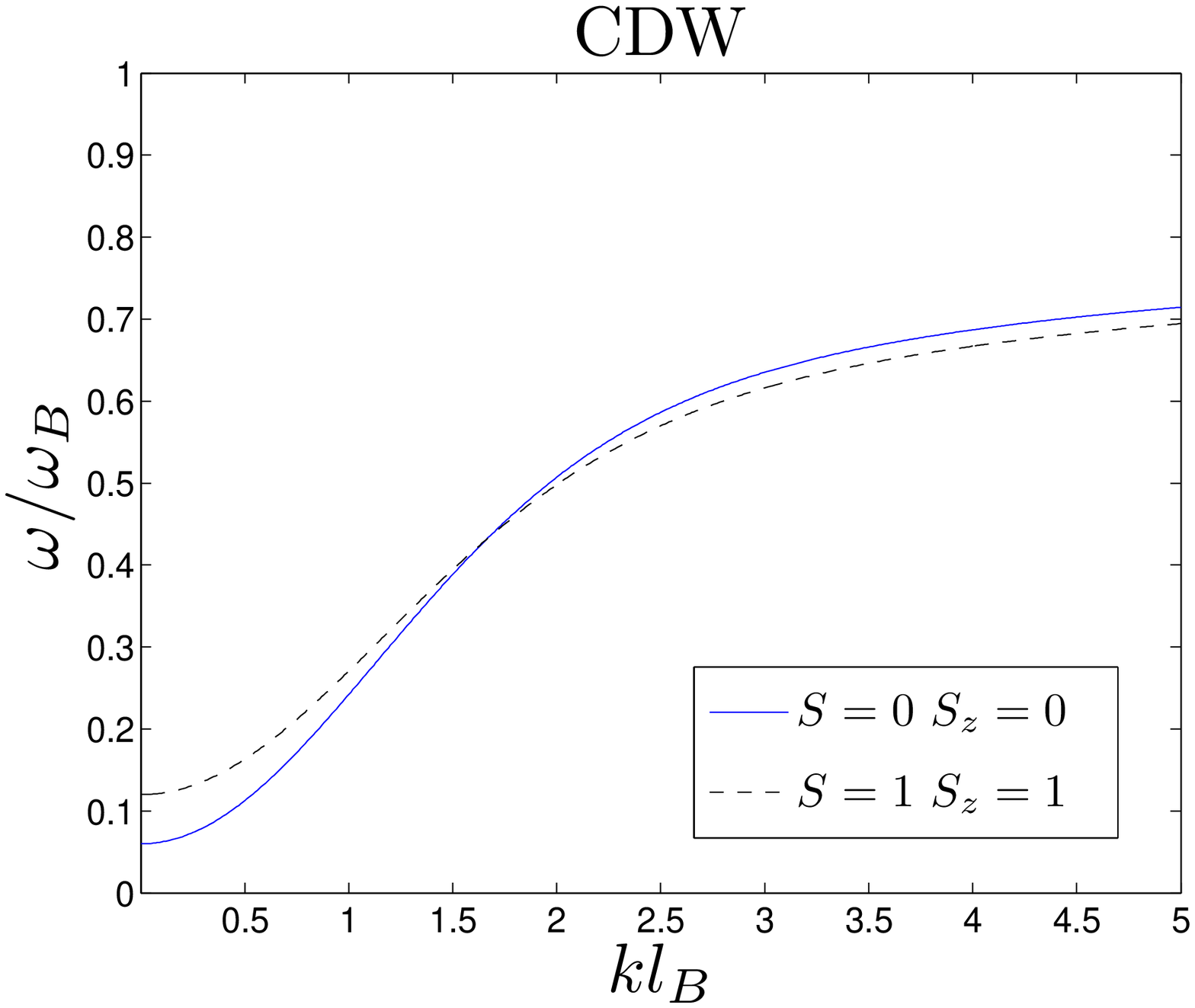} \\
    \includegraphics[width=\columnwidth]{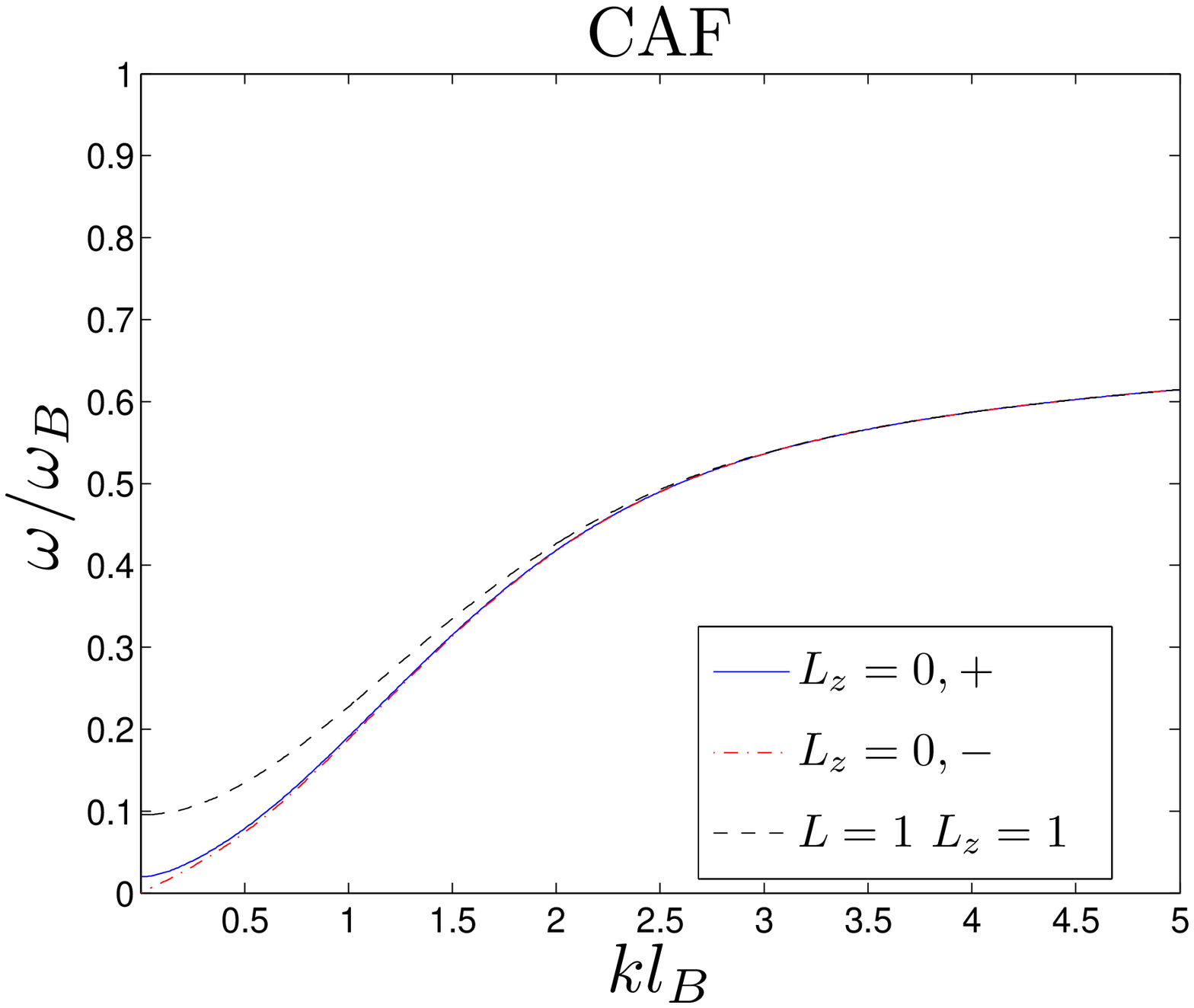} &
    \includegraphics[width=\columnwidth]{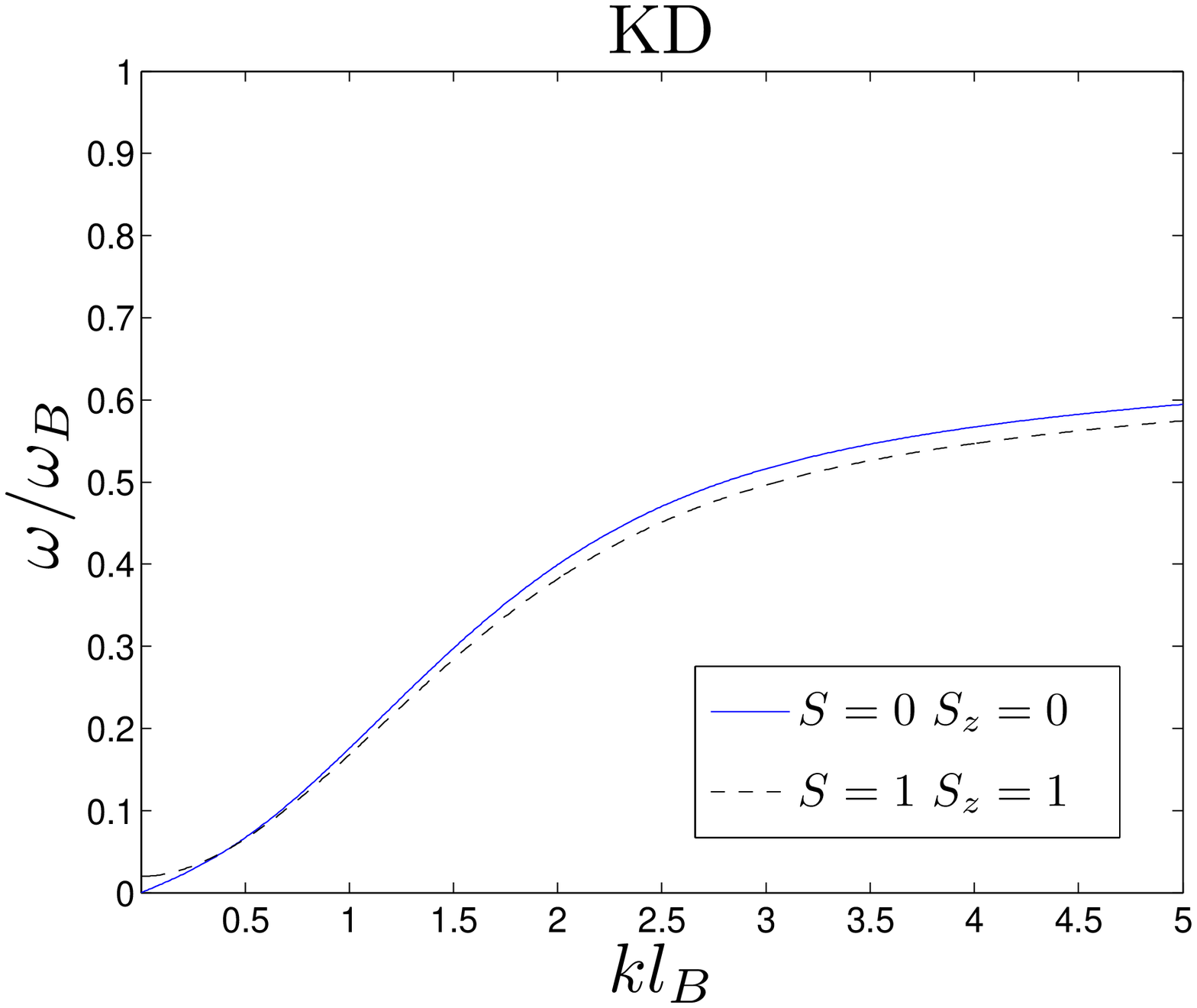} \\
\end{tabular}
\caption{Dispersion relation of the collective modes for the different phases of the $\nu=0$ QH state in monolayer graphene. In all the plots, the strength of the Coulomb interaction is set to $F_C=0.5\hbar\omega_B$. Upper left: dispersion relation of the F phase for $u_z=0.05\hbar\omega_B$, $u_{\perp}=-0.02\hbar\omega_B$ and $\epsilon_Z=0.045\hbar\omega_B$. Upper right: dispersion relation of the CDW phase for $u_z=-0.05\hbar\omega_B$, $u_{\perp}=-0.02\hbar\omega_B$ and $\epsilon_Z=0.01\hbar\omega_B$. Lower left: dispersion relation of the CAF phase for $u_z=0.05\hbar\omega_B$, $u_{\perp}=-0.02\hbar\omega_B$ and $\epsilon_Z=0.01\hbar\omega_B$. Lower right: dispersion relation of the KD phase for $u_z=0.01\hbar\omega_B$, $u_{\perp}=-0.02\hbar\omega_B$ and $\epsilon_Z=0.01\hbar\omega_B$.}
\label{fig:DispersionRelationMLG}
\end{figure*}

We now plot the dispersion relation for every phase. The results are shown in Fig. \ref{fig:DispersionRelationMLG}. Note that for the study of the CDW and KD phases we have to use unrealistic values for the coupling constants. The qualitative trends of the several dispersion relations are similar to those of the $N=0$ orbital modes in bilayer graphene; in fact, at $k=0$, they have the same formal expression as explained above. When they exist, the dispersion relation of the dynamical instabilities of the CAF and KD phases is qualitatively similar to that of the bilayer case (not shown).

We note that the neutral excitations of the F and CDW phases were computed in Ref. \cite{Doretto2007} using a bosonization approach and a different type of short-range interactions, nevertheless finding similar analytical expressions for the frequency of the modes. In addition, Ref. \cite{Murthy2016} obtained the dispersion relation for the F and CAF phases using the same valley-asymmetric interactions here considered but replacing the long-range Coulomb interaction by an effective short-range one in order to simulate the valley-spin stiffness of the waves.

\section{Effects of Landau-level mixing}\label{sec:renorm}

In this section, we consider the usual case where $F_C\gtrsim \hbar\omega_B$ and LL mixing cannot be neglected. One possible way to take it into account is through a static screening of the Coulomb interaction in the large-$N$ approximation \cite{Foster2008,Basko2008,Aleiner2007,Lemonik2010,Kharitonov2012}. Other approaches consider dynamical screening within the same approximation \cite{Gorbar2010,Gorbar2011,Gorbar2012} or allow for LL mixing in the TDHFA formalism \cite{Toke2013}. As we are mainly interested in the low-energy modes describing the phase transitions, static screening is expected to provide a good approximation in this limit; on the other hand, screening by LL mixing is not expected to describe correctly the dispersion relation near $k=0$ \cite{Toke2013}, which is precisely the most interesting region for our purposes.

In the large-$N$ approximation, the effective Coulomb interaction potential is obtained using a RPA-type screening
\begin{equation}\label{eq:RPACoulomb}
\bar{V}(\mathbf{k})=\frac{V_0(\mathbf{k})}{1-\Pi^{0}(\mathbf{k},0)V_0(\mathbf{k})},~V_0(\mathbf{k})=2\pi \frac{e^2_c}{\kappa|\mathbf{k}|},
\end{equation}
where $V_0(\mathbf{k})$ is the Fourier transform of the bare Coulomb interaction and $\Pi^{0}(\mathbf{k},\omega)$ is the non-interacting polarization for the $\nu=0$ QH state,
\begin{equation}\label{eq:freepolarization}
\hbar\Pi^{0}(\mathbf{k},\omega)=\sideset{}{'}\sum_{\substack{n_k,n_l\\ \alpha_k \alpha_l}}~\frac{\delta_{\alpha_k,\alpha_l}}{2\pi l^2_B}D^{0}_{nk\alpha_kn_l\alpha_l}(\omega)|A^{(2)}_{n_kn_l}(\mathbf{k})|^2~,
\end{equation}
with $n_k,n_l$ taking values for all integers except $-1$, as in Eq. (\ref{eq:fieldoperatorLL}). In the same fashion, $D^{0}_{nk\alpha_kn_l\alpha_l}(\omega)$ is the non-interacting two-particle propagator [see Eq. (\ref{eq:electronholefunction}) and related discussion for more details], with the non-interacting values for the energies and occupation numbers. Here, $A^{(2)}_{nn'}(\mathbf{k})$ is the bilayer graphene magnetic form factor:
\begin{eqnarray}\label{eq:magneticFFBilayer}
\nonumber A^{(2)}_{nn'}(\mathbf{k})&=&\frac{1}{2}\left[a^{-}_{n,n'}A_{|n|-2,|n'|-2}(\mathbf{k})\textrm{sgn}\ n~\textrm{sgn}\ n'\right.\\
&+&\left.a^{+}_{n,n'}A_{|n|,|n|'}(\mathbf{k})\right]\\
\nonumber a^{\pm}_{n,n'}&=&\sqrt{1\pm(\delta_{n0}+\delta_{n1})}\sqrt{1\pm(\delta_{n'0}+\delta_{n'1})}
\end{eqnarray}
$A_{nn'}(\mathbf{k})$ being the usual magnetic form factor, given by Eq. (\ref{eq:magneticFF}). The polarization $\Pi^{0}(\mathbf{k},\omega)$ is the Fourier transform of the non-interacting density-density correlation function and is obtained following an analog calculation to that described in Appendix \ref{subsec:dyson} but using the bare vertex and allowing for LL mixing.

After neglecting small corrections due to the Zeeman effect and the layer voltage, the free static polarization reads:
\begin{eqnarray}\label{eq:polarizationseries}
\Pi^{0}(\mathbf{k},0)&=&-\frac{N}{2\pi l^2_B}\frac{f(kl_B)}{\hbar\omega_B}\\
\nonumber f(kl_B)&=&\sum^{\infty}_{n_k=0}\sum^{\infty}_{n_l=2}~\frac{2|A^{(2)}_{n_k,-n_l}(\mathbf{k})|^2}{\sqrt{n_k(n_k-1)}+\sqrt{n_l(n_l-1)}}
\end{eqnarray}
with $N=4$ the number of the valley-spin components of the field, which is expected to be a sufficiently large value to provide a good approximation \cite{Basko2008,Lemonik2010,Kharitonov2012}. The function $f(x)$ is dimensionless as $|A^{(2)}_{n_kn_l}(\mathbf{k})|^2$ only depends on the momentum through the quantity $kl_B$. The effective screened Coulomb interaction can be then rewritten as:
\begin{equation}\label{eq:Screened}
\bar{V}(k)=\frac{V_0(k)}{1+N\frac{F_C}{\hbar\omega_B}\frac{f(kl_B)}{kl_B}}
\end{equation}
where we have made explicit the rotational invariance of all the magnitudes. For $F_C\ll \hbar\omega_B$, the screened interaction reduces to the bare Coulomb interaction used in the previous calculations.

\begin{figure}[bt]
\includegraphics[width=\columnwidth]{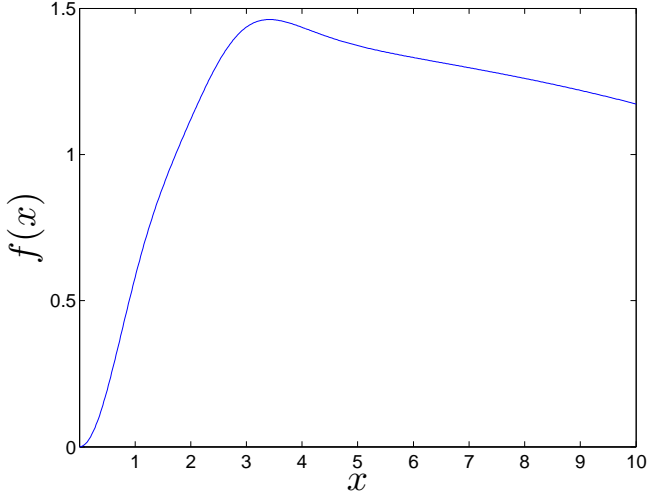}
\caption{Plot of the function $f(x)$ defined in Eq. (\ref{eq:polarizationseries}), where we have imposed a finite cutoff in the series $N_C=150$ for the computation.}
\label{fig:dimensionlessfunction}
\end{figure}

The dimensionless function $f(x)$ is plotted in Fig. \ref{fig:dimensionlessfunction}, in perfect agreement with the results of Refs. \cite{Gorbar2010,Gorbar2012}. For small $x$, it behaves as $f(x)\simeq\alpha x^2$,
\begin{eqnarray}\label{eq:dimensionlessfunction}
\alpha&=&\frac{1}{2}\sum_{n=2}^{\infty}\frac{\left(1-\sqrt{1-\frac{2}{n}}\right)^2}{\sqrt{1-\frac{1}{n}}\left(1+\sqrt{1-\frac{2}{n}}\right)}\\
\nonumber &=&
2\sum_{n=2}^{\infty}\frac{1}{n^2\sqrt{1-\frac{1}{n}}\left(1+\sqrt{1-\frac{2}{n}}\right)^3}\simeq 0.8771
\end{eqnarray}
and then $\bar{V}(k)\simeq V_0(k)$ for low momentum. On the other hand, $f(x)\sim1$ for $x\gtrsim 1$ so for relevant momentum $kl_B\sim 1$ the screened potential satisfies:
\begin{equation}\label{eq:ScreenedShortRange}
\bar{V}(k)\sim \frac{4\pi \hbar^2}{m}\frac{1}{N}
\end{equation}
The dimensionless strength of the screened Coulomb interaction is then of order $N^{-1}=0.25\ll 1$ and thus we can now safely neglect LL mixing and restrict once more ourselves to the ZLL.

The other effect that results from allowing LL mixing is the renormalization of the coupling constants \cite{Kharitonov2012,Kharitonov2012PRL}, that amounts to replace the bare coupling constants by their renormalized values, $\bar{g}_{\perp},\bar{g}_{z}$. However, this process is only strong for low values of the magnetic field \cite{Kharitonov2012PRL} so we do not study its effect in detail here, treating again the renormalized coupling constants as inputs for the computations.

Taking into account the previous observations, we consider the following effective Hamiltonian for the ZLL
\begin{widetext}
\begin{eqnarray}\label{eq:EffectiveHamiltonianScreened}
\nonumber\hat{H}^{(0)}&=&\int\mathrm{d}^2\mathbf{x}~\hat{\psi}^{\dagger}(\mathbf{x})(-\epsilon_V\tau_z-\epsilon_Z\sigma_z)\hat{\psi}(\mathbf{x})+
\frac{1}{2}\int\mathrm{d}^2\mathbf{x}~\mathrm{d}^2\mathbf{x'}:[\hat{\psi}^{\dagger}(\mathbf{x})\hat{\psi}(\mathbf{x})]\bar{V}(\mathbf{x}-\mathbf{x'})[\hat{\psi}^{\dagger}(\mathbf{x'})\hat{\psi}(\mathbf{x'})]:\\
&+&\sum_{i}\frac{1}{2}\int\mathrm{d}^2\mathbf{x}~\bar{g}_{i}:[\hat{\psi}^{\dagger}(\mathbf{x})\tau_{i}\hat{\psi}(\mathbf{x})]^2:
\end{eqnarray}
\end{widetext}
instead of that of Eq. (\ref{eq:EffectiveHamiltonian}). We see that the result of considering LL mixing is reduced to
replacing the Coulomb interaction and the coupling constants by their effective values $\bar{V}$ and $\bar{g}_{i}$. Note that, since the excitations within the ZLL change valley or spin, the direct contribution of the Coulomb interaction vanishes and then there is no double-counting in this approximation (see Appendix \ref{app:TDHFA} for more details about the diagrammatic formalism of the TDHFA).

This kind of effective renormalized Hamiltonian has already been used for studying the $\nu=0$ state or fractional QH effect in the ZLL \cite{Kharitonov2012,Kharitonov2012b,Abanin2013}. As the valley-spin structure of the Hamiltonian is preserved, the solutions to the corresponding HF equations are the same as before and thus the phase diagram is that of Fig. \ref{fig:PhaseDiagram} but replacing $u_{i}$ by $\bar{u}_i$. The same is true for the HF energies, obtained from the screened values $\bar{F}_{nm}$ [see Eq. (\ref{eq:FockFouriermatrixelementscreened})], finding numerically that they satisfy the same relation $\bar{F}_{00}>\bar{F}_{10}>\bar{F}_{11}$ and then $\bar{F}_{0}>\bar{F}_{1}$, so the equation for the transport gap is given by the adapted version of Eq. (\ref{eq:transportgap}).

Regarding the collective modes, using the same reasoning, we find that all the orbital-valley-spin classification of Sec. \ref{sec:colmod} still holds, including the orbital pseudospin structure at $k=0$. Hence, only the quantitative values of the dispersion relation change but not the qualitative features and the resulting physical conclusions. In particular, the structure of the Goldstone modes and of the dynamical instabilities is preserved.

\begin{figure*}[bt]
\begin{tabular}{@{}cccc@{}}
    \includegraphics[width=0.5\columnwidth]{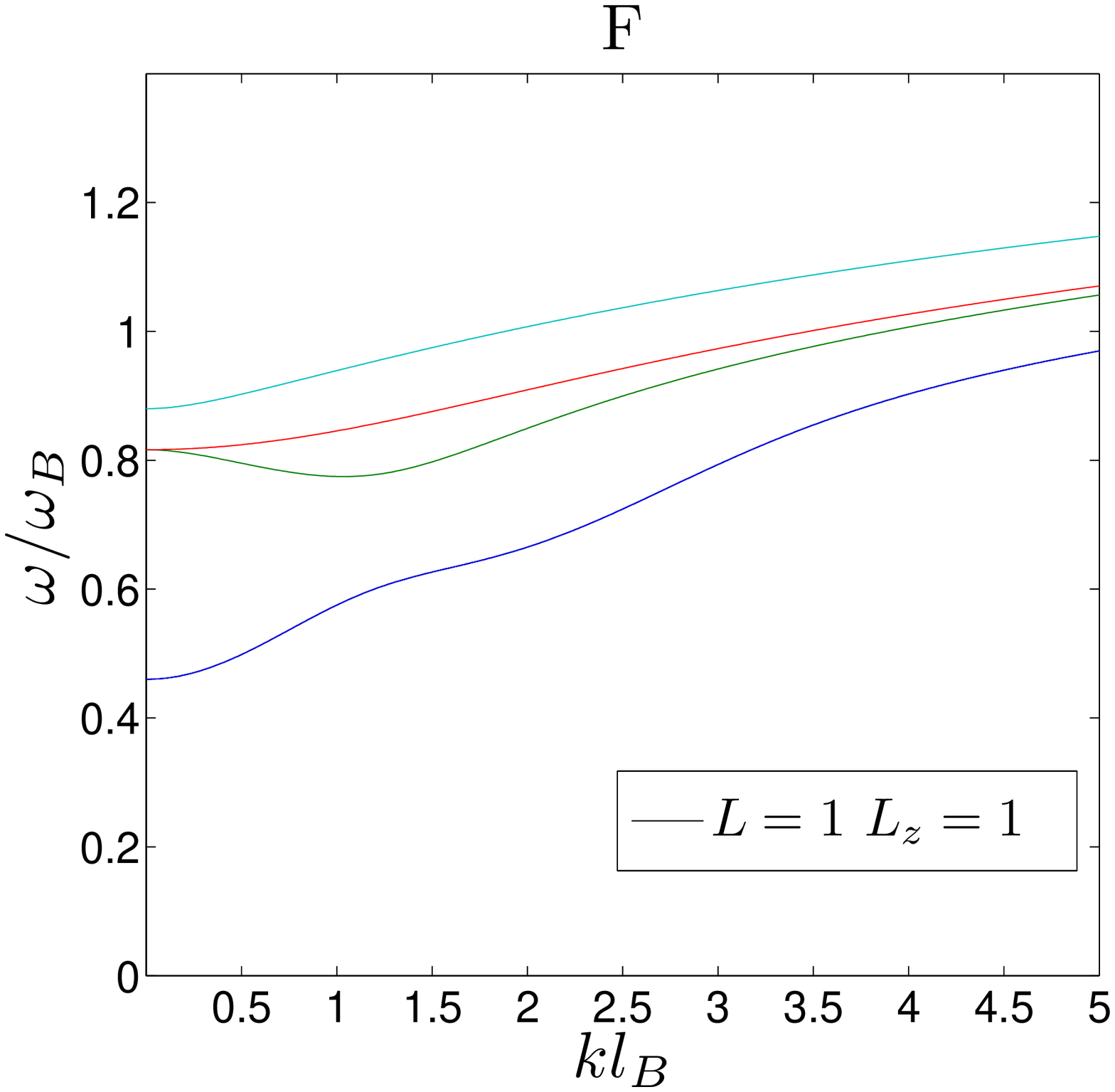} &
    \includegraphics[width=0.5\columnwidth]{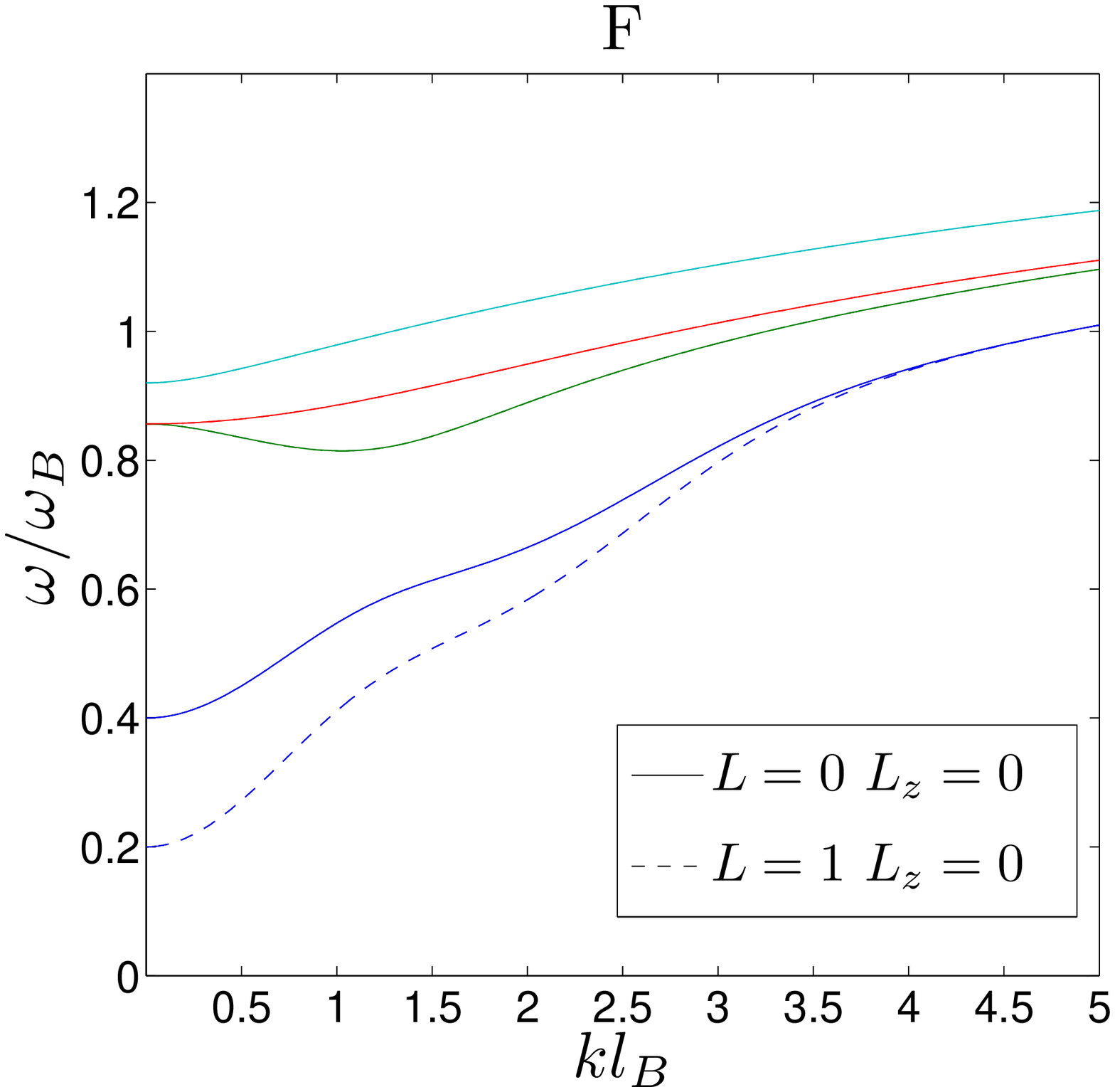} &
    \includegraphics[width=0.5\columnwidth]{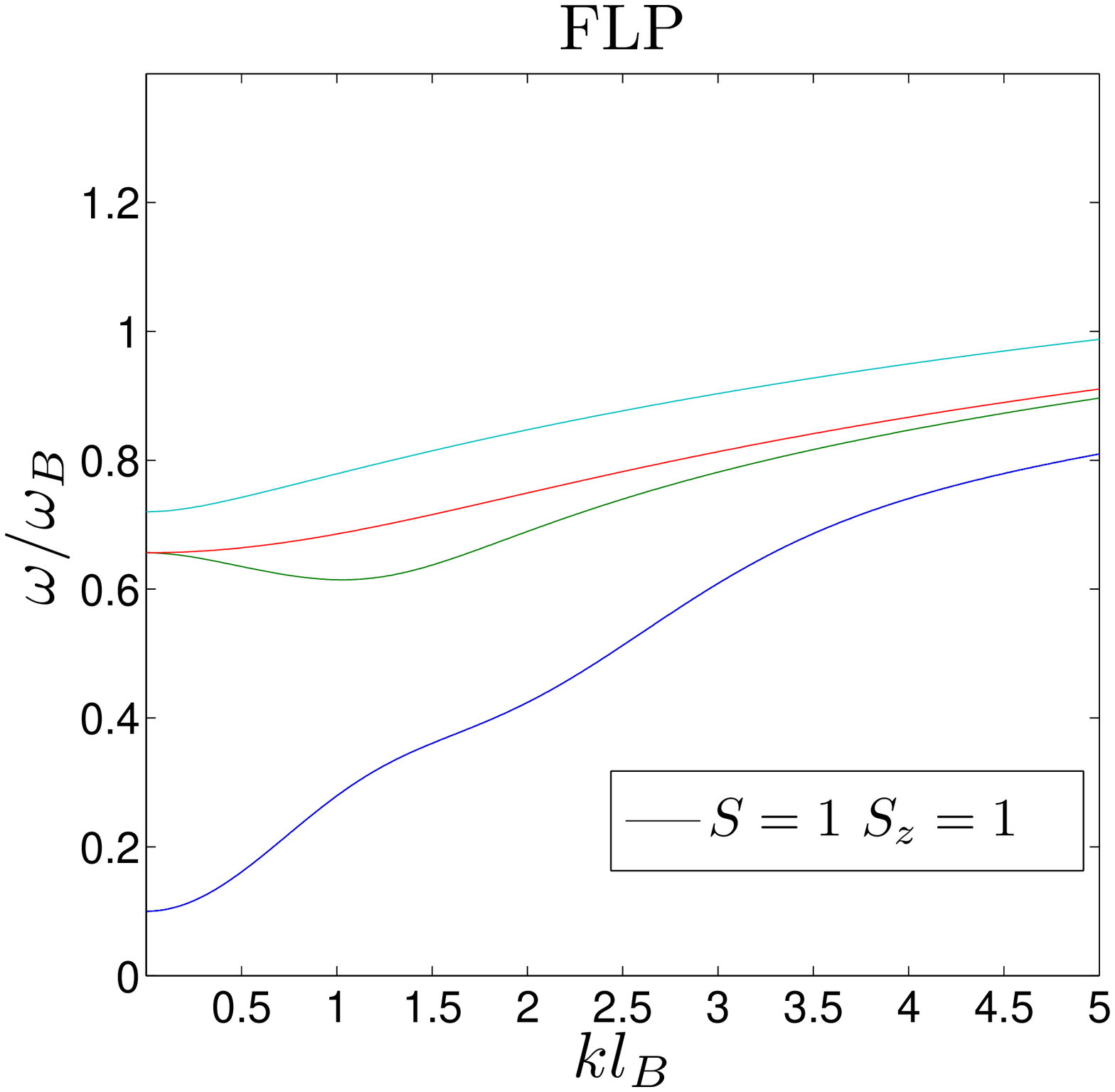} &
    \includegraphics[width=0.5\columnwidth]{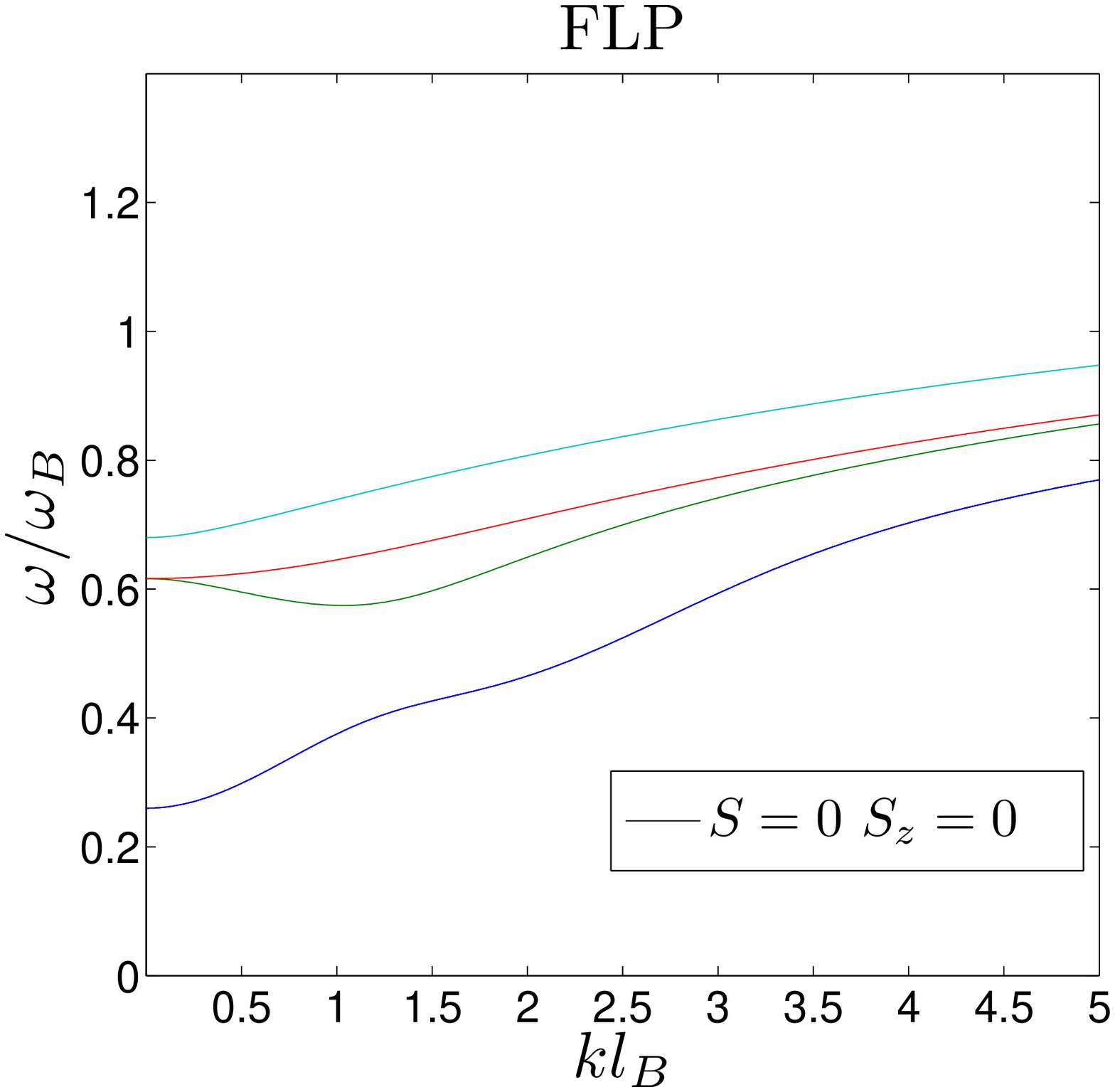} \\
    \includegraphics[width=0.5\columnwidth]{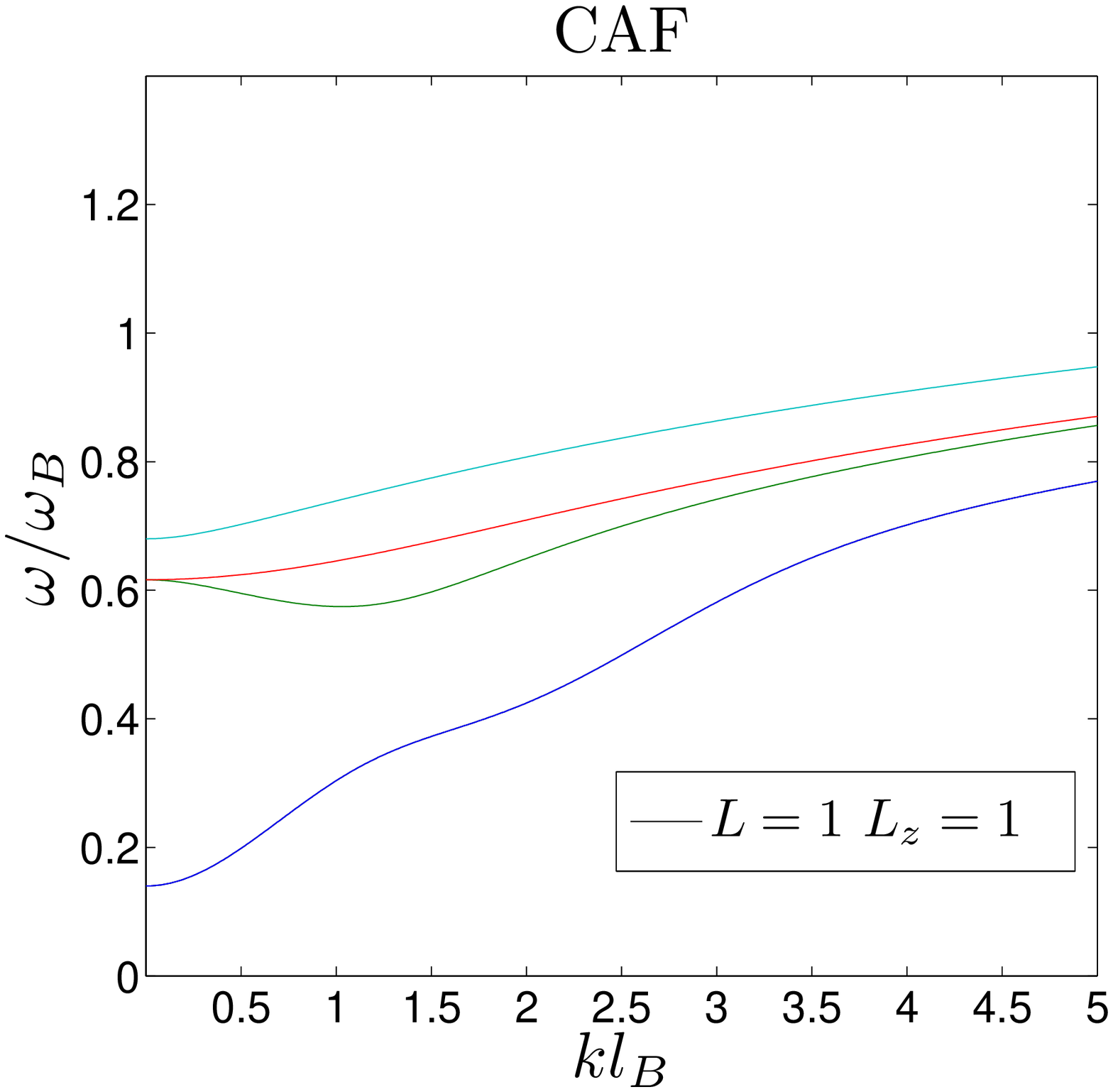} &
    \includegraphics[width=0.5\columnwidth]{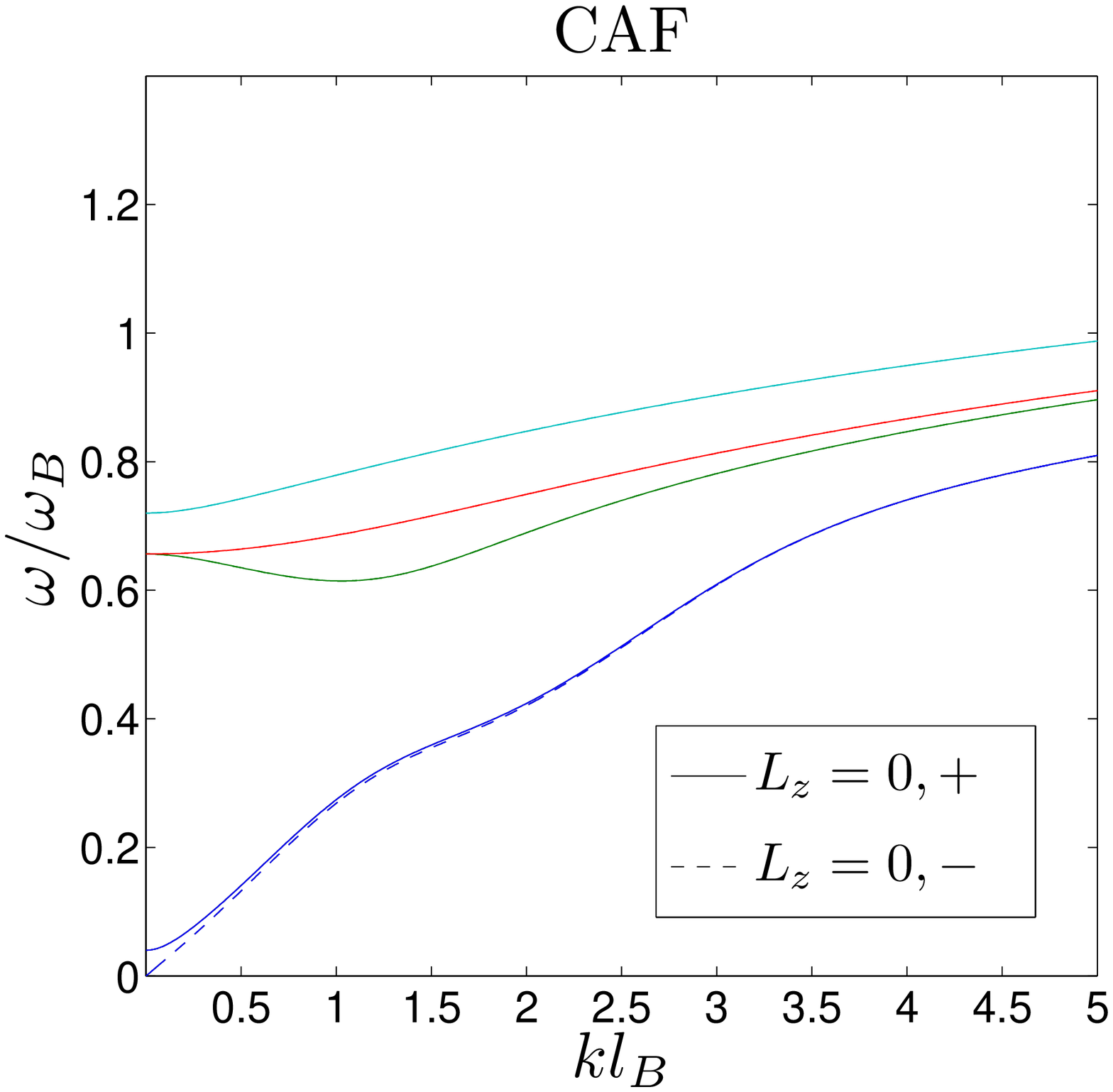} &
    \includegraphics[width=0.5\columnwidth]{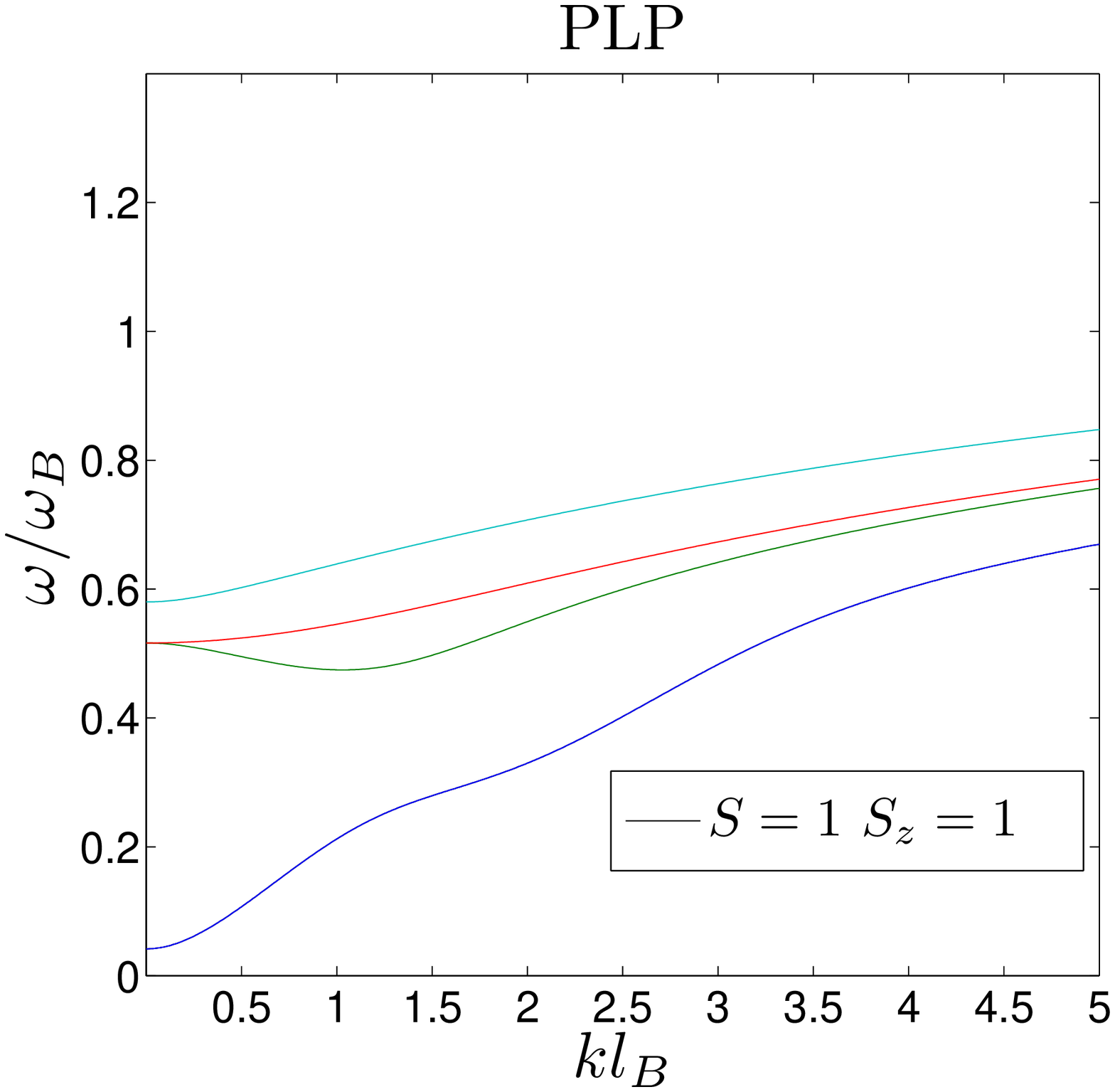} &
    \includegraphics[width=0.5\columnwidth]{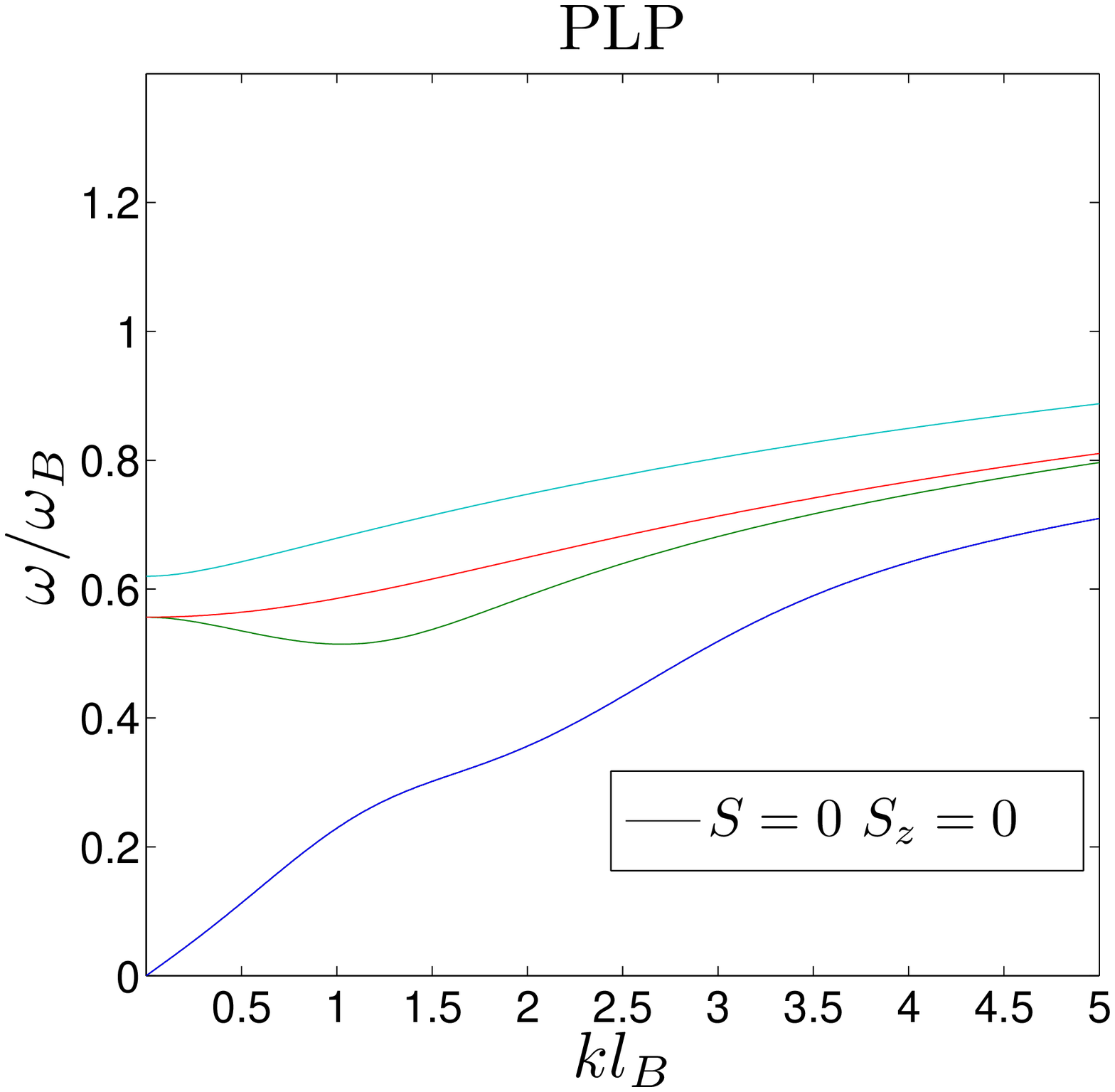} \\
\end{tabular}
\caption{Dispersion relation of the collective modes for the different phases of the $\nu=0$ QH state in bilayer graphene but now computed using the screened potential (\ref{eq:Screened}) with $F_C=4\hbar\omega_B$ while the rest of the parameters keep the same values of Fig. (\ref{fig:DispersionRelations}).}
\label{fig:DispersionRelationsRG}
\end{figure*}

\begin{figure}[bt]
\includegraphics[width=\columnwidth]{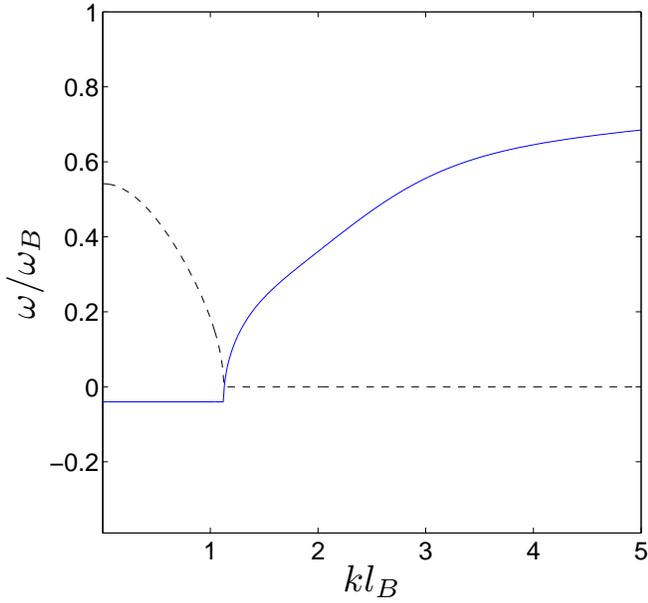}
\caption{Dispersion relation of the complex-frequency mode $\omega^{11}(k)$ for a PLP state, computed using the screened potential (\ref{eq:Screened}) with $F_C=4\hbar\omega_B$ and the rest of the parameters with the same values of right Fig. \ref{fig:ComplexMode}. The real (imaginary) part of the complex-frequency mode is plotted in solid (dashed) line.}
\label{fig:PLPRGexpmode}
\end{figure}

We now plot the numerical results for the computation of the dispersion relation using the screened Coulomb potential of Eq. (\ref{eq:Screened}) for a Coulomb interaction strength $F_C=4\hbar\omega_B$ and keeping the same values for the rest of parameters (including $\bar{u}_i=u_i$) in order to observe the effects of screening. In Fig. \ref{fig:DispersionRelationsRG}, we plot the dispersion relation of the modes for every phase and in Fig. \ref{fig:PLPRGexpmode} we represent the real and imaginary parts of the frequency of a dynamically unstable mode. We observe that the qualitative structure of the curves is similar to the unscreened case.

All the calculations developed in this section can be straightforwardly adapted to monolayer graphene, obtaining similar conclusions.

\section{Experimental remarks}\label{sec:remarksexps}

We make here some experimental remarks about the magnitudes computed in this work. First, we analyze the behavior of the transport gap, a quantity that can be measured, for instance, through local compressibility measurements \cite{Martin2010}, electronic transport measurements \cite{Zhao2010} or using the bias as a spectroscopic tool \cite{Velasco2012,Velasco2014,Shi2016}. As revealed by Eq. (\ref{eq:transportgap}), the transport gap has a contribution from Coulomb interactions, independent of the phase, and another contribution which actually depends on the valley-spin structure of the state. Using Eqs. (\ref{eq:Screened}), (\ref{eq:FockFouriermatrixelement}), it is straightforward to show that the screened value of the Coulomb contribution satisfies $\bar{F}_1=\hbar\omega_Bg\left(\frac{F_C}{\hbar\omega_B}\right)$, with $g(x)$ a dimensionless function that asymptotically behaves as [check Eq. (\ref{eq:dimensionlessgaptotal}) and related discussion for more details]
\begin{eqnarray}\label{eq:dimensionlessgap}
g(x)&\simeq& \frac{5}{4}\sqrt{\frac{\pi}{2}}x,~x\ll 1\\
\nonumber g(x)&\simeq&\frac{\ln x}{N\alpha}+0.668,~ x\gg 1
\end{eqnarray}
Hence, for low $F_C/\hbar\omega_B$ we recover the result for bare Coulomb interactions $\bar{F}_1\simeq F_1\propto \sqrt{B_{\perp}}$ while for strongly screened Coulomb interactions, due to the mild behavior of the logarithm, we are close to a linear behavior $\bar{F}_1\sim B_{\perp}$, already predicted in Ref. \cite{Gorbar2012}. On the other hand, the short-range energies also scale linearly with the perpendicular magnetic field, $u_{\perp},u_z\sim B_{\perp}$, at least for moderate fields $B_{\perp}\gtrsim 2~\text{T}$ \cite{Kharitonov2012PRL}. Hence, the gap is expected to grow linearly with $B_{\perp}$, as confirmed by experiments \cite{Martin2010,Velasco2012}; only for very high fields and large values of the dielectric constant $\kappa$ the unscreened regime $\Delta_{HF}\sim \sqrt{B_{\perp}}$ can be reached (note that even in this regime the Coulomb contribution is still dominant over those of short-range interactions). In fact, this behavior has been reported for transport gaps of other integer and fractional QH states \cite{Zhao2010,Shi2016}.

Interestingly, more information about the short-range energies can be obtained by comparing the transport gap of two different phases. Suppose that the transport gap for a F state with $\epsilon_V=0$ and a relatively high Zeeman energy $\epsilon_{Z,F}$ is measured and the process is repeated, keeping constant $B_{\perp}$, for a FLP state with voltage energy $\epsilon_{V,FLP}$ and negligible Zeeman term $\epsilon_Z\simeq0$. Then, as the Coulomb contribution is expected to be the same, by subtracting both gaps one finds
\begin{eqnarray}
\Delta_{HF,\text{F}}-\Delta_{HF,\text{FLP}}&=&\Delta^{bc}_{F}-\Delta^{bc}_{FLP}\\
\nonumber &=&4(u_z+u_{\perp})+2(\epsilon_{Z,F}-\epsilon_{V,FLP})
\end{eqnarray}
For sufficiently high in-plane magnetic field for the F phase, $\epsilon_{Z,F}$ becomes independent of $B_{\perp}$ and
hence one can obtain the value of $u_z+u_{\perp}$ from a linear fit of $\Delta_{HF,\text{F}}-\Delta_{HF,\text{FLP}}$ versus $B_{\perp}$. We note that, in order to determine the transport gap, this magnetoexciton gap has to be compared with the skyrmionic gap. According to Ref. \cite{Toke2011}, skyrmions are expected to be energetically favorable only for very large magnetic fields $B_{\perp}\gtrsim30~\text{T}$.

Regarding the detection of the collective modes, it is well known that their frequencies are the poles of the response functions to external fields $\hat{H}_{\rm{ext}}(\mathbf{x},t)$ (see Appendix \ref{app:TDHFA} for a detailed explanation). We now present a study of the required valley-spin structure of the operator $\hat{H}_{\rm{ext}}$ in order to detect the various modes. For instance, a magnetic field in the $i$ direction couples to the spin density, proportional to the operator $\hat{S}_i$ while a layer voltage difference couples to the charge-density difference between the layers, proportional to the $\hat{L}_z$ operator. A recent promising work has shown that the layer polarization can be measured in detail in capacitive measurements, arising as a powerful tool to characterize quantum Hall states in bilayer graphene samples \cite{Hunt2016}. Unfortunately, the operators $\hat{L}_{x,y}$, corresponding to the effective ladder operators between the layers in the ZLL, are not so easily translated to real physical observables. Compound operators of the form $\hat{S}_{i}\hat{L}_{z}$ can be interpreted as a spin-density difference between the layers although it is not clear how they could be measured. Note that in the projected model considered in this work, the modes do not couple to density perturbations (proportional to a scalar in valley-spin space) since the excitations are valley-spin waves that change necessarily one of these two quantum numbers \cite{Murthy2016}.

In this way, for the F phase, the modes with $L=0$ couple to just the $\hat{S}_{-}$ operator, with $\hat{S}_{\pm}=\hat{S}_{x}\pm i\hat{S}_{y}$, while the $L=1, L_z=0,\pm 1$ modes couple to the compound operator $\hat{S}_{-}\hat{L}_{z,\pm}$, respectively. Thus, in principle, only the $L=0$ modes can be detected using their coupling to a magnetic field in the $x,y$ direction. For the FLP phase, the discussion is exactly the same but changing the role of spin and layer operators, which makes much harder their detection as they all couple to the $\hat{L}_{-}$ operator.

On the other hand, for the phases with spontaneously broken symmetries, the situation is much easier due to the lower number of conserved quantities. For instance, for the CAF phase, in the sector $L_z=0$, the modes $+$ couple to the $\hat{S}_{\pm}$ operators while the modes $-$, including the Goldstone mode, couple to $\hat{S}_{z}$: as pointed out in Ref. \cite{Murthy2016}, the coupling of the CAF phase with the $\hat{S}_{z}$ operator could be used in principle to unambiguously distinguish this phase from the others. The modes with $L=1, L_z=\pm 1$ couple to the $\hat{S}_{z}\hat{L}_{\pm}$ operators. For the PLP phase, the modes with $S=0$ couple to just the $\hat{L}_{z}$ operator while the modes with $S=1, S_z=0,\pm 1$ couple to the operators $\hat{L}_{z}\hat{S}_{z,\pm}$, respectively. Then, in a similar way to the CAF phase, the coupling of the modes to a layer voltage provides a unique feature of the PLP phase that can be used to clearly characterize it.

An alternative way to study the collective-mode spectra through measurements of thermal transport has been proposed in a very recent work \cite{Pientka2017ArXiv}.

\begin{figure*}[tb!]
\begin{tabular}{@{}cc@{}}
    \includegraphics[width=\columnwidth]{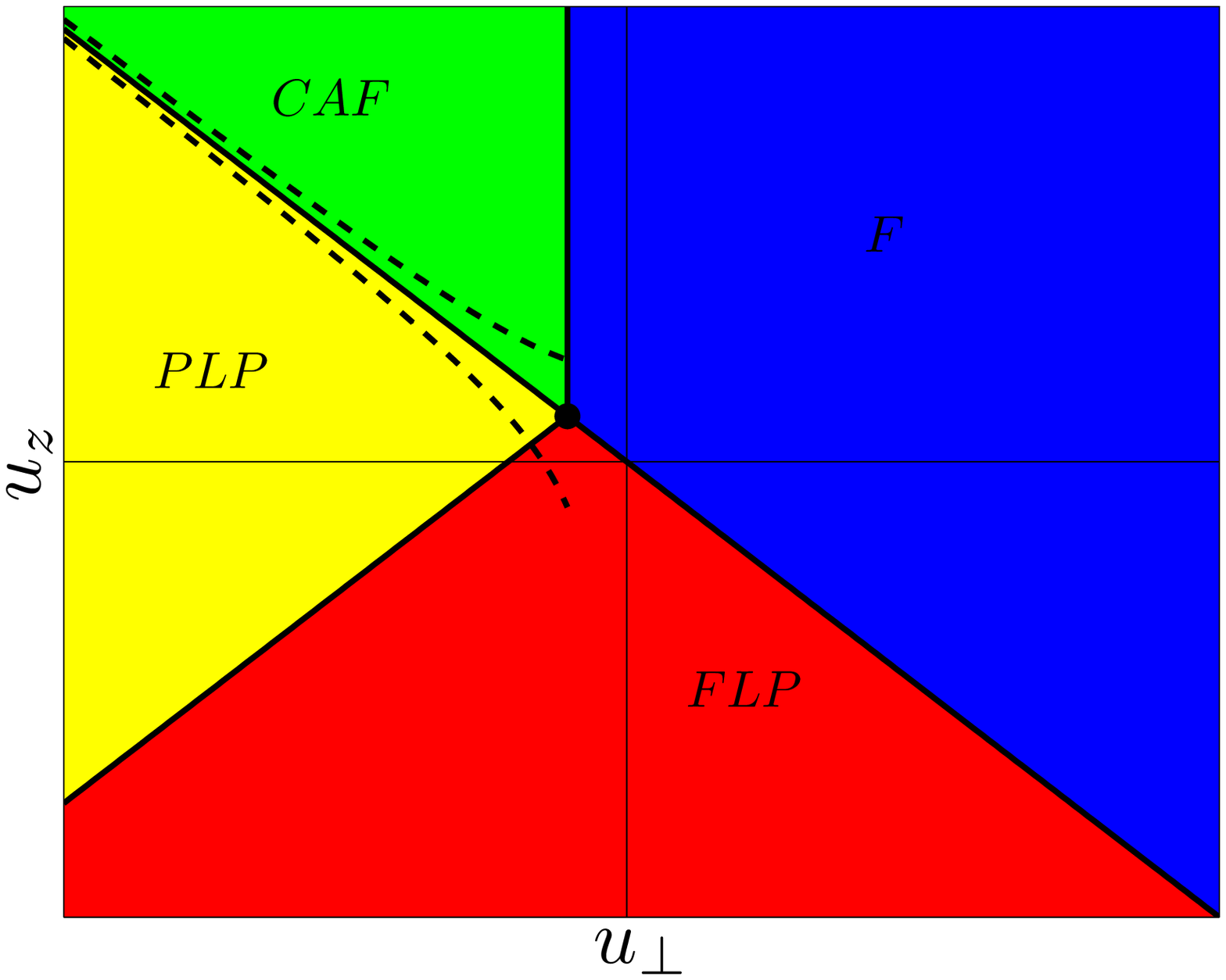} &
    \includegraphics[width=\columnwidth]{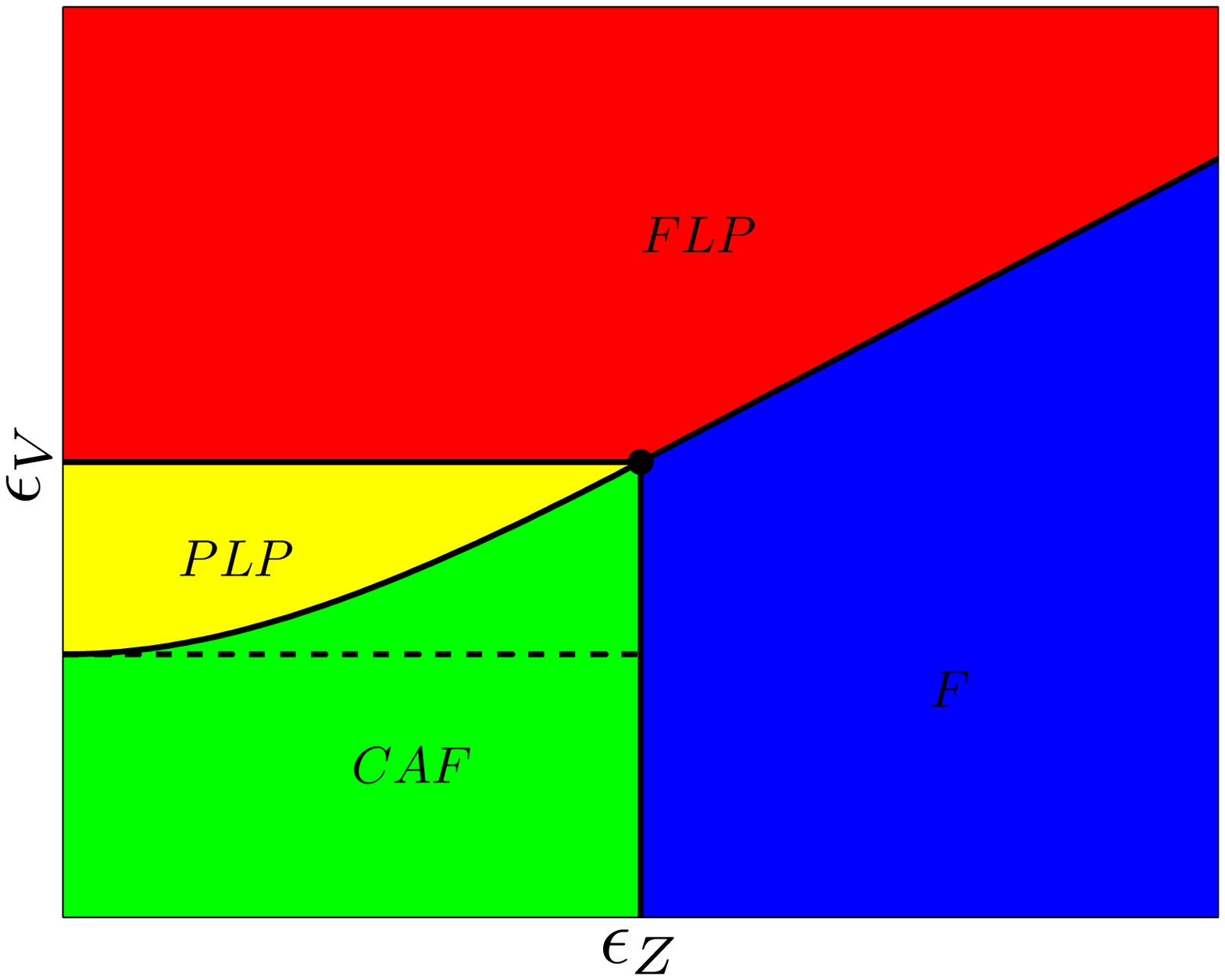} \\
\end{tabular}
\caption{Phase diagram of the $\nu=0$ QH state in parameter space $u_{\perp},u_{z}$ (left) and $\epsilon_Z,\epsilon_V$ (right), including the boundaries of the dynamically stable regions, represented as dashed lines.}
\label{fig:PhaseDiagramIns}
\end{figure*}

Regarding the detection of the dynamical instabilities, from Eqs. (\ref{eq:CAFinstabilitycondition}), (\ref{eq:PLPinstabilitycondition}) it is seen that the conditions for the appearance of unstable modes for both CAF and PLP phases are
\begin{eqnarray}\label{eq:grapheneinstability}
\frac{\epsilon^2_Z}{2u_{\perp}}&>&u_{\perp}+u_{z},~\text{(CAF)}\\
\nonumber u^2_{z}&>&u^2_{\perp}+\epsilon^2_Z,~\text{(PLP)}
\end{eqnarray}
Both conditions of instability are depicted in Fig. \ref{fig:PhaseDiagramIns}. Interestingly, the saturation of the two previous inequalities correspond to the boundary between the PLP and CAF phases for $\epsilon_V=0$ ($\epsilon_Z=0$). The first condition is unlikely to be achieved for the expected values of the coupling constants. The second condition is at least compatible with those values, but corresponds to a region where the PLP is not the actual ground state, as can be seen in the right plot. Naturally, the system is not dynamically unstable when starting from an equilibrium state. However, these potentially unstable dynamics could be explored in out of equilibrium situations.

For monolayer graphene, it is well known that the transport gap is governed by skyrmions \cite{Yang2006}, scaling as $\sim\sqrt{B_{\perp}}$ for both bare and screened interactions \cite{Kharitonov2012}, in agreement with the experimental results \cite{Jiang2007}. The same scaling occurs for the mean-field transport gap since the screened Coulomb contribution also behaves in monolayer graphene as $\bar{F}_{00}=\hbar\omega_Bh\left(\frac{F_C}{\hbar\omega_B}\right)$, with $h$ some dimensionless function. However, as now $F_C/\hbar\omega_B$ is independent of $B_{\perp}$, the gap always scales as $\sim\hbar\omega_B\sim\sqrt{B_{\perp}}$ for any value of $B_{\perp}$. With respect to the detection of the collective modes, as the valley-spin structure is formally equal to the bilayer scenario, the above discussion still holds although, unfortunately, there is not a direct equivalent of the layer voltage term. Alternative candidates for the job are distortions in the lattice \cite{Abanin2007,Fuchs2007} but they seem to be much less controllable from an experimental point of view.


\section{Conclusions} \label{sec:QHFMConclusions}

In this work, we have studied the $\nu=0$ QH state within a mean-field Hartree-Fock approach projected onto the zero-energy Landau level.
We have reproduced the mean-field phase diagram of the $\nu=0$ QH state of Ref. \cite{Kharitonov2012} by solving the self-consistent Hartree-Fock equations and computed the different mean-field energies and transport gaps.

After that, using the time-dependent Hartree-Fock approximation, we have computed the corresponding collective modes, that in the limit of zero momentum can be obtained analytically. In particular, we have provided a complete classification of the rich orbital-valley-spin structure of the modes and their symmetries. As a result, we have identified the singlet modes in orbital pseudospin at zero momentum as those characterizing the phase transitions since they are the lowest energy modes and the only ones coupled to the many-body contributions of short-range interactions at low momentum.

With respect to the valley-spin structure of the modes, at the boundary between the ferromagnetic and the fully layer-polarized phases, there is a gapless mode, resulting from a mean-field symmetry that can be regarded as a weak extension of the complete $SO(5)$ symmetry described in Ref. \cite{Wu2014}; the same phenomenon appears at the boundary between the canted anti-ferromagnetic and the partially layer-polarized phases. On the other hand, Goldstone modes appear at the boundaries between the ferromagnetic and canted anti-ferromagnetic phases and the fully and partially layer-polarized phases due to the spontaneous symmetry breakings. It is remarkable that, while the former Goldstone mode corresponds to a triplet mode in valley pseudospin, the latter is a spin-singlet mode. This contrast arises due to the different nature of both symmetry breaking mechanisms. A complementary study of the phase transitions could be provided in terms of the edge modes, extending the work presented in Refs. \cite{Murthy2016,Tikhonov2016} for the ferromagnetic-canted anti-ferromagnetic transition in monolayer graphene.

Another interesting feature of the phases with spontaneously broken symmetries is that they are able to present dynamical instabilities within this projected model. The study of such unstable dynamics in out of equilibrium scenarios is of particular interest as both the in-plane magnetic field and the interlayer voltage are easily manipulable in the laboratory. For instance, one can try to explore the occurrence of dynamical instabilities for a state initially prepared in the phases with spontaneously broken symmetries by applying a sudden quench in the external fields \cite{deNova2016a}.

The performed calculations for bilayer graphene can be straightforwardly adapted to monolayer graphene, due to the formal analogy of the Hamiltonian. We have found that most of the conclusions of the previous paragraphs still hold in the monolayer scenario. We have also analyzed the effects of LL mixing and we have accounted them by screening the long-range Coulomb interaction and by renormalizing the coupling constants of the short-range interactions. The resulting effective Hamiltonian is formally similar to that previously considered and hence, the phase diagram and the collective modes present the same structure discussed above.

Finally, we have made some experimental remarks about the transport gap and the detection of the instabilities. We have found that, within the current detection schemes, one can only aim at characterizing a restricted number of modes. Specifically, among all the modes describing the phase transitions, only those related to the F-CAF and FLP-PLP transitions couple to physical observables in the side of the spontaneously broken symmetry phase. Hence, the detection of the collective modes motivates the study and design of new possible experimental methods able to provide effective observables that couple to valley-pseudospin waves and measure its dispersion relation. In this sense, the recent work of Ref. \cite{Hunt2016} provides a promising line of research as it shows measurements of the layer polarization of the states.

%
%

\acknowledgments
We would like to specially thank E. Demler for his collaboration at the beginning the project. We also thank E. Shimsoni, P. Tikhonov, L. Brey, F. Guinea, F. Sols and T. Stauber for useful discussions. This work has been supported by MINECO (Spain) through grants FIS2010-21372 and FIS2013-41716-P, Comunidad de Madrid through grant MICROSERES-CM (S2009/TIC-1476) and also partially by the Israel Science Foundation.

\appendix

\section{Solutions of the Hartree-Fock equations}\label{app:magneticFF}

We compute in this Appendix the solutions to the Hartree-Fock equations for the $\nu=0$ QH state. We start by reviewing the case of the 2D electron gas (2DEG), as many of the developed techniques are also used for the actual computation for graphene, presented later.

\subsection{Basic results of the 2DEG}

As a first step, we define the magnetic form factors
\begin{equation}\label{eq:defmagneticFF}
A_{nn'}\left(\mathbf{k}\right)\equiv\int\mathrm{d}x\, \phi_n\left(x-\frac{k_yl^2_B}{2}\right)\phi_{n'}\left(x+\frac{k_yl^2_B}{2}\right)e^{-ik_xx}
\end{equation}
where the functions $\phi_n(x)$ are the usual harmonic oscillator wave functions
\begin{equation}\label{eq:oscillatorwavefunctions}
\phi_n(x)=\braket{x|n}=\frac{1}{\sqrt{2^nn!\sqrt{\pi}l_B}}H_n\left(\frac{x}{l_B}\right)e^{-\frac{x^2}{2l^2_B}},
\end{equation}
In order to compute the explicit expression of the magnetic form factors, we rewrite Eq. (\ref{eq:defmagneticFF}) as:
\begin{eqnarray}\label{eq:magneticFFelement}
\nonumber A_{nn'}\left(\mathbf{k}\right)&=&e^{-i\frac{k_xk_yl^2_B}{2}}\int\mathrm{d}x\, \phi_n\left(x\right)\phi_{n'}\left(x+k_yl^2_B\right)e^{-ik_xx}\\
&=&e^{-i\frac{k_xk_yl^2B}{2}}\braket{n|e^{-ik_xx}e^{i\frac{P_x}{\hbar}k_yl^2_B}|n'}
\end{eqnarray}
and we recall that $P_x$, the momentum operator in the $x$-direction, is the generator of translations in the $x$ direction. Rewriting the previous expression in terms of the usual harmonic oscillator destruction operator $a$ in the $x$-direction yields
\begin{eqnarray}\label{eq:destructors}
\nonumber A_{nn'}\left(\mathbf{k}\right)&=&\braket{n|e^{\frac{(-ik_x-k_y)}{\sqrt{2}}l_Ba^{\dagger}}e^{\frac{(-ik_x+k_y)}{\sqrt{2}}l_Ba}|n'}e^{-\frac{(kl_B)^2}{4}}\\
a&=&\frac{\frac{x}{l_B}+i\frac{p_xl_B}{\hbar}}{\sqrt{2}}
\end{eqnarray}
Then, after some straightforward manipulations, we obtain
\begin{widetext}
\begin{eqnarray}\label{eq:magneticFF}
\nonumber A_{nn'}\left(\mathbf{k}\right)&=&\sqrt{\frac{n'!}{n!}}\left[\frac{(-ik_x-k_y)l_B}{\sqrt{2}}\right]^{n-n'} L^{n-n'}_{n'}\left[\frac{(kl_B)^2}{2}\right]e^{-\frac{(kl_B)^2}{4}},~n\geq n'\\
L^{m}_n(x)&=&e^x \frac{x^{-m}}{n!} \frac{d^n}{dx^n}x^{n+m} e^{-x}=\sum^n_{j=0}\frac{(-1)^j(n+m)!}{j!(n-j)!(m+j)!}x^j,~m\geq0\\
\nonumber L^{m}_n(x)&=&(-1)^m x^{-m}\frac{(n+m)!}{n!}L^{-m}_{n+m}(x),~m<0
\end{eqnarray}
\end{widetext}
with $L^{m}_n(x)$ a generalized Laguerre polynomial. From the above relations it follows that $A_{nn'}\left(\mathbf{k}\right)=A^*_{n'n}\left(-\mathbf{k}\right)$. It is worth studying the dependence on $\mathbf{k}$ of the magnetic form factors. If we switch to the following polar coordinates, $(k_x,k_y)=k(\sin \varphi_{\mathbf{k}},\cos \varphi_{\mathbf{k}})$, we find that
\begin{equation}\label{eq:magneticFFspherical}
A_{nn'}\left(\mathbf{k}\right)=e^{i(n-n')\varphi_{\mathbf{k}}}A_{nn'}(k_x=0,k_y=k)\propto e^{i(n-n')\varphi_{\mathbf{k}}}
\end{equation}

As in practice we restrict to the ZLL, we give the explicit expression of the magnetic form factors involved in the calculations:
\begin{eqnarray}\label{eq:magneticFFZLL}
A_{00}\left(\mathbf{k}\right)&=&e^{-\frac{(kl_B)^2}{4}}\\
\nonumber A_{10}\left(\mathbf{k}\right)&=&\frac{-ik_x-k_y}{\sqrt{2}}e^{-\frac{(kl_B)^2}{4}}\\
\nonumber A_{01}\left(\mathbf{k}\right)&=&\frac{-ik_x+k_y}{\sqrt{2}}e^{-\frac{(kl_B)^2}{4}}\\
\nonumber A_{11}\left(\mathbf{k}\right)&=&\left[1-\frac{(kl_B)^2}{2}\right]e^{-\frac{(kl_B)^2}{4}}
\end{eqnarray}
Interestingly, in the context of the Wigner function, the magnetic form factors are the Moyal functions of the harmonic oscillator \cite{Schleich2001}.

For simplicity, we first review the usual case of integer QH states in the 2DEG, where the field operator only has two components corresponding to the spin polarizations $\xi=\pm$. The extension to the graphene scenario is discussed in the next subsection. The non-interacting eigenfunctions of the 2DEG are:
\begin{eqnarray}\label{eq:magneticwavefunctions}
\phi^0_{n,k,\xi}(\mathbf{x})&=&\phi_{n,k}(\mathbf{x})\chi_{\xi}\\
\nonumber \phi_{n,k}(\mathbf{x})&\equiv&\braket{\mathbf{x}|n,k}=\frac{e^{iky}}{\sqrt{L_y}}\phi_n(x+kl^2_B),~n=0,1,2\ldots
\end{eqnarray}
where $\phi_{n,k}(\mathbf{x})$ are the orbital wave functions of the magnetic levels and $\chi_{\xi}$ is the spin wave-function with polarization $\xi$. The non-interacting eigenvalues of these wave functions are
\begin{equation}
\epsilon^0_{n,\xi}=\left(n+\frac{1}{2}\right)\hbar\omega_B-\xi\epsilon_Z
\end{equation}
with $\omega_B$ and $\epsilon_Z$ the corresponding cyclotron and Zeeman energy.

We now suppose that the electrons interact through an arbitrary scalar potential $V$ (as it is, for instance, the Coulomb interaction $V_0$). Its spatial matrix elements are given in terms of the magnetic form factors:
\begin{widetext}
\begin{eqnarray}\label{eq:scalarmatrixelements}
\nonumber V^{n_ln_kn_jn_m}_{p_lp_kp_jp_m}&\equiv& \braket{n_lp_l~n_jp_j|V|n_kp_k~n_mp_m}=\int\mathrm{d}^2\mathbf{x}~\mathrm{d}^2\mathbf{x'}~\phi^*_{n_l,p_l}(\mathbf{x})\phi_{n_k,p_k}(\mathbf{x})V(\mathbf{x}-\mathbf{x'})\phi^*_{n_j,p_j}(\mathbf{x'})\phi_{n_m,p_m}(\mathbf{x'})\\
&=&\frac{1}{S}\sum_{\mathbf{q}}\delta_{p_k-p_l,-q_y}\delta_{p_m-p_j,q_y}e^{-iq_x(p_l-p_j-q_y)l^2_B}V(\mathbf{q})A_{n_ln_k}(-\mathbf{q})A_{n_jn_m}(\mathbf{q})
\end{eqnarray}
The HF equations associated to the previous interacting potential are (see Appendix \ref{subsec:stationaryrev} for a derivation of the HF equations)
\begin{eqnarray}\label{eq:2DHFeqs}
\nonumber \epsilon_{n,\xi}\phi_{n,k,\xi}(\mathbf{x})&=&\frac{\mathbf{\pi}^2}{2m}\phi_{n,k,\xi}(\mathbf{x})+\sum_{m,p,\xi'}\nu_{m,\xi'}\left(\left[\int\mathrm{d}^2\mathbf{x'}~V(\mathbf{x}-\mathbf{x'})\phi_{m,p,\xi'}^{\dagger}(\mathbf{x'})\phi_{m,p,\xi'}(\mathbf{x'}) \right]\phi_{n,k,\xi}(\mathbf{x})\right.\\
\nonumber &-&\left.\int\mathrm{d}^2\mathbf{x'}~V(\mathbf{x}-\mathbf{x'})\phi_{m,p,\xi'}(\mathbf{x})\phi_{m,p,\xi'}^{\dagger}(\mathbf{x'})\phi_{n,k,\xi}(\mathbf{x'})\right)-\epsilon_Z\sigma_z\phi_{n,k,\xi}(\mathbf{x})\\
\end{eqnarray}
with $\nu_{m,\xi'}$ the occupation number of each LL and the components of the vector $\mathbf{\pi}=(\pi_x,\pi_y)$ given after Eq. (\ref{eq:aniquilacion}). We analyze now the spatial structure of the previous equation. For that purpose, we define two associated mean-field potentials, denoted as $V^{H,F}$, that take into account the Hartree (direct) and Fock (exchange) contributions from the potential $V$, respectively. First, we address the {\it non-self-consistent} (NSC) problem, where the mean-field potentials are created by the bare (non-interacting) wave functions $\phi^0_{n,k,\xi}$. In that case, the above HF equations are:
\begin{equation}\label{eq:2DNSCHFeqs} \epsilon_{n,\xi}\phi_{n,k,\xi}(\mathbf{x})=\frac{\mathbf{\pi}^2}{2m}\phi_{n,k,\xi}(\mathbf{x})+\sum_{m,\xi'}\nu_{m,\xi'}\left[\int\mathrm{d}^2\mathbf{x'}~V_m^{H}(\mathbf{x},\mathbf{x'}) -V_m^{F}(\mathbf{x},\mathbf{x'})\chi_{\xi'}\chi^{\dagger}_{\xi'}\right]\phi_{n,k,\xi}(\mathbf{x'})-\epsilon_Z\sigma_z\phi_{n,k,\xi}(\mathbf{x})
\end{equation}
\end{widetext}
where $V_m^{H,F}$ are the contributions from each magnetic level to the mean-field Hartree and Fock potentials:
\begin{eqnarray}\label{eq:HFpotential}
\nonumber V_m^{H}(\mathbf{x},\mathbf{x'})&=&\int\mathrm{d}^2\mathbf{x''}~V(\mathbf{x}-\mathbf{x''})K_m(\mathbf{x''},\mathbf{x''})\delta(\mathbf{x}-\mathbf{x'})\\
V_m^{F}(\mathbf{x},\mathbf{x'})&=&V(\mathbf{x}-\mathbf{x'})K_m(\mathbf{x},\mathbf{x'})
\end{eqnarray}
The function $K_n(\mathbf{x},\mathbf{x'})\equiv K_{nn}(\mathbf{x},\mathbf{x'})$ is the spatial part of the Green's function, with
\begin{eqnarray}\label{eq:spatialGreenfunction}
K_{nn'}(\mathbf{x},\mathbf{x'})&=&\sum_p \phi_{n,p}(\mathbf{x})\phi^*_{n',p}(\mathbf{x'})
\end{eqnarray}
We can compute explicitly $K_{nn'}(\mathbf{x},\mathbf{x'})$ by transforming the discrete sum over $p$ into an integral and by making the transformation $x_p=X+pl^2_B$, with the center of mass and relative coordinates defined as $\mathbf{R}\equiv(\mathbf{x}+\mathbf{x'})/2$ and $\Delta \mathbf{x}\equiv\mathbf{x}-\mathbf{x'}$. The result is
\begin{equation}\label{eq:LLNonDiagonalGreenfunction}
K_{nn'}(\mathbf{x},\mathbf{x'})=\frac{e^{-i\frac{X\Delta y}{l^2_B}}}{2\pi l^2_B}A_{nn'}\left(-\frac{\Delta y}{l^2_B},-\frac{\Delta x}{l^2_B}\right)
\end{equation}
In particular,
\begin{equation}\label{eq:LLGreenfunction}
K_n(\mathbf{x},\mathbf{x'})=\frac{e^{-i\frac{X\Delta y}{l^2_B}}}{2\pi l^2_B}L_n\left(\frac{\Delta r^2}{2l^2_B}\right)e^{-\frac{\Delta r^2}{4 l^2_B}},
\end{equation}
with $\Delta r=|\Delta \mathbf{x}|$. Since $K_{nn'}(\mathbf{x},\mathbf{x})=\frac{\delta_{nn'}}{2\pi l^2_B}$, $(2\pi l^2_B)^{-1}$ being the homogeneous density corresponding to a completely filled magnetic level, the Hartree contribution is uniform:
\begin{equation}\label{eq:Hartreepotential}
V_m^{H}(\mathbf{x},\mathbf{x'})=\frac{V(0)}{2\pi l^2_B}\delta(\mathbf{x}-\mathbf{x'})~,
\end{equation}
$V(0)$ being the Fourier transform of the potential evaluated at $\mathbf{k}=0$. In the usual case of the Coulomb interaction, the Hartree potential is canceled by the positive charge background. For a short-range interaction, proportional to $\delta(\mathbf{x}-\mathbf{x'})$, the spatial part of the Hartree potential is equal to the Fock potential.

With respect to the Fock potential, for general purposes, we consider the following matrix elements
\begin{widetext}
\begin{eqnarray}\label{eq:generalFockelements}
\braket{n k|V_{mm'}^{F}|n'k'} &\equiv& \int\mathrm{d}^2\mathbf{x}\mathrm{d}^2\mathbf{x'}~ \phi^*_{n,k}(\mathbf{x})V(\mathbf{x}-\mathbf{x'})K_{mm'}(\mathbf{x},\mathbf{x'})\phi_{n',k'}(\mathbf{x'})
\end{eqnarray}
The case of the Fock potential in Eq. (\ref{eq:HFpotential}) is obtained for $m=m'$. To compute the previous integral, we switch to the center of mass and relative coordinates defined before Eq. (\ref{eq:LLGreenfunction}) and integrate along the center of mass coordinates $X,Y$, obtaining:
\begin{equation}\label{eq:Fockgeneralized}
\braket{n k|V_{mm'}^{F}|n' k'}=\frac{\delta_{k,k'}}{2\pi l^2_B}\int\mathrm{d}^2 \mathbf{x}~A_{n'n}\left(\frac{y}{l^2_B},\frac{x}{l^2_B}\right)A_{mm'}\left(-\frac{y}{l^2_B},-\frac{x}{l^2_B} \right)V(\mathbf{x})
\end{equation}
Now, we assume the typical situation where the potential $V(\mathbf{x})$ is rotationally invariant (as the Coulomb potential, for example). The polar dependence described in Eq. (\ref{eq:magneticFFspherical}) gives the result
\begin{eqnarray}\label{eq:FockNonDiagonalmatrixelements}
\braket{n k|V_{mm'}^{F}|n' k'}&=&\delta_{kk'}\delta_{n-n',m-m'}F_{nmm'}
\end{eqnarray}
with $F_{nmm'}$ the {\it on-shell} value of the previous integral. Specifically, for the Fock potential, $m=m'$, we get
\begin{eqnarray}\label{eq:Fockmatrixelements}
\braket{n k|V_{m}^{F}|n' k'}&=&\delta_{kk'}\delta_{nn'}F_{nm}\\
\nonumber F_{nm}&=&\frac{1}{l^2_B}\int_0^{\infty}~\mathrm{d}r~ r ~V(r)L_{n}\left(\frac{r^2}{2l^2_B}\right)L_{m}\left(\frac{r^2}{2l^2_B}\right)e^{-\frac{r^2}{2l^2_B}}
\end{eqnarray}
\end{widetext}
Note that $F_{mn}=F_{nm}$. Amazingly, the Hartree and Fock potentials of Eq. (\ref{eq:HFpotential}) are diagonal in the magnetic base of the eigenfunctions of Eq. (\ref{eq:magneticwavefunctions}), which implies that the self-consistent orbital wave functions are indeed the same as the non-interacting ones. Moreover, from Eq. (\ref{eq:2DNSCHFeqs}), it is easy to check that this result also holds for the spin part. Thus, the complete self-consistent wave functions are equal to the non-interacting ones.

It is useful to reproduce the previous results in Fourier space by writing the matrix elements of the Hartree and Fock potentials in terms of the matrix elements of the potential $V$ in Eq. (\ref{eq:scalarmatrixelements}). For instance, for the Hartree potential we trivially find:
\begin{equation}
\braket{n k|V_{m}^{H}|n' k'}=\sum_p V^{nn'mm}_{kk'pp}=\frac{V(0)}{2\pi l^2_B}\delta_{kk'}\delta_{nn'}
\end{equation}
On the other hand, for the generalized Fock potential of Eq. (\ref{eq:Fockgeneralized}), we obtain
\begin{eqnarray}\label{eq:HFPotentialFourier}
\braket{n k|V_{mm'}^{F}|n' k'}&=&\sum_p V^{nmm'n'}_{kppk'}\\
\nonumber &=&\frac{\delta_{kk'}}{(2\pi)^2}\int\mathrm{d}^2\mathbf{q}~V(\mathbf{q}) A^*_{mn}(\mathbf{q})A_{m'n'}(\mathbf{q})
\end{eqnarray}
Again, if we consider that the potential is rotationally invariant, its Fourier transform is also rotationally invariant and hence we recover the result of Eq. (\ref{eq:FockNonDiagonalmatrixelements}). In particular, for $m=m'$ we find
\begin{equation}\label{eq:FockFouriermatrixelement}
F_{nm}=\frac{1}{(2\pi)^2}\int\mathrm{d}^2\mathbf{q}~|A_{mn}(\mathbf{q})|^2V(\mathbf{q})
\end{equation}
Equations (\ref{eq:Fockmatrixelements}) and (\ref{eq:FockFouriermatrixelement}) are related to each other through the identity:
\begin{align}\label{eq:FFB}
& \int\frac{\mathrm{d}^2\mathbf{q}}{(2\pi)^2}~e^{i(q_xk_y-q_yk_x)l^2_B}A^*_{n_kn_j}(\mathbf{q})A_{n_ln_m}(\mathbf{q})\\
\nonumber & =\frac{1}{2\pi l^2_B}A^*_{n_kn_l}(\mathbf{k})A_{n_jn_m}(\mathbf{k})
\end{align}
which can be proven by inserting the definition of the magnetic form factors, Eq. (\ref{eq:defmagneticFF}).

\subsection{Diagonalization of the Hartree-Fock equations in graphene}\label{app:SCHF}

After the previous training, we obtain the HF solutions for the $\nu=0$ QH in bilayer graphene (the monolayer case is just a trivial extension of this calculation). We start by computing the eigenfunctions of the single-particle Hamiltonian (\ref{eq:spBHamiltonian}). Its matrix elements, $\braket{n k\alpha|H_0|n'k'\alpha'}$, with $\braket{\mathbf{x}|n k\alpha}=\Psi^{0}_{n,k,\alpha}(\mathbf{x})$ the wave functions of Eqs. (\ref{eq:Landaueigenfunctions})-(\ref{eq:ZLL}), are given by:
\begin{widetext}
\begin{eqnarray}\label{eq:spmnmn}
\nonumber\braket{n k\alpha|H_0|n'k'\alpha'}&=& \epsilon_n\delta_{nn'}\delta_{kk'}\delta_{\alpha\alpha'}-\epsilon_V\delta_{n,-n'}\delta_{kk'}(\tau_z)_{\alpha\alpha'}-\epsilon_Z\delta_{nn'}\delta_{kk'}(\sigma_z)_{\alpha\alpha'},~|n|\neq0,1\\
\braket{n k\alpha|H_0|n'k'\alpha'}&=&-\epsilon_V\delta_{nn'}\delta_{kk'}(\tau_z)_{\alpha\alpha'}-\epsilon_Z\delta_{nn'}\delta_{kk'}(\sigma_z)_{\alpha\alpha'},~n=0,1
\end{eqnarray}
\end{widetext}
with $(\tau_i)_{\alpha\alpha'}=\chi^{\dagger}_{\alpha}\tau_i\chi_{\alpha'}$. We see that the previous Hamiltonian is diagonal within the ZLL while it mixes the LLs $\pm n$ due to the layer voltage. The reason is that, outside the ZLL, the wave functions are not localized on one specific layer and thus they are able to experiment the effect of the voltage. However, as long as $\epsilon_V\ll \hbar\omega_B$, it is a good approximation to consider that $H_0$ is diagonal in every LL so their corresponding eigenfunctions are still given by $\Psi^{0}_{n,p,\alpha}(\mathbf{x})$. The associated eigenvalues $H_0\ket{nk\alpha}=\hbar\omega^{0}_{n,\alpha}\ket{nk\alpha}$ are then
\begin{eqnarray}\label{eq:spmnmneigenvalues}
\hbar\omega^{0}_{n,\alpha}&\simeq&\epsilon_n-\epsilon_Z\xi,~|n|\neq 0,1\\
\nonumber  \hbar\omega^{0}_{n,\alpha}&=&-\epsilon_Z\xi-\epsilon_V \lambda,~|n|= 0,1
\end{eqnarray}
where $\lambda,\xi$ label the valley and spin polarizations, $\lambda=\pm 1$ corresponding to $K,K'$ valleys. The non-interacting spinors are given by all possible orthogonal combinations of valley-spin polarizations, $\chi_{\alpha}^0=\ket{\pm n_z}\otimes\ket{\pm s_z}$; see Eq. (\ref{eq:Fphase}) for the notation of the wave-functions in valley-spin space. For the $\nu=0$ QH state, all the states with $n\leq-2$ are filled and empty for $n\geq 2$ while the ZLL is half-filled. As the non-interacting energy only depends on the polarization in valley-spin space, the two magnetic levels are filled in the same way in valley-spin space. Thus, following the notation of the main text, we label the occupied non-interacting spinors as $\chi^0_{a,b}$ and the empty ones as $\chi^0_{c,d}$. In particular, for $\epsilon_V<\epsilon_Z$, the ZLL is filled in the same F configuration of Eq. (\ref{eq:Fphase}) and for $\epsilon_V>\epsilon_Z$ it is filled in the FLP configuration of Eq. (\ref{eq:FLPphase}). Hence, in the non-interacting problem, these are the only two possible phases for the $\nu=0$ QH state.

We now address the effect of the interactions, given by Eqs. (\ref{eq:coulombHamiltonian}), (\ref{eq:srHamiltonian}). The resulting HF equations for the $\nu=0$ QH state are:

\begin{widetext}
\begin{eqnarray}\label{eq:HFeqsComplete}
\epsilon_{n,\alpha}\Psi_{n,k,\alpha}(\mathbf{x})&=&H_0\Psi_{n,k,\alpha}(\mathbf{x})-\sideset{}{'}\sum_{m=-\infty}^{\infty}\sum_{p,\beta}\nu_{m,\beta}\int\mathrm{d}^2\mathbf{x'}~V_0(\mathbf{x}-\mathbf{x'})\Psi_{m,p,\beta}(\mathbf{x})\Psi_{m,p,\beta}^{\dagger}(\mathbf{x'})\Psi_{n,k,\alpha}(\mathbf{x'})\\
\nonumber&+&\sideset{}{'}\sum_{i,j}g_{ij}\sideset{}{'}\sum_{m=-\infty}^{\infty}\sum_{p,\beta}\nu_{m,\beta}\left([\Psi_{m,p,\beta}^{\dagger}(\mathbf{x})T_{ij}\Psi_{m,p,\beta}(\mathbf{x})]T_{ij}\Psi_{n,k,\alpha}(\mathbf{x})- T_{ij}\Psi_{m,p,\beta}(\mathbf{x})\Psi_{m,p,\beta}^{\dagger}(\mathbf{x})T_{ij}\Psi_{n,k,\alpha}(\mathbf{x})\right)
\end{eqnarray}
\end{widetext}
where the Hartree term of the Coulomb potential is canceled by the positive charge background. As in the 2DEG case, we first consider the NSC problem in order to understand the structure of the equations.

\subsubsection{Non-self consistent problem}

The spatial part of the non-interacting Green's function is given by a $2\times 2$ matrix in the $\bar{A}\bar{B}$ subspace
\begin{equation}\label{eq:GreenBarefunctionSpatialDef}
K^{(2)}_n(\mathbf{x},\mathbf{x'})\equiv\sum_{p}\Psi^{0}_{n,p}(\mathbf{x})\Psi^{0\dagger}_{n,p}(\mathbf{x'})~.
\end{equation}
Operating in the same fashion as in the 2DEG case, we obtain:
\begin{widetext}
\begin{eqnarray}\label{eq:GreenBarefunctionSpatial}
K^{(2)}_n(\mathbf{x},\mathbf{x'})&=&\frac{1}{2}\left[\begin{array}{cc}
K_{|n|-2,|n|-2}(\mathbf{x},\mathbf{x'})& \textrm{sgn}\ n\ K_{|n|-2,|n|}(\mathbf{x},\mathbf{x'})\\
\textrm{sgn}\ n\  K_{|n|,|n|-2}(\mathbf{x},\mathbf{x'}) & K_{|n|,|n|}(\mathbf{x},\mathbf{x'})
\end{array}\right],~|n|\neq 0,1\\
\nonumber K^{(2)}_n(\mathbf{x},\mathbf{x'})&=&\left[\begin{array}{cc}
0 & 0\\
0 & K_{n}(\mathbf{x},\mathbf{x'})
\end{array}\right],~n=0,1
\end{eqnarray}
\end{widetext}
In particular, for $\mathbf{x}=\mathbf{x'}$,  $K^{(2)}_n(\mathbf{x},\mathbf{x'})=(4\pi l^2_B)^{-1}\text{diag}[1,1]$ for $|n|\neq 0,1$ and $K^{(2)}_n(\mathbf{x},\mathbf{x'})=(2\pi l^2_B)^{-1}P_{\bar{B}}$ for $n=0,1$, $P_{\bar{B}}=\text{diag}[0,1]$ being the projector onto the subspace $\bar{B}$.

The matrix $K^{(2)}_n(\mathbf{x},\mathbf{x'})$ is key to understand the orbital structure of the HF equations. With that finality, we write the NSC version of the HF equations (\ref{eq:HFeqsComplete}) as:
\begin{equation}\label{eq:meanfieldHamiltonian}
\epsilon_{n,\alpha}\Psi_{n,k,\alpha}(\mathbf{x})=\int\mathrm{d}^2\mathbf{x'}~H_{NSCHF}(\mathbf{x},\mathbf{x'})\Psi_{n,k,\alpha}(\mathbf{x'})
\end{equation}
and separate the different contributions to the {\it non-self-consistent} mean-field Hamiltonian $H_{NSCHF}$:
\begin{equation}\label{eq:meanfieldDSHamiltonian}
H_{NSCHF}=H_0+H_{NSCDS}+H_{NSCZLL}
\end{equation}
where $H_0$ is just the non-interacting Hamiltonian. The term $H_{NSCDS}$ represents the NSC mean-field potential created by the Dirac sea, formed by all the non-interacting states with $n\leq -2$,
\begin{eqnarray}\label{eq:DiracZLLHFHamiltonian}
\nonumber H_{NSCDS}(\mathbf{x},\mathbf{x'})&=&-\sum^{\infty}_{m=2}\left[(V_0)^{(2)}_{-m}(\mathbf{x},\mathbf{x'})+\bar{u}\delta(\mathbf{x}-\mathbf{x'})\right]\\
\nonumber (V_0)^{(2)}_{m}(\mathbf{x},\mathbf{x'})&\equiv&V_0(\mathbf{x}-\mathbf{x'})K^{(2)}_m(\mathbf{x},\mathbf{x'})\\
\bar{u}&=&\sideset{}{'}\sum_{i,j}\frac{g_{ij}}{4\pi l^2_B}
\end{eqnarray}
Since all valley-spin polarizations are occupied in the Dirac sea, the Hartree term of the short-range interactions vanishes and the corresponding Fock term is just an scalar. The last term of Eq. (\ref{eq:meanfieldDSHamiltonian}) corresponds to the NSC mean-field potential created by the filled states of the ZLL:
\begin{widetext}
\begin{equation}\label{eq:NSCZLLHFHamiltonian}
H_{NSCZLL}(\mathbf{x},\mathbf{x'})=-\sum_{m=0,1}(V_0)^{(2)}_{m}(\mathbf{x},\mathbf{x'})P^0+\left[\sum_{i}u_{i}\left([\text{tr}(P^0\tau_{i})]\tau_{i}-\tau_{i}P^0\tau_{i}\right)P_{\bar{B}}
-v_{i}\tau_{i}P^0\tau_{i}P_{\bar{A}}\right]\delta(\mathbf{x}-\mathbf{x'})
\end{equation}
with $P^0=\chi^0_{a}\chi^{0\dagger}_{a}+\chi^0_{b}\chi^{0\dagger}_{b}$ the analog of the matrix $P$ [defined after Eq. (\ref{eq:HFenergy})] for the non-interacting spinors, $P_{\bar{A}}=\text{diag}[1,0]$ the projector onto the subspace $\bar{A}$ and $v_{i}=(g_{ix}+g_{iy})/\pi l^2_B$. The matrix elements of the Fock contribution of the Coulomb potential are
\begin{eqnarray}\label{eq:GraphenegeneralFockelements}
\nonumber \braket{n k|(V_0)^{(2)}_{m}|n'k'} &\equiv& \int\mathrm{d}^2\mathbf{x}\mathrm{d}^2\mathbf{x'}~ \Psi^{0\dagger}_{n,k}(\mathbf{x})(V_0)^{(2)}_{m}(\mathbf{x},\mathbf{x'})\Psi^0_{n',k'}(\mathbf{x'})=\left[C_{nm}\delta_{n,n'}+D_{nm}\delta_{n,-n'}\right]\delta_{kk'},~|n|\neq 0,1\\
\braket{n k|(V_0)^{(2)}_{m}|n'k'}&=&\frac{1+\delta_{m,0}+\delta_{m,1}}{2}F_{nm}\delta_{nn'}\delta_{kk'},~n=0,1~,
\end{eqnarray}
\end{widetext}
with $C_{nm},D_{nm}$ some coefficients that can be expressed in terms of the values of the generalized matrix elements of Eq. (\ref{eq:FockNonDiagonalmatrixelements}); their particular expression is not interesting for the current purposes. As we see, the Fock term of the Coulomb potential mixes $n$ with $-n$ outside the ZLL and it is diagonal within the ZLL, in the same fashion of the non-interacting Hamiltonian. It is easy to show that this behavior also persists in the matrix elements of the short-range terms; therefore, we conclude that the NSC Hamiltonian $H_{NSCHF}$ only couples the LLs $\pm n$ and leaves invariant the ZLL. Moreover, even when the mean-field potentials are self-consistently renormalized, this structure is preserved. Hence, the orbital part of the self-consistent wave functions $\Psi_{n,k,\alpha}$ is constructed from the non-interacting orbital wave functions $\Psi^0_{\pm n,k}$. In the following, we neglect the coupling between $\pm n$ LLs as we are assuming that interactions are weak enough to neglect LL mixing. A nice discussion about the coupling of the $\pm n$ LLs for strong Coulomb interactions is presented in Ref. \cite{Toke2013}. 

\subsubsection{Self-consistent problem}

We now switch to the actual {\it self-consistent} problem. We rewrite Eq. (\ref{eq:HFeqsComplete}) as previously done for the NSC problem
\begin{equation}\label{eq:meanfieldHamiltonian}
\epsilon_{n,\alpha}\Psi_{n,k,\alpha}(\mathbf{x})=\int\mathrm{d}^2\mathbf{x'}~H_{HF}(\mathbf{x},\mathbf{x'})\Psi_{n,k,\alpha}(\mathbf{x'})
\end{equation}
and separate the different contributions along the same lines of Eq. (\ref{eq:meanfieldDSHamiltonian})
\begin{equation}\label{eq:meanfieldDSHamiltonianSC}
H_{HF}=H_0+H_{DS}+H_{ZLL}
\end{equation}
As we are neglecting LL mixing, we can regard the Dirac sea as frozen and hence the self-consistent HF mean-field potential created by the Dirac sea, $H_{DS}$, is the same as the NSC one of Eq. (\ref{eq:DiracZLLHFHamiltonian}), $H_{DS}=H_{NSCDS}$. With respect to the self-consistent mean-field potential created by the filled states of the ZLL, Eq. (\ref{eq:NSCZLLHFHamiltonian}), we only have to replace the NSC projector $P^0$ by the self-consistent one, $P$, as the filling of the ZLL is of the same form [see discussion after Eq. (\ref{eq:HFeqs})]. Hence, we focus on the ZLL in order to determine the {\it self-consistent} valley-spin wave functions of the ZLL.

For that purpose, since $H_{HF}$ leaves invariant the ZLL, we only need to take the components of $H_{HF}$ in the sublattice $\bar{B}$, obtaining the projected HF Hamiltonian $H^{(0)}_{HF}$ for the ZLL, $H^{(0)}_{HF}=P_{\bar{B}}H_{HF}P_{\bar{B}}=H^{(0)}_{0}+H^{(0)}_{DS}+H^{(0)}_{ZLL}$. $H^{(0)}_{0}$ corresponds to just the layer-voltage and Zeeman terms, $H^{(0)}_{0}=-\epsilon_V\tau_z-\epsilon_Z\sigma_z$, as the kinetic energy is zero. The interaction of the ZLL with the inert Dirac sea can be taken into account by a scalar non-local potential $V_{DS}(\mathbf{x},\mathbf{x'})$,
\begin{widetext}
\begin{eqnarray}\label{eq:HFprojected}
V_{DS}(\mathbf{x},\mathbf{x'})&\equiv&H^{(0)}_{DS}(\mathbf{x},\mathbf{x'})=P_{\bar{B}}H_{DS}(\mathbf{x},\mathbf{x'})P_{\bar{B}}=-\sum^{\infty}_{m=2}~\left[\frac{1}{2}(V_0)^{F}_{m}(\mathbf{x},\mathbf{x'})+\bar{u}\delta(\mathbf{x}-\mathbf{x'})\right]
\end{eqnarray}
while the self-consistent interaction inside the ZLL gives rise to
\begin{eqnarray}
H^{(0)}_{ZLL}(\mathbf{x},\mathbf{x'})&=&-\sum_{m=0,1}(V_0)^{F}_{m}(\mathbf{x},\mathbf{x'})P+\sum_{i}u_{i}\left([\text{tr}(P\tau_{i})]\tau_{i}-\tau_{i}P\tau_{i}\right)\delta(\mathbf{x}-\mathbf{x'})
\end{eqnarray}
Note that, after projecting, the interaction with the Dirac sea corresponds to a one-body operator.

Then, the resulting HF equation for the ZLL
\begin{equation}\label{eq:meanfieldHamiltonian}
\epsilon_{n,\alpha}\Psi_{n,k,\alpha}(\mathbf{x})=\int\mathrm{d}^2\mathbf{x'}~H^{(0)}_{HF}(\mathbf{x},\mathbf{x'})\Psi_{n,k,\alpha}(\mathbf{x'})
\end{equation}
is completely equivalent to the HF equation (\ref{eq:HFeqs}) derived from the effective Hamiltonian (\ref{eq:EffectiveHamiltonian}) and gives
\begin{eqnarray}\label{eq:HFanalytical}
\nonumber\epsilon_{n,\alpha}\Psi_{n,k,\alpha}(\mathbf{x})&=&-\frac{1}{2}\sum^{\infty}_{m=2}\int\mathrm{d}^2\mathbf{x'}(V_0)^{F}_{m}(\mathbf{x},\mathbf{x'})\Psi_{n,k,\alpha}(\mathbf{x'})
-\sum_{m=0,1}\int\mathrm{d}^2\mathbf{x'}(V_0)^{F}_{m}(\mathbf{x},\mathbf{x'})P\Psi_{n,k,\alpha}(\mathbf{x'})\\
\nonumber &+&\sum_{i}u_{i}\left([\text{tr}(P\tau_{i})]\tau_{i}-\tau_{i}P\tau_{i}\right)\Psi_{n,k,\alpha}(\mathbf{x})
-\epsilon_V\tau_z\Psi_{n,k,\alpha}(\mathbf{x})-\epsilon_Z\sigma_z\Psi_{n,k,\alpha}(\mathbf{x})\\
\end{eqnarray}
where we have reabsorbed in the Hamiltonian the infinite energy shift provided by the sum of the term $\bar{u}\delta(\mathbf{x}-\mathbf{x'})$ in the expression of $V_{DS}(\mathbf{x},\mathbf{x'})$. Since the mean-field HF potentials are invariant under unitary transformations that leave invariant the subspace formed by the filled states [see Eq. (\ref{eq:TheoryHFEquations}) and related discussion], the self-consistent problem is still diagonal in the orbital part as the two magnetic levels $n=0,1$ are filled exactly in the same way. Thus, the self-consistent wave functions have the form $\Psi_{n,k,\alpha}(\mathbf{x})=\Psi^0_{n,k}(\mathbf{x})\chi_{\alpha}$.

In order to arrive at an equation for the spinor $\chi_{\alpha}$, we multiply by $\Psi^{0,\dagger}_{n,k}(\mathbf{x})$ and integrate in Eq. (\ref{eq:HFanalytical}), obtaining:
\begin{eqnarray}\label{eq:HFenergyDirac}
\epsilon_{n,\alpha}\chi_{\alpha}&=&-\frac{1}{2}\sum_{m=2}^{\infty}F_{nm}\chi_{\alpha}-(F_{n0}+F_{n1})P\chi_{\alpha}+\sum_{i}u_{i}\left([\text{tr}(P\tau_{i})]\tau_{i}-\tau_{i}P\tau_{i}\right)\chi_{\alpha}-\epsilon_V\tau_z\chi_{\alpha}-\epsilon_Z\sigma_z\chi_{\alpha}
\end{eqnarray}
\end{widetext}
with $F_{nm}$ the eigenvalues of the Fock potential associated to the Coulomb interaction, Eq. (\ref{eq:Fockmatrixelements}). The contribution from the mean-field interaction with the Dirac sea, which is the origin of the analog of the Lamb shift, corresponds to the first term at the r.h.s. of the above equation. Following the regularization procedure of Ref. \cite{Shizuya2012}, we rearrange the associated series as:
\begin{equation}\label{eq:Diraccontribution}
\sum_{m=2}^{\infty}F_{nm}=\sum_{m=0}^{\infty}F_{nm}-F_{n0}-F_{n1}
\end{equation}
The completeness relation
\begin{equation}\label{eq:completenessrelation}
\sum_{m=0}^{\infty}|A_{nm}(\mathbf{k})|^2=1~,
\end{equation}
can be proven from the definition of the magnetic form factors Eq. (\ref{eq:defmagneticFF}) and the completeness relation of the harmonic oscillator wave functions. Then, using Eqs. (\ref{eq:FockFouriermatrixelement}), (\ref{eq:completenessrelation}), we find that
\begin{equation}
\sum_{m=0}^{\infty}F_{nm}=\frac{1}{(2\pi)^2}\int\mathrm{d}^2\mathbf{q}~V_0(\mathbf{q})=V_0(\mathbf{x}=0)~,
\end{equation}
which represents a constant (infinite) energy shift that can be absorbed in the Hamiltonian, as that arising from the short-range interactions [see discussion below Eq. (\ref{eq:HFprojected})]. After this regularization, and by defining $F_{n}\equiv F_{n0}+F_{n1}$ as in the main text, we finally get Eq. (\ref{eq:HFenergy}).

The formalism developed in this section is also valid for screened Coulomb interactions if we just replace $V_0(k)$ by $\bar{V}(k)$ of Eq. (\ref{eq:Screened}). The expression for the screened Fock energies $\bar{F}_{nm}$ then takes the form:
\begin{equation}\label{eq:FockFouriermatrixelementscreened}
\bar{F}_{nm}=\frac{1}{(2\pi)^2}\int\mathrm{d}^2\mathbf{q}~|A_{mn}(\mathbf{q})|^2\bar{V}(\mathbf{q})=\hbar\omega_Bg_{nm}\left(\frac{F_C}{\hbar\omega_B}\right)
\end{equation}
with $g_{nm}(x)$ the dimensionless functions obtained from the integral
\begin{equation}
g_{nm}(x)=\int_0^{\infty}\mathrm{d}z~\frac{|A_{mn}\left(0,\frac{z}{l_B}\right)|^2}{\frac{1}{x}+N\frac{f(z)}{z}}
\end{equation}
In particular, the dimensionless function $g(x)$ associated to the screened Coulomb contribution to the transport gap, $\bar{F}_1=\bar{F}_{10}+\bar{F}_{11}$ [see Eq. (\ref{eq:dimensionlessgap})], reads:
\begin{equation}\label{eq:dimensionlessgaptotal}
g(x)=g_{10}(x)+g_{11}(x)=\int_0^{\infty}\mathrm{d}z~\frac{1-\frac{z^2}{2}+\frac{z^4}{4}}{\frac{1}{x}+N\frac{f(z)}{z}}e^{-\frac{z^2}{2}}
\end{equation}
For small $x$, one recovers the unscreened result for $g(x)$ while for large $x$, $g(x)$ diverges logarithmically since $f(z)\simeq \alpha z^2$ for small $z$.

\section{Time-dependent Hartree-Fock approximation}\label{app:TDHFA}

We review in this Appendix the basic theory of the TDHFA. First, we consider a variational approach for the wave function \cite{Thouless1961} and then we connect the results with a diagrammatic calculation of the correlation functions \cite{Kallin1984,Wang2002}, which gives identical results for the computation of the collective modes. Alternative approaches can also be seen in Refs. \cite{Giuliani2005,Negele2008}.

\subsection{Variational formalism}\label{subsec:TDHFAwavefunction}

\subsubsection{Stationary situation}\label{subsec:stationaryrev}

We start by reviewing the stationary HF equations in the usual case of a system of $N$ fermions governed by a second-quantization Hamiltonian of the form \cite{Fetter2003}:
\begin{equation}\label{eq:2ndHamiltonian}
\hat{H}=\sum_{l,k}(H_{sp})_{lk}\hat{c}^{\dagger}_l\hat{c}_k+\frac{1}{2}\sum_{l,k,j,m}V_{lk,jm}\hat{c}^{\dagger}_l\hat{c}^{\dagger}_j\hat{c}_m\hat{c}_k
\end{equation}
where the eigenfunctions of the single-particle Hamiltonian $H_{sp}$ are known. The indices $l,k,j,m$ label the states of an orthonormal basis of the single-particle Hilbert space. The stationary HF equations are obtained by looking for a Slater determinant that minimizes the expectation value of the Hamiltonian
\begin{equation}\label{eq:Slater}
\ket{\Psi_0}=\prod_{\lambda=1}^N\hat{c}^{\dagger}_{\lambda}\ket{0}~,
\end{equation}
$\lambda=1\ldots N$ being $N$ orthogonal states. Now, we consider small particle-hole perturbations around the single-particle states in the so-called Thouless parametrization \cite{Thouless1961}:
\begin{eqnarray}\label{eq:Slaterperturbed}
\ket{\Psi}&=&\prod_{\lambda=1}^N\left(\hat{c}^{\dagger}_{\lambda}+\sum_{\lambda}w_{\Lambda\lambda}\hat{c}^{\dagger}_{\Lambda}\right)\ket{0}\\
\nonumber &=&\prod_{\lambda=1}^N\left(\hat{c}^{\dagger}_{\lambda}+\sum_{\Lambda}w_{\Lambda\lambda}\hat{c}^{\dagger}_{\Lambda}\hat{c}_{\lambda}\hat{c}^{\dagger}_{\lambda}\right)\ket{0}\\
\nonumber &=&e^{\sum_{\Lambda\lambda}w_{\Lambda\lambda}\hat{c}^{\dagger}_{\Lambda}\hat{c}_{\lambda}}\ket{\Psi_0}
\end{eqnarray}
where the sum in $\Lambda$ runs over all the unoccupied states. Along this section, we will use lower Greek indices $\lambda,\sigma$ for labeling occupied levels and upper Greek indices $\Lambda,\Sigma$ for empty levels, while Latin indices $l,k,j,m$ will label arbitrary (filled or empty) states.

The condition for $\ket{\Psi_0}$ to be an extreme of the energy, $\bra{\Psi_0}\hat{H}\ket{\Psi_0}$, reads:
\begin{equation}\label{eq:Extreme}
\bra{\Psi_0}\hat{c}^{\dagger}_{\lambda}\hat{c}_{\Lambda}\hat{H}\ket{\Psi_0}=0
\end{equation}
which implies that $H^{HF}_{\Lambda\lambda}=0$, with
\begin{eqnarray}\label{eq:TheoryHartreeFockHamiltonianGeneral}
\hat{H}^{HF}&=&\sum_{l,k}H^{HF}_{lk}\hat{c}^{\dagger}_l\hat{c}_k\\
\nonumber H^{HF}_{lk}&=&(H_{sp})_{lk}+\sum_{\lambda}\left(V_{lk,\lambda\lambda}-V_{l\lambda,\lambda k}\right)~,
\end{eqnarray}
Thus, if $\ket{\Psi_0}$ is an extreme, the mean-field Hartree-Fock Hamiltonian $\hat{H}^{HF}$ created by $\ket{\Psi_0}$ cannot connect occupied and empty states. Moreover, as $\hat{H}^{HF}$ is invariant under unitary transformations that leave invariant the subspace spanned by the occupied and empty states, we can choose an orthonormal basis of the Hilbert space that diagonalizes $\hat{H}^{HF}$ so
\begin{eqnarray}\label{eq:TheoryHFEquations}
H^{HF}_{lk}&=&\epsilon_k\delta_{lk}\\
\nonumber \epsilon_k&=&(H_{sp})_{kk}+\sum_{\lambda}\left(V_{kk,\lambda\lambda}-V_{k\lambda,\lambda k}\right)
\end{eqnarray}
which corresponds to the usual HF equations for the self-consistent wave functions written in terms of matrix elements. The Hartree-Fock Hamiltonian is then
\begin{equation}
\hat{H}^{HF}=\sum_{k}\epsilon_k\hat{c}^{\dagger}_k\hat{c}_k=\sum_{\lambda}\epsilon_\lambda\hat{c}^{\dagger}_\lambda\hat{c}_\lambda+\sum_{\Lambda}\epsilon_\Lambda\hat{c}^{\dagger}_\Lambda\hat{c}_\Lambda
\end{equation}
The total energy of the state $\ket{\Psi_0}$ is:
\begin{equation}\label{eq:TheoryEnergyHF}
E_{HF}=\bra{\Psi_0}\hat{H}\ket{\Psi_0}=\sum_{\lambda}(H_{sp})_{\lambda\lambda}+\frac{1}{2}\sum_{\lambda,\sigma}\left(V_{\sigma\sigma,\lambda\lambda}-V_{\sigma\lambda,\lambda\sigma}\right)
\end{equation}
In order to check if $\ket{\Psi_0}$ is a true minimum of the expectation value of $\hat{H}$, we consider fluctuations around $\ket{\Psi_0}$ [as given by Eq. (\ref{eq:Slaterperturbed})] and keeping only terms up to second order in the coefficients $w_{\Lambda \lambda}$, we get the quadratic form
\begin{widetext}
\begin{eqnarray}\label{eq:TheoryLandauHF}
\bra{\Psi}(\hat{H}-E_{HF})\ket{\Psi}&=&\frac{1}{2}\sum_{\Lambda,\lambda,\Sigma,\sigma}w^{*}_{\Lambda\lambda}X_{\lambda\Lambda,\sigma\Sigma}w_{\Sigma\sigma}+w^{*}_{\Lambda\lambda}X_{\lambda\Lambda,\Sigma\sigma}w^{*}_{\Sigma\sigma}+
w_{\Lambda\lambda}X_{\Lambda\lambda,\sigma\Sigma}w_{\sigma\Sigma}+w_{\Lambda\lambda}X_{\Lambda\lambda,\Sigma\sigma}w^{*}_{\Sigma\sigma}\\
\nonumber &\equiv&\frac{1}{2}\mathbf{W}^{\dagger}X\mathbf{W}
\end{eqnarray}
with $\mathbf{W}$ being a column vector containing the coefficients $w_{\Lambda\lambda},w^{*}_{\Lambda\lambda}$,
\begin{equation}
\mathbf{W}=\left[\begin{array}{c}w_{\Lambda\lambda} \\ w^{*}_{\Lambda\lambda} \end{array}\right]
\end{equation}
The elements of the matrix $X$ are
\begin{eqnarray}\label{eq:TheoryXmatrixelementsParticleHole}
\nonumber X_{\lambda\Lambda,\sigma\Sigma}&=&\bra{\Psi_0}\hat{c}^{\dagger}_{\lambda}\hat{c}_{\Lambda}(\hat{H}-E_{HF})\hat{c}^{\dagger}_{\Sigma}\hat{c}_{\sigma}\ket{\Psi_0}=(\epsilon_{\Lambda}-\epsilon_{\lambda})\delta_{\lambda\sigma,\Lambda\Sigma}+V_{\Lambda\lambda,\sigma\Sigma}-V_{\sigma\lambda,\Lambda\Sigma}\\
\nonumber X_{\lambda\Lambda,\Sigma\sigma}&=&\bra{\Psi_0}\hat{c}^{\dagger}_{\lambda}\hat{c}_{\Lambda}\hat{c}^{\dagger}_{\sigma}\hat{c}_{\Sigma}(\hat{H}-E_{HF})\ket{\Psi_0}
=\bra{\Psi_0}\hat{c}^{\dagger}_{\lambda}\hat{c}_{\Lambda}\hat{c}^{\dagger}_{\sigma}\hat{c}_{\Sigma}\hat{H}\ket{\Psi_0}
=V_{\Lambda\lambda,\Sigma\sigma}-V_{\Sigma\lambda,\Lambda\sigma}\\
\nonumber X_{\Lambda\lambda,\sigma\Sigma}&=&X^{*}_{\sigma\Sigma,\Lambda\lambda}=X^{*}_{\lambda\Lambda,\Sigma\sigma}=\bra{\Psi_0}\hat{H}\hat{c}^{\dagger}_{\Lambda}\hat{c}_{\lambda}\hat{c}^{\dagger}_{\Sigma}\hat{c}_{\sigma}\ket{\Psi_0}\\
X_{\Lambda\lambda,\Sigma\sigma}&=&X^{*}_{\lambda\Lambda,\sigma\Sigma}
\end{eqnarray}
Then, if the matrix $X$ is positive definite, $\ket{\Psi_0}$ corresponds to a true local minimum of the expectation value of the Hamiltonian and hence it is an energetically stable solution. We will refer to the elements $X_{\lambda\Lambda,\sigma\Sigma}$, $X_{\Lambda\lambda,\Sigma\sigma}$ as normal and to the elements $X_{\lambda\Lambda,\Sigma\sigma}$, $X_{\Lambda\lambda,\sigma\Sigma}$ as anomalous since they are not given by a matrix elements between two particle-hole excitations but rather between the ground state and an excited state with two pairs of particle-holes.

For notational convenience, we rewrite the matrix $X$ in terms of arbitrary states $kl,jm$ as:
\begin{eqnarray}\label{eq:TheoryXmatrixelements}
X_{kl,jm}&=&(\nu_{k}-\nu_{l})(\epsilon_{l}-\epsilon_{k})\delta_{kj}\delta_{lm}+V_{lk,jm}-V_{jk,lm}~,
\end{eqnarray}
\end{widetext}
where $\nu_{k},\nu_{l}$ represent the number occupation of the state $k,l$. We keep in mind that the only valid matrix elements of $X$ in Eq. (\ref{eq:TheoryXmatrixelements}) are those with the pair index $kl$ corresponding to one level filled and one level empty, so $\nu_{k}-\nu_{l}=\pm 1$ if $k$ is filled (empty) and $l$ is empty (filled); this also applies for the pair $jm$. In the same fashion, we rewrite the components of the vector $\mathbf{W}$ as $W_{kl}=w_{lk}$ for $\nu_{k}-\nu_{l}=1$ and $W_{kl}=w^*_{kl}$ for $\nu_{k}-\nu_{l}=-1$.

Using this notation, we can rewrite the quadratic form of Eq. (\ref{eq:TheoryLandauHF}) in a more compact way:
\begin{equation}\label{eq:TheoryCD}
\mathbf{W}^{\dagger}X\mathbf{W}=\sideset{}{'}\sum_{kl}\sideset{}{'}\sum_{jm}W^{*}_{kl}X_{kl,jm}W_{jm}
\end{equation}
where $'$ denotes now that we only sum over the proper values of the pair indices $kl,jm$.

\subsubsection{Time-dependent situation}

The Schrödinger equation for the time evolution of the wave function can also be derived from a variational principle. In particular, a solution of the Schrödinger equation, $\ket{\Psi(t)}$, must be an extreme of the functional
\begin{equation}\label{eq:TDHFAFormalism}
L(t)=\bra{\Psi(t)}\hat{H}-i\hbar\frac{d}{dt}\ket{\Psi(t)}
\end{equation}
The equations for the TDHFA arise when imposing a time-dependent solution of the same form of Eq. (\ref{eq:Slaterperturbed}),
\begin{equation}\label{eq:TDHFAansatz}
\ket{\Psi(t)}=f(t)e^{-i\frac{E_{HF}}{\hbar}t}e^{\sum_{\Lambda\lambda}w_{\Lambda\lambda}(t)\hat{c}^{\dagger}_{\Lambda}\hat{c}_{\lambda}}\ket{\Psi_0}
\end{equation}
In that case, we find
\begin{eqnarray}\label{eq:TDHFAFunctional}
L(t)&\simeq&\frac{1}{2}\mathbf{W}^{\dagger}(t)X\mathbf{W}(t)|f(t)|^2\\
\nonumber &-&i\hbar f^*(t)\frac{d f}{d t}\left(1+\sum_{\Lambda\lambda}|w_{\Lambda\lambda}(t)|^2\right)\\
\nonumber &-&|f(t)|^2\sum_{\Lambda\lambda} w^*_{\Lambda\lambda}(t)i\hbar\frac{d w_{\Lambda\lambda}}{d t}
\end{eqnarray}
where we have only kept terms up to second order in $w_{\Lambda\lambda}$. The resulting equation of motion for $w_{\Lambda\lambda}$ is:
\begin{eqnarray}\label{eq:TDHFAEqMotion0}
\nonumber &~&|f|^2\sum_{\Sigma,\sigma}\left(X_{\lambda\Lambda,\sigma\Sigma}w_{\Sigma\sigma}+X_{\lambda\Lambda,\Sigma\sigma}w^{*}_{\Sigma\sigma}\right)=\\
&~&i\hbar f^*\frac{d f}{d t}w_{\Lambda\lambda}+|f|^2 i\hbar\frac{d w_{\Lambda\lambda}}{d t}
\end{eqnarray}
Combining this equation of motion with that of $f(t)$, it is straightforward to show that:
\begin{eqnarray}\label{eq:TDHFASuppEqs}
\frac{d}{dt}\left[|f|^2\left(1+\sum_{\Lambda\lambda}|w_{\Lambda\lambda}|^2\right)\right]&=&0\\
\nonumber f^*\frac{d f}{d t}-f\frac{d f^*}{d t}&=&0
\end{eqnarray}
Thus, $f^*\frac{d f}{d t}$ is quadratic in the coefficients $w_{\Lambda\lambda}$ and since we must consistently keep only the lowest order terms for the equation of motion for $w_{\Lambda\lambda}$, Eq. (\ref{eq:TDHFAEqMotion0}) reduces to the linear $f$-independent equation
\begin{equation}\label{eq:TDHFAEqMotion1}
\sum_{\Sigma,\sigma}X_{\lambda\Lambda,\sigma\Sigma}w_{\Sigma\sigma}+X_{\lambda\Lambda,\Sigma\sigma}w^{*}_{\Sigma\sigma}=i\hbar\frac{d w_{\Lambda\lambda}}{d t}
\end{equation}

If we now perform a linear expansion in modes for $w_{\Lambda\lambda}$,
\begin{eqnarray}\label{eq:BdGmodesexpansionLarge}
w_{\Lambda\lambda}(t)&=&\sum_n \gamma_n u_{n,\Lambda\lambda}e^{-i\omega_nt}+\gamma^{*}_nv^*_{n,\Lambda\lambda}e^{i\omega_nt},~
\end{eqnarray}
$\gamma_n$ being the amplitude of each mode, we finally arrive at the equations for the TDHFA
\begin{eqnarray}\label{eq:TDHFABdGLarge}
\nonumber \sum_{\Sigma,\sigma}X_{\lambda\Lambda,\sigma\Sigma}u_{n,\Sigma\sigma}+X_{\lambda\Lambda,\Sigma\sigma}v_{n,\Sigma\sigma}&=&\hbar\omega_n u_{n,\Lambda\lambda}\\
\nonumber -\sum_{\Sigma,\sigma}X_{\Lambda\lambda,\sigma\Sigma}u_{n,\Sigma\sigma}+X_{\Lambda\lambda,\Sigma\sigma}v_{n,\Sigma\sigma}&=&\hbar\omega_n v_{n,\Lambda\lambda}\\
\end{eqnarray}
In matrix form, the previous equation simply reads as
\begin{equation}\label{eq:TDHFATontos}
\left[\begin{array}{c|c}
N & A \\
\hline
-A^{*} & - N^{*}
\end{array}\right]\left[\begin{array}{c}u_{n}\\ v_{n}\end{array}\right]=\hbar\omega\left[\begin{array}{c}u_{n}\\ v_{n}\end{array}\right]
\end{equation}
with $N,A$ denoting the normal, anomalous sectors of the matrix $X$. As $N$ is an Hermitian matrix, whenever the anomalous elements are non-zero, Eq. (\ref{eq:TDHFATontos}) is a non-Hermitian eigenvalue equation and thus, it can present complex eigenvalues. We also remark that the $u$ and $v$ components are mixed only for non-vanishing anomalous elements. It is worth noting the strong formal analogy of the above equations with the bosonic Bogoliubov-de Gennes equations \cite{Pitaevskii2003,Pethick2008}.

For convenience, we rewrite the above equations using the compact notation developed in Eqs. (\ref{eq:TheoryXmatrixelements}),(\ref{eq:TheoryCD}). In this way, Eq. (\ref{eq:TDHFAEqMotion1}) can be put as
\begin{equation}\label{eq:TDHFACompactTD}
\tilde{X}\mathbf{W}=i\hbar\frac{d\mathbf{W}}{dt},~\tilde{X}=\tilde{T}X
\end{equation}
with the elements of the matrix $\tilde{T}$ given by $\tilde{T}_{kl,jm}=(\nu_{k}-\nu_{l})\delta_{kj}\delta_{lm}$. The expansion in modes of Eq. (\ref{eq:BdGmodesexpansionLarge}) reads:
\begin{eqnarray}\label{eq:BdGmodesexpansion}
\mathbf{W}(t)&=&\sum_n \gamma_n \mathbf{Z}_ne^{-i\omega_nt}+\gamma^{*}_n\bar{\mathbf{Z}}_ne^{i\omega_nt}\\
\nonumber \mathbf{Z}_n&=&\left[\begin{array}{c}u_{n,\Lambda\lambda}\\ v_{n,\Lambda\lambda}\end{array}\right],\bar{\mathbf{Z}}_n=\left[\begin{array}{c}v^*_{n,\Lambda\lambda}\\ u^*_{n,\Lambda\lambda}\end{array}\right]
\end{eqnarray}
and the equations of the TDHFA, Eq. (\ref{eq:TDHFABdGLarge}), are simply
\begin{eqnarray}\label{eq:TDHFABdG}
\tilde{X}\mathbf{Z}_n=\hbar\omega_n\mathbf{Z}_n
\end{eqnarray}

Exploiting the analogies with the bosonic BdG equations, we find some interesting properties of the TDHFA equations. For instance, if $\mathbf{Z}_n$ is a mode with frequency $\omega_n$, then the conjugate $\bar{\mathbf{Z}}_n$ is as well a mode with eigenvalue $-\omega^*_n$.
\begin{eqnarray}\label{eq:TDHFAconjugate}
\tilde{X}\bar{\mathbf{Z}}_n=-\hbar\omega^*_n\bar{\mathbf{Z}}_n
\end{eqnarray}
The TDHFA equations also have an associated Klein-Gordon type scalar product given by:
\begin{equation}\label{eq:TDHFAKG}
(\mathbf{Z}_n|\mathbf{Z}_m)\equiv\mathbf{Z}^{\dagger}_n\tilde{T}\mathbf{Z}_m
\end{equation}
In particular, if $\mathbf{Z}_n,\mathbf{Z}_m$ are two eigenmodes of Eq. (\ref{eq:BdGmodesexpansion}), then
\begin{eqnarray}\label{eq:TDHFAKGProperty}
(\omega_m-\omega^*_n)(\mathbf{Z}_n|\mathbf{Z}_m)=0
\end{eqnarray}
from which follows that two modes with different eigenvalues are orthogonal according to this scalar product. Another important property is that the previous scalar product is not positive definite so the norm of a given solution $\mathbf{Z}_n$, defined as $(\mathbf{Z}_n|\mathbf{Z}_n)$, can be positive, negative or zero. In particular, the norm of the conjugate $\bar{\mathbf{Z}}_n$ has the opposite sign to that of $\mathbf{Z}_n$. It is immediately deduced from Eq. (\ref{eq:TDHFAKGProperty}) that any mode with complex frequency has zero norm. Also, as $\tilde{T}\tilde{X}^{\dagger}\tilde{T}=\tilde{X}$, the complex-frequency modes appear in pairs $(\omega_n,\omega_n^*)$.

Interestingly, we can further relate the energetic stability of the HF solution $\ket{\Psi_0}$ (discussed at the end of Sec. \ref{subsec:stationaryrev}) with the frequencies of the collective modes computed in the TDHFA through
\begin{equation}\label{eq:TDHFALandauDynSta}
\mathbf{Z}^{\dagger}_nX\mathbf{Z}_n=\hbar\omega_n(\mathbf{Z}_n|\mathbf{Z}_n)
\end{equation}

If the state $\ket{\Psi_0}$ truly minimizes the expectation value of the Hamiltonian, the matrix $X$ is positive definite and then there are no dynamical instabilities since that would imply the presence of a mode with zero norm, $(\mathbf{Z}_n|\mathbf{Z}_n)=0$. Therefore, the presence of dynamical instabilities is only possible if the state is energetically unstable.

Also, if $\ket{\Psi_0}$ is energetically stable, the modes with positive (negative) frequency have positive (negative) normalization. Thus, the presence of a mode with positive (negative) normalization and negative (positive) frequency reveals that the system is energetically unstable. Hence, in practice, we only need to compute the modes with positive norm since the rest of the modes are obtained through the conjugation $\mathbf{Z}_n\rightarrow \bar{\mathbf{Z}}_n$. Indeed, they are the two sides of the same coin as they both give rise to the same wave function.

More interesting properties appear in the presence of a continuous symmetry in the Hamiltonian. A continuous symmetry arises when the Hamiltonian is invariant under a continuous transformation of the form $\hat{U}^{\dagger}(\varphi)\hat{H}\hat{U}(\varphi)=\hat{H}$, with $\hat{U}(\varphi)=e^{-i\varphi \hat{L}}$, $\hat{L}$ being a Hermitian operator of the form
\begin{equation}\label{eq:Generator}
\hat{L}=\sum_{l,k}L_{lk}\hat{c}^{\dagger}_l\hat{c}_k
\end{equation}

As well known, the presence of such a symmetry gives rise to the appearance of a gapless Goldstone mode when the symmetry is spontaneously broken by the ground state. This result is recovered in the TDHFA by taking into account that the ground state is a solution of the HF equations. In that case, if we rewrite Eq. (\ref{eq:Extreme}) as
\begin{equation}\label{eq:ExtremeGoldstone}
\bra{\Psi_0}\hat{c}^{\dagger}_{\lambda}\hat{c}_{\Lambda}\hat{U}^{\dagger}(\varphi)\hat{H}\hat{U}(\varphi)\ket{\Psi_0}=0
\end{equation}
and expand the exponential for small $\varphi$, we find that a mode with zero frequency arises, $X\mathbf{Z}_G=0$, with
\begin{equation}\label{eq:TDHFAGoldstone}
\mathbf{Z}_G=\left[\begin{array}{c}iL_{\Lambda\lambda} \\ -iL^{*}_{\Lambda\lambda} \end{array}\right]
\end{equation}
$L_{\Lambda\lambda}$ being the matrix elements of $\hat{L}$ that connect filled and empty states. This gapless mode disappears whenever all the $L_{\Lambda\lambda}$ are zero; in that case, the unitary transformation given by $\hat{U}(\varphi)$ leaves invariant the state $\ket{\Psi_0}$ so the symmetry is not spontaneously broken.

The above argument also predicts the existence of such a gapless mode whenever the system presents a continuous {\it mean-field} symmetry, by which we denote a continuous transformation that is not a exact symmetry of the Hamiltonian, $\hat{U}^{\dagger}(\varphi)\hat{H}\hat{U}(\varphi)\neq\hat{H}$, but it satisfies Eq. (\ref{eq:ExtremeGoldstone}), which means that $\hat{U}(\varphi)\ket{\Psi_0}$ is also a mean-field ground state with the same energy of $\ket{\Psi_0}$.

Using the developed formalism, it is straightforward to compute the response function to a small perturbation introduced by an external field $H_{\rm{ext}}(t)$ within the TDHFA. Adding $H_{\rm{ext}}$ to the equilibrium Hamiltonian $\hat{H}$ in the functional of Eq. (\ref{eq:TDHFAFormalism}), using the same ansatz of Eq. (\ref{eq:TDHFAansatz}) and retaining only the lowest order terms in the amplitude of the external field gives an inhomogeneous version of Eq. (\ref{eq:TDHFACompactTD})
\begin{equation}\label{eq:TDHFACompactResponseFunction}
i\hbar\frac{d\mathbf{W}}{dt}=\tilde{X}\mathbf{W}+\mathbf{h},~\mathbf{h}_{kl}=(\nu_k-\nu_l)(H_{\rm{ext}})_{lk}
\end{equation}
Thus, to first order, the system only responds to the particle-hole matrix elements of the external field, perturbing only the expectation value of particle-hole operators. Then, if we consider the change to lowest order in the expectation value of a particular observable $A$ due to the coupling with an external field $H_{\rm{ext}}=B$, $\Delta A(t)=\sideset{}{'}\sum_{kl} A_{kl}W_{kl}(t)$, we find by taking the Fourier transform in the above expression that
\begin{equation}\label{eq:TDHFAMotherResponse}
\Delta A(\omega)=\mathbf{A}^{\dagger}\chi(\omega)\mathbf{B}(\omega),~\chi(\omega)=(\hbar\omega\tilde{T}-X)^{-1}
\end{equation}
with $\mathbf{A},\mathbf{B}$ vectors containing the particle-hole matrix elements of $A,B$, $\mathbf{A}_{kl}=A^{*}_{kl}$ and the same for $\mathbf{B}$. The function $\chi(\omega)$ is called the mother of all response functions \cite{Giuliani2005}. Its poles are given by the condition
\begin{equation}\label{eq:detcollmod}
\det~[X-\hbar\omega\tilde{T}]=0
\end{equation}
which after a trivial transformation gives the TDHFA eigenvalue equation $\det~[\tilde{X}-\hbar\omega]=0$, completely equivalent to Eq. (\ref{eq:TDHFABdG}). Hence, the poles of the different response functions are indeed the frequencies of the collective modes, as well known from the general theory of the response function \cite{Fetter2003}.

\subsubsection{TDHFA for the $\nu=0$ quantum Hall state in graphene}\label{app:TDHFAnu0}

We now apply the previous formalism to the study of the $\nu=0$ QH state in graphene, although many of the results presented proceed from the general theory of integer QH states. In that case, we take the Hamiltonian of Eq. (\ref{eq:2ndHamiltonian}) as the effective Hamiltonian of Eq. (\ref{eq:EffectiveHamiltonian}). The states of the Hilbert space are labeled by the dummy index $k$, which represents all the quantum numbers $(n_k,p_k,\alpha_k)$ at the same time, with $n_k$ the magnetic number, $p_k$ the $y$-momentum and $\alpha_k$ the polarization in valley-spin space. In particular, in our ZLL projected Hamitonian, $n_k=0,1$ and $\alpha_k=a,b,c,d$. The single-particle Hamiltonian $H_{sp}$ corresponds to the layer voltage, the Zeeman term and the mean-field interaction with the Dirac sea while the total effective potential corresponds to the Coulomb and short-range interactions. Since we are restricting to the ZLL, where the orbital part of the wave functions is proportional to the magnetic wave function $\phi_n$, the matrix elements of the total effective interaction potential [$V_{lk,jm}$ in Eq. (\ref{eq:2ndHamiltonian})] can be straightforwardly calculated with the help of Eq. (\ref{eq:scalarmatrixelements}) as
\begin{widetext}
\begin{equation}\label{eq:matrixelementsexplicit}
V_{lk,jm}=(V_0)^{n_ln_kn_jn_m}_{p_lp_kp_jp_m}\delta_{\alpha_l\alpha_k}\delta_{\alpha_j\alpha_m}
+(V_{sr})^{n_ln_kn_jn_m}_{p_lp_kp_jp_m}\sum_i g_i  (\tau_i)_{\alpha_l\alpha_k}(\tau_i)_{\alpha_j\alpha_m},~V_{sr}(\mathbf{x})\equiv\delta(\mathbf{x})
\end{equation}
\end{widetext}
The corresponding HF equations (\ref{eq:TheoryHFEquations}) read in terms of the self-consistent wave functions as Eq. (\ref{eq:HFeqs}). As discussed in the main text and in Appendix \ref{eq:2DHFeqs}, the orbital part of the self-consistent wave functions is equal to that of the non-interacting wave functions and the ground state $\ket{\Psi_0}$ is given by Eq. (\ref{eq:2ndHFsolution}).

With respect to the TDHFA equations (\ref{eq:TDHFABdG}), since the occupation number does not depend on the $y$-momentum for integer QH states, we perform the following transformation in momentum space for an arbitrary vector $\mathbf{Z}$:
\begin{eqnarray}\label{eq:magnetoexcitontransformation}
Z_{kl}(\mathbf{k})&\equiv& Z_{n_k\alpha_kn_l\alpha_l}(\mathbf{k})\\
\nonumber &=&\sqrt{\frac{1}{N_B}}\sum_{p_l,p_k}e^{i\frac{p_l+p_k}{2}k_xl^2_B}\delta_{p_l-p_k,k_y}Z_{kl},
\end{eqnarray}
so we get rid of the momentum coordinates and obtain a discrete matrix equation only in the magnetic and valley-spin indices:
\begin{equation}\label{eq:TDHFAmagnetoexciton}
\tilde{X}(\mathbf{k})\mathbf{Z}(\mathbf{k})=\hbar\omega(\mathbf{k})\mathbf{Z}(\mathbf{k}), ~\tilde{X}(\mathbf{k})=\tilde{T}X(\mathbf{k})
\end{equation}
$\mathbf{Z}(\mathbf{k})$ being a vector with components $Z_{kl}(\mathbf{k})$ and $\tilde{T}$ the correspondent version of the same matrix in only magnetic and valley-spin indices. The elements of the matrix $X(\mathbf{k})$, $X_{kl,jm}(\mathbf{k})\equiv X_{n_k\alpha_kn_l\alpha_l,n_j\alpha_jn_m\alpha_m}(\mathbf{k})$, are computed from
\begin{widetext}
\begin{equation}\label{eq:magnetoexcitondispersionmatrix}
\frac{1}{N_B}\sum_{p_l,p_k}\sum_{p_j,p_m}e^{i\frac{p_l+p_k}{2}k_xl^2_B}\delta_{p_l-p_k,k_y}X_{kl,jm}e^{-i\frac{p_j+p_m}{2}k'_xl^2_B}\delta_{p_m-p_j,k'_y}=X_{kl,jm}(\mathbf{k})\delta_{\mathbf{k},\mathbf{k'}}
\end{equation}
which gives:
\begin{eqnarray}\label{eq:ladderRPA}
X_{kl,jm}(\mathbf{k})&=&(\nu_k-\nu_l)(\epsilon_l-\epsilon_k)\delta_{kj}\delta_{lm}+W_{lk,jm}(\mathbf{k}),~W_{lk,jm}(\mathbf{k})=U^{RPA}_{lk,jm}(\mathbf{k})-U^{LAD}_{jk,lm}(\mathbf{k})\\
\nonumber U^{RPA}_{lk,jm}(\mathbf{k})&=&\frac{1}{2\pi l^2_B}A_{n_ln_k}(-\mathbf{k})A_{n_jn_m}(\mathbf{k})U_{\alpha_l\alpha_k,\alpha_j\alpha_m}(\mathbf{k})\\
\nonumber U^{LAD}_{jk,lm}(\mathbf{k})&=&
\int\frac{\mathrm{d}^2\mathbf{q}}{(2\pi)^2}~e^{i(q_xk_y-q_yk_x)l^2_B}A_{n_jn_k}(-\mathbf{q})A_{n_ln_m}(\mathbf{q})U_{\alpha_j\alpha_k,\alpha_l\alpha_m}(\mathbf{q})\\
\nonumber U_{\alpha_l\alpha_k,\alpha_j\alpha_m}(\mathbf{k})&=&
V_0(\mathbf{k})\delta_{\alpha_l\alpha_k}\delta_{\alpha_j\alpha_m}+\frac{4\pi\hbar^2}{m}\sum_i g_i (\tau_i)_{\alpha_l\alpha_k}(\tau_i)_{\alpha_j\alpha_m},
\end{eqnarray}
\end{widetext}
The energies $U^{RPA},U^{LAD}$ result from direct (exchange) interactions; see also discussion after Eq. (\ref{eq:calculationladderRPA}). The matrix $X(\mathbf{k})$ satisfies the properties:
\begin{eqnarray}
\label{eq:hermiticX}X_{kl,jm}(\mathbf{k})&=&X^{*}_{jm,kl}(\mathbf{k})\\
\label{eq:symmetryminusomega} X_{kl,jm}(\mathbf{k})&=&X^{*}_{lk,mj}(-\mathbf{k})
\end{eqnarray}
Following Ref. \cite{Wang2002}, we will refer to the matrix $X(\mathbf{k})$ as the dispersion matrix.

As a result, the transformation given by Eq. (\ref{eq:magnetoexcitondispersionmatrix}) diagonalizes the TDHFA equations so the wave vector $\mathbf{k}$ becomes a good quantum number that represents the momentum of the magnetoexciton wave function, Eq. (\ref{eq:magnetoexcitonwavefunction}). By defining $\hat{M}^{\dagger}_{kl}\equiv\hat{M}^{\dagger}_{n_{k}\alpha_{k}n_{l}\alpha_{l}}(\mathbf{k})$, the matrix elements of $X(\mathbf{k})$ can be easily expressed in terms of magnetoexciton matrix elements as
\begin{widetext}
\begin{eqnarray}\label{eq:magnetoexcitonmatrixelements}
\nonumber X_{\lambda\Lambda,\sigma\Sigma}(\mathbf{k})&=&\bra{\Psi_0}\hat{M}_{\Lambda\lambda}(\mathbf{k})\left(\hat{H}-E_{HF}\right)\hat{M}^{\dagger}_{\Sigma\sigma}(\mathbf{k})\ket{\Psi_0}\\
X_{\lambda\Lambda,\Sigma\sigma}(\mathbf{k})&=&\bra{\Psi_0}\hat{M}_{\Lambda\lambda}(\mathbf{k})\hat{M}_{\Sigma\sigma}(-\mathbf{k})\hat{H}\ket{\Psi_0}\\
\nonumber X_{\Lambda\lambda,\sigma\Sigma}(\mathbf{k})&=&X^{*}_{\sigma\Sigma,\Lambda\lambda}(\mathbf{k})=X^{*}_{\lambda\Lambda,\Sigma\sigma}(-\mathbf{k})=\bra{\Psi_0}\hat{H}\hat{M}^{\dagger}_{\Lambda\lambda}(-\mathbf{k})\hat{M}^{\dagger}_{\Sigma\sigma}(\mathbf{k})\ket{\Psi_0}\\
\nonumber X_{\Lambda\lambda,\Sigma\sigma}(\mathbf{k})&=&X^{*}_{\lambda\Lambda,\sigma\Sigma}(-\mathbf{k})=\bra{\Psi_0}\hat{M}_{\Sigma\sigma}(-\mathbf{k})\left(\hat{H}-E_{HF}\right)\hat{M}^{\dagger}_{\Lambda\lambda}(-\mathbf{k})\ket{\Psi_0}
\end{eqnarray}
with $X_{\Lambda\lambda,\sigma\Sigma}(\mathbf{k})$, $X_{\lambda\Lambda,\Sigma\sigma}(\mathbf{k})$ the anomalous matrix elements.

The wave functions associated to the collective modes computed from Eq. (\ref{eq:TDHFAmagnetoexciton}), $\ket{\Psi(\mathbf{k},t)}=\hat{M}^{\dagger}(\mathbf{k},t)\ket{\Psi_0}$, are characterized by the particle-hole operators $\hat{M}^{\dagger}(\mathbf{k},t)$, which are given in terms of linear combinations of magnetoexcitons
\begin{eqnarray}\label{eq:magnetoexcitontransformationwavefunctionTDHFA}
\hat{M}^{\dagger}(\mathbf{k},t)&=&\sum_{\Lambda,\lambda}u_{\Lambda\lambda}(\mathbf{k})\hat{M}^{\dagger}_{\Lambda\lambda}(\mathbf{k})e^{-i\omega(\mathbf{k})t}+
v^*_{\Lambda\lambda}(\mathbf{k})\hat{M}^{\dagger}_{\Lambda\lambda}(-\mathbf{k})e^{i\omega(\mathbf{k})t}\\
\nonumber u_{\Lambda\lambda}(\mathbf{k})&=&
\sqrt{\frac{1}{N_B}}\sum_{p_{\Lambda},p_{\lambda}}e^{i\frac{p_{\Lambda}+p_{\lambda}}{2}k_xl^2_B}\delta_{p_{\Lambda}-p_{\lambda},k_y}u_{\Lambda\lambda},~
v_{\Lambda\lambda}(\mathbf{k})=
\sqrt{\frac{1}{N_B}}\sum_{p_{\Lambda},p_{\lambda}}e^{i\frac{p_{\Lambda}+p_{\lambda}}{2}k_xl^2_B}\delta_{p_{\lambda}-p_{\Lambda},k_y}v_{\Lambda\lambda},
\end{eqnarray}
as can be seen from Eqs. (\ref{eq:BdGmodesexpansionLarge}), (\ref{eq:magnetoexcitontransformation}).
\end{widetext}

If all the $v_{\Lambda\lambda}$ coefficients are zero, the excitation is a pure magnetoexciton with momentum $\mathbf{k}$ while for non-vanishing $v_{\Lambda\lambda}$ the excitations are combinations of magnetoexcitons with momentum $\pm \mathbf{k}$. This can be understood from the fact that, as the $v$ components arise from the anomalous matrix elements (connecting the ground state with a state with two magnetoexcitons created), in order to conserve total momentum the two magnetoexcitons must sum total momentum zero, see Eq. (\ref{eq:magnetoexcitonmatrixelements}).

With respect to the response function, after inverting relation (\ref{eq:magnetoexcitondispersionmatrix}), one gets from Eq. (\ref{eq:TDHFAMotherResponse}) that
\begin{equation}\label{eq:TDHFAMotherResponseMagnetoexciton}
\Delta A(\omega)=\sum_{\mathbf{k}}\mathbf{A}^{\dagger}(\mathbf{k})\chi(\mathbf{k},\omega)\mathbf{B}(\mathbf{k},\omega)\\
\end{equation}
with the elements $\mathbf{A}_{kl}(\mathbf{k}),\mathbf{B}_{kl}(\mathbf{k},\omega)$ defined as in Eq. (\ref{eq:magnetoexcitontransformation}). The mother of all response functions is now diagonalized in momentum and reads $\chi(\mathbf{k},\omega)=\left[\hbar\omega\tilde{T}-X(\mathbf{k})\right]^{-1}$. Of special interest is the case where the operators $A,B$ are  of the form:
\begin{eqnarray}\label{eq:OperatorDensity}
\nonumber A&=&A(\mathbf{x})=\hat{\psi}^{\dagger}(\mathbf{x})\theta_A^{\dagger}\hat{\psi}(\mathbf{x})\\
B&=&B(\mathbf{x}')=A^{\dagger}(\mathbf{x}')
\end{eqnarray}
with $\theta_A$ some operator in valley-spin space, as they characterize the response of the system to the introduction of a perturbation in the charge, spin or interlayer density. After some manipulations, it is seen that
\begin{eqnarray}\label{eq:TDHFAMotherTotalResponse}
\nonumber \Delta A(\mathbf{x},\omega)&=&\int\frac{\mathrm{d}^2\mathbf{k}}{(2\pi)^2}~e^{i\mathbf{k}(\mathbf{x}-\mathbf{x}')}\Delta A(\mathbf{k},\omega)\\
\Delta A(\mathbf{k},\omega)&=&\frac{1}{2\pi l^2_B}
\mathbf{a}^{\dagger}(\mathbf{k})\chi(\mathbf{k},\omega)\mathbf{a}(\mathbf{k})\\
\nonumber \mathbf{a}_{kl}(\mathbf{k})&=&A^{*}_{n_kn_l}(\mathbf{k})\theta_{\alpha_l\alpha_k},~\theta_{\alpha_l\alpha_k}\equiv\chi^{\dagger}_{\alpha_l}\theta_A\chi_{\alpha_k}
\end{eqnarray}

\subsection{Diagrammatic formalism}\label{subsec:dyson}

\begin{figure*}
\includegraphics[width=\columnwidth]{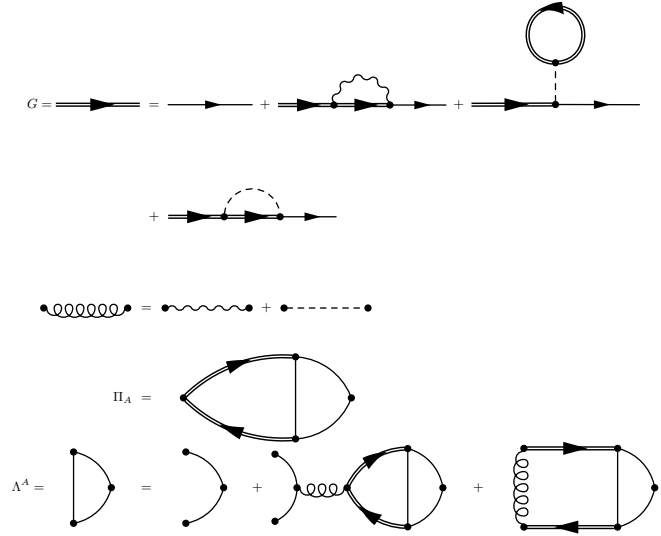}
\caption{Top: diagrammatic representation of the HF equations. The double (single) lines represent the dressed (bare) Green's function. The wiggly line represents the Coulomb interaction and the dashed line represents the short-range interactions. The direct (Hartree) term of the Coulomb interaction is suppressed by the uniform positive charge background.
Middle: diagrammatic representation of the total potential and the correlation function $\Pi_A$, Eq. (\ref{eq:corrfunc}).
Bottom: diagrammatic representation of the vertex equation in the TDHFA, Eq. (\ref{eq:vertexequation}).}\label{fig:Diagramas}
\end{figure*}

Following Refs. \cite{Kallin1984,Wang2002}, we rederive the results of the previous section using a diagrammatic expansion. For definiteness, we restrict from the beginning to the particular case of the $\nu=0$ QH state (although the extrapolation of the results for general integer QH states is immediate). The first step is to compute the self-consistent Green's function, whose equation is given by the diagrammatic representation of the first line of Fig. \ref{fig:Diagramas}. As well known \cite{Fetter2003}, this equation leads to the HF equations for the self-consistent wave functions, Eq. (\ref{eq:HFeqs}), in terms of which the Green's functions is written:
\begin{eqnarray}\label{eq:GreenZLL}
G(\mathbf{x},\mathbf{x'},\omega)&=&\sum_{k}\Psi_{k}(\mathbf{x})\Psi^{\dagger}_{k}(\mathbf{x'})G_{k}(\omega)\\
\nonumber G_k(\omega)&=&\frac{1-\nu_{k}}{\omega-\omega_{k}+i\eta}+\frac{\nu_{k}}{\omega-\omega_{k}-i\eta}
\end{eqnarray}
with $\eta=0^{+}$ and $\hbar\omega_{k}=\epsilon_{k}$ the HF energy. We follow here the notation introduced in Sec. \ref{app:TDHFAnu0} so the dummy index $k$ labels all the quantum numbers $(n_k,p_k,\alpha_k)$.

Since the corresponding formalism is closely related to that of the response function \cite{Fetter2003}, the collective-mode frequencies can also obtained from the poles of the set of correlation functions:
\begin{equation}\label{eq:corrfunc}
\Pi_A(x,x')=-i\braket{T\{\Delta[\hat{\psi}^{\dagger}(x)\theta_A^{\dagger}\hat{\psi}(x)]\Delta[\hat{\psi}^{\dagger}(x')\theta_A\hat{\psi}(x')]\}}
\end{equation}
where $x=(\mathbf{x},t)$, $T$ denotes time-ordering, all the expectation values are evaluated in the ground state of the system in the Heisenberg picture and $\Delta \hat{O}=\hat{O}-\braket{\hat{O}}$ denotes the fluctuations of an operator around its mean value.

The Fourier transform of the correlation function gives:
\begin{eqnarray}\label{eq:FTcorrfunc}
\nonumber \Pi_A(\mathbf{k},\omega)&=&\int\mathrm{d}^3x~e^{-ikx}\Pi_A(x,0)
\\ \mathrm{d}^3x&\equiv& \mathrm{d}^2\mathbf{x}\mathrm{d}t,~kx=\mathbf{k}\mathbf{x}-\omega t
\end{eqnarray}
after taking into account that the correlation function is invariant under time translations (due to the fact that the Hamiltonian is time independent) and also under spatial translations (this last property is shown explicitly later).
\begin{widetext}
We now proceed to compute the previous correlation functions within the diagrammatic version of the TDHFA, in which the equation for $\Pi_A(\mathbf{k},\omega)$ reads
\begin{equation}\label{eq:FTcorrvertex}
\Pi_A(\mathbf{k},\omega)=-\frac{i}{\hbar}\sum_{k,l}\int\frac{d\omega'}{2\pi}~\frac{1}{S}\sum_{\mathbf{k'}}\Lambda^{A(0)*}_{kl}(\mathbf{k},\omega)G_k(\omega')G_l(\omega+\omega')\Lambda^A_{kl}(\mathbf{k'},\omega)
\end{equation}
where we have introduced the dressed vertex function $\Lambda^A_{kl}(\mathbf{k'},\omega)$. In the above equation, we have made use of the identity $\sum_{\mathbf{k'}}e^{i\mathbf{k'}\mathbf{x'}}=S\delta(\mathbf{x'})$. We remark that the index $k$ in Eq. (\ref{eq:FTcorrvertex}) labels the quantum numbers $(n_k,p_k,\alpha_k)$ and not momentum. The expression for the non-interacting vertex $\Lambda^{A(0)}_{kl}(\mathbf{k},\omega)$ is:
\begin{equation}\label{eq:barevertex}
\Lambda^{A(0)}_{kl}(\mathbf{k},\omega)=\int \mathrm{d}^2\mathbf{x}~\Psi^{\dagger}_{l}(\mathbf{x})e^{i\mathbf{k}\mathbf{x}}\theta_A\Psi_{k}(\mathbf{x})=
e^{-ik_x\frac{p_k+p_l}{2}l^2_B}\delta_{p_l-p_k,k_y}A^*_{n_kn_l}(\mathbf{k})\theta_{\alpha_l\alpha_k}
\end{equation}
with $\theta_{\alpha_l\alpha_k}$ defined in Eq. (\ref{eq:TDHFAMotherTotalResponse}).

In the TDHFA, the Dyson's equation for the dressed vertex function is given by:
\begin{equation}\label{eq:vertexequation}
\Lambda^A_{kl}(\mathbf{k'},\omega)=\Lambda^{A(0)}_{kl}(\mathbf{k'},\omega)-\frac{i}{\hbar}\sum_{j,m}\left[V_{lk,jm}-V_{jk,lm}\right]\int\frac{d\omega'}{2\pi}~G_j(\omega')G_m(\omega+\omega')\Lambda^A_{jm}(\mathbf{k'},\omega)\\
\end{equation}
with $V_{lk,jm}$ the matrix elements of the total interaction potential, see Eq. (\ref{eq:matrixelementsexplicit}).

The diagrams for the total bare interaction, the correlation function and the vertex equation in the TDHFA are shown in Fig. \ref{fig:Diagramas}. Hence, within the TDHFA, the dressed vertex function is the bare vertex (first term) plus the series corresponding to bubble (second term) and ladder (third term) diagrams. The RPA approximation would correspond to just keep only the bubble diagrams. By defining the two-particle propagator
\begin{equation}\label{eq:electronholefunction}
i D_{kl}(\omega)\equiv \int\frac{d\omega'}{2\pi}~G_k(\omega')G_l(\omega+\omega')=i\left[\frac{\nu_k(1-\nu_l)}{\omega+\omega_k-\omega_l+i\eta}-\frac{\nu_l(1-\nu_k)}{\omega+\omega_k-\omega_l-i\eta}\right]~,
\end{equation}
we simplify Eqs. (\ref{eq:FTcorrvertex}), (\ref{eq:vertexequation}) as
\begin{eqnarray}\label{eq:FTvertex}
\hbar\Pi_A(\mathbf{k},\omega)&=&\frac{1}{S}\sum_{k,l}\sum_{\mathbf{k'}}\Lambda^{A(0)*}_{kl}(\mathbf{k},\omega)D_{kl}(\omega)\Lambda^A_{kl}(\mathbf{k'},\omega)\\
\nonumber \Lambda^A_{kl}(\mathbf{k'},\omega)&=&\Lambda^{A(0)}_{kl}(\mathbf{k'},\omega)+\frac{1}{\hbar}\sum_{j,m}\left[V_{lk,jm}-V_{jk,lm}\right]D_{jm}(\omega)\Lambda^A_{jm}(\mathbf{k'},\omega)
\end{eqnarray}
We note that the two-particle propagator $D_{kl}(\omega)$ does not depend on the momenta $p_k,p_l$ and it is only non-zero whenever the pair index $kl$ represents one filled level and one empty.

Further simplifications can be made by summing over all the momenta
\begin{eqnarray}\label{eq:Pidiagonalization}
\hbar\Pi_A(\mathbf{k},\omega)&=&\sum_{\substack{n_k,n_l\\ \alpha_k \alpha_l}}\sum_{\mathbf{k'}}~\theta^*_{\alpha_l\alpha_k}A_{n_kn_l}(\mathbf{k})D_{kl}(\omega)L^A_{kl}(\mathbf{k},\mathbf{k'},\omega)\\
\label{eq:vertexdiagonalization}
L^A_{kl}(\mathbf{k},\mathbf{k'},\omega)&\equiv&\frac{1}{S}\sum_{p_l,p_k}e^{i\frac{p_l+p_k}{2}k_xl^2_B}\delta_{p_l-p_k,k_y}\Lambda^A_{kl}(\mathbf{k'},\omega)
\end{eqnarray}
The function $L^A_{kl}$ only depends in $k,l$ through the magnetic level and the valley-spin polarization so, thanks to these manipulations, we get rid of the momentum indices and simplify the vertex equation. For instance, for the bare vertex, $\Lambda^{A(0)}_{kl}(\mathbf{k'},\omega)$, one finds
\begin{equation}\label{eq:zeroL}
L^{(0)A}_{kl}(\mathbf{k},\mathbf{k'},\omega)=\frac{1}{S}\sum_{p_l,p_k}e^{i\frac{p_l+p_k}{2}k_xl^2_B}\delta_{p_l-p_k,k_y} \Lambda^{A(0)}_{kl}(\mathbf{k'},\omega)=\frac{1}{2\pi l^2_B}\delta_{\mathbf{k},\mathbf{k'}}A^*_{n_kn_l}(\mathbf{k})\theta_{\alpha_l\alpha_k}
\end{equation}
while for the ladder and bubble diagrams, after using the expression for the matrix elements given in Eq. (\ref{eq:matrixelementsexplicit}),
\begin{equation}\label{eq:calculationladderRPA}
\frac{1}{S}\frac{1}{\hbar}\sum_{j,m}\sum_{p_l,p_k}e^{i\frac{p_l+p_k}{2}k_xl^2_B}\delta_{p_l-p_k,k_y}
\left[V_{lk,jm}-V_{jk,lm}\right]D_{jm}(\omega)\Lambda^A_{jm}(\mathbf{k'},\omega)=
\frac{1}{\hbar}\sum_{\substack{n_j,n_m\\ \alpha_j \alpha_m}}W_{lk,jm}(\mathbf{k})D_{jm}(\omega)L^A_{jm}(\mathbf{k},\mathbf{k'},\omega)
\end{equation}
\end{widetext}
with $W_{lk,jm}(\mathbf{k})$ given in Eq. (\ref{eq:ladderRPA}). From the above expression, it is easy to understand the notation of the quantities $U^{RPA},U^{LAD}$: they take into account the energy contribution from bubble and ladder diagrams, respectively \cite{Wang2002}. Since we have performed the summation in all the momentum dummy variables, from now on we use the index $k$ to label just the quantum numbers $(n_k,\alpha_k)$.

After inserting the expression for $\Lambda^A_{kl}(\mathbf{k},\omega)$ of Eq. (\ref{eq:FTvertex}) in Eq. (\ref{eq:vertexdiagonalization}) and using Eqs. (\ref{eq:zeroL})-(\ref{eq:calculationladderRPA}), one realizes that $L^A_{kl}(\mathbf{k},\mathbf{k'},\omega)$ is diagonal in $(\mathbf{k},\mathbf{k'})$. Hence, after defining a new dressed vertex function:
\begin{equation}\label{eq:newvertex}
L^A_{kl}(\mathbf{k},\mathbf{k'},\omega)\equiv \frac{1}{2\pi l^2_B}\delta_{\mathbf{k},\mathbf{k'}}\tilde{\Lambda}^{A}_{kl}(\mathbf{k},\omega)
\end{equation}
the original integral equation (\ref{eq:vertexequation}) for the dressed vertex is transformed into a discrete matrix equation:
\begin{eqnarray}\label{eq:diagonalvertexequation}
\tilde{\Lambda}^{A}_{kl}(\mathbf{k},\omega)&=&A^*_{n_kn_l}(\mathbf{k})\theta_{\alpha_l\alpha_k}\\
\nonumber &+&\frac{1}{\hbar}\sum_{jm}W_{lk,jm}(\mathbf{k})D_{jm}(\omega)\tilde{\Lambda}^{A}_{jm}(\mathbf{k},\omega)
\end{eqnarray}
where we remark that the labels $l,k,j,m$ now only represent the pair index corresponding to the magnetic level $n$ and the valley-spin polarization $\alpha$. The correlation function then reads:
\begin{equation}\label{eq:PiAfinal}
\hbar\Pi_A(\mathbf{k},\omega)=\frac{1}{2\pi l^2_B}\sum_{kl}~\theta^*_{\alpha_l\alpha_k}A_{n_kn_l}(\mathbf{k})D_{kl}(\omega)\tilde{\Lambda}^{A}_{kl}(\mathbf{k},\omega)
\end{equation}
The $\delta_{\mathbf{k},\mathbf{k'}}$ factor appearing in Eq. (\ref{eq:newvertex}) ensures the translational invariance of the correlation function previously assumed. In order to solve the matrix equation (\ref{eq:diagonalvertexequation}) we define:
\begin{eqnarray}
\label{eq:Polvector}\Pi^A_{kl}(\mathbf{k},\omega)&\equiv&\frac{1}{\hbar}D_{kl}(\omega)\tilde{\Lambda}^{A}_{kl}(\mathbf{k},\omega)\\\label{eq:dispersionmatrix}
X_{kl,jm}(\mathbf{k},\omega)&\equiv&W_{lk,jm}(\mathbf{k})-\delta_{kj}\delta_{lm} \hbar D^{-1}_{kl}(\omega)
\end{eqnarray}
and then
\begin{equation}
\sum_{jm} X_{kl,jm}(\mathbf{k},\omega)\Pi^A_{jm}(\mathbf{k},\omega)=-\mathbf{a}_{kl}(\mathbf{k})
\end{equation}
with $\mathbf{a}_{kl}$ given by Eq. (\ref{eq:TDHFAMotherTotalResponse}). According to Eqs. (\ref{eq:electronholefunction}), (\ref{eq:Polvector}) and (\ref{eq:dispersionmatrix}), the only valid matrix elements of $X_{kl,jm}(\mathbf{k},\omega)$ are those where the index pair $kl$ correspond to one level occupied and the other level empty (the same goes for $jm$). Therefore, $X_{kl,jm}(\mathbf{k},\omega=0)=X_{kl,jm}(\mathbf{k})$ is exactly the same of Eq. (\ref{eq:ladderRPA}).

Indeed, after rewriting Eq. (\ref{eq:PiAfinal}) as
\begin{eqnarray}\label{eq:Pi}
\nonumber \Pi_A(\mathbf{k},\omega)&=&-\frac{1}{2\pi l^2_B}\sideset{}{'}\sum_{kl}\sideset{}{'}\sum_{jm}~\mathbf{a}^{*}_{kl}(\mathbf{k})X^{-1}_{kl,jm}(\mathbf{k},\omega)\mathbf{a}_{jm}(\mathbf{k})\\
&=&-\frac{1}{2\pi l^2_B}\mathbf{a}^{\dagger}(\mathbf{k})X^{-1}(\mathbf{k},\omega)\mathbf{a}(\mathbf{k})
\end{eqnarray}
and following the discussion leading to Eq. (\ref{eq:detcollmod}), we find that the poles of the correlation function also give the collective-mode energies, as they are given by the condition $\det~[\tilde{X}(\mathbf{k})-\hbar\omega]=0$, which is the same eigenvalue problem of Eq. (\ref{eq:TDHFAmagnetoexciton}). In particular, note that Eqs. (\ref{eq:TDHFAMotherTotalResponse}), (\ref{eq:Pi}) are equivalent as $X^{-1}(\mathbf{k},\omega)=-\chi(\mathbf{k},\omega)$, revealing the close link between the correlation and the response functions mentioned at the beginning of the section.

Although the diagrammatic formalism may be more robust from a fundamental point of view, the form of the TDHFA equations obtained in Sec. \ref{app:TDHFAnu0} provides a simpler and clearer physical picture of the TDHFA since it allows us to understand the collective modes in terms of wave functions.

\section{Computation of the dispersion relation in the TDHFA}\label{app:analyticalTDHFA}

We devote this section to the computation of the solutions to the TDHFA equations for the $\nu=0$ QH state within the projected model considered in this work.

\subsection{Analytical results for the dispersion matrix}

We start by computing analytically the elements of the dispersion matrix, $X_{kl,jm}(\mathbf{k})$, given by Eq. (\ref{eq:ladderRPA}). In particular, we focus on the non-trivial many-body contribution arising from interactions, $W_{lk,jm}(\mathbf{k})$. In order to give its explicit expression, we separate the Coulomb from the short-ranged terms
\begin{equation}\label{eq:interactingcontributions}
W_{lk,jm}(\mathbf{k})=W^{C}_{lk,jm}(\mathbf{k})+W^{sr}_{lk,jm}(\mathbf{k})
\end{equation}
The Coulomb term, $W^{C}_{lk,jm}(\mathbf{k})$, has no direct (RPA) contribution as it vanishes when considering the proper elements of the dispersion matrix because in our model the empty levels have different valley-spin polarization from that of the filled ones. Thus, we only have to consider the exchange (ladder) contribution, which yields:

\begin{widetext}
\begin{eqnarray}\label{eq:Coulombmatrices}
\nonumber W^{C}_{lk,jm}(\mathbf{k})&=&-U^{C}_{n_jn_k,n_ln_m}(\mathbf{k})\delta_{\alpha_j\alpha_k}\delta_{\alpha_l\alpha_m}\\
U^{C}_{n_jn_k,n_ln_m}(\mathbf{k})&=&\int\frac{\mathrm{d}^2\mathbf{q}}{(2\pi)^2}~e^{i(q_xk_y-q_yk_x)l^2_B}A_{n_jn_k}(-\mathbf{q})A_{n_ln_m}(\mathbf{q})V_0(\mathbf{q})
\end{eqnarray}
These energies are given in terms of modified Bessel functions, as seen by computing the element $U^{C}_{00,00}(\mathbf{k})$. After changing to polar coordinates, taking into account that only the real part of the complex exponential survives, performing the integral over the radial coordinate and  using the integral representation of the modified Bessel functions,
\begin{equation}\label{eq:BesselModifiedintegral}
I_n(x)=\frac{1}{2\pi}\int_0^{2\pi}\mathrm{d}\varphi ~e^{x\cos\varphi}e^{-in\varphi}~,
\end{equation}
one finds that:
\begin{equation}\label{eq:Coulomb0000}
U^{C}_{00,00}(\mathbf{k})=F_{00}\frac{1}{2\pi}\int_0^{2\pi}\mathrm{d}\varphi ~e^{-\frac{(kl_B)^2\sin^2\varphi}{2}}=F_{00}I_0\left[\frac{(kl_B)^2}{4}\right]e^{-\frac{(kl_B)^2}{4}}
\end{equation}
with $F_{00}$ given by Eq. (\ref{eq:Fockeigenvalues}).

The other elements $U^{C}_{n_jn_k,n_ln_m}(\mathbf{k})$ can also be computed analytically. Instead of using the complicated expression found in the tables, we simply take into account that $A_{n_ln_m}(\mathbf{q})$ is a polynomial in $q_x,q_y$ multiplied by $e^{-\frac{(ql_B)^2}{4}}$. Then,
\begin{equation}
U^{C}_{n_jn_k,n_ln_m}(\mathbf{k})=\int\frac{\mathrm{d}^2\mathbf{q}}{(2\pi)^2}~P_{n_jn_k,n_ln_m}(q_x,q_y)e^{i(q_xk_y-q_yk_x)l^2_B}V_0(\mathbf{q})e^{-\frac{(ql_B)^2}{2}}
\end{equation}
with $P_{n_jn_k,n_ln_m}(q_x,q_y)$ some polynomial. For $U^{C}_{00,00}(\mathbf{k})$, $P_{00,00}(q_x,q_y)=1$. Therefore, from the usual properties of Fourier transforms, one has that
\begin{eqnarray}\label{eq:Coulombmatricesanalytical}
U^{C}_{n_jn_k,n_ln_m}(\mathbf{k})=P_{n_jn_k,n_ln_m}\left(-i\frac{\partial}{\partial k_y},i\frac{\partial}{\partial k_x}\right)U^{C}_{00,00}(\mathbf{k})
\end{eqnarray}
Since the Bessel functions are solutions of a second order differential equation, their higher order derivatives can always be put in terms of themselves and their first derivatives. For $I_0(x)$, $\frac{dI_0}{dx}=I_1(x)$, which means that the elements $U^{C}_{n_jn_k,n_ln_m}(\mathbf{k})$ are expressed through combinations of $I_0,I_1$ and polynomials in $k_x,k_y$. For instance, from Eq. (\ref{eq:magneticFFZLL}), we obtain
\begin{equation}
U^{C}_{10,00}=\frac{\frac{\partial}{\partial k_y}+i\frac{\partial}{\partial k_x}}{\sqrt{2}}U^{C}_{00,00}(\mathbf{k})=F_{00}\frac{k_y+ik_x}{2\sqrt{2}}\left(I_1\left[\frac{(kl_B)^2}{4}\right]-I_0\left[\frac{(kl_B)^2}{4}\right]\right)e^{-\frac{(kl_B)^2}{4}}
\end{equation}
and so on. For large $x$, a saddle-point approximation of the integral in Eq. (\ref{eq:BesselModifiedintegral}) gives $I_0(x)\sim\frac{e^x}{\sqrt {x}}$. Then, for large wave vector $k$, $kl_B\gg 1$, $U^{C}_{00,00}(\mathbf{k})\sim (kl_B)^{-1}$. As the other elements are obtained from derivatives of $U^{C}_{00,00}(\mathbf{k})$, the total matrix $U^C$ decays at least as $(kl_B)^{-1}$ for large $k$.

We now turn our attention to the short-range interactions. Using the property (\ref{eq:FFB}), one finds that the orbital part of RPA and ladder contributions is the same and then:
\begin{eqnarray}\label{eq:SRmatrices}
W^{sr}_{lk,jm}(\mathbf{k})&=&\frac{1}{2}A_{n_ln_k}(-\mathbf{k})A_{n_jn_m}(\mathbf{k})u^{\alpha_k\alpha_l,\alpha_j\alpha_m},
~u^{\alpha_k\alpha_l,\alpha_j\alpha_m}\equiv\sum_iu_i\left[(\tau_i)_{\alpha_l\alpha_k}(\tau_i)_{\alpha_j\alpha_m}
-(\tau_i)_{\alpha_j\alpha_k}(\tau_i)_{\alpha_l\alpha_m}\right]
\end{eqnarray}
Thus, all the valley-spin structure of the short-range interactions is captured by the effective couplings $u^{\alpha_k\alpha_l,\alpha_j\alpha_m}$, which present the following properties:
\begin{eqnarray}\label{eq:Melements} u^{\alpha_k\alpha_l,\alpha_j\alpha_m}&=&(u^{\alpha_j\alpha_m,\alpha_k\alpha_l})^*=(u^{\alpha_l\alpha_k,\alpha_m\alpha_j})^*=-u^{\alpha_k\alpha_j,\alpha_l\alpha_m}
\end{eqnarray}

In order to simplify the notation, we define a matrix $C(\mathbf{k})$ containing the exchange Coulomb energies with the following index ordering in orbital space:
\begin{equation}\label{eq:Coulombmagneticmatrix}
C(\mathbf{k})=\left[\begin{array}{cccc}
C_{00,00}&C_{00,01}&C_{00,10}&C_{00,11}\\
C_{01,00}&C_{01,01}&C_{01,10}&C_{01,11}\\
C_{10,00}&C_{10,01}&C_{10,10}&C_{10,11}\\
C_{11,00}&C_{11,01}&C_{11,10}&C_{11,11}\\
\end{array}\right],~C_{n_kn_l,n_jn_m}(\mathbf{k})\equiv U^{C}_{n_jn_k,n_ln_m}(\mathbf{k})
\end{equation}
We also define an analog matrix $R(\mathbf{k})$ for the short-range interactions
\begin{equation}\label{eq:SRmagneticmatrices}
R_{n_kn_l,n_jn_m}(\mathbf{k})\equiv \frac{1}{2}A_{n_ln_k}(-\mathbf{k})A_{n_jn_m}(\mathbf{k})= \frac{1}{2}A^*_{n_kn_l}(\mathbf{k})A_{n_jn_m}(\mathbf{k})
\end{equation}

Then, after taking into account that the only valid matrix elements of $X_{kl,jm}(\mathbf{k})$ are those with the pair indices $kl,jm$ corresponding to one level filled and one level empty and that the occupation number for the $\nu=0$ QH state only depends on the valley-spin polarization, it is seen that the matrix $\tilde{X}(\mathbf{k})$ of the TDHFA equations (\ref{eq:TDHFAmagnetoexciton}) is a $32\times 32$ matrix of the form:
\begin{equation}\label{eq:XYgen32}
\tilde{X}(\mathbf{k})=\left[\begin{array}{rrrr|rrrr}
Y^{ac,ac} & Y^{ac,ad} & Y^{ac,bc} & Y^{ac,bd} & Y^{ac,ca} & Y^{ac,da} & Y^{ac,cb} & Y^{ac,db}\\
Y^{ad,ac} & Y^{ad,ad} & Y^{ad,bc} & Y^{ad,bd} & Y^{ad,ca} & Y^{ad,da} & Y^{ad,cb} & Y^{ad,db}\\
Y^{bc,ac} & Y^{bc,ad} & Y^{bc,bc} & Y^{bc,bd} & Y^{bc,ca} & Y^{bc,da} & Y^{bc,cb} & Y^{bc,db}\\
Y^{bd,ac} & Y^{bd,ad} & Y^{bd,bc} & Y^{bd,bd} & Y^{bd,ca} & Y^{bd,da} & Y^{bd,cb} & Y^{bd,db}\\
\hline
- Y^{ca,ac} & - Y^{ca,ad} & - Y^{ca,bc} & - Y^{ca,bd} & - Y^{ca,ca} & - Y^{ca,da} & - Y^{ca,cb} & - Y^{ca,db}\\
- Y^{da,ac} & - Y^{da,ad} & - Y^{da,bc} & - Y^{da,bd} & - Y^{da,ca} & - Y^{da,da} & - Y^{da,cb} & - Y^{da,db}\\
- Y^{cb,ac} & - Y^{cb,ad} & - Y^{cb,bc} & - Y^{cb,bd} & - Y^{cb,ca} & - Y^{cb,da} & - Y^{cb,cb} & - Y^{cb,db}\\
- Y^{db,ac} & - Y^{db,ad} & - Y^{db,bc} & - Y^{db,bd} & - Y^{db,ca} & - Y^{db,da} & - Y^{db,cb} & - Y^{db,db}
\end{array}\right]~.
\end{equation}
where the $4\times 4$ matrices $Y^{\alpha_k\alpha_l,\alpha_j\alpha_m}(\mathbf{k})$ are the building blocks of $\tilde{X}(\mathbf{k})$ and represent the different valley-spin sectors. As discussed after Eq. (\ref{eq:TDHFALandauDynSta}), computing just the modes with positive norm is sufficient for characterizing the collective modes.

After some straightforward manipulations, the matrices $Y^{\alpha_k\alpha_l,\alpha_j\alpha_m}(\mathbf{k})$ can be written as:
\begin{eqnarray}\label{eq:pseudospinmatrices}
Y^{\alpha_k\alpha_l,\alpha_j\alpha_m}(\mathbf{k})&=&[\Delta^{\alpha_k\alpha_l}I+F(\mathbf{k})]\delta_{\alpha_k\alpha_j}\delta_{\alpha_l\alpha_m}+R(\mathbf{k})u^{\alpha_k\alpha_l,\alpha_j\alpha_m}\\
\nonumber \Delta^{\alpha_k\alpha_l}&=&(\nu_{\alpha_k}-\nu_{\alpha_l})(\epsilon_{\alpha_l}-\epsilon_{\alpha_k})=|\epsilon_{\alpha_l}-\epsilon_{\alpha_k}|,~
F(\mathbf{k})\equiv\text{diag}\left[F_0,\frac{F_{0}+F_{1}}{2},\frac{F_{0}+F_{1}}{2},F_{1}\right]-C(\mathbf{k})
\end{eqnarray}
We remind that the occupied states ($\nu_{\alpha}=1$) correspond to polarizations $\alpha=a,b$ while the empty ones ($\nu_{\alpha}=0$) correspond to $\alpha=c,d$ and that $I$ is the $4\times 4$ identity matrix. Between the square brackets of Eq. (\ref{eq:pseudospinmatrices}), $\Delta^{\alpha_k\alpha_l}$ represents the valley-spin part of the mean-field energy gap [see Eq. (\ref{eq:HFenergystructure})] while the matrix $F(\mathbf{k})$ contains all the terms involving Coulomb interactions, both mean-field and many-body contributions. At the same time, they are all multiplied by a diagonal tensor with respect to the valley-spin indices, so the different valley-spin sectors of $\tilde{X}$ are only connected through terms arising from short-range interactions, proportional to $u^{\alpha_k\alpha_l,\alpha_j\alpha_m}$. This fact simplifies notably the calculations since many of the effective couplings $u^{\alpha_k\alpha_l,\alpha_j\alpha_m}$ are related to each other through Eq. (\ref{eq:Melements}) and also some of them vanish due to symmetry considerations (see next section) and consequently we do not need to take into account the full $32\times 32$ problem. In particular, using Eq. (\ref{eq:Melements}), it is shown that the only independent and non-vanishing element in the anomalous sector [the off-diagonal boxes in Eq. (\ref{eq:XYgen32})] is $u^{ac,db}$.

Some analytical properties of the dispersion relation can be obtained from the above results. First, as explained in the main text, at $\mathbf{k}=0$ the total orbital pseudospin $O$ and its $z$-component $O_z$ are good quantum numbers. We can check this fact by making a unitary transformation in the matrix $\tilde{X}$ that switches to the singlet and triplet magnetoexciton base, $\hat{M}^{\dagger}_{OO_z,\alpha\alpha'}(\mathbf{k})$,
\begin{eqnarray}\label{eq:magnetoexcitontripletorbital}
\hat{M}^{\dagger}_{11,\alpha\alpha'}(\mathbf{k})&=&\hat{M}^{\dagger}_{1\alpha0\alpha'}(\mathbf{k})\\
\nonumber \hat{M}^{\dagger}_{10,\alpha\alpha'}(\mathbf{k})&=&\frac{1}{\sqrt{2}}\left[\hat{M}^{\dagger}_{1\alpha1\alpha'}(\mathbf{k})
-\hat{M}^{\dagger}_{0\alpha0\alpha'}(\mathbf{k})\right]\\
\nonumber \hat{M}^{\dagger}_{1-1,\alpha\alpha'}(\mathbf{k})&=&\hat{M}^{\dagger}_{0\alpha1\alpha'}(\mathbf{k})\\
\nonumber \hat{M}^{\dagger}_{00,\alpha\alpha'}(\mathbf{k})&=&\frac{1}{\sqrt{2}}\left[\hat{M}^{\dagger}_{1\alpha1\alpha'}(\mathbf{k})
+\hat{M}^{\dagger}_{0\alpha0\alpha'}(\mathbf{k})\right]
\end{eqnarray}
This transformation yields a simplified form at $\mathbf{k}=0$ for the matrices $Y^{\alpha_k\alpha_l,\alpha_j\alpha_m}$,
\begin{eqnarray}\label{eq:zerofrequencyequation}
Y^{\alpha_k\alpha_l,\alpha_j\alpha_m}(\mathbf{k}=0)&=&\left(\Delta^{\alpha_k\alpha_l}I+F\right)\delta_{\alpha_k\alpha_j}\delta_{\alpha_l\alpha_m}
+Ru^{\alpha_k\alpha_l,\alpha_j\alpha_m}\\
\nonumber F&=&\text{diag}\left[0,\frac{F_{00}+F_{11}}{2},\frac{F_{00}+F_{11}}{2},2F_{01}\right],~R=\text{diag}[1,0,0,0],
\end{eqnarray}
\end{widetext}
where the order of the magnetic indices for rows and columns corresponds to $OO_z=00,11,1-1,10$. All matrices are now diagonal in this new basis. In particular, the energy of the singlet $OO_z=00$ mode is independent of the Coulomb strength and is the only one that presents a non-trivial structure in valley-spin due to many-body contributions from short-range interactions, represented by the matrix $R$. This contrasts to the case of orbital-triplet modes, where the many-body corrections due to short-range interactions vanish and the collective-mode frequencies are immediately obtained from the diagonal of $\Delta^{\alpha_k\alpha_l}I+F$.

For $\mathbf{k}\neq 0$, we show that the resulting dispersion relation is isotropic. The key point is the polar structure of the magnetic form factors, shown in Eq. (\ref{eq:magneticFFspherical}). As $(k_x,k_y)=k(\sin \varphi_\mathbf{k},\cos \varphi_\mathbf{k})$,
\begin{equation}\label{eq:Rsphericalstructure}
R_{n_kn_l,n_jn_m}(\mathbf{k})=e^{i(n_l-n_k)\varphi_\mathbf{k}}e^{i(n_j-n_m)\varphi_\mathbf{k}}R_{n_kn_l,n_jn_m}(k)
\end{equation}
with $R_{n_kn_l,n_jn_m}(k)$ a function that only depends on $k=|\mathbf{k}|$ [not to be confused with the subindex $k$ that labels the quantum numbers $(n_k,\alpha_k)$]. For the Coulomb interaction, we switch to the following polar coordinates in the integral of Eq. (\ref{eq:Coulombmatrices}), $\mathbf{q}=q(\sin \varphi_\mathbf{q},\cos \varphi_\mathbf{q})$. This gives
\begin{widetext}
\begin{equation}
U^{C}_{n_jn_k,n_ln_m}(\mathbf{k})=\int_0^{\infty}\mathrm{d}q~\left(\frac{1}{2\pi}\int_0^{2 \pi}\mathrm{d}\varphi_\mathbf{q}~e^{iqkl^2_B\sin(\varphi_\mathbf{q}-\varphi_\mathbf{k})}e^{-i(n_k+n_m-n_j-n_l)\varphi_\mathbf{q}}\right)w_{n_jn_k,n_ln_m}(q)
\end{equation}
with $w_{n_jn_k,n_ln_m}(q)$ some function that only depends on $q$. The polar integral gives
\begin{equation}
\frac{1}{2\pi}\int_0^{2\pi}\mathrm{d}\varphi_\mathbf{q}~
e^{iqkl^2_B\sin(\varphi_\mathbf{q}-\varphi_\mathbf{k})}e^{-i(n_k+n_m-n_j-n_l)\varphi_\mathbf{q}}=e^{-i(n_k+n_m-n_j-n_l)\varphi_\mathbf{k}}J_{n_k+n_m-n_j-n_l}(qkl^2_B)
\end{equation}
\end{widetext}
This implies that the matrix $C(\mathbf{k})$ has the same dependence on the polar angle of Eq. (\ref{eq:Rsphericalstructure}). On the other hand, for the diagonal elements $n_kn_l=n_jn_m$ and the factor $e^{i(n_l-n_k)\varphi_\mathbf{k}}e^{i(n_j-n_m)\varphi_\mathbf{k}}$ becomes the unity. Hence, $\tilde{X}_{kl,jm}(\mathbf{k})$ is of the form
\begin{equation}\label{eq:isotropyphase}
\tilde{X}_{kl,jm}(\mathbf{k})=e^{-i(n_k-n_l)\varphi_\mathbf{k}}\tilde{X}_{kl,jm}(k)e^{i(n_j-n_m)\varphi_\mathbf{k}}
\end{equation}
with $X_{kl,jm}(k)$ depending only on $k=|\mathbf{k}|$. Then, after making an appropriated phase transformation in the magnetic indices we get rid of the dependence on the polar angle of the momentum and obtain a matrix $\tilde{X}_{kl,jm}(k)$ which explicitly depends solely on $k=|\mathbf{k}|$. Therefore, the resulting dispersion relation is isotropic, $\omega(\mathbf{k})=\omega(k)$.

We remark that the previous proof only relies on the fact that the Coulomb interaction is rotationally invariant and not on its particular form. Thus, this result applies for any rotationally invariant interaction potential; in particular, it holds for the screened interaction considered in Sec. \ref{sec:renorm}.

For completeness, we give the expression of the rotationally invariant matrices $F(k),R(k)$ that arise from the matrix $\tilde{X}(k)$ in Eq. (\ref{eq:isotropyphase}). For Coulomb interactions,
\begin{widetext}
\begin{eqnarray}\label{eq:Ckisotropic}
\nonumber C(k)&=&F_{00}e^{-\frac{(kl_B)^2}{4}}\left[\begin{array}{cccc}I_0 & \frac{kl_B}{2\sqrt{2}}\left(R_{01}\right) & -\frac{kl_B}{2\sqrt{2}}R_{01} & \frac{I_0}{2}-\frac{(kl_B)^2}{4}R_{01} \\
\frac{kl_B}{2\sqrt{2}}R_{01} &
\frac{I_0}{2}+\frac{(kl_B)^2}{4}R_{01} & \frac{I_1}{2}-\frac{(kl_B)^2}{4}R_{01} & \frac{kl_B}{4\sqrt{2}}\left[S_{01}-(kl_B)^2R_{01}\right] \\-\frac{kl_B}{2\sqrt{2}}R_{01} & \frac{I_1}{2}-\frac{(kl_B)^2}{4}R_{01} & \frac{I_0}{2}+\frac{(kl_B)^2}{4}R_{01}  & -\frac{kl_B}{4\sqrt{2}}\left[S_{01}-(kl_B)^2R_{01}\right] \\
\frac{I_0}{2}-\frac{(kl_B)^2}{4}R_{01} & \frac{kl_B}{4\sqrt{2}}\left[S_{01}-(kl_B)^2R_{01}\right] & -\frac{kl_B}{4\sqrt{2}}\left[S_{01}-(kl_B)^2R_{01}\right] & \frac{3}{4}I_0-\frac{(kl_B)^2}{4}I_0+\frac{(kl_B)^4}{8}R_{01} \\
\end{array}\right]\\
I_{0,1}&\equiv& I_{0,1}\left[\frac{(kl_B)^2}{4}\right],S_{01}\equiv I_0+I_1~,~R_{01}\equiv I_0-I_1
\end{eqnarray}
and $F(k)$ obtained through Eq. (\ref{eq:pseudospinmatrices}), while for short-range interactions
\begin{equation}\label{eq:Rkisotropic}
R(k)=r(k)a(k)a^{\dagger}(k),~a(k)\equiv\frac{1}{\sqrt{2\left(1+\frac{(kl_B)^4}{8}\right)}}\left[\begin{array}{c} 1\\ \frac{kl_B}{\sqrt{2}} \\ -\frac{kl_B}{\sqrt{2}} \\ 1-\frac{(kl_B)^2}{2}\\ \end{array}\right],~r(k)=e^{-\frac{(kl_B)^2}{2}}\left(1+\frac{(kl_B)^4}{8}\right)
\end{equation}
\end{widetext}
Note that the matrix $R(k)$ is a matrix of rank $1$ as it is proportional to the projector on the vector $a(k)$. Interestingly, from the above expressions it is immediate to check that the vector $[0,1,1,0]^T$ is an eigenvector of the matrices $F(k),R(k)$; in particular, it is orthogonal to $a(k)$. Hence, the following linear combination of magnetoexcitons with $O_z=\pm1$ and $\nu_{\alpha_k}-\nu_{\alpha_l}=1$,
\begin{equation}\label{eq:N2magneto} \hat{M}^{\dagger}_{\alpha_l\alpha_k}(\mathbf{k})=\frac{1}{\sqrt{2}}\left[e^{i\varphi_\mathbf{k}}\hat{M}^{\dagger}_{1\alpha_l,0\alpha_k}(\mathbf{k})+
e^{-i\varphi_\mathbf{k}}\hat{M}^{\dagger}_{0\alpha_l,1\alpha_k}(\mathbf{k})\right]
\end{equation}
is always an eigenmode of the TDHFA equations, which describes the $N=2$ orbital modes, with frequency
\begin{equation}\label{eq:N2mode}
\hbar\omega_2^{\alpha_l\alpha_k}(k)=\Delta^{\alpha_k\alpha_l}+F_{00}\left[\frac{11}{8}-\frac{I_0+I_1}{2}e^{-\frac{(kl_B)^2}{4}}\right]
\end{equation}
Note that the only dependence on the short-range interactions of this mode is contained in the valley-spin contribution of the single-particle gap, $\Delta^{\alpha_k\alpha_l}$, so the dependence in $k$ solely involves Coulomb interactions.

Finally, we note that considering screened Coulomb interactions only amounts to replace the matrix $F(k)$ by its screened version, $\bar{F}(k)$.

\subsection{Computation of the collective modes in bilayer graphene}\label{app:technical}

We now discuss the details of the computation of the collective modes for the different states of Eqs. (\ref{eq:Fphase}),(\ref{eq:FLPphase}),(\ref{eq:CAFphase}), (\ref{eq:PLPphase}). For all the phases, we give the independent non-vanishing coefficients $u^{\alpha_k\alpha_l,\alpha_j\alpha_m}$ that characterize the valley-spin structure, discuss the symmetries of the modes and the associated eigenvalue problem and compute the stiffness coefficients for all the modes.

\subsubsection{Ferromagnetic phase}

\begin{eqnarray}\label{eq:characterizationF}
\nonumber u^{ac,ac}&=&u^{bd,bd}=-u_z,~u^{ac,bd}=-2u_{\perp}\\
u^{bc,bc}&=&u^{ad,ad}=u_z
\end{eqnarray}
The problem is diagonalized in valley-spin space by considering magnetoexcitons with well defined valley pseudospin numbers, $\hat{M}^{\dagger}_{nn',LL_z}(\mathbf{k})$:
\begin{eqnarray}\label{eq:FMESIM}
\hat{M}^{\dagger}_{nn',11}(\mathbf{k})&=&\hat{M}^{\dagger}_{ncn'b}(\mathbf{k})\\
\nonumber \hat{M}^{\dagger}_{nn',1-1}(\mathbf{k})&=&\hat{M}^{\dagger}_{ndn'a}(\mathbf{k})\\
\nonumber \hat{M}^{\dagger}_{nn',10}(\mathbf{k})&=&\frac{1}{\sqrt{2}}\left[\hat{M}^{\dagger}_{ncn'a}(\mathbf{k})
-\hat{M}^{\dagger}_{ndn'b}(\mathbf{k})\right]\\
\nonumber \hat{M}^{\dagger}_{nn',00}(\mathbf{k})&=&\frac{1}{\sqrt{2}}\left[\hat{M}^{\dagger}_{ncn'a}(\mathbf{k})
+\hat{M}^{\dagger}_{ndn'b}(\mathbf{k})\right]
\end{eqnarray}
Since the ground state has well-defined quantum number for $S,S_z,L,L_z$ and all magnetoexcitons have $S=1,S_z=-1$, the anomalous matrix element vanish due to spin conservation and hence dynamical instabilities cannot appear in the ferromagnetic state. As a consequence, for a fixed value of $LL_z$, the frequency and the orbital structure of the modes (corresponding to the $N=0,1,2,3$ orbital modes) are computed from the $4\times4$ eigenvalue equation
\begin{eqnarray}\label{eq:FCollectivemodeequation}
Y^{LL_z}(k)Z_{LL_z}(k)&=&\hbar\omega^{LL_z}(k)Z_{LL_z}(k),\\
\nonumber Z_{LL_z}&=&\left[u_{00,LL_z},u_{10,LL_z},u_{01,LL_z},u_{11,LL_z}\right]^T
\end{eqnarray}
where we have made use of the isotropic expressions (\ref{eq:isotropyphase})-(\ref{eq:Rkisotropic}). All the matrices $Y^{\mu=LL_z}(k)$ present the same structure:
\begin{equation}\label{eq:TDHFAstructure44}
Y^{\mu}(k)=\Delta^{\mu}I+F(k)+u^{\mu}R(k)
\end{equation}
where the valley-spin gaps $\Delta^{\mu}$ and effective coupling energies $u^{\mu}$ are
\begin{eqnarray}\label{eq:TDHFAValleySpinGapsCouplings}
\nonumber \Delta^{00}&=&\Delta^{10}=\Delta^{ac},~\Delta^{11}=\Delta^{bc},~\Delta^{1-1}=\Delta^{ad}\\
\nonumber u^{L0}&=&u^{ac,ac}+(-1)^Lu^{ac,bd},~u^{11}=u^{1-1}=u^{bc,bc}\\
\end{eqnarray}
The previous relations imply that the dispersion relation for the triplet modes with $L_z=\pm 1$ is the same, only shifted by the layer voltage, $\omega^{1-1}(k)=\omega^{11}(k)+4\epsilon_V$, and that $Z_{1-1}(k)=Z_{11}(k)$. The wave function of each mode is created by the operator
\begin{equation}\label{eq:FME}
\hat{M}_{LL_z}^{\dagger}(\mathbf{k},t)=\sum_{n,n'}e^{i(n-n')\varphi_\mathbf{k}}u_{nn',LL_z}(k)\hat{M}^{\dagger}_{nn',LL_z}(\mathbf{k})e^{-i\omega(k)t}
\end{equation}

Using standard techniques it is straightforward to compute from Eq. (\ref{eq:TDHFAstructure44}) the stiffness coefficients of the different modes:
\begin{eqnarray}\label{eq:stiffnessexact}
\nonumber \rho^{\mu}_0&=&\left(\frac{23}{32}F_{00}-u^{\mu}-\frac{\left(\frac{3}{4}F_{00}-u^{\mu}\right)^2}{\frac{7}{4}F_{00}-2u^{\mu}}\right)l^2_B\\
\nonumber \rho^{\mu}_1&=&\left(-\frac{9}{16}F_{00}+\frac{u^{\mu}}{2}+\frac{\left(\frac{3}{4}F_{00}-u^{\mu}\right)^2}{\frac{7}{4}F_{00}-2u^{\mu}}\right)l^2_B\\
\rho^{\mu}_2&\equiv&\rho_2=\frac{1}{16}F_{00}l^2_B,~\rho^{\mu}_3\equiv\rho_3=\frac{7}{32}F_{00}l^2_B
\end{eqnarray}
The expression for the stiffness of the $N=2$ modes can also be obtained directly from Eq. (\ref{eq:N2mode}). Note that the stiffness coefficients of the modes $N=2,3$ are independent of short-range interactions and hence of the valley-spin symmetry of the mode; this result holds for all the phases. Remarkably, $\rho^{\mu}_1<0$ for every mode as the leading contribution goes as
\begin{equation}\label{eq:stiffness1leading}
\rho^{\mu}_1=-\left[\frac{27}{112}F_{00}-\frac{3}{294}u^{\mu}+O\left(\frac{u^{\mu}}{F_{00}}\right)\right]l^2_B\simeq-\frac{27}{112}F_{00}l^2_B
\end{equation}

\subsubsection{Full layer-polarized phase}\label{subsec:FLPcol}

\begin{eqnarray}\label{eq:characterizationFLP}
\nonumber u^{ac,ac}&=&u^{bd,bd}=2u_{\perp}+u_z,~u^{ac,bd}=2u_{\perp}\\
u^{bc,bc}&=&u^{ad,ad}=u_z
\end{eqnarray}

As this phase is the analog of the F phase in valley space, the results for the FLP phase are formally analog to those of the F phase and they are obtained by replacing $LL_z$ by $SS_z$ in Eqs. (\ref{eq:FMESIM})-(\ref{eq:stiffnessexact}). As mentioned in the main text, all triplet modes present the same dispersion relation, only shifted by the Zeeman energy, $\omega^{1\pm1}(k)=\omega^{10}(k)\mp2\epsilon_Z$ and $Z_{11}(k)=Z_{10}(k)=Z_{1-1}(k)$.

\subsubsection{Canted anti-ferromagnetic phase}\label{subsec:CAFcol}

\begin{eqnarray}\label{eq:characterizationCAF}
\nonumber u^{ac,ac}&=&u^{bd,bd}=-u_z,~u^{ac,bd}=-2u_{\perp}\cos^2 \theta_S\\
\nonumber u^{bc,bc}&=&u^{ad,ad}=u_z+2u_{\perp}\sin^2 \theta_S,\\
u^{ac,db}&=&2u_{\perp}\sin^2 \theta_S
\end{eqnarray}

Since the CAF state has only well defined value of the $z$ valley pseudospin, $L_z=0$, the anomalous element $u^{ac,db}$ becomes non-zero, so the collective modes are combinations of magnetoexcitons with $\pm \mathbf{k}$, see Eq. (\ref{eq:magnetoexcitontransformationwavefunctionTDHFA}) and related discussion. For the same reason, complex-frequency modes can appear.

The eigenvalue problem is split in sectors with fixed $L_z$. First, we consider the subspace with $L_z=0$, where we define:
\begin{eqnarray}\label{eq:CAFMESIM}
\nonumber \hat{M}^{\dagger}_{nn',+}(\mathbf{k})&=&\frac{1}{\sqrt{2}}\left[\hat{M}^{\dagger}_{ncn'a}(\mathbf{k})
+\hat{M}^{\dagger}_{ndn'b}(\mathbf{k})\right]\\
\nonumber \hat{M}^{\dagger}_{nn',-}(\mathbf{k})&=&\frac{1}{\sqrt{2}}\left[\hat{M}^{\dagger}_{ncn'a}(\mathbf{k})
-\hat{M}^{\dagger}_{ndn'b}(\mathbf{k})\right]\\
\end{eqnarray}
Note that, due to the different spin orientations of $s_a,s_b$, these modes have no well-defined value of $L$; only in the limit $\theta_S=0$ they match the proper $L=0,1$ modes. In this new basis, the eigenvalue problem is diagonalized in valley-spin space and reduced to two independent $8\times 8$ blocks:
\begin{widetext}
\begin{eqnarray}\label{eq:CAFCollectiveModeEquation}
\tilde{Y}^{\pm}(k)Z_{\pm}(k)&=&\hbar\omega^{\pm}(k)Z_{\pm}(k)\\
\nonumber Z_{\pm}&=&\left[u_{00,\pm},u_{10,\pm},u_{01,\pm},u_{11,\pm},v_{00,\pm},v_{01,\pm},v_{10,\pm},v_{11,\pm}\right]^T\\
\nonumber \tilde{Y}^{\pm}(k)&=&\left[\begin{array}{cc}
Y^{\pm}(k)  & \pm A(k) \\
\mp A(k) & -Y^{\pm}(k)   \\
\end{array}\right],
\end{eqnarray}
where the reader is advised to pay attention to the interchange of magnetic indices in the $v$ components of the vector $Z$. The matrices $Y^{\pm}(k)$ are given by Eq. (\ref{eq:TDHFAstructure44}) with the values $\Delta^{\pm}=\Delta^{ac}$ and $u^{\pm}=u^{ac,ac}\pm u^{ac,bd}$ and the matrix $A(k)$ characterizes the anomalous matrix elements of $\tilde{X}(k)$, $A(k)\equiv u^{ac,db} R(k)$.

The velocity of the Goldstone mode is computed, after some tedious but straightforward algebra [note that the eigenvalue problem of Eq. (\ref{eq:CAFCollectiveModeEquation}) at $k=0$ gives a non-diagonal Jordan canonical form for the orbital-singlet modes rather than the usual diagonal matrix], as
\begin{equation}\label{eq:Goldstonevelocity}
\frac{\hbar v_G}{l_B}=\sqrt{2A\left[\frac{23}{32}F_{00}+\Delta-\frac{\left(\frac{3}{4}F_{00}+\Delta\right)^2}{\frac{7}{4}F_{00}+2\Delta}\right]},~\Delta\equiv\Delta^{ac},~A\equiv|u^{ac,db}|
\end{equation}

The wave function of the modes is created by the operator:
\begin{eqnarray}\label{eq:CAFLz0magnetoexciton}
\nonumber \hat{M}_{\pm}^{\dagger}(\mathbf{k},t)&=&\sum_{n,n'}e^{i(n-n')\varphi_\mathbf{k}}\left[u_{nn',\pm}(k)\hat{M}^{\dagger}_{nn',\pm}(\mathbf{k})e^{-i\omega(k)t}+
v^*_{nn',\pm}(k)\hat{M}^{\dagger}_{nn',\pm}(-\mathbf{k})e^{i\omega(k)t}\right]\\
\end{eqnarray}

For the components with $L_z=\pm 1$, the eigenvalue problem reads:
\begin{eqnarray}\label{eq:CAFCollectiveModeEquationTriplet}
\nonumber \tilde{Y}^{1\pm1}(k)Z_{1\pm1}(k)&=&\hbar\omega^{1\pm1}(k)Z_{1\pm1}(k),\\
\nonumber Z_{1\pm1}&=&\left[u_{00,1\pm1},u_{10,1\pm1},u_{01,1\pm1},u_{11,1\pm1},v_{00,1\mp1},v_{01,1\mp1},v_{10,1\mp1},v_{11,1\mp1}\right]^T\\
\tilde{Y}^{1\pm 1}(k)&=&\left[\begin{array}{cc}
Y^{1\pm 1}(k) & -A(k)\\
A(k) & -Y^{1\mp 1}(k)
\end{array}\right],
\end{eqnarray}
where the expressions for $Y^{1\pm 1}(k)$ are the same as in the F state, and the operator associated to the wave function of the collective modes is:
\begin{eqnarray}\label{eq:CAFMETriplet}
\hat{M}_{1\pm1}^{\dagger}(\mathbf{k},t)&=&\sum_{n,n'}e^{i(n-n')\varphi_\mathbf{k}}\left[u_{nn',1\pm1}(k)\hat{M}^{\dagger}_{nn',1\pm1}(\mathbf{k})e^{-i\omega(k)t}+
v^*_{nn',1\mp1}(k)\hat{M}^{\dagger}_{nn',1\mp1}(-\mathbf{k})e^{i\omega(k)t}\right]
\end{eqnarray}
\end{widetext}
which we see that mixes magnetoexcitons with $\pm\mathbf{k},~L_z=\pm 1$ in order to conserve both total momentum and total value of $L_z$ in the anomalous matrix element. As in the F phase, $\omega^{1-1}_N(k)=\omega^{11}_N(k)+4\epsilon_V$ and $Z_{1-1}(k)=Z_{11}(k)$.

The expression of the stiffness coefficients for the orbital-singlet modes (when $\mu$ does not correspond to the Goldstone one) is:
\begin{widetext}
\begin{eqnarray}\label{eq:stiffnessanomalous}
\nonumber \rho^{\mu}_0&=&\left[(c^2_{\mu}+d^2_{\mu})\left(\frac{23}{32}F_{00}-u^{\mu}\right)-2c_{\mu}d_{\mu} A-
\frac{\left(c_{\mu}\left[\frac{3}{4}F_{00}-u^{\mu}\right]-d_{\mu}A\right)^2}{\frac{7}{4}F_{00}+2\Delta-2\sqrt{(u^\mu+\Delta)^2-A^2}}
-\frac{\left(d_{\mu}\left[\frac{3}{4}F_{00}-u^{\mu}\right]-c_{\mu}A\right)^2}{\frac{7}{4}F_{00}+2\Delta+2\sqrt{(u^\mu+\Delta)^2-A^2}}\right]l^2_B,\\
c_{\mu}&=&\frac{1}{2}\left[\left(\frac{u^\mu+\Delta-A}{u^\mu+\Delta+A}\right)^{\frac{1}{4}}+\left(\frac{u^\mu+\Delta+A}{u^\mu+\Delta-A}\right)^{\frac{1}{4}}\right],
~d_{\mu}=\frac{1}{2}\left[\left(\frac{u^\mu+\Delta-A}{u^\mu+\Delta+A}\right)^{\frac{1}{4}}-\left(\frac{u^\mu+\Delta+A}{u^\mu+\Delta-A}\right)^{\frac{1}{4}}\right]
\end{eqnarray}
For the $N=1$ modes, the expression varies if there is a Goldstone mode:
\begin{eqnarray}\label{eq:stiffnessnoGoldstone}
\nonumber \rho^{\mu}_1&=&\left[-\frac{9}{16}F_{00}+\frac{u^{\mu}}{2}+\frac{\left(c_{\mu}\left[\frac{3}{4}F_{00}-u^{\mu}\right]-d_{\mu}A\right)^2}{\frac{7}{4}F_{00}+2\Delta-2\sqrt{(u^\mu+\Delta)^2-A^2}}
-\frac{\left(d_{\mu}\left[\frac{3}{4}F_{00}-u^{\mu}\right]-c_{\mu}A\right)^2}{\frac{7}{4}F_{00}+2\Delta+2\sqrt{(u^\mu+\Delta)^2-A^2}}\right]l^2_B,~\mu=+,11,1-1\\
\label{eq:stiffnessGoldstone}\rho^{\mu}_1&=&\left[-\frac{9}{16}F_{00}+\frac{u^{\mu}}{2}+\frac{\left(\frac{3}{4}F_{00}-u^{\mu}\right)^2-A^2}{\frac{7}{4}F_{00}+2\Delta}
+2A\frac{\left(\frac{3}{4}F_{00}-u^{\mu}-A\right)^2}{\left(\frac{7}{4}F_{00}+2\Delta\right)^2}\right]l^2_B,~\mu=-,
\end{eqnarray}
\end{widetext}
We note that the leading contribution in the Coulomb interaction strength for $\rho^{\mu}_1$ still satisfies Eq. (\ref{eq:stiffness1leading}).


\subsubsection{Partially layer-polarized phase}\label{subsec:PLPcol}

\begin{eqnarray}\label{eq:characterizationPLP}
\nonumber u^{ac,ac}&=&u^{bd,bd}=u_z+2u_{\perp}\\
\nonumber u^{ac,bd}&=&2u_{\perp}+(u_z-u_{\perp})\sin^2\theta_V\\
\nonumber u^{bc,bc}&=&u^{ad,ad}=u_z-(u_z-u_{\perp})\sin^2\theta_V\\
u^{ac,db}&=&(u_z-u_{\perp})\sin^2\theta_V
\end{eqnarray}
The valley-spin structure of the modes is similar to that of the FLP phase and the problem is diagonalized by considering modes with well defined spin number $S,S_z$.
The main difference is that, as the Hamiltonian does not commute with the operator $\hat{L}_n$, the anomalous matrix element is non-zero.
\begin{widetext}
For the modes with $S_z=0$, we find that the collective modes are given by:
\begin{eqnarray}\label{eq:PLPCollectiveModeEquation}
\nonumber \tilde{Y}^{S0}(k)Z_{S0}(k)&=&\hbar\omega^{S0}(k)Z_{S0}(k),\\
\nonumber Z_{S0}&=&\left[u_{00,S0},u_{10,S0},u_{01,S0},u_{11,S0},v_{00,S0},v_{01,S0},v_{10,S0},v_{11,S0}\right]^T\\
\tilde{Y}^{S0}(k)&=&\left[\begin{array}{cc}
Y^{S0}(k)  &  (-1)^S A(k) \\
-(-1)^S A(k) & -Y^{S0}(k)   \\
\end{array}\right],
\end{eqnarray}
\end{widetext}
The corresponding wave functions are those of Eq. (\ref{eq:CAFLz0magnetoexciton}) but changing $\pm$ by $SS_z=00,10$, respectively. The exact expression for the velocity of the Goldstone mode, $\omega^{00}_0(k)\simeq v_Gk$, is also given by Eq. (\ref{eq:Goldstonevelocity}). For the modes with $S_z=\pm 1$, the results are formally analog to those of Eqs. (\ref{eq:CAFCollectiveModeEquationTriplet}), (\ref{eq:CAFMETriplet}), but with $SS_z$ playing the role of $LL_z$. The values of $\Delta^{SS_z},~u^{SS_z}$ characterizing the matrices $Y^{SS_z}(k)$ are computed in the same way as in the FLP case, also finding that $\omega^{1\pm1}(k)=\omega^{10}(k)\mp2\epsilon_Z$ and $Z_{11}(k)=Z_{10}(k)=Z_{1-1}(k)$.

The expression of the stiffness coefficient for the orbital-singlet mode $\rho^{10}_0$ is given by Eq. (\ref{eq:stiffnessanomalous}) and those of $\rho^{10}_1,\rho^{00}_1$ by the first and second line of Eq. (\ref{eq:stiffnessGoldstone}), respectively.

\subsection{Computation of the collective modes in monolayer graphene}\label{app:technicalMLG}

The TDHFA previously developed for computing the collective modes of the $\nu=0$ QH state of bilayer graphene is also valid for the monolayer. In particular, the dispersion relation is also obtained from the eigenvalues of the matrix $\tilde{X}(\mathbf{k})$ of Eq. (\ref{eq:XYgen32}) but now there is only one possible value for the magnetic index, $n=0$. Thus, as the structure in the valley-spin space is the same as for the bilayer problem, the building blocks of the dispersion matrix $\tilde{X}$, $Y^{\alpha_k\alpha_l,\alpha_j\alpha_m}(\mathbf{k})$, are now scalars instead of $4\times 4$ matrices:
\begin{widetext}
\begin{eqnarray}\label{eq:MLGpseudospinmatrices}
Y^{\alpha_k\alpha_l,\alpha_j\alpha_m}(k)&=&\left[\Delta^{\alpha_k\alpha_l}+F(k)\right]\delta_{\alpha_k\alpha_j}\delta_{\alpha_l\alpha_m}+R(k)u^{\alpha_k\alpha_l,\alpha_j\alpha_m}\\
\nonumber F(k)&\equiv& F_{00}-C_{00,00}(k)=F_{00}\left[1-I_0\left[\frac{(kl_B)^2}{4}\right]e^{-\frac{(kl_B)^2}{4}}\right]\\
\nonumber R(k)&\equiv& 2R_{00,00}(k)=e^{-\frac{(kl_B)^2}{2}}
\end{eqnarray}
\end{widetext}
where we have made explicit the isotropy of the dispersion relation. All the coefficients $\Delta^{\alpha_k\alpha_l}$ and $u^{\alpha_k\alpha_l,\alpha_j\alpha_m}$ have the same value as in the bilayer case but with $\epsilon_V=0$; the same holds for the magnitudes $\Delta^{\mu}$ and $u^{\mu}$. Hence, the dispersion relation is computed in the same way as in the previous section but replacing all the matrices $Y$ by the corresponding scalars, notably simplifying the calculations.

We now give the explicit analytical results for the dispersion relation of the collective modes $\omega^{\mu}(k)$
and the associated stiffness $\rho^{\mu}$ for every phase, characterized by the same valley-spin symmetries $\mu$ as in the bilayer case. Note that the components $u_{nn'},v_{nn'}$ of the magnetoexciton wave functions are reduced now to scalars as the orbital structure is trivial for monolayer graphene.

\subsubsection{Ferromagnetic and charge-density wave phases}

\begin{eqnarray}\label{eq:MLGspinflip}
\hbar\omega^{\mu}(k)&=&Y^{\mu}(k)\equiv\Delta^{\mu}+F(k)+R(k)u^{\mu}\\
\nonumber \rho^{\mu}&=&\left(\frac{F_{00}}{4}-\frac{u^\mu}{2}\right)l^2_B
\end{eqnarray}
As $\epsilon_V=0$, for the F phase $\omega^{1-1}(k)=\omega^{11}(k)$. Note the strong similarity in the formula for the stiffness with
that of bilayer graphene for dominant Coulomb interactions, Eq. (\ref{eq:stiffnesspro}).

\subsubsection{Canted anti-ferromagnetic and Kekul\'e distortion phases}\label{subsec:MLGCAF}

\begin{equation}\label{eq:MLGCAFsp}
\hbar\omega^{\mu}(k)=\sqrt{\left[Y^{\mu}(k)\right]^2 -\left[AR(k)\right]^2}
\end{equation}
Similarly, $\omega^{1-1}(k)=\omega^{11}(k)$ in the CAF phase.

The velocity of the Goldstone modes is simply:
\begin{equation}\label{eq:GoldstonevelocityMLG}
\frac{\hbar v_G}{l_B}=\sqrt{A\left[\frac{F_{00}}{2}+\Delta^{G}\right]}
\end{equation}
with $\Delta^{G}=\Delta^{-}$ for the CAF phase and $\Delta^{G}=\Delta^{00}$ for the KD phase, while the stiffness coefficients are:
\begin{equation}\label{eq:MLGCAFstiffness}
\rho^{\mu}=\frac{(u^{\mu}+\Delta^{\mu})\left(\frac{F_{00}}{4}-\frac{u^{\mu}}{2}\right)+\frac{A^2}{2}}{\sqrt{(u^{\mu}+\Delta^{\mu})^2-A^2}}l^2_B
\end{equation}
Both results can be compared with those for the bilayer for dominant Coulomb interactions, Eqs. (\ref{eq:Goldstonevelocitypro}), (\ref{eq:stiffnessproCAF}).

\bibliographystyle{apsrev4-1}
\bibliography{QHFM}

\begin{thebibliography}{68}%
\makeatletter
\providecommand \@ifxundefined [1]{%
 \@ifx{#1\undefined}
}%
\providecommand \@ifnum [1]{%
 \ifnum #1\expandafter \@firstoftwo
 \else \expandafter \@secondoftwo
 \fi
}%
\providecommand \@ifx [1]{%
 \ifx #1\expandafter \@firstoftwo
 \else \expandafter \@secondoftwo
 \fi
}%
\providecommand \natexlab [1]{#1}%
\providecommand \enquote  [1]{``#1''}%
\providecommand \bibnamefont  [1]{#1}%
\providecommand \bibfnamefont [1]{#1}%
\providecommand \citenamefont [1]{#1}%
\providecommand \href@noop [0]{\@secondoftwo}%
\providecommand \href [0]{\begingroup \@sanitize@url \@href}%
\providecommand \@href[1]{\@@startlink{#1}\@@href}%
\providecommand \@@href[1]{\endgroup#1\@@endlink}%
\providecommand \@sanitize@url [0]{\catcode `\\12\catcode `\$12\catcode
  `\&12\catcode `\#12\catcode `\^12\catcode `\_12\catcode `\%12\relax}%
\providecommand \@@startlink[1]{}%
\providecommand \@@endlink[0]{}%
\providecommand \url  [0]{\begingroup\@sanitize@url \@url }%
\providecommand \@url [1]{\endgroup\@href {#1}{\urlprefix }}%
\providecommand \urlprefix  [0]{URL }%
\providecommand \Eprint [0]{\href }%
\providecommand \doibase [0]{http://dx.doi.org/}%
\providecommand \selectlanguage [0]{\@gobble}%
\providecommand \bibinfo  [0]{\@secondoftwo}%
\providecommand \bibfield  [0]{\@secondoftwo}%
\providecommand \translation [1]{[#1]}%
\providecommand \BibitemOpen [0]{}%
\providecommand \bibitemStop [0]{}%
\providecommand \bibitemNoStop [0]{.\EOS\space}%
\providecommand \EOS [0]{\spacefactor3000\relax}%
\providecommand \BibitemShut  [1]{\csname bibitem#1\endcsname}%
\let\auto@bib@innerbib\@empty
\bibitem [{\citenamefont {Klitzing}\ \emph {et~al.}(1980)\citenamefont
  {Klitzing}, \citenamefont {Dorda},\ and\ \citenamefont
  {Pepper}}]{Klitzing1980}%
  \BibitemOpen
  \bibfield  {author} {\bibinfo {author} {\bibfnamefont {K.~v.}\ \bibnamefont
  {Klitzing}}, \bibinfo {author} {\bibfnamefont {G.}~\bibnamefont {Dorda}}, \
  and\ \bibinfo {author} {\bibfnamefont {M.}~\bibnamefont {Pepper}},\ }\href
  {\doibase 10.1103/PhysRevLett.45.494} {\bibfield  {journal} {\bibinfo
  {journal} {Phys. Rev. Lett.}\ }\textbf {\bibinfo {volume} {45}},\ \bibinfo
  {pages} {494} (\bibinfo {year} {1980})}\BibitemShut {NoStop}%
\bibitem [{\citenamefont {Tsui}\ \emph
  {et~al.}(1982{\natexlab{a}})\citenamefont {Tsui}, \citenamefont {Gossard},
  \citenamefont {Field}, \citenamefont {Cage},\ and\ \citenamefont
  {Dziuba}}]{Tsui1982a}%
  \BibitemOpen
  \bibfield  {author} {\bibinfo {author} {\bibfnamefont {D.~C.}\ \bibnamefont
  {Tsui}}, \bibinfo {author} {\bibfnamefont {A.~C.}\ \bibnamefont {Gossard}},
  \bibinfo {author} {\bibfnamefont {B.~F.}\ \bibnamefont {Field}}, \bibinfo
  {author} {\bibfnamefont {M.~E.}\ \bibnamefont {Cage}}, \ and\ \bibinfo
  {author} {\bibfnamefont {R.~F.}\ \bibnamefont {Dziuba}},\ }\href {\doibase
  10.1103/PhysRevLett.48.3} {\bibfield  {journal} {\bibinfo  {journal} {Phys.
  Rev. Lett.}\ }\textbf {\bibinfo {volume} {48}},\ \bibinfo {pages} {3}
  (\bibinfo {year} {1982}{\natexlab{a}})}\BibitemShut {NoStop}%
\bibitem [{\citenamefont {Tsui}\ \emph
  {et~al.}(1982{\natexlab{b}})\citenamefont {Tsui}, \citenamefont {Stormer},\
  and\ \citenamefont {Gossard}}]{Tsui1982}%
  \BibitemOpen
  \bibfield  {author} {\bibinfo {author} {\bibfnamefont {D.~C.}\ \bibnamefont
  {Tsui}}, \bibinfo {author} {\bibfnamefont {H.~L.}\ \bibnamefont {Stormer}}, \
  and\ \bibinfo {author} {\bibfnamefont {A.~C.}\ \bibnamefont {Gossard}},\
  }\href {\doibase 10.1103/PhysRevLett.48.1559} {\bibfield  {journal} {\bibinfo
   {journal} {Phys. Rev. Lett.}\ }\textbf {\bibinfo {volume} {48}},\ \bibinfo
  {pages} {1559} (\bibinfo {year} {1982}{\natexlab{b}})}\BibitemShut {NoStop}%
\bibitem [{\citenamefont {Stormer}\ \emph {et~al.}(1983)\citenamefont
  {Stormer}, \citenamefont {Chang}, \citenamefont {Tsui}, \citenamefont
  {Hwang}, \citenamefont {Gossard},\ and\ \citenamefont
  {Wiegmann}}]{Stormer1983}%
  \BibitemOpen
  \bibfield  {author} {\bibinfo {author} {\bibfnamefont {H.~L.}\ \bibnamefont
  {Stormer}}, \bibinfo {author} {\bibfnamefont {A.}~\bibnamefont {Chang}},
  \bibinfo {author} {\bibfnamefont {D.~C.}\ \bibnamefont {Tsui}}, \bibinfo
  {author} {\bibfnamefont {J.~C.~M.}\ \bibnamefont {Hwang}}, \bibinfo {author}
  {\bibfnamefont {A.~C.}\ \bibnamefont {Gossard}}, \ and\ \bibinfo {author}
  {\bibfnamefont {W.}~\bibnamefont {Wiegmann}},\ }\href {\doibase
  10.1103/PhysRevLett.50.1953} {\bibfield  {journal} {\bibinfo  {journal}
  {Phys. Rev. Lett.}\ }\textbf {\bibinfo {volume} {50}},\ \bibinfo {pages}
  {1953} (\bibinfo {year} {1983})}\BibitemShut {NoStop}%
\bibitem [{\citenamefont {Laughlin}(1983)}]{Laughlin1983}%
  \BibitemOpen
  \bibfield  {author} {\bibinfo {author} {\bibfnamefont {R.~B.}\ \bibnamefont
  {Laughlin}},\ }\href {\doibase 10.1103/PhysRevLett.50.1395} {\bibfield
  {journal} {\bibinfo  {journal} {Phys. Rev. Lett.}\ }\textbf {\bibinfo
  {volume} {50}},\ \bibinfo {pages} {1395} (\bibinfo {year}
  {1983})}\BibitemShut {NoStop}%
\bibitem [{\citenamefont {McCann}\ and\ \citenamefont
  {Fal'ko}(2006)}]{McCann2006}%
  \BibitemOpen
  \bibfield  {author} {\bibinfo {author} {\bibfnamefont {E.}~\bibnamefont
  {McCann}}\ and\ \bibinfo {author} {\bibfnamefont {V.~I.}\ \bibnamefont
  {Fal'ko}},\ }\href {\doibase 10.1103/PhysRevLett.96.086805} {\bibfield
  {journal} {\bibinfo  {journal} {Phys. Rev. Lett.}\ }\textbf {\bibinfo
  {volume} {96}},\ \bibinfo {pages} {086805} (\bibinfo {year}
  {2006})}\BibitemShut {NoStop}%
\bibitem [{\citenamefont {Yang}\ \emph {et~al.}(2006)\citenamefont {Yang},
  \citenamefont {Das~Sarma},\ and\ \citenamefont {MacDonald}}]{Yang2006}%
  \BibitemOpen
  \bibfield  {author} {\bibinfo {author} {\bibfnamefont {K.}~\bibnamefont
  {Yang}}, \bibinfo {author} {\bibfnamefont {S.}~\bibnamefont {Das~Sarma}}, \
  and\ \bibinfo {author} {\bibfnamefont {A.~H.}\ \bibnamefont {MacDonald}},\
  }\href {\doibase 10.1103/PhysRevB.74.075423} {\bibfield  {journal} {\bibinfo
  {journal} {Phys. Rev. B}\ }\textbf {\bibinfo {volume} {74}},\ \bibinfo
  {pages} {075423} (\bibinfo {year} {2006})}\BibitemShut {NoStop}%
\bibitem [{\citenamefont {Iyengar}\ \emph {et~al.}(2007)\citenamefont
  {Iyengar}, \citenamefont {Wang}, \citenamefont {Fertig},\ and\ \citenamefont
  {Brey}}]{Iyengar2007}%
  \BibitemOpen
  \bibfield  {author} {\bibinfo {author} {\bibfnamefont {A.}~\bibnamefont
  {Iyengar}}, \bibinfo {author} {\bibfnamefont {J.}~\bibnamefont {Wang}},
  \bibinfo {author} {\bibfnamefont {H.~A.}\ \bibnamefont {Fertig}}, \ and\
  \bibinfo {author} {\bibfnamefont {L.}~\bibnamefont {Brey}},\ }\href {\doibase
  10.1103/PhysRevB.75.125430} {\bibfield  {journal} {\bibinfo  {journal} {Phys.
  Rev. B}\ }\textbf {\bibinfo {volume} {75}},\ \bibinfo {pages} {125430}
  (\bibinfo {year} {2007})}\BibitemShut {NoStop}%
\bibitem [{\citenamefont {Doretto}\ and\ \citenamefont
  {Smith}(2007)}]{Doretto2007}%
  \BibitemOpen
  \bibfield  {author} {\bibinfo {author} {\bibfnamefont {R.~L.}\ \bibnamefont
  {Doretto}}\ and\ \bibinfo {author} {\bibfnamefont {C.~M.}\ \bibnamefont
  {Smith}},\ }\href {\doibase 10.1103/PhysRevB.76.195431} {\bibfield  {journal}
  {\bibinfo  {journal} {Phys. Rev. B}\ }\textbf {\bibinfo {volume} {76}},\
  \bibinfo {pages} {195431} (\bibinfo {year} {2007})}\BibitemShut {NoStop}%
\bibitem [{\citenamefont {Bychkov}\ and\ \citenamefont
  {Martinez}(2008)}]{Bychkov2008}%
  \BibitemOpen
  \bibfield  {author} {\bibinfo {author} {\bibfnamefont {Y.~A.}\ \bibnamefont
  {Bychkov}}\ and\ \bibinfo {author} {\bibfnamefont {G.}~\bibnamefont
  {Martinez}},\ }\href {\doibase 10.1103/PhysRevB.77.125417} {\bibfield
  {journal} {\bibinfo  {journal} {Phys. Rev. B}\ }\textbf {\bibinfo {volume}
  {77}},\ \bibinfo {pages} {125417} (\bibinfo {year} {2008})}\BibitemShut
  {NoStop}%
\bibitem [{\citenamefont {Jung}\ and\ \citenamefont
  {MacDonald}(2009)}]{Jung2009}%
  \BibitemOpen
  \bibfield  {author} {\bibinfo {author} {\bibfnamefont {J.}~\bibnamefont
  {Jung}}\ and\ \bibinfo {author} {\bibfnamefont {A.~H.}\ \bibnamefont
  {MacDonald}},\ }\href {\doibase 10.1103/PhysRevB.80.235417} {\bibfield
  {journal} {\bibinfo  {journal} {Phys. Rev. B}\ }\textbf {\bibinfo {volume}
  {80}},\ \bibinfo {pages} {235417} (\bibinfo {year} {2009})}\BibitemShut
  {NoStop}%
\bibitem [{\citenamefont {Abanin}\ \emph {et~al.}(2009)\citenamefont {Abanin},
  \citenamefont {Parameswaran},\ and\ \citenamefont {Sondhi}}]{Abanin2009}%
  \BibitemOpen
  \bibfield  {author} {\bibinfo {author} {\bibfnamefont {D.~A.}\ \bibnamefont
  {Abanin}}, \bibinfo {author} {\bibfnamefont {S.~A.}\ \bibnamefont
  {Parameswaran}}, \ and\ \bibinfo {author} {\bibfnamefont {S.~L.}\
  \bibnamefont {Sondhi}},\ }\href {\doibase 10.1103/PhysRevLett.103.076802}
  {\bibfield  {journal} {\bibinfo  {journal} {Phys. Rev. Lett.}\ }\textbf
  {\bibinfo {volume} {103}},\ \bibinfo {pages} {076802} (\bibinfo {year}
  {2009})}\BibitemShut {NoStop}%
\bibitem [{\citenamefont {C\^ot\'e}\ \emph {et~al.}(2010)\citenamefont
  {C\^ot\'e}, \citenamefont {Lambert}, \citenamefont {Barlas},\ and\
  \citenamefont {MacDonald}}]{Cote2010}%
  \BibitemOpen
  \bibfield  {author} {\bibinfo {author} {\bibfnamefont {R.}~\bibnamefont
  {C\^ot\'e}}, \bibinfo {author} {\bibfnamefont {J.}~\bibnamefont {Lambert}},
  \bibinfo {author} {\bibfnamefont {Y.}~\bibnamefont {Barlas}}, \ and\ \bibinfo
  {author} {\bibfnamefont {A.~H.}\ \bibnamefont {MacDonald}},\ }\href {\doibase
  10.1103/PhysRevB.82.035445} {\bibfield  {journal} {\bibinfo  {journal} {Phys.
  Rev. B}\ }\textbf {\bibinfo {volume} {82}},\ \bibinfo {pages} {035445}
  (\bibinfo {year} {2010})}\BibitemShut {NoStop}%
\bibitem [{\citenamefont {Shizuya}(2010)}]{Shizuya2010}%
  \BibitemOpen
  \bibfield  {author} {\bibinfo {author} {\bibfnamefont {K.}~\bibnamefont
  {Shizuya}},\ }\href {\doibase 10.1103/PhysRevB.81.075407} {\bibfield
  {journal} {\bibinfo  {journal} {Phys. Rev. B}\ }\textbf {\bibinfo {volume}
  {81}},\ \bibinfo {pages} {075407} (\bibinfo {year} {2010})}\BibitemShut
  {NoStop}%
\bibitem [{\citenamefont {Gorbar}\ \emph {et~al.}(2010)\citenamefont {Gorbar},
  \citenamefont {Gusynin},\ and\ \citenamefont {Miransky}}]{Gorbar2010}%
  \BibitemOpen
  \bibfield  {author} {\bibinfo {author} {\bibfnamefont {E.~V.}\ \bibnamefont
  {Gorbar}}, \bibinfo {author} {\bibfnamefont {V.~P.}\ \bibnamefont {Gusynin}},
  \ and\ \bibinfo {author} {\bibfnamefont {V.~A.}\ \bibnamefont {Miransky}},\
  }\href {\doibase 10.1103/PhysRevB.81.155451} {\bibfield  {journal} {\bibinfo
  {journal} {Phys. Rev. B}\ }\textbf {\bibinfo {volume} {81}},\ \bibinfo
  {pages} {155451} (\bibinfo {year} {2010})}\BibitemShut {NoStop}%
\bibitem [{\citenamefont {Gorbar}\ \emph {et~al.}(2011)\citenamefont {Gorbar},
  \citenamefont {Gusynin}, \citenamefont {Jia},\ and\ \citenamefont
  {Miransky}}]{Gorbar2011}%
  \BibitemOpen
  \bibfield  {author} {\bibinfo {author} {\bibfnamefont {E.~V.}\ \bibnamefont
  {Gorbar}}, \bibinfo {author} {\bibfnamefont {V.~P.}\ \bibnamefont {Gusynin}},
  \bibinfo {author} {\bibfnamefont {J.}~\bibnamefont {Jia}}, \ and\ \bibinfo
  {author} {\bibfnamefont {V.~A.}\ \bibnamefont {Miransky}},\ }\href {\doibase
  10.1103/PhysRevB.84.235449} {\bibfield  {journal} {\bibinfo  {journal} {Phys.
  Rev. B}\ }\textbf {\bibinfo {volume} {84}},\ \bibinfo {pages} {235449}
  (\bibinfo {year} {2011})}\BibitemShut {NoStop}%
\bibitem [{\citenamefont {Gorbar}\ \emph {et~al.}(2012)\citenamefont {Gorbar},
  \citenamefont {Gusynin}, \citenamefont {Miransky},\ and\ \citenamefont
  {Shovkovy}}]{Gorbar2012}%
  \BibitemOpen
  \bibfield  {author} {\bibinfo {author} {\bibfnamefont {E.~V.}\ \bibnamefont
  {Gorbar}}, \bibinfo {author} {\bibfnamefont {V.~P.}\ \bibnamefont {Gusynin}},
  \bibinfo {author} {\bibfnamefont {V.~A.}\ \bibnamefont {Miransky}}, \ and\
  \bibinfo {author} {\bibfnamefont {I.~A.}\ \bibnamefont {Shovkovy}},\ }\href
  {\doibase 10.1103/PhysRevB.85.235460} {\bibfield  {journal} {\bibinfo
  {journal} {Phys. Rev. B}\ }\textbf {\bibinfo {volume} {85}},\ \bibinfo
  {pages} {235460} (\bibinfo {year} {2012})}\BibitemShut {NoStop}%
\bibitem [{\citenamefont {Kim}\ \emph {et~al.}(2011)\citenamefont {Kim},
  \citenamefont {Lee},\ and\ \citenamefont {Tutuc}}]{Kim2011}%
  \BibitemOpen
  \bibfield  {author} {\bibinfo {author} {\bibfnamefont {S.}~\bibnamefont
  {Kim}}, \bibinfo {author} {\bibfnamefont {K.}~\bibnamefont {Lee}}, \ and\
  \bibinfo {author} {\bibfnamefont {E.}~\bibnamefont {Tutuc}},\ }\href
  {\doibase 10.1103/PhysRevLett.107.016803} {\bibfield  {journal} {\bibinfo
  {journal} {Phys. Rev. Lett.}\ }\textbf {\bibinfo {volume} {107}},\ \bibinfo
  {pages} {016803} (\bibinfo {year} {2011})}\BibitemShut {NoStop}%
\bibitem [{\citenamefont {Shizuya}(2011)}]{Shizuya2011}%
  \BibitemOpen
  \bibfield  {author} {\bibinfo {author} {\bibfnamefont {K.}~\bibnamefont
  {Shizuya}},\ }\href {\doibase 10.1103/PhysRevB.84.075409} {\bibfield
  {journal} {\bibinfo  {journal} {Phys. Rev. B}\ }\textbf {\bibinfo {volume}
  {84}},\ \bibinfo {pages} {075409} (\bibinfo {year} {2011})}\BibitemShut
  {NoStop}%
\bibitem [{\citenamefont {Shizuya}(2012)}]{Shizuya2012}%
  \BibitemOpen
  \bibfield  {author} {\bibinfo {author} {\bibfnamefont {K.}~\bibnamefont
  {Shizuya}},\ }\href {\doibase 10.1103/PhysRevB.86.045431} {\bibfield
  {journal} {\bibinfo  {journal} {Phys. Rev. B}\ }\textbf {\bibinfo {volume}
  {86}},\ \bibinfo {pages} {045431} (\bibinfo {year} {2012})}\BibitemShut
  {NoStop}%
\bibitem [{\citenamefont {Kharitonov}(2012{\natexlab{a}})}]{Kharitonov2012}%
  \BibitemOpen
  \bibfield  {author} {\bibinfo {author} {\bibfnamefont {M.}~\bibnamefont
  {Kharitonov}},\ }\href {\doibase 10.1103/PhysRevB.85.155439} {\bibfield
  {journal} {\bibinfo  {journal} {Phys. Rev. B}\ }\textbf {\bibinfo {volume}
  {85}},\ \bibinfo {pages} {155439} (\bibinfo {year}
  {2012}{\natexlab{a}})}\BibitemShut {NoStop}%
\bibitem [{\citenamefont {Kharitonov}(2012{\natexlab{b}})}]{Kharitonov2012PRL}%
  \BibitemOpen
  \bibfield  {author} {\bibinfo {author} {\bibfnamefont {M.}~\bibnamefont
  {Kharitonov}},\ }\href {\doibase 10.1103/PhysRevLett.109.046803} {\bibfield
  {journal} {\bibinfo  {journal} {Phys. Rev. Lett.}\ }\textbf {\bibinfo
  {volume} {109}},\ \bibinfo {pages} {046803} (\bibinfo {year}
  {2012}{\natexlab{b}})}\BibitemShut {NoStop}%
\bibitem [{\citenamefont {Kharitonov}(2012{\natexlab{c}})}]{Kharitonov2012a}%
  \BibitemOpen
  \bibfield  {author} {\bibinfo {author} {\bibfnamefont {M.}~\bibnamefont
  {Kharitonov}},\ }\href {\doibase 10.1103/PhysRevB.86.195435} {\bibfield
  {journal} {\bibinfo  {journal} {Phys. Rev. B}\ }\textbf {\bibinfo {volume}
  {86}},\ \bibinfo {pages} {195435} (\bibinfo {year}
  {2012}{\natexlab{c}})}\BibitemShut {NoStop}%
\bibitem [{\citenamefont {Kharitonov}(2012{\natexlab{d}})}]{Kharitonov2012b}%
  \BibitemOpen
  \bibfield  {author} {\bibinfo {author} {\bibfnamefont {M.}~\bibnamefont
  {Kharitonov}},\ }\href {\doibase 10.1103/PhysRevB.86.075450} {\bibfield
  {journal} {\bibinfo  {journal} {Phys. Rev. B}\ }\textbf {\bibinfo {volume}
  {86}},\ \bibinfo {pages} {075450} (\bibinfo {year}
  {2012}{\natexlab{d}})}\BibitemShut {NoStop}%
\bibitem [{\citenamefont {Lambert}\ and\ \citenamefont
  {C\^ot\'e}(2013)}]{Lambert2013}%
  \BibitemOpen
  \bibfield  {author} {\bibinfo {author} {\bibfnamefont {J.}~\bibnamefont
  {Lambert}}\ and\ \bibinfo {author} {\bibfnamefont {R.}~\bibnamefont
  {C\^ot\'e}},\ }\href {\doibase 10.1103/PhysRevB.87.115415} {\bibfield
  {journal} {\bibinfo  {journal} {Phys. Rev. B}\ }\textbf {\bibinfo {volume}
  {87}},\ \bibinfo {pages} {115415} (\bibinfo {year} {2013})}\BibitemShut
  {NoStop}%
\bibitem [{\citenamefont {T\ifmmode~\mbox{\H{o}}\else
  \H{o}\fi{}ke}(2013)}]{Toke2013}%
  \BibitemOpen
  \bibfield  {author} {\bibinfo {author} {\bibfnamefont {C.}~\bibnamefont
  {T\ifmmode~\mbox{\H{o}}\else \H{o}\fi{}ke}},\ }\href {\doibase
  10.1103/PhysRevB.88.241411} {\bibfield  {journal} {\bibinfo  {journal} {Phys.
  Rev. B}\ }\textbf {\bibinfo {volume} {88}},\ \bibinfo {pages} {241411}
  (\bibinfo {year} {2013})}\BibitemShut {NoStop}%
\bibitem [{\citenamefont {Wu}\ \emph {et~al.}(2014)\citenamefont {Wu},
  \citenamefont {Sodemann}, \citenamefont {Araki}, \citenamefont {MacDonald},\
  and\ \citenamefont {Jolicoeur}}]{Wu2014}%
  \BibitemOpen
  \bibfield  {author} {\bibinfo {author} {\bibfnamefont {F.}~\bibnamefont
  {Wu}}, \bibinfo {author} {\bibfnamefont {I.}~\bibnamefont {Sodemann}},
  \bibinfo {author} {\bibfnamefont {Y.}~\bibnamefont {Araki}}, \bibinfo
  {author} {\bibfnamefont {A.~H.}\ \bibnamefont {MacDonald}}, \ and\ \bibinfo
  {author} {\bibfnamefont {T.}~\bibnamefont {Jolicoeur}},\ }\href {\doibase
  10.1103/PhysRevB.90.235432} {\bibfield  {journal} {\bibinfo  {journal} {Phys.
  Rev. B}\ }\textbf {\bibinfo {volume} {90}},\ \bibinfo {pages} {235432}
  (\bibinfo {year} {2014})}\BibitemShut {NoStop}%
\bibitem [{\citenamefont {Knothe}\ and\ \citenamefont
  {Jolicoeur}(2015)}]{Knothe2015}%
  \BibitemOpen
  \bibfield  {author} {\bibinfo {author} {\bibfnamefont {A.}~\bibnamefont
  {Knothe}}\ and\ \bibinfo {author} {\bibfnamefont {T.}~\bibnamefont
  {Jolicoeur}},\ }\href {\doibase 10.1103/PhysRevB.92.165110} {\bibfield
  {journal} {\bibinfo  {journal} {Phys. Rev. B}\ }\textbf {\bibinfo {volume}
  {92}},\ \bibinfo {pages} {165110} (\bibinfo {year} {2015})}\BibitemShut
  {NoStop}%
\bibitem [{\citenamefont {Murthy}\ \emph {et~al.}(2016)\citenamefont {Murthy},
  \citenamefont {Shimshoni},\ and\ \citenamefont {Fertig}}]{Murthy2016}%
  \BibitemOpen
  \bibfield  {author} {\bibinfo {author} {\bibfnamefont {G.}~\bibnamefont
  {Murthy}}, \bibinfo {author} {\bibfnamefont {E.}~\bibnamefont {Shimshoni}}, \
  and\ \bibinfo {author} {\bibfnamefont {H.~A.}\ \bibnamefont {Fertig}},\
  }\href {\doibase 10.1103/PhysRevB.93.045105} {\bibfield  {journal} {\bibinfo
  {journal} {Phys. Rev. B}\ }\textbf {\bibinfo {volume} {93}},\ \bibinfo
  {pages} {045105} (\bibinfo {year} {2016})}\BibitemShut {NoStop}%
\bibitem [{\citenamefont {Tikhonov}\ \emph {et~al.}(2016)\citenamefont
  {Tikhonov}, \citenamefont {Shimshoni}, \citenamefont {Fertig},\ and\
  \citenamefont {Murthy}}]{Tikhonov2016}%
  \BibitemOpen
  \bibfield  {author} {\bibinfo {author} {\bibfnamefont {P.}~\bibnamefont
  {Tikhonov}}, \bibinfo {author} {\bibfnamefont {E.}~\bibnamefont {Shimshoni}},
  \bibinfo {author} {\bibfnamefont {H.~A.}\ \bibnamefont {Fertig}}, \ and\
  \bibinfo {author} {\bibfnamefont {G.}~\bibnamefont {Murthy}},\ }\href
  {\doibase 10.1103/PhysRevB.93.115137} {\bibfield  {journal} {\bibinfo
  {journal} {Phys. Rev. B}\ }\textbf {\bibinfo {volume} {93}},\ \bibinfo
  {pages} {115137} (\bibinfo {year} {2016})}\BibitemShut {NoStop}%
\bibitem [{\citenamefont {Knothe}\ and\ \citenamefont
  {Jolicoeur}(2016)}]{Knothe2016}%
  \BibitemOpen
  \bibfield  {author} {\bibinfo {author} {\bibfnamefont {A.}~\bibnamefont
  {Knothe}}\ and\ \bibinfo {author} {\bibfnamefont {T.}~\bibnamefont
  {Jolicoeur}},\ }\href {\doibase 10.1103/PhysRevB.94.235149} {\bibfield
  {journal} {\bibinfo  {journal} {Phys. Rev. B}\ }\textbf {\bibinfo {volume}
  {94}},\ \bibinfo {pages} {235149} (\bibinfo {year} {2016})}\BibitemShut
  {NoStop}%
\bibitem [{\citenamefont {Jia}\ \emph {et~al.}(2017)\citenamefont {Jia},
  \citenamefont {Pyatkovskiy}, \citenamefont {Gorbar},\ and\ \citenamefont
  {Gusynin}}]{Jia2017}%
  \BibitemOpen
  \bibfield  {author} {\bibinfo {author} {\bibfnamefont {J.}~\bibnamefont
  {Jia}}, \bibinfo {author} {\bibfnamefont {P.~K.}\ \bibnamefont
  {Pyatkovskiy}}, \bibinfo {author} {\bibfnamefont {E.~V.}\ \bibnamefont
  {Gorbar}}, \ and\ \bibinfo {author} {\bibfnamefont {V.~P.}\ \bibnamefont
  {Gusynin}},\ }\href {\doibase 10.1103/PhysRevB.95.045410} {\bibfield
  {journal} {\bibinfo  {journal} {Phys. Rev. B}\ }\textbf {\bibinfo {volume}
  {95}},\ \bibinfo {pages} {045410} (\bibinfo {year} {2017})}\BibitemShut
  {NoStop}%
\bibitem [{\citenamefont {T\ifmmode~\mbox{\H{o}}\else \H{o}\fi{}ke}\ and\
  \citenamefont {Fal'ko}(2011)}]{Toke2011}%
  \BibitemOpen
  \bibfield  {author} {\bibinfo {author} {\bibfnamefont {C.}~\bibnamefont
  {T\ifmmode~\mbox{\H{o}}\else \H{o}\fi{}ke}}\ and\ \bibinfo {author}
  {\bibfnamefont {V.~I.}\ \bibnamefont {Fal'ko}},\ }\href {\doibase
  10.1103/PhysRevB.83.115455} {\bibfield  {journal} {\bibinfo  {journal} {Phys.
  Rev. B}\ }\textbf {\bibinfo {volume} {83}},\ \bibinfo {pages} {115455}
  (\bibinfo {year} {2011})}\BibitemShut {NoStop}%
\bibitem [{\citenamefont {S\'ari}\ and\ \citenamefont
  {T\ifmmode~\mbox{\H{o}}\else \H{o}\fi{}ke}(2013)}]{Sari87}%
  \BibitemOpen
  \bibfield  {author} {\bibinfo {author} {\bibfnamefont {J.}~\bibnamefont
  {S\'ari}}\ and\ \bibinfo {author} {\bibfnamefont {C.}~\bibnamefont
  {T\ifmmode~\mbox{\H{o}}\else \H{o}\fi{}ke}},\ }\href {\doibase
  10.1103/PhysRevB.87.085432} {\bibfield  {journal} {\bibinfo  {journal} {Phys.
  Rev. B}\ }\textbf {\bibinfo {volume} {87}},\ \bibinfo {pages} {085432}
  (\bibinfo {year} {2013})}\BibitemShut {NoStop}%
\bibitem [{\citenamefont {Foster}\ and\ \citenamefont
  {Aleiner}(2008)}]{Foster2008}%
  \BibitemOpen
  \bibfield  {author} {\bibinfo {author} {\bibfnamefont {M.~S.}\ \bibnamefont
  {Foster}}\ and\ \bibinfo {author} {\bibfnamefont {I.~L.}\ \bibnamefont
  {Aleiner}},\ }\href {\doibase 10.1103/PhysRevB.77.195413} {\bibfield
  {journal} {\bibinfo  {journal} {Phys. Rev. B}\ }\textbf {\bibinfo {volume}
  {77}},\ \bibinfo {pages} {195413} (\bibinfo {year} {2008})}\BibitemShut
  {NoStop}%
\bibitem [{\citenamefont {Basko}\ and\ \citenamefont
  {Aleiner}(2008)}]{Basko2008}%
  \BibitemOpen
  \bibfield  {author} {\bibinfo {author} {\bibfnamefont {D.~M.}\ \bibnamefont
  {Basko}}\ and\ \bibinfo {author} {\bibfnamefont {I.~L.}\ \bibnamefont
  {Aleiner}},\ }\href {\doibase 10.1103/PhysRevB.77.041409} {\bibfield
  {journal} {\bibinfo  {journal} {Phys. Rev. B}\ }\textbf {\bibinfo {volume}
  {77}},\ \bibinfo {pages} {041409} (\bibinfo {year} {2008})}\BibitemShut
  {NoStop}%
\bibitem [{\citenamefont {Aleiner}\ \emph {et~al.}(2007)\citenamefont
  {Aleiner}, \citenamefont {Kharzeev},\ and\ \citenamefont
  {Tsvelik}}]{Aleiner2007}%
  \BibitemOpen
  \bibfield  {author} {\bibinfo {author} {\bibfnamefont {I.~L.}\ \bibnamefont
  {Aleiner}}, \bibinfo {author} {\bibfnamefont {D.~E.}\ \bibnamefont
  {Kharzeev}}, \ and\ \bibinfo {author} {\bibfnamefont {A.~M.}\ \bibnamefont
  {Tsvelik}},\ }\href {\doibase 10.1103/PhysRevB.76.195415} {\bibfield
  {journal} {\bibinfo  {journal} {Phys. Rev. B}\ }\textbf {\bibinfo {volume}
  {76}},\ \bibinfo {pages} {195415} (\bibinfo {year} {2007})}\BibitemShut
  {NoStop}%
\bibitem [{\citenamefont {Lemonik}\ \emph {et~al.}(2010)\citenamefont
  {Lemonik}, \citenamefont {Aleiner}, \citenamefont {Toke},\ and\ \citenamefont
  {Fal'ko}}]{Lemonik2010}%
  \BibitemOpen
  \bibfield  {author} {\bibinfo {author} {\bibfnamefont {Y.}~\bibnamefont
  {Lemonik}}, \bibinfo {author} {\bibfnamefont {I.~L.}\ \bibnamefont
  {Aleiner}}, \bibinfo {author} {\bibfnamefont {C.}~\bibnamefont {Toke}}, \
  and\ \bibinfo {author} {\bibfnamefont {V.~I.}\ \bibnamefont {Fal'ko}},\
  }\href {\doibase 10.1103/PhysRevB.82.201408} {\bibfield  {journal} {\bibinfo
  {journal} {Phys. Rev. B}\ }\textbf {\bibinfo {volume} {82}},\ \bibinfo
  {pages} {201408} (\bibinfo {year} {2010})}\BibitemShut {NoStop}%
\bibitem [{\citenamefont {Sondhi}\ \emph {et~al.}(1993)\citenamefont {Sondhi},
  \citenamefont {Karlhede}, \citenamefont {Kivelson},\ and\ \citenamefont
  {Rezayi}}]{Sondhi1993}%
  \BibitemOpen
  \bibfield  {author} {\bibinfo {author} {\bibfnamefont {S.~L.}\ \bibnamefont
  {Sondhi}}, \bibinfo {author} {\bibfnamefont {A.}~\bibnamefont {Karlhede}},
  \bibinfo {author} {\bibfnamefont {S.~A.}\ \bibnamefont {Kivelson}}, \ and\
  \bibinfo {author} {\bibfnamefont {E.~H.}\ \bibnamefont {Rezayi}},\ }\href
  {\doibase 10.1103/PhysRevB.47.16419} {\bibfield  {journal} {\bibinfo
  {journal} {Phys. Rev. B}\ }\textbf {\bibinfo {volume} {47}},\ \bibinfo
  {pages} {16419} (\bibinfo {year} {1993})}\BibitemShut {NoStop}%
\bibitem [{\citenamefont {Moon}\ \emph {et~al.}(1995)\citenamefont {Moon},
  \citenamefont {Mori}, \citenamefont {Yang}, \citenamefont {Girvin},
  \citenamefont {MacDonald}, \citenamefont {Zheng}, \citenamefont {Yoshioka},\
  and\ \citenamefont {Zhang}}]{Moon1995}%
  \BibitemOpen
  \bibfield  {author} {\bibinfo {author} {\bibfnamefont {K.}~\bibnamefont
  {Moon}}, \bibinfo {author} {\bibfnamefont {H.}~\bibnamefont {Mori}}, \bibinfo
  {author} {\bibfnamefont {K.}~\bibnamefont {Yang}}, \bibinfo {author}
  {\bibfnamefont {S.~M.}\ \bibnamefont {Girvin}}, \bibinfo {author}
  {\bibfnamefont {A.~H.}\ \bibnamefont {MacDonald}}, \bibinfo {author}
  {\bibfnamefont {L.}~\bibnamefont {Zheng}}, \bibinfo {author} {\bibfnamefont
  {D.}~\bibnamefont {Yoshioka}}, \ and\ \bibinfo {author} {\bibfnamefont
  {S.-C.}\ \bibnamefont {Zhang}},\ }\href {\doibase 10.1103/PhysRevB.51.5138}
  {\bibfield  {journal} {\bibinfo  {journal} {Phys. Rev. B}\ }\textbf {\bibinfo
  {volume} {51}},\ \bibinfo {pages} {5138} (\bibinfo {year}
  {1995})}\BibitemShut {NoStop}%
\bibitem [{\citenamefont {C\^ot\'e}\ and\ \citenamefont
  {Barrette}(2013)}]{Cote2013}%
  \BibitemOpen
  \bibfield  {author} {\bibinfo {author} {\bibfnamefont {R.}~\bibnamefont
  {C\^ot\'e}}\ and\ \bibinfo {author} {\bibfnamefont {M.}~\bibnamefont
  {Barrette}},\ }\href {\doibase 10.1103/PhysRevB.88.245445} {\bibfield
  {journal} {\bibinfo  {journal} {Phys. Rev. B}\ }\textbf {\bibinfo {volume}
  {88}},\ \bibinfo {pages} {245445} (\bibinfo {year} {2013})}\BibitemShut
  {NoStop}%
\bibitem [{\citenamefont {Mayorov}\ \emph {et~al.}(2011)\citenamefont
  {Mayorov}, \citenamefont {Elias}, \citenamefont {Mucha-Kruczynski},
  \citenamefont {Gorbachev}, \citenamefont {Tudorovskiy}, \citenamefont
  {Zhukov}, \citenamefont {Morozov}, \citenamefont {Katsnelson}, \citenamefont
  {Fal'ko}, \citenamefont {Geim},\ and\ \citenamefont
  {Novoselov}}]{Mayorov2011}%
  \BibitemOpen
  \bibfield  {author} {\bibinfo {author} {\bibfnamefont {A.~S.}\ \bibnamefont
  {Mayorov}}, \bibinfo {author} {\bibfnamefont {D.~C.}\ \bibnamefont {Elias}},
  \bibinfo {author} {\bibfnamefont {M.}~\bibnamefont {Mucha-Kruczynski}},
  \bibinfo {author} {\bibfnamefont {R.~V.}\ \bibnamefont {Gorbachev}}, \bibinfo
  {author} {\bibfnamefont {T.}~\bibnamefont {Tudorovskiy}}, \bibinfo {author}
  {\bibfnamefont {A.}~\bibnamefont {Zhukov}}, \bibinfo {author} {\bibfnamefont
  {S.~V.}\ \bibnamefont {Morozov}}, \bibinfo {author} {\bibfnamefont {M.~I.}\
  \bibnamefont {Katsnelson}}, \bibinfo {author} {\bibfnamefont
  {V.}~\bibnamefont {Fal'ko}}, \bibinfo {author} {\bibfnamefont {A.~K.}\
  \bibnamefont {Geim}}, \ and\ \bibinfo {author} {\bibfnamefont {K.~S.}\
  \bibnamefont {Novoselov}},\ }\href {\doibase 10.1126/science.1208683}
  {\bibfield  {journal} {\bibinfo  {journal} {Science}\ }\textbf {\bibinfo
  {volume} {333}},\ \bibinfo {pages} {860} (\bibinfo {year}
  {2011})}\BibitemShut {NoStop}%
\bibitem [{\citenamefont {Hofstadter}(1976)}]{Hofstadter1976}%
  \BibitemOpen
  \bibfield  {author} {\bibinfo {author} {\bibfnamefont {D.~R.}\ \bibnamefont
  {Hofstadter}},\ }\href {\doibase 10.1103/PhysRevB.14.2239} {\bibfield
  {journal} {\bibinfo  {journal} {Phys. Rev. B}\ }\textbf {\bibinfo {volume}
  {14}},\ \bibinfo {pages} {2239} (\bibinfo {year} {1976})}\BibitemShut
  {NoStop}%
\bibitem [{\citenamefont {Yang}\ \emph {et~al.}(2016)\citenamefont {Yang},
  \citenamefont {Couturaud}, \citenamefont {Desrat}, \citenamefont {Consejo},
  \citenamefont {Kazazis}, \citenamefont {Yakimova}, \citenamefont
  {Syv\"aj\"arvi}, \citenamefont {Goiran}, \citenamefont {B\'eard},
  \citenamefont {Frings}, \citenamefont {Pierre}, \citenamefont {Cresti},
  \citenamefont {Escoffier},\ and\ \citenamefont {Jouault}}]{Yang2016}%
  \BibitemOpen
  \bibfield  {author} {\bibinfo {author} {\bibfnamefont {M.}~\bibnamefont
  {Yang}}, \bibinfo {author} {\bibfnamefont {O.}~\bibnamefont {Couturaud}},
  \bibinfo {author} {\bibfnamefont {W.}~\bibnamefont {Desrat}}, \bibinfo
  {author} {\bibfnamefont {C.}~\bibnamefont {Consejo}}, \bibinfo {author}
  {\bibfnamefont {D.}~\bibnamefont {Kazazis}}, \bibinfo {author} {\bibfnamefont
  {R.}~\bibnamefont {Yakimova}}, \bibinfo {author} {\bibfnamefont
  {M.}~\bibnamefont {Syv\"aj\"arvi}}, \bibinfo {author} {\bibfnamefont
  {M.}~\bibnamefont {Goiran}}, \bibinfo {author} {\bibfnamefont
  {J.}~\bibnamefont {B\'eard}}, \bibinfo {author} {\bibfnamefont
  {P.}~\bibnamefont {Frings}}, \bibinfo {author} {\bibfnamefont
  {M.}~\bibnamefont {Pierre}}, \bibinfo {author} {\bibfnamefont
  {A.}~\bibnamefont {Cresti}}, \bibinfo {author} {\bibfnamefont
  {W.}~\bibnamefont {Escoffier}}, \ and\ \bibinfo {author} {\bibfnamefont
  {B.}~\bibnamefont {Jouault}},\ }\href {\doibase
  10.1103/PhysRevLett.117.237702} {\bibfield  {journal} {\bibinfo  {journal}
  {Phys. Rev. Lett.}\ }\textbf {\bibinfo {volume} {117}},\ \bibinfo {pages}
  {237702} (\bibinfo {year} {2016})}\BibitemShut {NoStop}%
\bibitem [{\citenamefont {Shizuya}(2013)}]{Shizuya2013}%
  \BibitemOpen
  \bibfield  {author} {\bibinfo {author} {\bibfnamefont {K.}~\bibnamefont
  {Shizuya}},\ }\href {\doibase 10.1103/PhysRevB.87.085413} {\bibfield
  {journal} {\bibinfo  {journal} {Phys. Rev. B}\ }\textbf {\bibinfo {volume}
  {87}},\ \bibinfo {pages} {085413} (\bibinfo {year} {2013})}\BibitemShut
  {NoStop}%
\bibitem [{\citenamefont {{Maher}}\ \emph {et~al.}(2013)\citenamefont
  {{Maher}}, \citenamefont {{Dean}}, \citenamefont {{Young}}, \citenamefont
  {{Taniguchi}}, \citenamefont {{Watanabe}}, \citenamefont {{Shepard}},
  \citenamefont {{Hone}},\ and\ \citenamefont {{Kim}}}]{Maher2013}%
  \BibitemOpen
  \bibfield  {author} {\bibinfo {author} {\bibfnamefont {P.}~\bibnamefont
  {{Maher}}}, \bibinfo {author} {\bibfnamefont {C.~R.}\ \bibnamefont {{Dean}}},
  \bibinfo {author} {\bibfnamefont {A.~F.}\ \bibnamefont {{Young}}}, \bibinfo
  {author} {\bibfnamefont {T.}~\bibnamefont {{Taniguchi}}}, \bibinfo {author}
  {\bibfnamefont {K.}~\bibnamefont {{Watanabe}}}, \bibinfo {author}
  {\bibfnamefont {K.~L.}\ \bibnamefont {{Shepard}}}, \bibinfo {author}
  {\bibfnamefont {J.}~\bibnamefont {{Hone}}}, \ and\ \bibinfo {author}
  {\bibfnamefont {P.}~\bibnamefont {{Kim}}},\ }\href {\doibase
  10.1038/nphys2528} {\bibfield  {journal} {\bibinfo  {journal} {Nature
  Physics}\ }\textbf {\bibinfo {volume} {9}},\ \bibinfo {pages} {154} (\bibinfo
  {year} {2013})},\ \Eprint {http://arxiv.org/abs/1212.3846} {1212.3846}
  \BibitemShut {NoStop}%
\bibitem [{\citenamefont {Young}\ \emph {et~al.}(2014)\citenamefont {Young},
  \citenamefont {Sanchez-Yamagishi}, \citenamefont {Hunt}, \citenamefont
  {Choi}, \citenamefont {Watanabe}, \citenamefont {Taniguchi}, \citenamefont
  {Ashoori},\ and\ \citenamefont {Jarillo-Herrero}}]{Young2014}%
  \BibitemOpen
  \bibfield  {author} {\bibinfo {author} {\bibfnamefont {A.~F.}\ \bibnamefont
  {Young}}, \bibinfo {author} {\bibfnamefont {J.}~\bibnamefont
  {Sanchez-Yamagishi}}, \bibinfo {author} {\bibfnamefont {B.}~\bibnamefont
  {Hunt}}, \bibinfo {author} {\bibfnamefont {S.~H.}\ \bibnamefont {Choi}},
  \bibinfo {author} {\bibfnamefont {K.}~\bibnamefont {Watanabe}}, \bibinfo
  {author} {\bibfnamefont {T.}~\bibnamefont {Taniguchi}}, \bibinfo {author}
  {\bibfnamefont {R.}~\bibnamefont {Ashoori}}, \ and\ \bibinfo {author}
  {\bibfnamefont {P.}~\bibnamefont {Jarillo-Herrero}},\ }\href@noop {}
  {\bibfield  {journal} {\bibinfo  {journal} {Nature}\ }\textbf {\bibinfo
  {volume} {505}},\ \bibinfo {pages} {528} (\bibinfo {year}
  {2014})}\BibitemShut {NoStop}%
\bibitem [{\citenamefont {Kallin}\ and\ \citenamefont
  {Halperin}(1984)}]{Kallin1984}%
  \BibitemOpen
  \bibfield  {author} {\bibinfo {author} {\bibfnamefont {C.}~\bibnamefont
  {Kallin}}\ and\ \bibinfo {author} {\bibfnamefont {B.~I.}\ \bibnamefont
  {Halperin}},\ }\href {\doibase 10.1103/PhysRevB.30.5655} {\bibfield
  {journal} {\bibinfo  {journal} {Phys. Rev. B}\ }\textbf {\bibinfo {volume}
  {30}},\ \bibinfo {pages} {5655} (\bibinfo {year} {1984})}\BibitemShut
  {NoStop}%
\bibitem [{\citenamefont {Abanin}\ \emph {et~al.}(2013)\citenamefont {Abanin},
  \citenamefont {Feldman}, \citenamefont {Yacoby},\ and\ \citenamefont
  {Halperin}}]{Abanin2013}%
  \BibitemOpen
  \bibfield  {author} {\bibinfo {author} {\bibfnamefont {D.~A.}\ \bibnamefont
  {Abanin}}, \bibinfo {author} {\bibfnamefont {B.~E.}\ \bibnamefont {Feldman}},
  \bibinfo {author} {\bibfnamefont {A.}~\bibnamefont {Yacoby}}, \ and\ \bibinfo
  {author} {\bibfnamefont {B.~I.}\ \bibnamefont {Halperin}},\ }\href {\doibase
  10.1103/PhysRevB.88.115407} {\bibfield  {journal} {\bibinfo  {journal} {Phys.
  Rev. B}\ }\textbf {\bibinfo {volume} {88}},\ \bibinfo {pages} {115407}
  (\bibinfo {year} {2013})}\BibitemShut {NoStop}%
\bibitem [{\citenamefont {Martin}\ \emph {et~al.}(2010)\citenamefont {Martin},
  \citenamefont {Feldman}, \citenamefont {Weitz}, \citenamefont {Allen},\ and\
  \citenamefont {Yacoby}}]{Martin2010}%
  \BibitemOpen
  \bibfield  {author} {\bibinfo {author} {\bibfnamefont {J.}~\bibnamefont
  {Martin}}, \bibinfo {author} {\bibfnamefont {B.~E.}\ \bibnamefont {Feldman}},
  \bibinfo {author} {\bibfnamefont {R.~T.}\ \bibnamefont {Weitz}}, \bibinfo
  {author} {\bibfnamefont {M.~T.}\ \bibnamefont {Allen}}, \ and\ \bibinfo
  {author} {\bibfnamefont {A.}~\bibnamefont {Yacoby}},\ }\href {\doibase
  10.1103/PhysRevLett.105.256806} {\bibfield  {journal} {\bibinfo  {journal}
  {Phys. Rev. Lett.}\ }\textbf {\bibinfo {volume} {105}},\ \bibinfo {pages}
  {256806} (\bibinfo {year} {2010})}\BibitemShut {NoStop}%
\bibitem [{\citenamefont {Zhao}\ \emph {et~al.}(2010)\citenamefont {Zhao},
  \citenamefont {Cadden-Zimansky}, \citenamefont {Jiang},\ and\ \citenamefont
  {Kim}}]{Zhao2010}%
  \BibitemOpen
  \bibfield  {author} {\bibinfo {author} {\bibfnamefont {Y.}~\bibnamefont
  {Zhao}}, \bibinfo {author} {\bibfnamefont {P.}~\bibnamefont
  {Cadden-Zimansky}}, \bibinfo {author} {\bibfnamefont {Z.}~\bibnamefont
  {Jiang}}, \ and\ \bibinfo {author} {\bibfnamefont {P.}~\bibnamefont {Kim}},\
  }\href {\doibase 10.1103/PhysRevLett.104.066801} {\bibfield  {journal}
  {\bibinfo  {journal} {Phys. Rev. Lett.}\ }\textbf {\bibinfo {volume} {104}},\
  \bibinfo {pages} {066801} (\bibinfo {year} {2010})}\BibitemShut {NoStop}%
\bibitem [{\citenamefont {Velasco~Jr}\ \emph {et~al.}(2012)\citenamefont
  {Velasco~Jr}, \citenamefont {Jing}, \citenamefont {Bao}, \citenamefont {Lee},
  \citenamefont {Kratz}, \citenamefont {Aji}, \citenamefont {Bockrath},
  \citenamefont {Lau}, \citenamefont {Varma}, \citenamefont {Stillwell} \emph
  {et~al.}}]{Velasco2012}%
  \BibitemOpen
  \bibfield  {author} {\bibinfo {author} {\bibfnamefont {J.}~\bibnamefont
  {Velasco~Jr}}, \bibinfo {author} {\bibfnamefont {L.}~\bibnamefont {Jing}},
  \bibinfo {author} {\bibfnamefont {W.}~\bibnamefont {Bao}}, \bibinfo {author}
  {\bibfnamefont {Y.}~\bibnamefont {Lee}}, \bibinfo {author} {\bibfnamefont
  {P.}~\bibnamefont {Kratz}}, \bibinfo {author} {\bibfnamefont
  {V.}~\bibnamefont {Aji}}, \bibinfo {author} {\bibfnamefont {M.}~\bibnamefont
  {Bockrath}}, \bibinfo {author} {\bibfnamefont {C.}~\bibnamefont {Lau}},
  \bibinfo {author} {\bibfnamefont {C.}~\bibnamefont {Varma}}, \bibinfo
  {author} {\bibfnamefont {R.}~\bibnamefont {Stillwell}},  \emph {et~al.},\
  }\href@noop {} {\bibfield  {journal} {\bibinfo  {journal} {Nature
  Nanotechnology}\ }\textbf {\bibinfo {volume} {7}},\ \bibinfo {pages} {156}
  (\bibinfo {year} {2012})}\BibitemShut {NoStop}%
\bibitem [{\citenamefont {Velasco}\ \emph {et~al.}(2014)\citenamefont
  {Velasco}, \citenamefont {Lee}, \citenamefont {Zhao}, \citenamefont {Jing},
  \citenamefont {Kratz}, \citenamefont {Bockrath},\ and\ \citenamefont
  {Lau}}]{Velasco2014}%
  \BibitemOpen
  \bibfield  {author} {\bibinfo {author} {\bibfnamefont {J.}~\bibnamefont
  {Velasco}}, \bibinfo {author} {\bibfnamefont {Y.}~\bibnamefont {Lee}},
  \bibinfo {author} {\bibfnamefont {Z.}~\bibnamefont {Zhao}}, \bibinfo {author}
  {\bibfnamefont {L.}~\bibnamefont {Jing}}, \bibinfo {author} {\bibfnamefont
  {P.}~\bibnamefont {Kratz}}, \bibinfo {author} {\bibfnamefont
  {M.}~\bibnamefont {Bockrath}}, \ and\ \bibinfo {author} {\bibfnamefont
  {C.}~\bibnamefont {Lau}},\ }\href {http://dx.doi.org/10.1021/nl4043399}
  {\bibfield  {journal} {\bibinfo  {journal} {Nano Letters}\ }\textbf {\bibinfo
  {volume} {14}},\ \bibinfo {pages} {1324} (\bibinfo {year} {2014})},\ \Eprint
  {http://arxiv.org/abs/http://dx.doi.org/10.1021/nl4043399}
  {http://dx.doi.org/10.1021/nl4043399} \BibitemShut {NoStop}%
\bibitem [{\citenamefont {Shi}\ \emph {et~al.}(2016)\citenamefont {Shi},
  \citenamefont {Lee}, \citenamefont {Che}, \citenamefont {Pi}, \citenamefont
  {Espiritu}, \citenamefont {Stepanov}, \citenamefont {Smirnov}, \citenamefont
  {Lau},\ and\ \citenamefont {Zhang}}]{Shi2016}%
  \BibitemOpen
  \bibfield  {author} {\bibinfo {author} {\bibfnamefont {Y.}~\bibnamefont
  {Shi}}, \bibinfo {author} {\bibfnamefont {Y.}~\bibnamefont {Lee}}, \bibinfo
  {author} {\bibfnamefont {S.}~\bibnamefont {Che}}, \bibinfo {author}
  {\bibfnamefont {Z.}~\bibnamefont {Pi}}, \bibinfo {author} {\bibfnamefont
  {T.}~\bibnamefont {Espiritu}}, \bibinfo {author} {\bibfnamefont
  {P.}~\bibnamefont {Stepanov}}, \bibinfo {author} {\bibfnamefont
  {D.}~\bibnamefont {Smirnov}}, \bibinfo {author} {\bibfnamefont {C.~N.}\
  \bibnamefont {Lau}}, \ and\ \bibinfo {author} {\bibfnamefont
  {F.}~\bibnamefont {Zhang}},\ }\href {\doibase 10.1103/PhysRevLett.116.056601}
  {\bibfield  {journal} {\bibinfo  {journal} {Phys. Rev. Lett.}\ }\textbf
  {\bibinfo {volume} {116}},\ \bibinfo {pages} {056601} (\bibinfo {year}
  {2016})}\BibitemShut {NoStop}%
\bibitem [{\citenamefont {{Hunt}}\ \emph {et~al.}(2016)\citenamefont {{Hunt}},
  \citenamefont {{Li}}, \citenamefont {{Zibrov}}, \citenamefont {{Wang}},
  \citenamefont {{Taniguchi}}, \citenamefont {{Watanabe}}, \citenamefont
  {{Hone}}, \citenamefont {{Dean}}, \citenamefont {{Zaletel}}, \citenamefont
  {{Ashoori}},\ and\ \citenamefont {{Young}}}]{Hunt2016}%
  \BibitemOpen
  \bibfield  {author} {\bibinfo {author} {\bibfnamefont {B.~M.}\ \bibnamefont
  {{Hunt}}}, \bibinfo {author} {\bibfnamefont {J.~I.~A.}\ \bibnamefont {{Li}}},
  \bibinfo {author} {\bibfnamefont {A.~A.}\ \bibnamefont {{Zibrov}}}, \bibinfo
  {author} {\bibfnamefont {L.}~\bibnamefont {{Wang}}}, \bibinfo {author}
  {\bibfnamefont {T.}~\bibnamefont {{Taniguchi}}}, \bibinfo {author}
  {\bibfnamefont {K.}~\bibnamefont {{Watanabe}}}, \bibinfo {author}
  {\bibfnamefont {J.}~\bibnamefont {{Hone}}}, \bibinfo {author} {\bibfnamefont
  {C.~R.}\ \bibnamefont {{Dean}}}, \bibinfo {author} {\bibfnamefont
  {M.}~\bibnamefont {{Zaletel}}}, \bibinfo {author} {\bibfnamefont {R.~C.}\
  \bibnamefont {{Ashoori}}}, \ and\ \bibinfo {author} {\bibfnamefont {A.~F.}\
  \bibnamefont {{Young}}},\ }\href@noop {} {\bibfield  {journal} {\bibinfo
  {journal} {ArXiv e-prints}\ } (\bibinfo {year} {2016})},\ \Eprint
  {http://arxiv.org/abs/1607.06461} {arXiv:1607.06461 [cond-mat.mes-hall]}
  \BibitemShut {NoStop}%
\bibitem [{\citenamefont {{Pientka}}\ \emph {et~al.}(2017)\citenamefont
  {{Pientka}}, \citenamefont {{Waissman}}, \citenamefont {{Kim}},\ and\
  \citenamefont {{Halperin}}}]{Pientka2017ArXiv}%
  \BibitemOpen
  \bibfield  {author} {\bibinfo {author} {\bibfnamefont {F.}~\bibnamefont
  {{Pientka}}}, \bibinfo {author} {\bibfnamefont {J.}~\bibnamefont
  {{Waissman}}}, \bibinfo {author} {\bibfnamefont {P.}~\bibnamefont {{Kim}}}, \
  and\ \bibinfo {author} {\bibfnamefont {B.~I.}\ \bibnamefont {{Halperin}}},\
  }\href@noop {} {\bibfield  {journal} {\bibinfo  {journal} {ArXiv e-prints}\ }
  (\bibinfo {year} {2017})},\ \Eprint {http://arxiv.org/abs/1703.01235}
  {arXiv:1703.01235 [cond-mat.mes-hall]} \BibitemShut {NoStop}%
\bibitem [{\citenamefont {Jiang}\ \emph {et~al.}(2007)\citenamefont {Jiang},
  \citenamefont {Zhang}, \citenamefont {Stormer},\ and\ \citenamefont
  {Kim}}]{Jiang2007}%
  \BibitemOpen
  \bibfield  {author} {\bibinfo {author} {\bibfnamefont {Z.}~\bibnamefont
  {Jiang}}, \bibinfo {author} {\bibfnamefont {Y.}~\bibnamefont {Zhang}},
  \bibinfo {author} {\bibfnamefont {H.~L.}\ \bibnamefont {Stormer}}, \ and\
  \bibinfo {author} {\bibfnamefont {P.}~\bibnamefont {Kim}},\ }\href {\doibase
  10.1103/PhysRevLett.99.106802} {\bibfield  {journal} {\bibinfo  {journal}
  {Phys. Rev. Lett.}\ }\textbf {\bibinfo {volume} {99}},\ \bibinfo {pages}
  {106802} (\bibinfo {year} {2007})}\BibitemShut {NoStop}%
\bibitem [{\citenamefont {Abanin}\ \emph {et~al.}(2007)\citenamefont {Abanin},
  \citenamefont {Lee},\ and\ \citenamefont {Levitov}}]{Abanin2007}%
  \BibitemOpen
  \bibfield  {author} {\bibinfo {author} {\bibfnamefont {D.~A.}\ \bibnamefont
  {Abanin}}, \bibinfo {author} {\bibfnamefont {P.~A.}\ \bibnamefont {Lee}}, \
  and\ \bibinfo {author} {\bibfnamefont {L.~S.}\ \bibnamefont {Levitov}},\
  }\href {\doibase 10.1103/PhysRevLett.98.156801} {\bibfield  {journal}
  {\bibinfo  {journal} {Phys. Rev. Lett.}\ }\textbf {\bibinfo {volume} {98}},\
  \bibinfo {pages} {156801} (\bibinfo {year} {2007})}\BibitemShut {NoStop}%
\bibitem [{\citenamefont {Fuchs}\ and\ \citenamefont
  {Lederer}(2007)}]{Fuchs2007}%
  \BibitemOpen
  \bibfield  {author} {\bibinfo {author} {\bibfnamefont {J.-N.}\ \bibnamefont
  {Fuchs}}\ and\ \bibinfo {author} {\bibfnamefont {P.}~\bibnamefont
  {Lederer}},\ }\href {\doibase 10.1103/PhysRevLett.98.016803} {\bibfield
  {journal} {\bibinfo  {journal} {Phys. Rev. Lett.}\ }\textbf {\bibinfo
  {volume} {98}},\ \bibinfo {pages} {016803} (\bibinfo {year}
  {2007})}\BibitemShut {NoStop}%
\bibitem [{\citenamefont {de~Nova}\ \emph {et~al.}( )\citenamefont {de~Nova},
  \citenamefont {Zapata},\ and\ \citenamefont {Demler}}]{deNova2016a}%
  \BibitemOpen
  \bibfield  {author} {\bibinfo {author} {\bibfnamefont {J.~R.~M.}\
  \bibnamefont {de~Nova}}, \bibinfo {author} {\bibfnamefont {I.}~\bibnamefont
  {Zapata}}, \ and\ \bibinfo {author} {\bibfnamefont {E.}~\bibnamefont
  {Demler}},\ }\href@noop {} {\bibfield  {journal} {\bibinfo  {journal} {To be
  published}\ } (\bibinfo {year} {~})}\BibitemShut {NoStop}%
\bibitem [{\citenamefont {Schleich}(2001)}]{Schleich2001}%
  \BibitemOpen
  \bibfield  {author} {\bibinfo {author} {\bibfnamefont {W.}~\bibnamefont
  {Schleich}},\ }\href {https://books.google.es/books?id=AjhCLtDbaFAC} {\emph
  {\bibinfo {title} {Quantum Optics in Phase Space}}}\ (\bibinfo  {publisher}
  {Wiley-VCH},\ \bibinfo {year} {2001})\BibitemShut {NoStop}%
\bibitem [{\citenamefont {Thouless}(1961)}]{Thouless1961}%
  \BibitemOpen
  \bibfield  {author} {\bibinfo {author} {\bibfnamefont {D.}~\bibnamefont
  {Thouless}},\ }\href {https://books.google.co.il/books?id=W\_9QAAAAMAAJ}
  {\emph {\bibinfo {title} {The quantum mechanics of many-body systems}}},\
  Pure and applied physics\ (\bibinfo  {publisher} {Academic Press},\ \bibinfo
  {year} {1961})\BibitemShut {NoStop}%
\bibitem [{\citenamefont {Wang}\ \emph {et~al.}(2002)\citenamefont {Wang},
  \citenamefont {Das~Sarma}, \citenamefont {Demler},\ and\ \citenamefont
  {Halperin}}]{Wang2002}%
  \BibitemOpen
  \bibfield  {author} {\bibinfo {author} {\bibfnamefont {D.-W.}\ \bibnamefont
  {Wang}}, \bibinfo {author} {\bibfnamefont {S.}~\bibnamefont {Das~Sarma}},
  \bibinfo {author} {\bibfnamefont {E.}~\bibnamefont {Demler}}, \ and\ \bibinfo
  {author} {\bibfnamefont {B.~I.}\ \bibnamefont {Halperin}},\ }\href {\doibase
  10.1103/PhysRevB.66.195334} {\bibfield  {journal} {\bibinfo  {journal} {Phys.
  Rev. B}\ }\textbf {\bibinfo {volume} {66}},\ \bibinfo {pages} {195334}
  (\bibinfo {year} {2002})}\BibitemShut {NoStop}%
\bibitem [{\citenamefont {Giuliani}\ and\ \citenamefont
  {Vignale}(2005)}]{Giuliani2005}%
  \BibitemOpen
  \bibfield  {author} {\bibinfo {author} {\bibfnamefont {G.}~\bibnamefont
  {Giuliani}}\ and\ \bibinfo {author} {\bibfnamefont {G.}~\bibnamefont
  {Vignale}},\ }\href {https://books.google.co.il/books?id=kFkIKRfgUpsC} {\emph
  {\bibinfo {title} {Quantum Theory of the Electron Liquid}}},\ Masters Series
  in Physics and Astronomy\ (\bibinfo  {publisher} {Cambridge University
  Press},\ \bibinfo {year} {2005})\BibitemShut {NoStop}%
\bibitem [{\citenamefont {Negele}\ and\ \citenamefont
  {Orland}(2008)}]{Negele2008}%
  \BibitemOpen
  \bibfield  {author} {\bibinfo {author} {\bibfnamefont {J.}~\bibnamefont
  {Negele}}\ and\ \bibinfo {author} {\bibfnamefont {H.}~\bibnamefont
  {Orland}},\ }\href {https://books.google.es/books?id=mx5CfeeEkm0C} {\emph
  {\bibinfo {title} {Quantum Many-particle Systems}}},\ Advanced Books
  Classics\ (\bibinfo  {publisher} {Westview Press},\ \bibinfo {year}
  {2008})\BibitemShut {NoStop}%
\bibitem [{\citenamefont {Fetter}\ and\ \citenamefont
  {Walecka}(2003)}]{Fetter2003}%
  \BibitemOpen
  \bibfield  {author} {\bibinfo {author} {\bibfnamefont {A.}~\bibnamefont
  {Fetter}}\ and\ \bibinfo {author} {\bibfnamefont {J.}~\bibnamefont
  {Walecka}},\ }\href {https://books.google.es/books?id=0wekf1s83b0C} {\emph
  {\bibinfo {title} {Quantum Theory of Many-particle Systems}}},\ Dover Books
  on Physics\ (\bibinfo  {publisher} {Dover Publications},\ \bibinfo {year}
  {2003})\BibitemShut {NoStop}%
\bibitem [{\citenamefont {Pitaevskii}\ and\ \citenamefont
  {Stringari}(2003)}]{Pitaevskii2003}%
  \BibitemOpen
  \bibfield  {author} {\bibinfo {author} {\bibfnamefont {L.}~\bibnamefont
  {Pitaevskii}}\ and\ \bibinfo {author} {\bibfnamefont {S.}~\bibnamefont
  {Stringari}},\ }\href@noop {} {\emph {\bibinfo {title} {{Bose-Einstein
  Condensation}}}}\ (\bibinfo  {publisher} {Clarendon Press},\ \bibinfo
  {address} {Oxford},\ \bibinfo {year} {2003})\BibitemShut {NoStop}%
\bibitem [{\citenamefont {Pethick}\ and\ \citenamefont
  {Smith}(2008)}]{Pethick2008}%
  \BibitemOpen
  \bibfield  {author} {\bibinfo {author} {\bibfnamefont {C.~J.}\ \bibnamefont
  {Pethick}}\ and\ \bibinfo {author} {\bibfnamefont {H.}~\bibnamefont
  {Smith}},\ }\href@noop {} {\emph {\bibinfo {title} {{Bose-Einstein
  condensation in dilute gases}}}}\ (\bibinfo  {publisher} {Cambridge
  University Press},\ \bibinfo {address} {Cambridge},\ \bibinfo {year}
  {2008})\BibitemShut {NoStop}%
\end{thebibliography}%

\end{document}